# THE PROPERTIES OF LOW REDSHIFT
# INTERGALACTIC O VI ABSORBERS DETERMINED FROM
# HIGH S/N OBSERVATIONS OF 14 QSOs
# WITH THE COSMIC ORIGINS SPECTROGRAPH[1]

B. D. Savage T.-S. Kim[2], B. P. Wakker
Department of Astronomy, University of Wisconsin, 475 N. Charter St., Madison, WI 53706, USA

B. Keeney, J. M. Shull, J. T. Stocke & J. C. Green
Department of Astrophysical and Planetary Sciences, University of Colorado, 389 UCB Boulder, CO 80309, USA

## ABSTRACT

We report on the observed properties of the plasma revealed through high signal-to-noise (S/N) observations of 54 intervening O VI absorption systems containing 85 O VI and 133 H I components in a blind survey of 14 QSOs observed at ~18 km s$^{-1}$ resolution with the Cosmic Origins Spectrograph (COS) over a redshift path of 3.52 at z < 0.5. Simple systems with one or two H I components and one O VI component comprise 50% of the systems. For a sample of 45 well-aligned absorption components where the temperature can be estimated, we find evidence for cool photoionized gas in 31 (69%) and warm gas (6 > log T > 5) in 14 (31%) of the components. The total hydrogen content of the 14 warm components can be estimated from the temperature and the measured value of log N(H I). The very large implied values of log N(H) range from 18.38 to 20.38 with a median of 19.35. The metallicity, [O/H], in the 6 warm components with log T > 5.45 ranges from -1.93 to 0.03 with a median value of -1.0 dex. Ground-based galaxy redshift studies reveal that most of the absorbers we detect sample gas in the IGM extending 200 to 600 kpc beyond the closest associated galaxy. For the warm aligned O VI absorbers, we estimate $\Omega_b$(O VI)$_{Warm}$ = 0.0019±0.0005 which corresponds to (4.1±1.1)% of the baryons at low z. The warm plasma traced by the aligned O VI and H I absorption contains nearly as many baryons as are found in galaxies.

## 1. INTRODUCTION

Most of the ordinary baryonic matter in the Universe lies in the very low-density intergalactic medium (IGM)[3]. The atoms in this low-density plasma spend most of their time in their ground

[1] Based on observations obtained with the NASA/ESA Hubble Space Telescope, which is operated by the Association of Universities for Research in Astronomy, Inc. under NASA contract NAS5-2655, and the NASA-CNES/ESA Far Ultraviolet Spectroscopic Explorer Mission operated by Johns Hopkins University, supported by NASA contract NAS 05-32885.

[2] T.-S. Kim has recently relocated to Osservatorio Astronomico di Trieste, Via G. B. Tiepolo 11, 34131 Trieste, Italy

electronic energy state.  They only emit energy following very rare particle collisions and are therefore difficult to detect in emission.   Most of the baryonic matter in the Universe is dark and is only easily revealed through absorption line spectroscopy of luminous background objects such as QSOs.  At high redshift, the IGM in the very distant Universe has been effectively studied during many years of absorption-line spectroscopy with ground based optical telescopes.  Those studies have revealed that nearly all of the baryons at $z \sim 2$ to 4 are situated in the cool (log T $\sim$ 3.5-4.5) photoionized H I forest of Lyman $\alpha$ absorbers (Rauch 1998).  The situation at low redshift is less clear because the accessible resonance absorption lines lie in the vacuum UV, EUV and X-ray regions of the spectrum and the space-based observatories so far available are only modest in size making the required photon starved absorption-line observations very time consuming.  Despite these difficulties it has been found that the Ly$\alpha$ forest of narrow absorbers at low z contains $\sim$30% of the baryons (Lehner et al. 2007; Danforth & Shull 2008, hereafter DS2008; Tilton et al. 2012)  while galaxies contain $\sim$7%.  The remainder of the baryons are believed to be situated in warm[3] ($5 <$ log T $< 6$) and hot[4] (log T = 6-7) highly ionized plasmas shock heated during the gravitational assembly of structures in the evolving Universe and by AGN/starburst outflows (Cen & Ostriker 1999; Dave et al. 2001; Tepper-Garcia et al. 2011).   The warm plasma can be detected via UV observations of highly ionized metal absorption lines such as O VI and Ne VIII along with weak but very broad Ly$\alpha$ (BLA)[5] absorption.   The detection of the hot plasma requires X-ray observations of O VII and/or O VIII absorption.  While there are many clear UV detections of warm gas, there are very few reliable X-ray detections of O VII/O VIII absorption in galaxy halos or IGM except at z = 0 associated with the disk and halo of the Milky Way.  Most of the claimed X-ray detections of O VII/O VIII beyond z = 0 are highly controversial with the exception of the 3$\sigma$ absorber at z = 0.03 toward H 2350-309 (Buote et al. 2009; Fang et al. 2010) that may be tracing halo gas of several foreground galaxies in the Sculptor wall (Williams et al. 2010, 2013). Yao et al. (2012) provide a recent update of the current state of the X-ray measurements.

---

[3] In this paper we adopt the following definitions for the names of the gaseous regions surrounding galaxies connecting to the very distant IGM situated well beyond the virial radii of galaxies. "Halo gas" refers to the gas in the virialized halos of galaxies extending to radial distances of up to $\sim$ 200 kpc around L* galaxies. We note that astronomers have begun to use "circumgalactic medium (CGM)" to refer to gas in the same region.  However, we prefer to use halo gas since the term has been in use ever since Bahcall & Spitzer (1969) proposed that intervening QSO absorption line systems are produced by highly extended gaseous galactic halos. There currently does not exist an accepted name for gas in the interface region between halo gas and the more distant undisturbed IGM, although in cosmological simulations that gas is commonly referred to as the 'filament gas' connecting to galaxies.

[4] We adopt the most common temperature naming convention for IGM research where cool, warm and hot refer to plasmas with log T = 3.5 to 4.5, 5 to 6, and > 6 , respectively.  The cool plasmas are most likely photoionized while the warm and hot plasmas are most likely collisionally ionized although photoionization may also contribute to the ionization.  The cool plasmas may be in ionization equilibrium where heating balances cooling. The warm and hot plasmas are most likely tracing cooling gas.  We note that these definitions are different for ISM and IGM research.

[5] We adopt the definition of broad Lyman $\alpha$ absorbers (BLAs) from Richter et al. (2006) to be H I lines with Doppler parameters b > 40 km $s^{-1}$.  A major challenge is separately determining the thermal and non-thermal contributions to the line broadening in BLAs.  If O VI is detected in the same absorbing gas, the 16x larger mass of O VI versus H I allows the separation as discussed in Section 3.3.  For log T = 5, 5.5, and 6, the thermal contribution to the Doppler parameter will be 41,  73 and 129 km $s^{-1}$, respectively.  Therefore, BLAs are possible tracers of the warm plasma in galaxy halos and the IGM.

The low z IGM is highly ionized. Understanding the origin of that ionization is crucial for determining elemental abundances in the gas and for ultimately understanding the total baryonic content of the gas. Because of its high frequency of occurrence and high state of ionization, O VI is an important tracer of very low density cool photoionized gas and warm collisionally ionized gas. In the cool gas the H I absorption related to the O VI is relatively easy to detect because it is strong and produces relatively narrow absorption lines. Therefore the studies evaluating the relations between H I and O VI absorption obtained with modest levels of S/N (Tripp et al. 2008, hereafter TSB2008; Thom & Chen 2008a hereafter TC2008a) revealed that a substantial percentage of the low z O VI absorbers were likely formed in cool photoionized gas. In the warm gas, the trace amount of H I associated with O VI is thermally broadened to large line widths and is very difficult to detect. However, moderately high S/N observations with the Cosmic Origins Spectrograph (COS) have begun to reveal broad BLA absorption associated with some of the O VI absorbers (Narayanan et al. 2010b; Savage et al. 2011a, 2011b, 2012) implying the existence of warm gas. Another way to detect the warm gas is to find Ne VIII absorption associated with the O VI absorption. Since it is difficult to produce Ne VIII through photoionization except at very low densites, the Ne VIII observations imply the detection of warm collisionally ionized gas (Savage et al. 2005, 2011a; Naryayanan et al. 2009, 2011, 2012; Tripp et al. 2011; Meiring et al. 2013).

Predicting the evolution of the highly ionized metals in gaseous halos and IGM requires complex hydrodynamical simulations along with crucial assumptions about the various processes heating the gas and the origins and mixing of the metals in the gas from galactic outflows and possibly population III stars. Simply treating the cooling of the gas properly is difficult because non-equilibrium cooling effects become very important for log T < 5.5 and the size of those effects depends strongly on the metallicity of the gas and its density evolution. With all the required assumptions and the computational complexity of properly treating the mixing, heating, and the non-equilibrium processes, it is not surprising that the recent simulations of the behavior of the plasma in the evolving universe reach completely different conclusions about the origin(s) of O VI in the IGM and gaseous halos. For example, the simulations of Kang et al. (2005), Oppenheimer & Dave (2009), Oppenheimer et al. (2012) predict that the low redshift O VI absorbers mostly arise from cold (log T ~ 4.3) photoionized gas while the simulations of Smith et al. (2011), Tepper-Garcia et al. (2011), and Cen (2012) conclude that collisional ionization plays a more important role in the creation of O VI. Smith et al. (2011) suggest that the differences in the predictions of the simulations mostly arise from the way the cooling is calculated. Not allowing for the non-equilibrium effects completely changes the outcome. Even when allowing for these effects, the precise manner in which they are included in the simulation will affect the outcome.

Another problem involves the reliability of the numerical techniques used in the different simulation codes. Simulations of the cooling of hot gas in the galactic halo using smoothed particle hydrodynamical (SPH) codes predict that thermal instabilities will lead to widespread cooling of the gas to form cold clumps (Sommer-Larsen 2006; Kaufmann et al. 2006) that rain down on the disk of the galaxy providing the fuel for star formation. More recent work reveals that the "classic" SPH codes incorrectly calculate the mixing of different gaseous phases because of numerical problems in the code (Agertz et al. 2007; Price 2008; Wadsley et al. 2008; Read et al. 2010). The recent cooling corona simulation of Hobbs et al. (2012) utilizing a new version of the SPH code such as the one described by Read & Hayfield (2012) reveals that the cold clumps seen in many classic SPH simulations are not present in the new simulations. Comparisons of the classic-SPH code results with those from moving-mesh simulations with the code AREPO performed by Sijacki et al. (2012), Vogelsberger et al. (2012), and Keres et al. (2012) also reveal major differences in the predictions of the evolution of the gas for the same assumed initial conditions. These authors provide plausible explanations for the origins of the differences.

The understanding of the complex set of processes affecting the gas phase baryons in the evolving Universe will probably require many years of theoretical research to better understand the thermal evolution of the gas and to better understand the numerical limitations of the various codes used to run the simulations. Turning to the observations we can ask: What clues are the observations providing? What is the relative mix of O VI absorbers tracing cool photoionized gas versus warm collisionally ionized gas? What is the total baryonic content of the absorbers? What information is there about the abundance of oxygen in the absorbers? Where are the most common O VI absorbers located relative to the distribution of galaxies? We will attempt to answer these observational questions in this paper.

A topic that has recently received a considerable amount of attention concerns the objects (galaxies, groups, filaments) associated with O VI absorbers. The general O VI absorber/galaxy association has been clearly established through a number of studies (Stocke et al. 2006; Wakker & Savage 2009; Chen & Mulchaey 2009; Prochaska et al. 2011a, 2011b; Stocke et al. 2113). For example, the study of the relationship of H I/O VI and nearby (z < 0.017) galaxies of Wakker & Savage (2009) extending to L ~0.1L* revealed that all of 13 O VI absorbers in the survey occur within 550 kpc of a L > 0.25L* galaxy. For an impact parameter of < 350 kpc, the covering factor of O VI around L > 0.1L* galaxies is 60-80% for field galaxies and 10-30% for group galaxies. For more luminous galaxies Tumlinson et al. (2011b) have shown that halos containing large column densities of O VI (log N(O VI) ~14.6) extending to 150 kpc are ubiquitous around star-forming galaxes and much less common around galaxies with little or no star formation. Stocke et al. (2013) found no H I or O VI absorption associated with absorption line galaxies.

In this paper we present the analysis of the properties of O VI absorption systems found in COS spectra of the 14 bright QSOs listed in Table 1. Our blind survey of QSOs provides a random set of sightlines through the low redshift IGM and thereby yields information about the most common O VI and H I absorbers. In contrast the COS-halo project of Tumlinson et al. (2011b) specifically targeted absorption by O VI close to L* star-forming galaxies with impact parameters of < 150 kpc.

The major goal of our study is to gain deeper insights about the origin of the ionization in the most common O VI systems so that the diagnostic potential of the O VI ion can be better exploited as a tracer of the cool and warm plasma in the IGM and galaxy halos.

Distances in this paper are physical distances assuming a $\Lambda$CDM cosmology with $\Omega_M = 0.3$, $\Lambda = 0.7$ and $H_0 = 70$ km s$^{-1}$ Mpc$^{-1}$ with $h_{70} = H_0/$ 70 km s$^{-1}$.

## 2. OBSERVATIONS

The primary observational data analyzed in the investigation are from COS. However, for many of the QSOs we also have archival observations from the Space Telescope Imaging Spectrograph (STIS) and the Far-Ultraviolet Spectrographic Explorer (FUSE). The STIS observations are valuable for viewing the absorption at higher resolution but lower S/N than seen in the COS observations. Figure 1 compares COS and STIS observations of H I $\lambda$1025 and O VI $\lambda\lambda$1032, 1037 in the z = 0.14231 system in the spectrum of PG 0953+415 over the wavelength range from 1165 to 1190 Å. Both spectra have been binned to 10 km s$^{-1}$ pixels to allow a direct comparison of the S/N achieved. The high efficiency of COS compared to STIS yields measurements with ~4 times higher S/N for this absorption system even though the STIS integration time was 5 times greater.

A cross-comparison of the COS and STIS observations helps to improve the wavelength accuracy of the COS results. The FUSE observations are valuable for searching for other IGM absorption lines in the O VI systems with redshifted wavelengths $\lambda < 1140$ Å. The FUSE measurements can be particularly valuable in searching for Ne VIII, O VI, O IV, O III, O II and C III in the absorption systems.

The QSOs observed are listed in Table 1 where they are ordered by increasing emission line redshift. $\lambda$(H I) and $\lambda$(O VI) indicate the observed redshifted wavelength range for H I and for O VI covered by the FUSE and COS observations out to within 5000 km s$^{-1}$ of the QSO emission line. Associated O VI absorption systems within 5000 km s$^{-1}$ of the QSO are not considered to avoid the proximity effect as recommended by TSB2008. z(O VI) indicates the redshift range for detecting intervening O VI absorption. For STIS, FUSE and COS, we also list the range of signal-to-noise (S/N) per resolution element in the observations produced by each instrument. All redshifts listed in this paper are in the heliocentric reference frame.

The object sample listed in Table 1 consists of many of the brightest low-to-moderate redshift QSOs for which it is possible to obtain high S/N spectra with COS in reasonable integration times. The sample provides a relatively unbiased view of the low z IGM in order to probe the most common types of O VI absorbers. In contrast other recent studies with COS have targeted QSOs known to be aligned with various types of foreground galaxies (Tumlinson et al. 2011b; Stocke et al. 2013) or have targeted higher redshift QSOs in order to study EUV absorption lines (Tripp et al. 2011; Meiring et al. 2013). In the latter case the redshifts are so large that H I $\lambda$1215 which is crucial for finding warm H I is not observed because it is shifted beyond the range of the COG G160M grating. The sample studied in this paper allows the study of O VI absorption not only situated within the virial radii of galaxy halos but also absorption occurring in the surrounding filamentary gas that connects to the distant IGM (see section 9).

## 2.1 COS

The COS observations analyzed in the paper were mostly obtained as part of the COS/GTO program (Green, PI) to study cool, warm and hot gas in the cosmic web and galaxy halos. Information about COS can be found in Froning & Green (2009), Green et al. (2012) and the COS HST Instrument Handbook (Holland et al. 2012). The inflight performance characteristics of COS are discussed in Osterman et al. (2011).

The QSO spectral integrations were obtained at different setup wavelengths with the G130M and G160M gratings in order to reduce the effects of detector fixed pattern noise (FPN) in the final combined spectra. For the G130M grating the central wavelengths used are 1291, 1300, 1309 and 1318 Å. For the G160M grating the central wavelengths used are 1589, 1600, 1611, and 1623 Å. The total integration time for each grating in ks is listed in Table 1.

The micro-channel plate delay-line detector was operated in the time-tag mode with the QSO centered in the 2.5" diameter primary science aperture. The spectra were processed with the CalCOS pipeline version listed in Table 1. The details of the methods the individual spectra are corrected for wavelength differences and combined into a single spectrum are discussed by Kim et al. (2014). We did not use the Danforth et al. (2010a) co-addition routine because it does not adequately allow for wavelength-dependent offsets between different exposures of the same target obtained with different grating setups. When STIS spectra were available, the COS absorption-line wavelengths in each integration were adjusted to agree with the STIS wavelengths which are estimated to have a velocity accuracy of ~1 km s$^{-1}$. In this case the COS spectra are estimated to have a velocity accuracy of 5 km s$^{-1}$. When STIS spectra are not available, the zero-point wavelength uncertainty in the COS observations was established via a cross comparison of ISM absorption lines between the COS and FUSE observations and with reference of the ISM absorption to 21 cm emission in the direction to each QSO. The radio spectra which are mostly from Kalberla et al. (2005) are displayed in Wakker et al. (2003) The resulting wavelength calibration of the COS observations in these cases is estimated to be accurate to within ~15 km s$^{-1}$. Improvements to ~10 km s$^{-1}$ can be made within particular absorption line systems that display higher order Lyman series absorption. Using COS spectra with reliable wavelength calibrations is very important when studying the presence or absence of H I in the O VI absorbing gas based on the velocity alignment of the H I and O VI absorption. The ability to detect broad Lyman $\alpha$ (BLA) absorption associated with the O VI absorption is crucial for understanding the ionization of the gas and for estimating its baryonic content.

For the COS spectra the iterative FPN correction scheme developed by Kim et al. (2013) was used to solve and correct for the FPN in the highest S/N spectra. A comparison between the FPN-corrected COS spectra and those with no correction reveals that the FPN introduces numerous features with rest frame equivalent widths of ~10 mÅ with some features being much stronger. A 10 mÅ feature if identified as H I would have a logarithmic column density of log N(H I) ~ 12.3. If the feature were instead identified as O VI, the column density would be log N(O VI ) ~ 12.9 because of the smaller O VI f-value (0.133) compared to that for H I (0.416).

Properly combined COS spectra that allow for the ~15 km s$^{-1}$ wavelength errors across the individual integrations yield combined spectra with a resolution of ~18 km s$^{-1}$ FWHM. However, the COS spectral line spread function has broad wings, with the strength of the wings increasing toward shorter wavelengths. When fitting line profiles to the COS observations obtained before 22 July 2012 we use the wavelength-dependent line spread function from Kriss (2011) that updates the original parameterization of Ghavamian et al. (2009). For the most recent observation of PKS 2155-304 obtained after the position of the spectrum on the detector was changed, we used the lifetime position 2 version of the LSF (see the appendix A2).

The COS LSF can introduce complications when trying to fit multicomponent absorbers. In some of the cases studied in this paper, the kinematical complexity of the absorber strongly limits attempts to determine good parameters for individual absorption components in the system.

The range of the final S/N per 18 km s$^{-1}$ resolution element achieved in the COS observations is listed in Table 1. The values for COS are 2 to 5 times larger than obtained with STIS per 7 km s$^{-1}$ resolution element for the same QSO. The higher S/N of the COS observations is important for defining the true properties of the O VI and the associated H I absorption. The ability to detect broad Lyman $\alpha$ (BLA) absorption associated with the O VI absorption is crucial for determining the temperature and ionization of warm gas and for estimating its baryonic content.

## 2.2 STIS

A number of the QSOs in our program were previously observed by STIS (see the reference list for Table 1 and the compilation by Tilton et al. 2012) using the medium-resolution FUV echelle mode E140M with the 0.2x0.2" aperture. The design and performance characteristics of STIS are described by Woodgate et al. (1998) and Kimble et al. (1998). The spectra extend from ~1150 Å to 1730 Å with a resolution of 7 km s$^{-1}$ and a wavelength uncertainty corresponding to a velocity uncertainty of ~1 km s$^{-1}$.

The spectra were processed using the calibrated files in the MAST archive as described by Wakker & Savage (2009). The STIS spectra have S/N levels per resolution element several times smaller than obtained with COS (see Table 1). The STIS observations permit an accurate wavelength calibration of the COS observations and more clearly reveal the multi-component structure of narrow low ionization metal lines. When the STIS observations are used in the profile fitting process we adopt the LSFs given in the STIS Handbooks (Bostroem & Proffitt 2011; Hernandez et al. 2012).

## 2.3 FUSE

The properties and inflight performance of the spectrographs aboard the FUSE satellite are discussed by Moos et al. (2000) and Sahnow et al. (2000). The satellite produced far-UV spectra from 912 to 1187 Å with a resolution of ~20-25 km s$^{-1}$ (FWHM). The FUSE observations were processed by the methods described by Wakker et al. (2003) and Wakker (2006) using version 2.4 of the FUSE calibration pipeline. The wavelength offsets of individual integrations on different detector segments were established with reference of the recorded UV ISM absorption lines including numerous H$_2$ lines to H I 21 cm emission in the direction of each QSO from various H I surveys before co-addition to produce a single spectrum. The H I spectra for most of the QSOs are shown in Wakker et al. (2003). For λ > 1000 Å the combined spectra only include the LiF channel integrations which have much higher throughput than the Si C channel integrations. The FUSE absolute velocity calibration when referenced to the H I absorption is estimated to be ±8 km s$^{-1}$. We assume a Gaussian spectral line spread function with FWHM = 25 km s$^{-1}$ when fitting line profiles to the FUSE observations unless stated otherwise in Appendix A.

The S/N per resolution element in the resulting spectra is several times larger for λ ~1000 to 1185 Å than for λ ~ 912 to 990 Å. The S/N range listed in Table 1 is for wavelengths near 1035 Å. When there is no entry for S/N in Table 1, FUSE observations of the QSO were not obtained.

## 3. ANALYSIS TECHNIQUES

In this section we discuss the various analysis techniques that have been applied to the observations presented in this paper. These techniques have previously been applied to the COS observations of particularly interesting absorption-line systems presented in our earlier papers (Naryayanan et al. 2010b, 2011, 2012; Savage et al. 2010, 2011a, 2011b, 2012).

## 3.1 Line Identifications

QSO absorption line spectra are less complex at low z than at high z because of the decreasing line density of the Lyα forest with decreasing redshift. Nevertheless, absorption line blending can be a problem. Therefore, to correctly identify and determine the properties of O VI absorption line systems it is necessary to make proper identifications and when necessary to correct for line blending produced by ISM absorption or other IGM absorption systems. For each of the QSOs listed in Table 1 we have identified all lines observed with significance > 3.5σ as determined from the profile fit code (see section 3.2). In the O VI absorption system velocity plots shown for each system in the appendix of this paper, absorbers from the ISM or from other systems are identified and shown with the green spectrum line fits in the panel for each absorber in the O VI system of interest. Most of the time those contaminating absorbers are well displaced from the absorber of interest and can be ignored. Occasionally they blend with the absorber of interest and the effects of the blending can be determined provided there are other detected lines of the same absorbing ion at other wavelengths that can be used to model the level of the contamination. When the blending is too strong, the properties of the particular IGM line of interest can't be reliably determined. The details of the blending and de-blending are provided in the appendix for each system.

The absorbers identified as O VI vary in quality. In the best cases both lines of the doublet are clearly identified and can be reliably fitted with one or more absorption components with no evidence of blending with other absorbers. In other cases only one line of the O VI λλ1031, 1037 doublet is observable because of either blending contamination in the other doublet line or because the weaker line of the doublet is too weak to detect. In these cases we have eliminated all plausible alternate identifications of the absorption we are attributing to O VI. When the candidate O VI absorber occurs at λ < 1215 Å we can rule out absorption by H I since the stronger H I lines in the same system would be observable at wavelengths longward of the candidate line if present. The O VI detections based on single lines of the O VI doublet occur at redshifts that are consistent with those for well-observed H I absorption systems. In a number of cases discussed in the appendix, a candidate O VI absorber associated with broad H I λ1215 absorption had a strength just below the 3σ acceptance criteria of our survey.

### 3.2. Profile Fits

We are interested in determining the physical state of the gas in the observed absorption line systems. Therefore information on absorption profile column densities, line widths and velocities are important for the spectroscopic diagnostic studies. We therefore have employed the VPFIT Code version 9.5 (http://www.ast.cam.ac.uk/~rfc/vpfit.html) discussed in Carswell et al. (2002) and Kim et al. (2007). For each absorption component we determine measures of the system redshift, z, the component velocity, v, the component Doppler parameter , b, and the column density, log N and their errors assuming the observed absorption can be fitted by single or multiple Voigt profile components. If the actual line shape is strongly affected by kinematic flows leading to line shapes not well characterized by Voigt profiles, this approach will lead to misleading results.

The error array provided by the pipeline COS processing yields errors a factor of 1.1 to 1.5 times larger than those actually measured in the continuum regions of the QSOs we have observed. Therefore, when fitting line profiles we rescaled the combined pipeline errors to agree with the actual errors observed in the continuum.

We choose the number of components by taking the minimum number required to give an acceptable fit for each absorber (i.e. $\chi_\nu^2 \sim 1$). In some cases the information on the narrower metal lines are used to determine the number of components likely present in the H I absorption lines, e.g. the z = 0.13850 absorber toward PG 1116+215 (see Appendix A). The absorption velocities, v, within a given absorption system are referenced to the system redshift which is taken to be the redshift of the strongest O VI component. For each O VI absorption system we show profile plots and the profile fits for the detected components in the Appendix. The fit results for each QSO are given in a table in the Appendix listing ion, λ, v ± error, b ± error and log N ± error, and a note column. The footnotes to each table contain detailed information about the fitting process including comments about potential line saturation, line blending problems, and marginally detected absorbers. Unless otherwise stated, all the different ions were fitted freely, i.e. the velocities of components were not fixed to agree with those for other absorbers.

The fit parameter errors derived by the VPFIT code allow for the statistical errors, the effects of absorber component overlap errors, and the continuum placement errors (when using VPFIT option <>). The continuum fitting process we have adopted is carefully described in section 2.2 of Kim et al. (2007). Most of the QSOs observed have relatively flat continua, and the continuum fitting errors are small. For BLAs aligned with H I absorption, we have carefully investigated the continuum fitting errors for HE 0153-4520 at z = 0.22600 (see Savage et al. 2011b) and 3C 263 at z = 0.14072 (see Savage et al. 2012).   In both of these cases, the continuum fitting errors were small compared to the statistical and absorber component overlap errors.

Note that the VPFIT errors do not allow for the possibility the absorber may have a more complex component structure than the one adopted.  This is a fundamental limitation of the profile fit process.  In some cases the associated systematic errors can be very large but difficult to quantify.

The detection significance of each absorber can be determined from the logarithmic errors listed for the measured logarithmic column densities, log N, in the tables of the appendix.  We note that logarithmic errors for isolated lines of 0.10, 0.14 and 0.18 dex correspond to detection significances of 5.0, 3.6, and $3.0\sigma$, respectively.  Footnotes to the appendix tables provide comments about the lower significance detections with logarithmic column density errors exceeding ~0.14 dex.

In addition to the S/N and continuum, finding BLAs and determining their properties is sensitive to the assumed LSF (see sections 2.1, 2.1, 2.3).  Fully addressing the validity of the COS LSF is beyond the scope of this paper.   However, the adopted COS LSF does an excellent job of describing strong single component H I Voigt profiles in the highest S/N measurements.   Kim et al. (2014) have compared H I Voigt profile fit results for COS and STIS for 3C 273. The COS and STIS observations both have very high S/N, so the comparison is not strongly affected by noise in the measurements when strong lines are fitted.   For strong single component H I lines revealed by the STIS 3C 273 observations, the line parameters inferred from COS are nearly identical to those inferred from STIS.  Examples include the H I lines at z = 0.04900, 0.06655, and 0.12004 toward 3C 273 where the Doppler parameters obtained for the COS and STIS observations, b(H I)$_{COS}$ and b(H I)$_{STIS}$ are 29.6±0.5 and 29.5±0.9;  35.2±0.3 and 35.6±0.7; and 22.7±0.04 and 22.2±0.07, respectively. The values of log N(H I) which range from 13.60 to 14.08 measured by COS and STIS for these three absorbers are identical within the small errors of < 0.01 dex.

Although we have detected relatively weak O VI absorbers with log N(O VI) from 13.0 to 13.3, our program has not been designed to do a proper assessment of the number density of O VI absorbers this weak.  Note that the O VI λ1031 equivalent width is W$_\lambda$ = 12.54 mÅ for log N(O VI) = 13.0 assuming no line saturation.   The presence of FPN impacts the study of weak lines.  Kim et al. (2014) have performed a careful study of dn/dz for H I absorbers in the very high quality COS and STIS observations of 3C 273 and MRK 876 where the COS FPN noise has been removed.  Their work reveals very substantial differences between the higher quality COS observations and the STIS observations when log N(H I) < 13.2.  The STIS measurements indicate 8 possible H I absorption features with <log N(H I)> = 12.54 ±0.22 that are not seen in the higher quality COS measurements. The Kim et al. (2014) completeness study of the higher quality COS observations shows that incompleteness for narrow H I absorption lines becomes very important for the COS 3C 273 observations with S/N ~ 100 for log N(H I) < 12.7.  We note the limit will be larger for broad H I absorption lines.   The column densities at which the problems are large for narrow H I absorption lines in the best observations available can be related to where similar problems would be expected to occur when trying to detect O VI in very high S/N observations.   The fλ-value for the O VI λ1031 line is 3.7 times or 0.57 dex smaller than for H I λ1215.  We would therefore expect to observe incompleteness and measuring problems when log N(O VI) < 13.3  in COS spectra with the S/N ~100.   Many of our COS observations with a large enough redshift to detect O VI at z > 0.11 have lower S/N from 40 to 90.  Therefore, our sample will have incompleteness and measuring problems when log N(O VI) < 13.5.

Determining a reliable measure of dn/dz for O VI for the column density range from log N(O VI) = 13.0 to 13.5  will require higher quality observations than we have obtained along with a better understanding of the effects of the COS fixed pattern noise which can introduce spurious O VI features with log N(O VI) ~ 13.0 if not removed from the spectra with an iterative FPN correction procedure. A full analysis of dn/dz at log N(O VI) < 13.5 is beyond the scope of this paper. Similar problems are associated with measuring O VI in FUSE observations when log N(O VI) < 13.5. FUSE also has FPN problems but their full impact on the measurement of weak O VI lines has not been carefully studied except over narrow wavelength ranges.

### 3.3. Determining T and $b_{NT}$ from Differences in Line Widths

An important technique for determining the temperature (and other properties) of metal absorption-line systems has been to identify metal absorbers for which the associated H I can also be observed. If the metal lines and the H I arise in the same absorbing region, a difference in the line width may arise because the large mass difference between the metal and H I causes the broadening of the H I to exceed that for the metal absorber. That difference in line width can be used to estimate the temperature and turbulent velocity of the gas. The measured line width, b, will have contributions from thermal, $b_T$, and non-thermal line broadening, $b_{NT}$. The thermal contribution to the broadening is $b_T = (2kT/m)^{1/2} = 0.129 \ (T/A)^{1/2}$ Km s$^{-1}$, where A is the atomic mass number and T is in K. In separating the non-thermal from the thermal line broadening it is usually assumed the non-thermal (turbulent) broadening has a Gaussian line shape in which case $b^2 = b_T^2 + b_{NT}^2$. An application of this equation to two species of very different mass (i.e. H I and O VI) permits the estimate of T and $b_{NT}$ in the absorbing region with the errors following from the standard propagation of errors using the measured b values and their fit errors.

The two absorbing species must be reasonably well aligned in velocity for the method to be valid. The velocity calibration uncertainty of the COS and FUSE observations of 5 to 15 km s$^{-1}$ implies that an alignment of < 10 km s$^{-1}$ within the fitting errors is a reasonable assumption for identifying the aligned absorbers in the COS and FUSE spectra. The O VI /H I absorbers do not have to be perfectly aligned to be approximately tracing a substantial amount of thermal broadening in a warm plasma. The analysis provides a measure of the thermal and non-thermal components to the broadening. If the detected plasma is isothermal and the O VI and H I co-exist in the same spatial region, the two absorbers should be well aligned if the non-thermal broadening is symmetrical. For fully developed turbulence the non-thermal broadening will be a symmetrical Gaussian function. If the non-thermal broadening is more complex, the use of the equations above for deriving the temperature is only approximate. If the absorbing gas is not isothermal and the non-thermal broadening is not symmetrical the O VI and H I profiles could have a velocity shift amounting to a modest fraction of the non-thermal value of the line broadening. In the simplest cases of co-spatial O VI and H I in isothermal plasmas with fully developed turbulence the analysis technique should provide an excellent estimate of the temperature. In the more complex physical situations the estimate will be less secure.

Even with good velocity alignments we need to use caution when interpreting values of T and $b_{NT}$ derived from this method. The two absorbing species may align in velocity but not be co-spatial. For example, the absorbing structure might have a cool interior and a hot exterior in which case the H I might trace the cooler gas and the O VI the hotter gas. Even in light of the many difficulties associated with the application of this method, it is valuable to search for absorbers where it appears the H I absorption is directly associated with the metal line absorption and to try to separate the thermal from the non-thermal contributions to the line broadening in order to estimate the temperature of the plasma.

# 4.  IONIZATION MECHANISMS

Understanding the origin of the ionization in QSO absorption-line systems is essential for using the systems to obtain information about the physical conditions in the absorbing gas and to determine the elemental abundances and baryonic content of the gas.  In the following subsections we briefly discuss the production of O VI and other observed ions including H I by photoionization, collisional ionization and the combination of the two ionization processes in different physical situations.  For a comprehensive discussion of the production of the lithium-like highly ionized species in various types of cool and warm plasmas in the Milky Way corona, see section 5 in Wakker et al. (2012).

## 4.1. Photoionization

We use the photoionization code Cloudy [versionC13.0,  Ferland et al. 1998] when investigating the possibility that absorption system components are created by photoionization balanced by recombination in time equilibrium.  We generally assume a one-zone slab model illuminated by the extragalactic background radiation field of Haardt & Madau (2001) incorporating photons from AGNs and starforming galaxies adjusted to the appropriate redshift of the absorber. The reference relative heavy elements abundances we adopt are from Asplund et al. (2009).  The level of the ionization in the absorber is determined by the ionization parameter $U = n_i/n_H$, where $n_i$ is the H I ionizing photon density and $n_H$ is the total hydrogen gas density.  We note that uncertainties in the assumed spectral shape of the ionizing extragalactic background radiation can introduce uncertainties as large as 0.5 dex in the derived metallicities of the photoionized absorbers (see Howk et al. 2009). In their recent study of the cosmic UV and X-ray background, Haardt & Madau (2012) emphasize that their new extragalactic background estimates are subject to a number of poorly determined parameters.

When studying the ionization of multi-phase systems it has been common to find that photoionization provides a good explanation for the cool gas in the system with log T ~ 4-4.7 but that alternate explanations are often required to understand the origin(s) of the more highly ionized species such as N V, O VI, and Ne VIII.   In most cases studied so far, these highly ionized species must exist in a much lower density cool photoionized region or in a warm (log T ~ 5-6)  region dominated by collisional ionization.

## 4.2. Collisional Ionization Equilibrium

In studying collisional ionization in its simplest case in this paper we adopt the collisional ionization equilibrium (CIE) calculations of Gnat & Sternberg (2007) and assume the solar relative heavy elemental abundances of Asplund et al. (2009).  In CIE the ionic conditions in the plasma only depend on temperature.   CIE curves of log N(X) vs log T from Gnat & Sternberg (2007) are shown as the solid and dashed lines in Figure 2 for C III,  C IV, N V, O III, O IV, O VI, O VII,  O VIII, Ne XIII and H I for an absorbing slab with log N(H) = 19 and [Z/H] = 0.

Note the relationship between the curves for O VI and H I in Figure 2.  In CIE O VI peaks in abundance at log T ~ 5.5 and at that temperature in gas with solar abundances N(O VI)/N(H I)  ~ 63 while at log T ~ 6,  N(O VI)/ N(H I) ~ 6.3.   For solar abundances in CIE there is significantly more O VI in the gas than H I and we would expect to find O VI absorption with very little associated H I. For 0.1 solar abundances the values of N(O VI)/N(H I) are 6.3 and 0.63, respectively, for the same temperature range.  The amount of O VI and H I in gas with 0.1 solar abundances and in CIE is roughly comparable for 5.8 < log T  < 6.2.  In a plasma with 0.1 solar abundances it should be possible to find associated H I absorption if the gas is in CIE.

CIE should be a good approximation for the ionization of H I for log T > 5 provided the density is high enough for electron collisional ionization to be more important than photoionization. In this case, log N(H) can be reliably determined from log N(H I) (see section 4.5) .

### 4.3. Static Non-Equilibrium Collisional Ionization in Radiatively Cooling Gas

Collisionally ionized gas containing high ionization stages of oxygen is likely heated by shocks from the gravitational assembly of matter or from shocks associated with galactic outflows. The warm or hot gas produced subsequently cools. Hot gas is quasi-stable because the cooling times are long. Warm gas can cool rapidly because the peak of the cooling curve occurs in the temperature range from log T = 5 to 5.5 where the principal coolants include O V, O VI, O IV, and O III. Therefore the O VI observed in warm gas of galaxy halos and IGM likely resides in cooling gaseous structures and the ionic state of the gas can be affected by non-equilibrium processes. In rapidly cooling gas the recombination of the high ions can lag behind the decrease in the temperature of the gas allowing more O VI to exist in the cooling gas at a given temperature than is predicted for gas in collisional ionization equilibrium. The non-equilibrium effects depend on the metallicity of the gas since the cooling rate of high metallicity gas is shorter than for low metallicity gas. Gnat & Sternberg (2007) have studied the non-equilibrium cooling of hot plasmas with the assumptions of cooling at constant pressure or cooling with constant volume with different assumed elemental abundances. In Figure 3 we show for O VI and H I the effects of the non-equilibrium isobaric cooling in plasmas with solar abundances and 0.1 solar abundances in a slab where log N(H) = 19. The CIE curves are solid while the non-CIE curves are dashed. The curve for H I is not affected by the non-equilibrium cooling. The H I/ H population ratio is strongly coupled to the electron velocity distribution which determines the temperature in the gas. The non-equilibrium effects for O VI are large for log T < 5.4 and relatively small for log T > 5.4. The non-equilibrium effects are greatly reduced as the elemental abundances are reduced since the primary coolants are the abundant heavy elements. While these non-equilibrium effects cause substantial problems in estimating log N(O) from log N(O VI) given the temperature of the gas, they do not influence the estimate of log N(H) from log N(H I) since the effects of non-equilibrium cooling does not modify the ionization of H I in the plasma. Therefore, an estimate of the baryonic content of the gas, log N(H), follows directly from the estimated temperature of the gas which is determined from the difference in the H I and O VI line widths. However, an estimate of the oxygen abundance in the gas must include the effects of the non-equilibrium cooling. The oxygen abundance will therefore be more uncertain than the estimate of the baryonic content of the plasma. We take advantage of this interesting situation when estimating the baryoinic content of the collisionally ionized absorbers for the aligned systems which imply log T > 5 for the plasma.

The behavior of the changing ionic ratios of O VI, C IV and Si IV in the CIE and non-CIE calculations of Gnat & Sternberg (2007) for solar CNO abundance ratios are shown as a function of log T with the black line (label: CIE) and magenta line (label: static cooling) in Figure 4. The figure does not allow for additional modifications to the high ion ratios that might be caused by the ionizing effects of the radiation from the extragalactic EUV background discussed in section 4.5 or for possible variations in the heavy element abundance ratios from one absorber to the next.

### 4.4. Non-Equilibrium Collisional Ionization in Radiatively Cooling Gas Flows

The non-equilibrium radiatively cooling model of Gnat & Sternberg (2007) described in section 4.3 does not account for the dynamical evolution of the gas. Allowing for the dynamical evolution changes the thermal evolution of the gas and alters the non-equilibrium behavior of the lithium like ions. One-dimensional radiative cooling flow models have been studied by Shapiro & Benjamin (1991, 1993), Benjamin & Shapiro (1993) and Benjamin (1994). Their models follow the thermal behavior and ionization of the gas in a planar steady state flow including the ionizing effects of the photons produced by the cooling hot gas. A full discussion of the model is found in the appendix of Wakker et al. (2012). The model ionic ratio column density predictions as a function of time integrated through the flow are shown as the blue line labeled 'cooling flow' in Figure 4 for solar abundance ratios. The thin (thick) lines are for flow velocities of 12-42 km s$^{-1}$ (20-30 km s$^{-1}$). The model of Shelton (1998) illustrated with the green line in Figure 4 also traces cooling halo gas that has been heated by old type Ia supernova occurring in the low halo. These models allow for the origin of Si IV in the cooling plasma by including the effects of the He$^+$ ionizing radiation emitting by the cooling gas.

### 4.5. Photo Ionization and Collisional Ionization in Warm Plasmas

Since all regions of the IGM and galaxy halos are penetrated by background radiation, photoionization can play a role in affecting the ionization of the gas in plasmas that are also ionized by electron collisions. The ionizing photons can come from the EUV background radiation or from ionizing sources near the absorber. Examples of combining photo and collisional ionization are found in the detailed discussions of particular absorption systems in the literature. For example, Narayanan et al. (2010b) found it necessary to include the effects of photoionization when analyzing the O VI/ H I absorber at z = 0.01203 toward MRK 290. In this absorber the presence of H I, O VI and absence of C IV was difficult to understand in gas with log T ~ 5.15 based on the difference in the H I and O VI line widths. The O VI can occur in such a gas at these temperatures if it is experiencing non-equilibrium cooling because the recombination of O VI into O V proceeds more slowly than the cooling. However, the absence of C IV in the cooling gas requires another explanation. Narayanan et al. (2010) found that by combining the non-equilibrium cooling of gas exposed to the EUV background radiation field with log U = -1.4 corresponding to a density $n_H$ ~ $4 \times 10^{-5}$ cm$^{-3}$ it was possible to reproduce the observed values of N(H I), N(O VI), and the limit on N(C IV) in a cooling plasma at log T = 5.15 with [O/H] = -1.7. Such a detailed study can be undertaken in particularly interesting situations where warm gas absorbers are not easily explained by either CIE or non-CIE processes.

Oppenheimer & Schaye (2013) have very recently presented an extensive theoretical study of non-equilibrium ionization and cooling of metal enriched gas in the presence of the extragalactic photo-ionizing background and provide guidelines regarding when the combined effects of photoionization from the background and collisional ionization in warm gas are important. Their study emphasized the effects of the photoionizing background at z = 1 where the ionizing intensity is ~10 times larger than at z ~ 0.2 which is appropriate for the systems studied in this investigation. A good fit to the H-photoionization rate produced by the extragalactic background is $\Gamma_H = (2.28 \times 10^{-14} \text{ s}^{-1}) (1+z)^{4.4}$ for 0 < z < 0.7 (Shull et al. 2012). Fully integrating the combined effects of non-equilibrium ionization in warm gas with the ionizing effects of the extragalactic background into a diagnostic study of QSO absorption line systems is an important goal for future studies but is beyond the scope of our paper. Such a program should also address the current very large uncertainties in the properties of the extragalactic background as discussed by Oppenheimer & Schaye (2013). At z = 1, the different versions of the extragalactic background differ by as much as 1.8 dex near the EUV energy of 114 eV required to produce O VI.

Danforth et al. (2010b) in their study of the baryonic content of BLAs have considered the combined effects of collisional ionization and photoionization by the extragalactic background on the H I.  At z = 0 they show that the critical density where the H I photoionization rate equals the e-collisional ionization rate is (8.9, 1.9 and 1.0) x$10^{-6}$ cm$^{-3}$  at log T = 5, 5.5 and 6 respectively, corresponding to low overdensities of δ < 10 at z ~ 0.  Therefore, high ionization parameters U are required for photoionization by the extragalactic background to become important when log T > 5.  Figure 1 of Danforth et al. (2010b) shows log (N(H I)/N(H)) as a function of temperature for log U = -1 and -2. The deviations from CIE for H I are insignificant for log U < -2 and log T > 5.  For log U = -1, the deviations from CIE increase from 0.1 dex at log T = 5.6 to 0.4 dex at log T = 5.  In the case of H I,  deviations from CIE only begin to become important for large values of log U at log T < 5.4.  At log T = 5 the deviations for H I from CIE  are < 0.18 dex provided log U < -1.5.

If an absorber is situated close to a luminous galaxy it is reasonable to consider if the radiation from the galaxy can affect the presence of O VI in the absorber.  A simple way of looking at this problem is to ask at what distance and for what galaxy properties does the radiation from the galaxy become as important as the extragalactic  background in producing O VI.  This problem has been carefully explored at z = 0 for the high velocity clouds in the Milky Way by Fox et al. (2005).  The first part of the problem is to develop a model for the emergent spectrum from the galaxy based on the known stellar sources of radiation and assumptions about the fraction of this radiation escaping from the disk.  The calculated spectrum of emergent radiation from the Milky Way is shown in Figure 8 of Fox et al. (2005).  The second part of the problem allows for the fact that the flux passing through an external cloud illuminated by the galactic radiation will depend on the orientation of the cloud with respect to the disk of the galaxy.  Figure 9 in Fox et al. (2005) addresses that issue.

Fox et al. (2005) in their Figure 8 show how radiation escaping from the Milky Way compares to that from the EUV background for various distances from the Milky Way for a line of sight with a galactic latitude b =+52 degrees.  For ionizing photons with E between 13.6 and 54 eV, the average density of the extragalactic background radiation field for AGNs from Haardt & Madau (1996) and that from the Milky Way  are the same at a distance of ~90 kpc.  The result would be nearly the same for the radiation field of AGNs and starforming galaxies since the galaxies do not contribute much radiation at >13.6 eV.   For E between 54 and 200 eV the radiation fields are roughly similar for distances  of 20 to 30 kpc.  The decrease is caused by the large drop in the flux of emergent radiation from galaxies at the He$^+$ 54 eV ionization edge.    These distance estimates are useful as a rough guide. The estimated H I ionization rate from the extragalactic background radiation increases by a factor of ~3  over the redshift range from 0 to 0.25 (see Table 3 in Haardt & Madau 2012).  Therefore at z = 0.25 the 90 kpc estimate from above reduces to 30 kpc.   As a rough guide, it is not necessary to worry about the effects of the hydrogen ionizing radiation from an intervening Milky Way like L* galaxy if the impact parameter exceeds  ~90 kpc.  This distance decreases by a factor of ~ 3 for harder radiation with E > 54 eV.   Of course a careful calculation must consider the true nature of the stellar content of the galaxy, whether or not the galaxy has an active nucleus, and the reliability of the estimated escape fraction of ionizing radiation. For the Milky Way the escape fraction is estimated to be ~ 6% from the intensity of Hα emission from Galactic high velocity clouds at known distances (Bland-Hawthorn & Maloney 1999; Putman et al. 2003) assuming the high velocity clouds are photoionized by radiation from the Galaxy.

### 4.6. O VI in Conductive Interfaces

Another example of a process whereby O VI can be created in the ISM or IGM is in the interface region between cool gas and hot gas.  In the interface the gas temperature must change from that associated with cool photoionized gas, log T ~ 4 to 4.5 to that associated with the hot gas with log T ~ 6 to 7.  The interface region will therefore contain gas with intermediate temperatures where O VI may be produced.  By following the non-equilibrium processes occurring in young evaporating interfaces or older condensing interfaces it is possible to estimate the expected column density of O VI in an individual interface as a function of the age of the interface.  The process has been modeled by Borkowski et al. (1990) and Gnat et al. (2010).

Support for the origin of O VI in the local interstellar medium in conductive interfaces is provided by the excellent observed alignment of O VI absorption and cool gas absorption by C II and O I in the clouds within 150 pc of the Sun (see Savage & Lehner 2006).  However, the situation in the Milky Way disk over greater distances is clearly more complicated since the interface origin for O VI does not explain the behavior in the dispersion in plots of log N(O VI) versus distance observed in the galactic plane in the O VI  disk gas survey of  Bowen et al. (2008).

A fundamental problem with the conductive interface model for the origin of O VI is a single interface is only expected to produce maximum O VI column densities ranging up to log N(O VI) ~ 13.0.  Therefore many interfaces would be required to produce the large O VI column densites often found in the IGM and galaxy halos.

Multiple conductive interfaces seem to be a reasonable explanation for the O VI found in Milky Way high velocity clouds (Sembach et al. 2003; Fox et al. 2005). In this case the absorbing structures are complex which in principal allows multiple interfaces to contribute to the total O VI column densities observed to be associated with HVCs which have an average log N(O VI ) = 13.95±0.34 (Sembach et al. 2003).

The lithium high ion ionic ratios predicted by the conductive interface model for solar CNO abundance ratios are illustrated in Figure 4 as the thin red lines (label: conductive interfaces) connecting to the large red circles.  The interfaces evolve rapidly along the thin red lines and spend most of their lifetimes in the region denoted by the red circles for a perpendicular and parallel orientation of the magnetic field.

## 4.7. O VI in Turbulent Interfaces

If a cool gaseous region is moving with respect to an exterior hot medium, turbulence at the boundary can mix the cool and hot gas and produce gas at intermediate temperatures containing O VI and other highly ionized species.  The temperature of the mixed gas changes rapidly through electron collisions but the ionization conditions in the gas respond more slowly so non-equilibrium ionization processes are important to determine the expected amount of O VI and other states of ionization in the gas.  The turbulent interface or turbulent mixing layer model was first proposed by Begelman & Fabian (1990) and further explored by Slavin et al. (1993),  Esquivel et al (2006) and Kwak & Shelton (2010).   The advantage of having a turbulent interface is that many interfaces can occur along a given line of sight allowing substantial column densities of O VI to be produced.

The turbulent mixing layer ion ratio predictions of Kwak & Shelton (2010) for solar CNO abundance ratios are shown in Figure 4 as the blue contours labled NEI TML.

A problem with the turbulent mixing layer process for the origin of O VI in the IGM and galaxy halos is the models generally predict C IV column densities that are larger than the O VI column densities.  Although the observations are limited, it is common to find O VI with little or no associated C IV at z ~ 0.

## 4.8. O VI in Shocked Plasmas

Electron collisional ionization in shock fronts with speeds of 200 to 500 km s$^{-1}$ are an excellent site for the production of the lithium like ions observed in the IGM and galaxy halo.  The shock models of Dopita & Sutherland (1996) for v = 200 to 300 km s$^{-1}$ with solar abundances predict large

O VI column densities and more than 10 times smaller column densities of C IV, N V, and Si IV. The predicted shock ionization high ion abundance ratios are shown in Figure 4 with the orange curves (label: shock heating) for different shock speeds

### 4.9. O VI Ionization Summary

There are many ways O VI can be created in the IGM and galaxy halos. Some involve photoionization in cool gas while others involve collisional ionization in warm/hot gas. The details can rapidly get complex if non-equilibrium processes are occurring and/or if there is a combination of photoionization and collisional ionization.

A primary and simple diagnostic of the possible production process involves estimating the temperature of the gas through the difference in line width of the O VI and H I when the absorption is aligned in velocity. If the temperature of the gas containing O VI can be estimated, the baryonic content of the gas is obtained from the total column density of hydrogen which can be estimated from N(H I) since for H I the value of $f_{H I} = N(H I)/N(H)$ should be approximately given by the CIE value if the inferred temperature is large enough for electron collisions to dominate the ionization in the plasma. This should be true for log T > 5 provided log U < -1.5.

A secondary diagnostic is to compare different observed ionic ratios such as N(Si IV)/N(C IV) versus N(C IV)/N(O VI) with the predictions of the various ionization models. The model predictions discussed above for gas with solar abundances are shown in Figure 4 along with measurements for gas in Milky Way HVCs and in damped Lyman alpha systems. In some cases a single measurement such as log[N(C IV)/N(O VI)] can be used to eliminate particular models since the predicted behavior of the ionic ratios change by very large factors from one model to the next (i.e. compare the curves for shock heating, turbulent mixing, and CIE in Figure 4). An important limitation of the ion ratio diagnostic is the intrinsic abundance ratios of the heavy elements must be assumed. Another limitation is the diversity of the possible models is extensive. A casual inspection of Figure 4 illustrates that problem. Therefore, a careful evaluation of the relative behavior of O VI and H I provides the most direct information on the ionization of O VI and the total baryonic content of the gas. Without the reference to H I, values of log N(H) and [O/H] can not be obtained.

### 5. General Properties of the O VI Absorbers

The O VI systems seen toward each QSO are discussed in detail in the Appendix and illustrated with velocity plots showing line profile fits to the ions detected. A table for each QSO given in the appendix lists the fit results for the systems detected along with notes about possible problems with the fitting process such as contamination from other absorbers.

We detect 54 O VI systems in the COS, FUSE and STIS observations of the 14 QSOs listed in Table 1 over a redshift range from 0.01 to 0.50. Tables 2 and 3 summarize the basic properties of the absorbers in the O VI systems. The total redshift path for the detection of O VI is 3.52. The unblocked redshift path for our sample will be approximately 4% smaller (DS2008) or 3.37. The 54 O VI systems contain 85 O VI components and 133 H I components. 54 of the 85 O VI components are aligned in velocity with the H I absorption.

Limiting our sample to the 48 O VI systems and 65 O VI components with log N(O VI) > 13.4 over an unblocked redshift path of 3.37 we obtain $(dn/dz)_{O VI} = 14\pm2$ and $19\pm3$, respectively. These values are in agreement with those presented by TSB2008 and Tilton et al. (2012) for $W_r > 30$ mÅ. Incompleteness and the effects of FPN strongly affects the measurements for log N(O VI) < 13.4 (see Section 3.1). Properly dealing with those issues and carefully assessing the effects of line blocking and the variable S/N is beyond the scope of our program that has been designed to probe the properties of the well-detected O VI systems and components.

Table 3 summarizes the component properties of the O VI absorbers. Simple systems with two or fewer H I components and one O VI component are very common representing 27 of 54 or 50% of the systems. Systems with three or more O VI components are rare, comprising 5 of 54 or 9% of the systems. We find an average of 1.57 O VI components per O VI system and 2.46 H I components per O VI system. The kinematical simplicity of the most common systems is an interesting observational result that should be used to constrain the hydrodynamical simulations of O VI in the IGM and galaxy halos.

Figure 5 is a profile plot summary of the different types of O VI systems observed. In this summary figure we only illustrate the observations and profile fits for O VI $\lambda 1031$ or $\lambda 1037$ and H I $\lambda 1215$ or $\lambda 1025$. The velocity plots in the appendix show profile fits to all of the observed ions in each system. Figures 5a and 5b shows two examples of O VI absorption aligned with H I absorption where the different line widths imply that the O VI occurs in warm ionized gas with log T > 5. Figures 5c and 5d shows two examples of O VI absorption aligned with H I absorption where the different line widths imply the O VI likely arises in cool photoionized gas. Figures 5e and 5f shows two examples where the O VI absorption is not aligned and clearly associated with H I absorption. For the PKS 0405-123 O VI system at z = 0.16716 the absorption is very complex. However, the O VI component at -278±2 km s$^{-1}$ has no associated H I absorption (see Savage et al. 2010). For PHL 1811 at z = 0.13280, the O VI component at 85±2 km s$^{-1}$ with log N(O VI) = 13.57±0.03 and b(O VI) = 23±3 km s$^{-1}$ is not clearly evident in H I $\lambda 1215$. There is possible H I absorption at 74±15 km s$^{-1}$ with log N(H I) = 12.65:±0.15: and b(H I) = 49:±13: with very uncertain properties. If the H I absorber is real N(H I)/N(O VI) = 0.12 (+0.05, -0.03) although the value should be considered and upper limit.

In order to understand the diverse properties of O VI in galaxy halos and IGM it is helpful to characterize the different types of O VI systems observed. Some very basic system properties useful for classifying the different types of absorber are the following: (1) Number of H I components in the system. (2) Number of O VI components in the system. (3) Presence or absence of a BLA. (4) Presence or absence of other metal lines. (5) Maximum velocity difference between the H I components in the system. (5) The velocity difference between the O VI components and the nearest H I component. For aligned components with $\Delta v < \sim 10$ km s$^{-1}$ it is possible the O VI and H I exist in the same plasma. (6) Column density of the strongest H I absorber. (7) Column density of the strongest O VI absorber. (7) For aligned O VI and H I components do the differences in b values imply cool (photoionized) gas or warm ( mostly collisionally ionized) gas? We consider components where the velocity difference is $< \sim 10$ km s$^{-1}$ to be aligned which implies the O VI and H I probably exist in the same plasma.

Table 2 lists the basic observed properties of the 54 O VI systems seen toward the 14 QSOs in our high S/N program. The table lists QSO, the redshift of the principal O VI absorber, Y for yes and N for no indicating the presence or absence of other detected metal lines, b values for the H I components, number of H I components, number of O VI components, $\Delta v$(H I), the maximum velocity difference between the H I components, $|\Delta v$(H I – O VI)$|_{min}$, the minimum velocity difference between H I and O VI for the different O VI components, log N(H I)$_{max}$ and log N(O VI)$_{max}$, the largest H I and O VI component column densities for the system, and comments about the alignment or not of the O VI and H I components (Y = yes aligned and N = not aligned).

Figures 6a-f summarizes the observed properties of O VI and H I in the O VI systems. Summary numbers for many quantities are provided in Table 3.

Figure 6a is a histogram of the redshift distribution for the 54 O VI systems detected. The systems range in redshift from 0.01 to 0.50 with a median redshift of 0.17.

Figure 6b is a histogram showing the number of H I and O VI components in each system. The simple systems with two or fewer H I and O VI components dominate the sample.

Figure 6c shows the number distribution of the maximum velocity difference between the H I absorption components in each of the 42 systems excluding the 12 systems with single H I components indicated with $\Delta v = 0$ km s$^{-1}$ in Table 2. There is a peak with $\Delta v < 20$ km s$^{-1}$ with a broad tail extending to 550 km s$^{-1}$. For 23 of 42 systems the H I components span a velocity range of $< 100$ km s$^{-1}$. For 11 of 42 systems the total velocity offset exceeds 200 km s$^{-1}$.

Figure 6d is a histogram of the minimum velocity difference between the H I and O VI components, $|\Delta v(\text{H I} - \text{O VI})|_{min}$, for 85 O VI components in 54 systems. We judge 54 of the 85 components to be aligned with $\Delta v < \sim 10$ km s$^{-1}$. The properties of the aligned components are given in Table 4.

Figure 6e shows the distributions of H I and O VI Doppler parameters, b, for all the components of H I and O VI observed in the 54 O VI systems. The median and average values of b for the H I components are 32 and 38.7 km s$^{-1}$, respectively. For O VI the median and average values are 27 and 28.8 km s$^{-1}$, respectively. The distribution for H I has a pronounced high velocity tail extending to 150 km s$^{-1}$. 29 of 133 H I components are BLAs with b > 50 km s$^{-1}$. 14 of these BLAs are aligned with a component of O VI. A similar high velocity tail has been seen at low redshift for samples including all H I absorbers (see Richter et al. 2006; Lehner et al. 2007; and Danforth et al. 2010b).

Figure 6f shows the distributions of log N(H I) and log N(O VI) for the components in all the systems. The O VI components range in column density from log N(O VI) = 13.00 to 14.59 with a median value of log N(O VI) = 13.68. In contrast, log N(H I) ranges from 12.45 to 16.61 with a median value of 13.64. The distributions are incomplete for log N(O VI) < 13.4 and log N(H I) < 13.0.

## 6. Properties of the O VI and H I Absorption in the Aligned Systems

When the O VI and H I absorption is aligned in velocity it is possible to derive physical information about the properties of the gas provided the O VI and H I are co-spatial. Table 4 lists the properties of O VI absorption in 54 components of the 85 total in our sample that are reasonably well aligned with H I absorption components. By well aligned we mean |v(O VI) – v(H I)| $\leq$ 10 km s$^{-1}$ within the errors. The table lists the QSO, the O VI system redshift and in parenthesis the velocity with respect to the system redshift of the component. For each component we give, QSO, z (v O VI component), log N(O VI), b(O VI), log N(H I), b(H I), the estimated values of log T and $b_{NT}$, the estimated values of log N(H) and [O/H], and a comment regarding the origin of the ionization with PI implying photoionization in cool gas and CI implying collisional ionization in warm gas. The estimated values of log T and $b_{NT}$ and their errors assume that the O VI and H I occur in the same plasma and the difference in the line widths for O VI and H I is due to thermal and non-thermal line broadening with $b^2 = (0.129)^2$ T/A + $b_{NT}^2$. Temperature estimates are given for 45 of the 54 components. In seven cases b(H I) < b(O VI) implying that the H I and O VI do not likely exist in the same gas and the H I that might be associated with the O VI is probably too weak to detect. In one case b(H I) = (8±2) b(O VI) which is also inconsistent with both species existing in the same plasma. For the PKS 2155-304 z = 0.05423 absorber, the H I errors are so large due to component overlap, a reliable temperature can not be estimated.

Figure 7 summarizes the observed and derived properties of the aligned O VI and H I components listed in Table 4. Figure 7a displays b(O VI) vs b(H I). The solid line shows where the data points would fall if the line broadening is dominated by thermal motions with $b^2 = 0.129T/A$ for both O VI and H I implying b(O VI)/b(H I) = 0.25. The dashed line shows where non-thermal motions dominate the broadening implying b(O VI)/b(H I) = 1. The blue diamond symbols identify absorbers where the derived temperature is low enough to be consistent with cool photoionized gas. The red square symbols identify absorbers where the derived temperature suggests warm gas where the ionization is likely dominated by collisional ionization. Black circles denote cases where b(H I) < b(O VI) and b(H I) = (8±2) b(O VI). In these cases the H I and O VI are unlikely to exist in the same plasma.

The solid line in Figure 7b displays a number versus temperature histogram for the 45 aligned components when the temperature can be estimated. 31(69%) of the aligned components have log T < 4.8 suggesting cool photoionized gas. 14 (31%) of the aligned components have log T > 5.0 suggesting warm gas. The observational bias that affects Figure 7b causes the number of systems with log T > 5.3 to be more difficult to detect because the associated BLAs are broad and weak. The red dashed line in Figure 7b displays the number versus temperature histogram from the STIS observations of TSB2008 for 28 aligned components. The lower S/N STIS observations are less sensitive to the BLAs associated with the O VI absorption (see Section 8).

Figure 7c displays log N(H I) versus log N(O VI) for the aligned absorbers. No correlation is evident. There is no clear separation of data points for the photoionized versus the collisionally ionized absorbers except the range of log N(H I) is smaller for the collisionally ionized absorbers.

Figure 7d displays log [N(O VI)/N(H I)] vs log N(H I) for the aligned absorbers. log [N(O IV)/N(H I)] appears to anti-correlate with log N(H I) for the photoionized absorbers. However, the apparent anti-correlation arises because log N(O VI) spans a relatively narrow range in column density while log N(H I) spans a large range. The anti- correlation therefore is mostly showing the behavior of -log [N(H I)] vs log N(H I). However, the plot does reveal that the collisionally ionized absorbers span a smaller range of log N(H I) than the photoionized absorbers as also seen in Figure 7c.

Figure 7e displays a histogram of the values for non-thermal broadening, $b_{NT}$, obtained for the warm absorbers (red dashed histogram) and for the cool photoionized absorbers (blue histogram). For the photoionized absorbers $b_{NT}$ ranges from 5 to 55 km s$^{-1}$ with a median of 23 km s$^{-1}$. For the collisionally ionized absorbers $b_{NT}$ ranges from <10 to 56 km s$^{-1}$ with a median of 29 km s$^{-1}$.

Figure 7f shows [O/H] for the 6 collisionally ionized absorbers with 5.4 < log T < 6.2. These abundance results for the warm gas are discussed in section 6.3. Reliable abundance estimates for warm gas in the IGM from absorption line data are only possible when the associated BLA can be detected and the value of log T is large enough (> 5.4) for non-equilibrium ionization effects to be small. [O/H] is observed to range from -1.93 to 0.03 with a median value of -1.03.

### 6.1. H I and O VI Aligned and b's Suggesting Photoionization in Cool Gas

The O VI absorbers discussed in Section 5 and listed in Table 4 reveal that 31of 45 O VI components that are well aligned with H I absorption components have line widths implying the detection of relatively cool gas with log T from 3.5 to 4.8. Gas with log T < 4.6 is too cool for collisional ionization to explain the origin of the O VI. This gas is probably photoionized and the derived temperatures are generally consistent with those expected for plasmas in photoionization equilibrium where the primary heating process is photoionization by the EUV background radiation field balanced by recombination. However, we note that for temperatures as large as log T = 4.7 to 4.8 are difficult to achieve from photoionization heating except in very low density plasmas. In these cases it is possible the O VI is tracing non-equilibrium ionization of O VI in cooling gas as discussed in the next section.

### 6.2. H I and O VI Aligned and b's implying either Photoionization or
### Non-Equilibrium Collisional Ionization

When log T determined from the difference in the H I and O VI line widths for aligned systems falls in the temperature range from log T ~ 4.7 to 5.0, the temperature is so high it is likely the gas is not in photoionization equilibrium since the UV photoionization heating is not adequate to produce such high temperatures unless the gas density is extremely low. In this situation the O VI may have originally been produced by collisional ionization but now exists in a plasma that has experienced non-equilibrium cooling where the recombination of O VI has lagged behind the recombination of H I as shown in Figure 3. This is commonly referred to as frozen-in ionization.

Among the 45 aligned systems listed in Table 4 with temperature estimates, four have temperatures ranging from log T = 4.7 to 4.8 including PG 1116+215 at z (v) = 0.13850 (0) with log T = 4.72 (+0.11, - 0.15); PG 1116+215 at z(v) = 0.16553(0) with log T = 4.71 (+0.07, -0.10); H 1821+643 at z(v) = 0.21329(0) with log T = 4.77 (+0.07, -0.06); and HE 0226-4410 at z(v) = 0.35523 (0) with log T = 4.83 (+0.05, -0.06). For the origin of the ionization we list PI/CI in Table 4 since the temperature is unusually high for photoionization to be the correct explanation. These systems may be tracing cooling collisionally ionized gas with delayed O VI recombination. However, in the summary plots of Figure 7 these four systems are included among the group of 31 systems believed to occur in photoionized gas.

### 6.3 H I and O VI Aligned and b's Implying the Detection of Warm Gas

The high S/N of the H I measurements in this paper has allowed us to detect a number of BLAs reasonably well aligned with O VI absorbing components. Table 4 contains 14 O VI components where the estimated value of log T from the line width comparison implies the detection of warm gas with log T > 5.0 implying the detection of warm gas. When log T > 5.4 origin of the ionization of O VI in these absorbers is most likely dominated by collisional ionization in warm gas since photoionization heating is unlikely to achieve such high temperatures except in extreme situations. For 5.0 < log T <5.5 the O VI ionization is probably from a combination of collisional ionization and photoionization (see section 4.5). When the temperatures are high enough (log T > 5.4) CIE should be a reasonably good approximation for determining N(O)/N(O VI). For log T < 5.4 non-equilibrium effects become important for understanding the ionization of O VI (see the curves in Fig. 3). However, in all cases the baryonic content of the systems, log N(H), can be estimated from the measured value of log N(H I) and the inferred value of log T since the value of the ionization correction, log [N(H)/N(H I)], is reliably given by the CIE value provided log U < -1.5 (see section 4.5) . If log U is as large as -1 the deviations from the CIE value are 0.37, 0.10 and 0.00 dex at log T = 5, 5.5 and 6, respectively. It is therefore only for the systems with log T < 5.4 that deviations from the CIE assumption for H I begin to become important. The effect of these deviations will be to cause the estimated total hydrogen column densities to increase by the deviation values given above. At log T > 5.5 the effect is insignificant even when log U = -1. At log T = 5 the correction is 0.05 and 0.37 dex at log U = -2 and -1, respectively.

The ionization corrections for the 14 components range from log [N(H)/N(H I)] = 5.00 to 6.18 with a median of 5.37 and a logarithmic dispersion of 0.63. The inferred values of log N(H) are listed in Table 3 for the warm absorbers. log N(H) ranges from 18.38 to 20.38 with a median of 19.35.

For six of the warm systems the temperature exceeds log T = 5.4. At such high temperatures the effects of non-equilibrium ionization should be relatively small for both O VI and H I (see Fig. 3) allowing an estimate of log [N(O)/N(O VI)] and log [N(H)/N(H I)] using the CIE ionization curves of Gnat and Sternberg (2007). With measures of log N(O VI) and log N(H I), the abundance of oxygen in the warm gas, [O/H], can be obtained. The six values are listed in Table 4 and plotted in Fig. 7f. The values range from [O/H] = -1.93 to 0.03 with a median of -1.03. This abundance study is biased toward finding absorption components with [O/H] < -0.7 dex because the H I is more difficult to detect in the higher metallicity systems given the relative behavior of O VI and H I shown in Figure 3. The O VI absorption components not aligned with H I probably trace the O VI absorbers in gas with higher oxygen abundances. The six values of [O/H] for the warm gas between the galaxies listed in Table 4 are not likely to increase very much for many years to come given the requirement for high S/N observations to reveal the BLAs associated with the O VI absorption.

For eight of the warm systems the temperature lies in the range from log T = 5.0 to 5.4 where non-equilibrium ionization effects are important for O VI. For the absorber toward MRK 290 at z = 0.01026 with log T = 5.07 (+0.13, -0.19) and log N(H) = 19.33 (+0.33, -0.56), Narayanan et al. (2010b) considered all the possible ionization processes and concluded the O VI is affected by non-equilibrium collisional ionization combined with the effects of photoionization from the EUV background radiation. They derived an oxygen abundance of [O/H] ~ -1.7. Such a detailed analysis for the other 8 warm systems with log T from 5 to 5.4 is beyond the scope of this paper. Once the temperature of the gas drops much below log T ~ 5.4, the delayed recombination of O VI in cooling gas can introduce very large uncertainties in the derivation of the oxygen abundance although reliable values of the total hydrogen column density can still be estimated.

The quality of the 14 BLA and O VI measurements discussed in this section varies. Table 5 gives an approximate ordering of the quality of the warm gas detections based on the comparison of the O VI absorption to the associated BLA. We list the 14 systems ordered by decreasing quality. Examples of the highest quality systems include HE 0153-4520 at z = 0.22600, MRK 290 at z = 0.01026, and 3C 263 at z = 0.14072. Examples of the lowest quality systems include PHL 1811 at 0.07773, H 1821+643 at z = 0.17036 and HE 0238-1904 at z = 0.47024. The quality ordering is based on our assessments of the reliability of the observed properties of the O VI and the BLA absorption line measurements.

Table 5 also lists the four systems with log T ~ 4.7-4.8 which could be tracing delayed recombination in a collisionally ionized cooling plasma rather than photoionized gas. Those systems are listed in the lower column under the heading possible collisionally ionized systems. Also listed under that heading are three systems where the origin of the ionization could be either photoionization in cool gas or collisional ionization in warm gas (see footnotes e, f and g). In the case of the 3C 263 system at z = 0.32567 it is clear the system is detecting warm gas due to the presence of Ne VIII. However, the error on the associated BLA implied a marginal detection significance of ~3σ.

The different sources of errors affecting the results listed in Tables 4 and 5 are discussed in various places in this paper. The basic data extraction has been performed with careful attention paid to removing fixed pattern noise and in determining accurate wavelengths of the individual spectra before the co-addition of specta. Standard extraction techniques will produce spectra with degraded resolution and inaccurate wavelengths. Therefore differences between our results and those obtained from observations using the standard pipeline processing can in part be explained as due to problems with the spectral extraction procedures. Fitting more components to the profiles would certainly produce different results. We have adopted the procedure to fit the minimum number of components that produces an acceptable $\chi^2$ to the multi-component absorption profile. The errors in the fit parameters v, log N, and b include the statistical errors and the errors associated with the separation of different components in a multi-component fit. For the aligned O VI and H I components the values of log T and their errors follow simply from the b values and their errors assuming the two absorbers co-exist in the same gas. That assumption will be only approximately correct and will introduce systematic errors that are difficult to quantify. Errors in derived quantities such as log N(H) and [O/H] are mostly influenced by the errors in log T through the temperature dependence of the ionization corrections for H I to H and O VI to O. The ionization corrections are more reliable for log T > 5.5 than for log T < 5.5. At the lower temperatures the ionization corrections for O VI to O are so large we do not try to estimate [O/H]. The derived values of log N(H) obtained when log T < 5.5 will increase over those listed when the density in the absorbing gas is low enough for photoionization to begin to play a role as discussed in section 4.5. We have been as careful as is reasonably possible in producing the results given in Tables 4 and 5. Observations with higher spectral resolution, higher S/N and better wavelength calibration will be required to improve upon the listed results. Unfortunately, we will need to wait many years for the deployment of a high efficiency far-UV spectrograph in space more capable than COS

## 7. O VI and H I in Non-Aligned Components

In 31 of 84 cases the O VI absorption components are not clearly aligned with H I absorption components. The most extreme cases are those in which there is no detectable H I at the velocity of the O VI. Several examples include: (1) The O VI at v = -278±2 km s$^{-1}$ in the z = 0.16716 Lyman limit absorber toward PKS 0405-123 where Savage et al. (2010) determined N(H I)/N(O VI) < 0.063. (2) The O VI at v = 343 km s$^{-1}$ in the z = 0.22497 system toward H 1821+643 where Narayanan et al. (2009) determined N(H I)/N(O VI) < 0.15. However, there are many other cases where the O VI absorption occurs over the velocity range of H I absorption but where there is little or no correspondence between the O VI and H I profiles. In these cases it is more difficult to establish a clear limit on the amount of H I that could be associated with the O VI because of the confusion with the other absorbing structures containing H I.

The non-aligned O VI absorbers can be created either by collisional ionization or by photoionization. In the case of collisional ionization equilibrium for solar abundances, Figure 2 reveals that N(H I)/N(O VI) is < 0.1 for log T from 5.35 to 5.85. For 0.1 solar abundances N(H I)/N(O VI) is < 0.2 only for a small temperature range near log T = 5.5. Therefore, it is easy to produce hydrogen-free O VI in collisionally ionized gas with solar abundances over a broad temperature range while it is difficult to hide the H I in collisionally ionized gas if the oxygen abundance is 0.1 solar. In the case of photoionization, obtaining N(H I)/N(O VI) < 0.1 requires a very high ionization parameter and therefore a relatively low gas density and a long path length. For example, for the absorber at v = -278 km s$^{-1}$ in the z = 0.16716 system toward PKS 0405-123, Savage et al. (2010, see their Fig. 10) showed that N(H I)/N(O VI) < 0.063 could be created by photoionization in gas with solar abundances with log U = -0.2, log $n_H$ = -5.5, log N(H) = 18.3, log T =4.7, P/k = 0.40 cm$^{-3}$K, and L = 130 kpc. However, the implied path length increases to 1.3 Mpc for 0.1 solar abundances. Therefore, the photoionization origin of the non-aligned O VI absorbers requires abundances near solar to avoid having path lengths that are so large that the expected line width from Hubble flow broadening would begin to exceed the observed line width. Of course, individual cases of non-aligned O VI absorbers might occur if there is a large enhancement of the local radiation field that preferentially destroys the H I.

In summary, PI or CI can explain the non-aligned O VI absorbers if the oxygen abundance is relatively large (N(O)/N(H) > ~0.5 solar). However, both PI and CI have problems explaining these absorbers if the oxygen abundance is ~ 0.1 solar or less. Therefore, it does not appear possible to clearly determine the physical state or ionization processes operating in the non-aligned hydrogen free O VI absorbers.

A crude temperature limit can be determined from the observed O VI line width assuming all the broadening is from thermal motions. For the two systems with the smallest limit on log [N(H I)/N(O VI)] including O VI at v = -278 km s$^{-1}$ in the z = 0.16716 absorber toward PKS 0405-123 and the O VI at v = 343 km s$^{-1}$ in the z = 0.22497 system toward H 1821+643, we obtain a 2σ limit on log T of < 6.5 and 5.5, respectively. These limits are consistent with either collisional ionization in warm or hot gas or with photoionization in cool gas where the broadening is dominated by non-thermal broadening. The turbulence would need to be fully developed to produce such broad, symmetrical and Gaussian-like optical depth in the O VI profile as seen in the system toward PKS 0405-123 analyzed by Savage et al. (2010) who concluded it is more reasonable to assume the broadening is dominated by thermal motions in a hot gas.

## 8. Comparision with Earlier Lower S/N Observations with STIS

When COS and STIS observations exist for the same QSO, the S/N in the COS spectra as listed in Table 1 is several times higher. Given the large difference in S/N, a detailed case by case study of the similarities and differences between all the COS and STIS measurements is not very informative except for someone interested in seeing how difficult it is to get reliable information from complex spectra when the S/N is low. Kim et al. (2013) discuss detailed intercomparisons of the COS and STIS observations of H I in the spectra of 3C 273 and MRK 876 for which the STIS observations are of high quality and the COS observations are of exceptional quality. For strong single component H I lines, the agreement is excellent for the derived values of b(H I) and log N(H I) (see section 2.1 and Kim et al. 2013). For blended lines and weaker lines the agreement is not as good although much of the difference is likely the result of the lower S/N of the STIS observations of these bright QSOs.

The greatly improved S/N for the COS spectra makes it easier to identify the component structure in the O VI and H I profiles and to search for weak broad H I λ1215 absorption. In cases where the STIS observation suggested the existence of a BLA, the higher S/N COS observations make the detection of the BLA convincing. The analysis of TSB2008 of STIS spectra revealed that for 28 aligned H I/ O VI absorbers 7 (25%) were associated with BLAs revealing gas with log T = 4.8 to 5.6 and 21 (75%) were associated with cooler H I implying gas with log T = 3.5 to 4.8. This result from TSB2008 is compared to the new COS result in Figure 7b with the red dashed and black histograms, respectively. However, we note that two of the warm plasma results of TSB2008 are shown in this paper to be questionable including the O VI system at z = 0.05885 toward PG 0953+414 and the z(v) = 0.13280 (85) absorber toward PHL 1811. The greater sensitivity of COS to weak absorbers has increased the detection rate of the BLAs associated with O VI to 31%.

It is interesting the STIS sample also reveals a very good quality well aligned O VI/BLA absorber toward PG 1444+407 at z = 0.22032 (see TSB2008, Fig. 41). This QSO was not included in the COS high S/N program. The absorber has b(H I) = 86±15 km s$^{-1}$ and b(O VI) = 36±8 km s$^{-1}$ and log N(H I) = 13.65±0.05 and log N(O VI) = 13.94±0.07. The different b values imply log T = 5.59 (0.16, -0.24). At such a high temperature the H I ionization should be well described by CIE implying log [N(H)/N(H I)] = 6.04 (0.24, -0.45), log N(H) = 19.68 (+0.25, -0.43), and [O/H] = -1.37 (+0.44, -0.19). These values are similar to those derived for many of the warm systems analyzed in this paper.

The most noteworthy scientific impact of the higher S/N is the clear demonstration from the new COS observations that there are cool and warm O VI absorbers traced by aligned O VI/H I components implying ~69% of the aligned absorbers are probably cool and photoionized and ~31% are warm  and probably mostly collisonally ionized .  The estimated temperatures for the aligned warm absorbers allows for estimates of log N(H) and [O/H] in the warm plasma.

## 9. Galaxies Associated with the Absorbers

H I and O VI absorbers are strongly associated with the extended surroundings of galaxies (Tripp et al. 1998, Stocke et al. 2006; Wakker & Savage 2009, Chen & Mulchaey 2009, Prochaska et al. 2011a, 2011b, Tumlinson et al. 2011b, Stocke et al. 2013, Johnson et al. 2013).  The very low redshift survey of Wakker & Savage (2009) extending to L ~ 0.1L* revealed that for impact parameters < 400 kpc the H I and O VI absorbers are estimated to contain 2 to 4 times the amount of baryons found in the associated galaxies.

The galaxies known to be associated with the O VI absorbers observed at high S/N in this paper are listed in Table 6 based on a search of the absorber/galaxy literature for galaxies with redshifts implying an absorber/galaxy rest-frame velocity difference $|\Delta v| < 400$ km s$^{-1}$.  For each absorber, the galaxy with the smallest known impact parameter, $\rho$, is listed along with other galaxies known to have  $\rho < 500$ kpc, except for the absorber at z = 0.00336 toward 3C 273 where we list only 4 of 17 known galaxies with  $\rho < 500$ kpc in the Virgo galaxy cluster.   The table is based on an inhomogeneous set of observations.  In some cases the galaxy/absorber association has been found through observations with 6-8 m ground based telescopes and have extended to small limiting luminosities.  In other cases the survey efforts are not as deep since they have been limited to spectroscopy with 4 m class telescopes.  For three QSOs containing 11 OVI systems (TON 236, HE 0153-4520, HE 0238-1904), even shallow survey observations do not exist.   The table lists the QSO, $z_{abs}$, $z_{gal}$, the total values of log N(H I) and log N(O VI) summed over all the components in each absorption system,  the galaxy name or coordinates, galaxy type, $\Delta v = v_{gal} - v_{abs}$, L/L*,  (L/L*)$_{Lim}$ , the detection limit for L/L* at the redshift of the absorber when available,   the impact parameter, $\rho$  in physical distance, and the reference.  Galaxies with $\rho$  < 1 Mpc possibly associated with the O VI systems in this paper have been identified for 40 of the 55 systems observed at high S/N.  For most of the 15 cases with no identified galaxies, deep survey measurements do not exit.  The listed values of L/L* and (L/L*)$_{Lim}$ are based on survey measurements made in many different photometric systems. We have generally adopted the estimates from the various papers.  Some of the references have done a good job of reporting (L/L*)$_{Lim}$ while others provide little information.  The listed values should be treated with caution.  The values of L and the survey limits reported in the more recent papers obtained with large ground based telescopes are usually well determined. In those cases the authors have generally adopted a non-evolving rest-frame absolute R-band magnitude for a 1.0L* galaxy of $M_{R*} = -21.17$ from Blanton et al. (2003) and k-corrections from Coleman et al. (1980) for Scd galaxies.  In the earlier papers the absolute B-band magnitude of $M_{B*} = -19.5$ from Loveday (1992) was adopted. For reference, L/L* ~ 0.2 for the Large Magellanic Cloud.

The galaxy types for the associated galaxies with the closest known impact parameter for  $\rho$ < 1000 kpc include 21- emission line and spiral galaxies and 11- absorption line galaxies with the type unknown for 8 galaxies.  While,  emission line galaxies dominate the sample, the presence of 11 absorption line galaxies imply that they also are associated with O VI absorption. Five of the 11 absorption line galaxies associated with absorbers have $\rho$ ranging from 124 to 281 kpc suggesting an actual physical association with the absorber.  The remaining six have ρ extending from 351 to 999 kpc.

The values of ρ for the galaxies with the smallest impact parameters associated with each O VI system range from 38 to 2009 kpc with a median value of 286 kpc.  There is only one O VI system listed in Table 5 where there is no evidence for an associated galaxy to large impact parameters when $L_{Lim} < 0.1L*$.  For the   z = 0.18285 absorber toward PKS 0405-123 Johnson et al. (2013), found no galaxies with L > 0.04L* for ρ < 250 kpc and L > 0.3L* for ρ < 1 Mpc. They suggest the absorber may lie in a galaxy void.  Both components of the O VI in the absorber appear to be photoionized. Since galaxies with L as small as 0.001L* are associated with other absorbers listed in Table 5, it would be important to obtain survey measurements for the z = 0.18285 system extending to smaller values of $L_{Lim}$.  Unfortunately, ruling out the possible existence of very faint galaxies associated with absorbers is observationally difficult even at z ∼ 0.2.

Figure 8a shows a histogram of the distribution of values of the impact parameters for the closest associated galaxies with ρ < 1200 kpc from Table 6.  When more than one galaxy is listed for a system, we use the galaxy with the smallest impact parameter.  The solid histogram is for all the closest associated galaxies.  The red dashed histogram is for associated galaxies with ρ determined from survey observations extending to L < 0.3L*.

Note that the observed distribution of ρ shown Figure 8a is relatively flat from ρ = 0 to 500 kpc. If we take 200 kpc as a representative virial radius for an L* galaxy, 12 of 37 systems are associated with a galaxy having an impact parameter smaller than this value while 20 of 37 lie from 200 to 650 kpc and 5 from 650 to 2009 kpc.   Therefore, most (25 of 37) of the absorbers sampled in our study of random QSO lines of sight probably refer to gas in the IGM beyond the virial radii of L* galaxies. The most common O VI absorber probably traces gas in the IGM far from luminous galaxies rather than gas in the virialized halos of galaxies that is found in the program of Tumlinson et al. (2011b) which specifically targeted luminous star-forming galaxies.

The circles in figures 8b and c show log N(O VI, total) and log N(H I, total) versus ρ for each system where the O VI and H I total column densities are summed over all the components in each system.  The circle size scales with the luminosity of the associated galaxy as described in the figure caption.  Filled circles are for values of ρ obtained from observations with $L_{Lim} < 0.3L*$.  Open circles are for less sensitive observations down to a luminosity only somewhat smaller than the luminosity of the detected galaxy as listed in Table 6.

 For our sample of O VI absorbers with associated galaxies, log N(O VI, total) ranges from 13.08 to 14.78 with a median value of  13.70.  There is a weak trend for the oxygen column density to increase ∼0.3 dex with decreasing impact parameter from 800 to 150 kpc with an additional   ∼0.8 dex increase for ρ  < 150 kpc.  Despite the inhomogeneous survey depth, it is clear that O VI absorbers with a mean log N(O VI) ∼ 13.7 are associated with both luminous and low- luminosity galaxies.

The behavior of log N(H I, total) for our sample with decreasing ρ is similar to that found for log N(O VI, total).  There is a slight increase with decreasing impact parameter from ρ = 800 kpc to 200 kpc with a ∼1 dex increase for ρ < 150 kpc.  A very clear difference between the behavior of O VI and H I vs impact parameter is the much smaller dispersion in the behavior of log N(O VI, total) vs  ρ  compared to log N(H I, total) vs  ρ. This result does not depend on the impact parameter itself since the same value has been used for the O VI and H I column densities.  The logarithmic dispersion of N(O VI, total) about the average is ∼ 0.3 dex while it is ∼ 0.6 dex for  N(H I, total).  The distribution of O VI in the IGM connecting to galaxies is much more uniform than the distribution of H I.

In order to compare the properties of the most common O VI systems measured in our blind survey to the O VI absorbers closely associated with galaxies, we also show in Figure 8b and 8c as the open blue squares the values of log N(O VI) and log N(H I) versus ρ for gas in the virialized halos of the L* star-forming galaxies with ρ ranging from 14 to 155 kpc studied by Tumlinson et al. (2011b) and Tumlinson et al. (2013). The Tumlinson et al. sample is based on COS observations of QSOs known to be closely aligned with foreground star-forming galaxies with small impact parameters. In contrast, our blind survey first identified the detectable O VI absorbers along random sight lines to bright QSOs and the associated galaxies were identified in subsequent galaxy redshift surveys.

The detected values of log N(O VI) in the Tumlinson et al. sample range from 14.3 to 15.1 with a median of 14.6. The detected values of log N(H I) range from 13.95 to >18.0. Note that approximately half of these H I measurements are lower limits because of line saturation. Profile fits to the observations by Tumlinson et al. (2013) imply the actual column densities may be ~ 1 dex larger on average.

Our sample of O VI systems merges with the Tumlinson et al. (2011b) sample for ρ < 200 kpc but is ~ 0.9 dex smaller for 200 < ρ < 650 kpc, revealing that O VI exists well beyond the virial radii of L* galaxies. Our sample of H I absorbers also merges with the Tumlinson et al. (2013) sample for ρ < 200 kpc. However, there is a very large increase in the value of log N(H I) near ρ = 150 kpc for the star forming galaxy sample implying a much larger value of N(H I)/N(O VI) in the Tumlinson et al. (2013) sample compared to what we measure at ρ > 200 kpc.

The panels in Figure 9 from bottom to top show log N(O VI), log N(H I) and log N(H) versus ρ for the aligned components in the O VI systems associated with galaxies. The different symbols denote CI absorbers (red squares), PI absorbers (blue diamonds) and absorbers where the ionization mechanism is unknown (black circles). The aligned component plots of log N(O VI) versus ρ and log N(H I) versus ρ in Figures 9c and 9b exhibit approximately the same behavior as for the total column density plots in Figures 8c and 8b. There is no clear segregation of the distribution of log N(O VI) or log N(H I) versus ρ for the PI absorbers (blue diamonds) or for the CI absorbers (red squares).

For the CI absorbers we can determine the total baryonic content from the temperature of the plasma. In Figure 9a we display the behavior of log N(H) versus ρ for the 14 CI absorbers for which impact parameters have been measured. This plot reveals that extremely large total H column densities trace gas in the IGM extending ~ 400 kpc away from the associated galaxies.

The top panel of Figure 9 reveals 6 of 12 collisionally ionized warm absorbers with good temperature estimates we have detected are situated close to galaxies with ρ < 200 kpc. However, 6 of 12 warm absorbers lie 230 to 800 kpc away from their associated galaxy. Although some of these could be associated with undetected dwarf galaxies, the galaxy redshift survey luminosity limits are relatively small (< 0.1L*) for many of the associated galaxies listed in Table 5. These absorbers situated far from galaxies therefore probably originate in the more extended IGM, which in simulations is organized as large-scale filaments, also known as the Cosmic Web.

It is interesting to explore the possible implications of the warm IGM plasma traced by the warm O VI/H I absorbers displayed in the top panel of Figure 9. The plasma is traced to impact parameters out to ~600 kpc. For ρ < 600 kpc the plasma appears to be relatively smoothly distributed implying a high covering factor. The median properties of the 14 warm absorbers are T =

$3.8 \times 10^5$ K and N(H) = $4.2 \times 10^{19}$ cm$^{-2}$.  If the typical absorbing path length is  L = 300 kpc,  the average total hydrogen particle density (= proton density) is ~$4 \times 10^{-5}$ (300 kpc/ L) cm$^{-3}$ and the pressure, P/k ~ 37 (300 kpc/L) cm$^{-3}$ K.

Such a widely distributed warm plasma could support imbedded photoionized clouds having a similar pressure. Stocke et al. (2013) have found that the cool PI absorbers at large impact parameters have values of P/k ranging from 3 cm$^{-3}$ K to 80 cm$^{-3}$ K with a median value of ~10 cm$^{-3}$ K. The warm plasma revealed by the O VI systems may represent the gas supporting the cool PI clouds.

## 10. Estimates of the Baryonic Content of Warm Gas in the Galaxy Halos and the IGM

There have been numerous estimates of the possible baryonic content of the O VI absorption-line systems. The estimates usually assume an average value of [O/H] = -1. In addition to the uncertain oxygen abundance, an assumption for the oxygen ionization correction must be made. It is common to assume an oxygen ionization correction N(O)/ N(O VI) ~ 5 which is valid for CIE at log T = 5.5 where O VI peaks in abundance  (see Fig. 2). The actual value of the oxygen ionization correction is expected to be larger (see the simulations of Shull et al. 2012). With log [N(H)/N(O)]$_O$ = 3.31 from Asplund et al. (2009)  and the assumptions above,  the value of log N(H) = log N(O VI) + 5.01.  If we apply this correction to the 14 warm absorbers listed in Table 4, we obtain values of log N(H) ranging from 18.21 to 19.77 with a median of 18.62. This median derived from the standard assumptions is 0.73 dex smaller than the median value of log N(H) = 19.25 listed in Table 3 for the warm absorbers obtained directly from measured value of log N(H I) and the inferred hydrogen ionization correction based on the derived temperature of the gas. The standard assumptions for estimating the baryonic content of individual O VI systems produces an estimate of N(H)  ~5.4 times smaller than for the warm absorber values listed in Table 4. Much of large difference arises from the assumption that the O VI in the IGM occurs near its temperature of peak abundance with log T ~ 5.5 where N(O)/N(O VI ) ~ 5.  However, a warm plasma cools very rapidly near this temperature.  It therefore is more likely to find O VI in warm plasmas at higher temperatures or in gas that has cooled to temperatures below log T = 5.  The rest of the difference could be due to the assumption that [O/H] ~ -1 in the O VI absorbers.  Although, Table 3 reveals that this assumption is consistent with the median observed value of [O/H] in the warm absorbers with log T > 5.4.

We can use the results of this paper to produce an improved estimate of the baryonic content of warm gas in galaxy halos and the IGM traced by O VI.  We only consider aligned O VI and H I absorbers where the line widths imply log T > 5.  For the 85 O VI absorption components included in this study, 45 are aligned and permit temperature estimates.  14 of these absorbers have temperatures implying the detection of warm gas with log T from 5.0 to 6.2.  Therefore, 14 of 85 or (16.5±4.4)% of the O VI components trace warm gas.  TSB2008 estimate that intervening O VI components at low redshift with W$_r$ > 30 mÅ have a number density per unit redshift of dn/dz = 21±3 . Using this value of dn/dz and the estimate that (16.5±4.4)% of O VI components trace warm gas we estimate that (dn/dz)$_{warm}$ = 3.5±0.8. Using the baryonic content estimate from Danforth et al. (2010b), the baryonic content of the O VI absorbers relative to the critical density, $\Omega_b$(O VI)$_{warm}$ = ($\mu m_H H_o/\rho_c c$) $\Sigma$N(H)/ $\Sigma\Delta X$, where the mean molecular weight including He, $\mu$ = 1.3,  m$_H$ is the mass of H, c is the speed of light, the critical density, $\rho_c$ = 3H$_o^2$/8$\pi$G, and $\Delta X$ = (1+z)$^2$ [$\Omega_M$(1+z)$^3$ +$\Omega_\Lambda$]$^{-1/2}$ $\Delta z$.   For the warm gas we take $\Sigma$N(H)/ $\Sigma\Delta X$ =  (dn/dz)$_{Warm}$ ($\Sigma\Delta z$ /$\Sigma\Delta X$) <N(H)$_{Warm}$>, and we obtain $\Omega_b$(O VI)$_{Warm}$ = 1.78×10$^{-23}$ h$_{70}^{-1}$ (dn/dz)$_{Warm}$ ($\Sigma\Delta z$ /$\Sigma\Delta X$) <N(H)$_{Warm}$> .  For the 14 O VI systems tracing warm gas listed in Table 4,  <N(H)$_{Warm}$> = 4.22 ×10$^{19}$ cm$^{-2}$ .  With these values  and  ($\Sigma\Delta z$ /$\Sigma\Delta X$) = 0.739 for the 14 QSOs we obtain a baryonic estimate for O VI systems containing warm gas of  $\Omega_b$(O

VI)$_{Warm}$ = 1.78x10$^{-23}$ h$_{70}^{-1}$ x 3.5 x 0.739x4.22x10$^{19}$ = (0.0019±0.0005)h$_{70}^{-1}$ , where the error estimate only includes the counting statistics in the estimate of (dn/dz)$_{Warm}$ for the O VI systems. Our estimate does not include possible contributions to the warm gas estimate from the 27 O VI components listed in Table 2 that are not aligned or for the 9 aligned components where the O VI and H I do not exist in the same gas. Some of these 36 absorbing components probably do trace warm or even hot gas as for example the very strong and very broad hydrogen free O VI absorber seen at -287 km s$^{-1}$ in the z = 0.16716 O VI system toward PKS 0405-123 (Savage et al 2010). The estimate ignores the three aligned O VI systems where log T ranges from 4.7 to 4.8 where the temperature is so high that photoionization seems unlikely but the temperature is below the traditional cutoff for warm gas at log T = 5. The estimate also does not include contributions to the warm gas estimate from BLAs for which associated O VI has not been detected. For example, we have omitted from Tables 2 and 3 two definite BLAs toward PG 1116+215 for which the possible presence of O VI is at the 2σ level (see A5 in the appendix). In the case of the BLAs with no detected metal lines, it is difficult to determine from the measured line width a temperature estimate because it is not clear how to separately determine the thermal and non-thermal contributions to the line broadening.

Table 7 summarizes various estimates for the different contributions to the baryonic content of the low redshift Universe. We have only listed values for those components for which the estimate is relatively secure and does not depend on very uncertain assumptions regarding the oxygen abundance in the gas or regarding the relative sizes of thermal and non-thermal broadening in BLAs. The total expected baryonic content, $\Omega_b$(total) = 0.0463±0.0024 is from Hinshaw et al. (2013) based on the analysis of the spectrum of the acoustic peaks in the cosmic microwave background spectrum obtained by the Wilkinson Microwave Anisotropy Probe.

The warm plasma estimate from this paper based on the 14 warm aligned O VI/H I absorbers listed in Table 4 includes contributions from gas in galaxy halos and the IGM. The estimate of $\Omega_b$(O VI)$_{Warm}$ = 0.0019±0.0005 corresponds to (4.1±1.1)% of all the expected baryons. In contrast (7±2)%, (4.0±1.5)%, and (28±15)% are found in galaxies, the hot intercluster medium, and the cool photoionized narrow H I absorbers in the Lyα forest according to the summary of Shull et al. (2012). The warm plasma traced by aligned O VI and H I absorption at low redshift contains nearly as many baryons as are found in galaxies.

Additional baryons exist in the non-aligned O VI absorbers. Some fraction of those will be in cool gas while the rest will be warm gas. However, without knowing the oxygen abundance or the origin of the ionization it is not possible to reliably estimate the baryonic content of the non-aligned absorbers. Therefore, the value (4.1±1.5)% listed above is a lower limit to the actual percentage of warm absorbers traced by O VI.

Additional warm baryons probably also exist in BLAs with no associated metal lines. Shull et al. (2012) estimate that the metal free BLAs may contribute (14±7)% to the baryon inventory based on very uncertain ionization corrections.

The actual detection of the hot IGM will require future X-ray satellites. However, even with X-ray observatories that could routinely measure galactic halo and IGM O VII and O VIII absorption, it will be necessary to assume values of [O/H] to convert measures of N(O VII) and N(O VIII) into estimates of N(H) unless the associated measure of N(H I) can also be obtained from a very broad and weak BLA at the same redshift. It is clear that astronomers will be debating the uncertainties assigned to the various estimates to the contributions to the baryonic components identified in Table 7 for many years to come.

## 11. Summary

We report on the observed properties of the gas revealed through high S/N observations of 54 intervening O VI absorption systems containing 85 O VI and 133 H I components in the spectra of 14 QSOs observed at ~18 km s$^{-1}$ resolution with COS at z < 0.5. Many of the QSOs have previously been observed with STIS at lower S/N. The total O VI redshift path corrected for ISM blockage is 3.52. The high S/N is especially crucial for determining the properties of the broad and weak H I absorption expected to be associated with the O VI absorption in gas with log T > 5. Special extraction techniques are used to improve the wavelength calibration of the individual spectra. Velocity plots and component fits to the absorption systems and discussions of their observed properties are found in the appendix. When the O VI and H I absorption components are aligned in velocity, their relative line widths can be used to estimate the temperature of the plasma containing O VI and determine other properties such as the oxygen abundance and total hydrogen column density. The principal results of this study are as follows:

1. The O VI system component structure ranges from very simple to complex. 50% of the systems have one O VI component and either one or two H I components. Only 10% of the systems have more than two O VI components. The average number of O VI (H I) components per O VI system is 1.57 (2.46).

2. For a sample of 45 O VI and H I absorption components well aligned in velocity for which temperature estimates are possible, we find evidence for cool photoionized gas in 69% of the components with narrow H I absorption and for warm gas in 31% of the components associated with broad H I absorption.

3. We show that, when the O VI absorption is not clearly associated with H I absorption in the components that are not aligned, it is not possible to determine the origin of the ionization of the O VI. It could either be produced in collisionally ionized or photoionized gas. However, in both cases the oxygen abundance must be relatively high, > ~0.3 times solar, in order to not produce detectable H I associated with the absorber.

4. The total hydrogen column density for each of the 14 warm components can be estimated from the temperature and the measured value of log N(H I) since CIE is a relatively good assumption for determining the ionization correction for hydrogen for warm gas provided log U < -1.5. The inferred values of log [N(H)/N(H I)] range from 4.77 to 6.80 with a median of 5.37. The values of log N(H) range from 18.38 to 20.38 with a median of 19.35. These are the only direct estimates of log N(H) in the warm IGM.

5. The abundance of oxygen can be determined in six warm components with 5.4 < log T < 6.2 for which the effects of non-CIE are relatively small for determining log [N(O)/N(O VI)]. In these components [O/H] ranges from -1.93 to +0.03 dex with a median value of -1.0 dex. These are the only direct estimates of [O/H] in the warm IGM.

6. The galaxies associated with many of the absorbers have been identified in previous studies and are listed in Table 5. The observed impact parameters have a median value of 310 kpc. 11 of 36 absorbers lie within projected distance ρ = 200 kpc of the associated galaxy while 19 of 36 lie in the IGM from 200 to 650 kpc. Most of the absorbers sampled in this study refer to gas beyond galaxies connecting to the IGM. The median total O VI column density for this gas is log N(O VI) = 13.7 which is ~0.9 dex smaller than the median O VI column density found in the virialized halos of L* star forming galaxies with impact parameters ρ < 150 kpc found by Tumlinson et al. (2011b). O VI is more uniformly distributed than H I in the gaseous regions connecting to galaxies sampled by our study.

7. We obtain $\Omega_b(O\ VI)_{Warm} = 0.0019 \pm 0.0005$ in the aligned O VI/H I systems which corresponds to $(4.1 \pm 1.1)\%$ of all the baryons with $\Omega_b(total) = 0.0463 \pm 0.0024$ from WMAP (Hinshaw et al. 2013). In contrast $(7 \pm 2)\%$, $(4 \pm 1.5)\%$, and $(28 \pm 15)\%$ of $\Omega_{b(total)}$ are found in galaxies, the hot intercluster medium, and the cool photoionized narrow H I absorbers in the Ly $\alpha$ forest. The warm plasma traced by aligned O VI and H I absorption at low redshift contains nearly as many baryons as are found in galaxies. Additional warm baryons exist in non-aligned O VI/ H I absorbers and in metal free BLAs.

8. The hydrodynamical simulations of the assembly of structures in the evolving universe must account for the formation of a mixture of oxygen enriched structures containing both cool photoionized gas and warm collisionally ionized gas located well beyond the virial radii of galaxies. The very simple kinematical nature of the most common absorber should be a product of the simulations.

We thank the many people involved with designing and building COS and determining its performance characteristics. We thank the referee for his/her help with improving a long and complex manuscript. Spectra were retrieved from the Multimission Archive (MAST) at the Space Telescope Science Institute. The study made use of the NASA/IPAC Extragalactic Data Base (NED) which is operated by the Jet Propulsion Laboratory, California Institute of Technology under contract with NASA. We acknowledge funding support to the University of Colorado, Boulder and the University of Wisconsin, Madison from NASA through the COS GTO NASA grants NNX08AC146 and NAS5-98043. T.-S. Kim has recently relocated at Osservatorio Astronomico di Trieste, Italy, and part of this work is supported by the European Research Council starting grant titled "Cosmology with the IGM" through grant GA- 257670.

*Facilities:* HST(COS), HST(STIS), FUSE.

Table 1
The  14 Bright QSOs Observed

| QSO[a] | $z_{em}$[b] | λ(Lyα) (Å) | λ(O VI) (Å) | z(O VI) | FUSE[c] (S/N) | STIS[c] (S/N) | COS[c] (S/N) | t (COS) ks G130M G160M | CalCOS | Date | Table[d] |
|---|---|---|---|---|---|---|---|---|---|---|---|
| MRK 290 | 0.030 | 1218-1232 | 1034-1046 | 0.002-0.013 | 19 | ... | 16-35 | 3.86 4.80 | 2.11 | 2009.10.28 | A1 |
| PKS 2155-304 | 0.117 | 1217-1335 | 1033-1133 | 0.002-0.098 | 31 | 12 | 70-110 | 4.56 ... | 2.17.3 | 2012.7.23 2012.7.28 | A2 |
| MRK 876 | 0.129 | 1218-1350 | 1034-1146 | 0.002-0.110 | 34 | 11 | 90-110 | 12.58 11.82 | 2.13.6 | 2010.4.8 2010.4.10 | A3 |
| 3C 273 | 0.157 | 1218-1383 | 1034-1174 | 0.002-0.138 | 27 | 27 | 110-155 | 4.00 ... | 2.17.3 | 2012.4.22 | A4 |
| PG 1116+215 | 0.177 | 1218-1407 | 1034-1194 | 0.002-0.157 | 25 | 12 | 68-90 | 4.68 5.53 | 2.13.6 | 2011.10.18 | A5 |
| PHL 1811 | 0.201[e] | 1218-1436 | 1034-1219 | 0.002-0.181 | 17 | 12 | 51-85 | 3.48 3.07 | 2.15.6 | 2010.10.29 | A6 |
| PG 0953+414 | 0.239 | 1217-1481 | 1033-1257 | 0.002-0.218 | 24 | 10 | 45-95 | 4.79 5.64 | 2.15.6 | 2011.10.18 | A7 |
| H 1821+643 | 0.297 | 1217-1550 | 1033-1316 | 0.002-0.275 | 29 | 13 | 33-90 | 12.0 0.53 | 2.17.3 | 2012.7.6 | A8 |
| TON 236 | 0.447 | 1218-1730 | 1135-1468 | 0.100-0.423 | 1[f] | ... | 17-41 | 6.56 9.39 | 2.15.6 | 2011.9.17 | A9 |
| HE 0153-4520 | 0.449 | 1218-1732 | 1135-1470 | 0.100-0.425 | 2[f] | ... | 27-58 | 5.23 5.89 | 2.15.6 | 2009.12.19 | A10 |
| HE 0226-4110 | 0.493 | 1218-1785 | 1034-1515 | 0.002-0.468 | 25 | 9 | 26-60 | 6.63 7.80 | 2.13.6 | 2010.2.3 | A11 |
| PKS 0405-123 | 0.573 | 1219-1824[g] | 1046-1548 | 0.014-0.500 | 19 | 7 | 55-70 | 22.17 11.06 | 2.17.3 | 2009.8.31[h] 2009.10.27-28 2009.12.21 | A12 |
| HE 0238-1904 | 0.631 | 1219-1795[i] | 1135-1523 | 0.100-0.476 | 13 | ... | 30-55 | 6.45 7.49 | 2.17.3 | 2009.12.31 2011.12.5 | A13 |
| 3C 263 | 0.652 | 1218-1795[i] | 1143-1523 | 0.108-0.476 | 14 | ... | 21-71 | 15.36 18.00 | 2.13.6 | 2010.01.01 | A14 |

Notes

[a] The previous analysis of the O VI systems in the bright QSOs listed above based on essentially the same set of FUSE and STIS observations have been presented in numerous investigations as listed at the end of the footnotes.

[b] $z_{em}$ is measured from the Ly-α emission line of the QSO.  The region of 5000 km s$^{-1}$ blueward of the QSO's Ly-α emission line is excluded due to the proximity effect (see TSB2008).

[c] The S/N per resolution element is listed.  For FUSE we list the S/N at 1030 Å per 25 km s$^{-1}$ while the S/N in the wavelength range from 1030Å -1187 Å is typically the same to within 20%.  The S/N for λ < 1000 Å is ~2 to 3 times smaller.  For STIS we list the S/N at 1250 Å per 7 km s$^{-1}$.  For COS we list the range of the S/N per 18 km s$^{-1}$ over the wavelength range from 1150 to 1750Å.

[d] The numbers for the tables and sections in the appendix where the absorption results for each QSO are reported.

[e] The redshift is uncertain due to the presence of the associated systems.

[f] The S/N is so low the observations are not useful for this program.

[g] The system at PKS 0405-123 O VI system at z = 0.49507 was included in the survey even though Lyα is redshifted off the wavelength range covered by the G160M grating.

[h] Only part of these early observations whose spectral shape is consistent with the spectra obtained later are included in the co-added observations.

[i] For λ > 1795Å H I λ1215  absorption redshifted off the end of the COS G160M grating.

Table 2. Basic Properties of the O VI Systems Observed at High S/N

| QSO | z | other metals | b(km s⁻¹) H I | # H I | # O VI | $\Delta v^a$ H I | \|Δv\| H I–O VI min. | max log N H I | max log N O VI | H I & O VI aligned[b] |
|---|---|---|---|---|---|---|---|---|---|---|
| MRK 290 | 0.01026ᶜ | N | 53, 32 | 2 | 1 | 113 | 3±7 | 14.38 | 13.65 | Y |
| PKS 2155-304 | 0.05423 | N | 28, 39, 21, 23 | 4 | 1 | 164 | 34±27 | 13.68 | 13.53 | N |
| PKS 2155-304 | 0.05722 | N | 50, 68, 18 | 3 | 1 | 151 | 25±4 | 14.32 | 13.65 | N |
| MRK 876 | 0.00315 | N | 58, 25 | 2 | 1 | 48 | 14±9 | 14.00 | 13.38 | Y |
| MRK 876 | 0.01171 | N | 22, 48 | 2 | 1 | 4 | 26±15 | 13.86 | 13.26 | N |
| 3C 273 | 0.00336 | N | 64, 31 | 2 | 1 | 7 | 3±7 | 13.54 | 13.39 | Y |
| 3C273 | 0.09022 | N | 48, 33 | 2 | 1 | 92 | 30±5 | 13.28 | 13.11 | N |
| 3C 273 | 0.12007 | N | 23 | 1 | 1 | 0 | 7±2 | 13.50 | 13.38 | Y |
| PG 1116+215 | 0.05897 | Y | 22, 33, 25 | 3 | 2 | 172 | 3±12, 24±4 | 13.57 | 13.49 | Y, N |
| PG 1116+215 | 0.13850 | Y | 14, 19, 46 | 3 | 1 | 14 | 0±1 | 15.95 | 13.81 | Y |
| PG 1116+215 | 0.16553 | N | 43, 34, 100, 17, 14, 43, 21 | 7 | 4 | 548 | 18±7, 26±14, 12±2, 4±6 | 14.53 | 13.93 | N, N, Y, Y |
| PHL 1811 | 0.07773 | Y | 82, 19 | 2 | 1 | 5 | 3±10 | 15.80 | 13.53 | Y |
| PHL 1811 | 0.13280 | N | 29, 81, 21, 49 | 4 | 4 | 207 | 0±2, 36±24, 1±4, 11±15:ᵈ | 14.67 | 13.74 | Y,N,Y, Nᵈ |
| PHL 1811 | 0.15785 | N | 97 | 1 | 2 | 0 | 80±8, 7±7 | 13.25 | 13.68 | N, Y |
| PHL 1811 | 0.17651 | Y | 26, 22 | 2 | 1 | 46 | 15±2 | 14.88 | 14.14 | N |
| PG 0953+414 | 0.00212 | N | 38 | 1 | 1 | 0 | 1±6 | 13.15 | 13.60 | Y |
| PG 0953+414 | 0.06808 | Y | 19, 37 | 2 | 1 | 34 | 2±5 | 14.34 | 14.29 | Y |
| PG 0953+414 | 0.14231 | Y | 20, 26, 29, 58 | 4 | 2 | 349 | 0±2, 4±3 | 13.57 | 14.09 | Y, Y |
| H 1821+643 | 0.02443 | N | 29 | 1 | 1 | 0 | 12±5 | 14.24 | 13.42 | Y |
| H 1821+643 | 0.12141 | N | 42, 82, 44 | 3 | 1 | 269 | 16±11 | 14.23 | 13.69 | Y |
| H 1821+643 | 0.17036 | N | 33, 54, 30 | 3 | 2 | 212 | 8±8, 33±15 | 13.86 | 13.69 | Y, N |
| H 1821+643 | 0.21329 | Y | 40 | 1 | 1 | 0 | 6±2 | 14.42 | 13.49 | Y |
| H 1821+643 | 0.22497ᵉ | Y | 35, 26, 95, 18, 58 | 5 | 3 | 396 | 23±17, 32±2, 50±3 | 15.28 | 14.25 | Y, N, N |
| H 1821+643 | 0.24531 | N | 36 | 1 | 1 | 0 | 1±5 | 13.08 | 13.76 | Y |
| H 1821+643 | 0.26656 | N | 46 | 1 | 1 | 0 | 4±3 | 13.64 | 13.61 | Y |
| TON 236 | 0.19452 | N | 49,16 | 2 | 1 | 82 | 0±4 | 14.05 | 14.03 | Y |
| TON 236 | 0.39944 | N | 45 | 1 | 1 | 0 | 16±21 | 14.10 | 13.74 | Y |
| HE 0153-4520 | 0.14887 | N | 35, 34 | 2 | 2 | 91 | 11±4, 11±3 | 13.34 | 14.02 | Y, Y |
| HE 0153-4520 | 0.17090 | N | 108, 34 | 2 | 2 | 32 | 56±3, 8±2 | 14.35 | 13.82 | N,Y |
| HE 0153-4520 | 0.22203 | N | 33, 53 | 2 | 2 | 12 | 19±2, 68±2 | 14.95 | 13.71 | N, N |
| HE 0153-4520 | 0.22600ᶠ | N | 28, 151 | 2 | 1 | 6 | 6±4 | 16.61 | 14.23 | Y |
| HE 0153-4520 | 0.29114 | N | 61, 20 | 2 | 2 | 28 | 24±4, 1±2 | 14.22 | 13.85 | N, Y |
| HE 0153-4520 | 0.40052 | Y | 42, 19, 24 | 3 | 3 | 129 | 1±6, 9±2ᵍ, 3±6 | 14.27 | 14.09 | Y, Nᵍ, Y |
| HE 0226-4110 | 0.01749 | N | 27 | 1 | 1 | 0 | 8±3 | 13.20 | 13.91 | Y |
| HE 0226-4110 | 0.20701ʰ | Y | 21, 100, 28 | 3 | 1 | 36 | 16±14 | 15.13 | 14.37 | Y |
| HE 0226-4110 | 0.22005 | N | 29 | 1 | 1 | 0 | 2±4 | 14.39 | 13.14 | Y |
| HE 0226-4110 | 0.34034 | Y | 52, 9, 85, 19 | 4 | 2 | 209 | 25±16, 5±3 | 13.47 | 13.90 | N, Y |
| HE 0226-4110 | 0.35523 | Y | 45, 16, 34 | 3 | 1 | 153 | 1±2 | 13.70 | 13.33 | Y |
| PKS 0405-123 | 0.09192 | Y | 36, 23 | 2 | 1 | 71 | 21±5 | 14.57 | 13.85 | N |
| PKS 0405-123 | 0.09657 | N | 29, 69 | 2 | 1 | 11 | 2±4 | 14.58 | 13.70 | Y |
| PKS 0405-123 | 0.16716ⁱ | Y | 54, 12, 27, 16, 26, 41 | 6 | 4 | 212 | 111±19, 1±2, 24±3, 1±3 | 16.45 | 14.45 | N, Y, N, Y |
| PKS 0405-123 | 0.18293 | Y | 18, 31, 32, 21 | 4 | 2 | 280 | 7±2, 10±2 | 14.71 | 13.84 | Y, Y |
| PKS 0405-123 | 0.29770 | N | 20, 61 | 2 | 1 | 22 | 9±3 | 13.82 | 13.60 | Y |
| PKS 0405-123 | 0.36160 | Y | 58, 29, 15, 32, 40 | 5 | 6 | 185 | 120±57, 38±57, 3±4, 27±10, 4±12, 25±12 | 14.96 | 13.69 | N, N, Y, N, Y, N |
| PKS 0405-123 | 0.36335 | Y | 26 | 1 | 1 | 0 | 11±2 | 13.22 | 13.48 | Y |
| PKS 0405-123 | 0.49507ʲ | Y | 51, 40 | 2 | 2 | 99 | 4±4, 46±4 | 14.14 | 14.31 | Y, N |
| HE 0238-1904 | 0.40107 | N | 31 | 1 | 1 | 0 | 1±2 | 14.93 | 13.47 | Y |
| HE 0238-1904 | 0.42430 | Y | 35, 24 | 2 | 2 | 70 | 15±2, 0±2 | 14.72 | 14.36 | N, Y |
| HE 0238-1904 | 0.47204 | N | 38, 47 | 2 | 2 | 148 | 14±4, 40±4 | 13.83 | 14.16 | Y, N |
| 3C 263 | 0.06342ᵏ | Y | 28, 45, 19, 21, 13 | 5 | 1 | 247 | 9±4 | 15.15 | 14.59 | Y |
| 3C 263 | 0.11389 | Y | 47, 71, 20 | 3 | 1 | 220 | 1±4 | 14.05 | 13.66 | Y |
| 3C 263 | 0.14072ᵏ | N | 28, 87 | 2 | 1 | 12 | 6±9 | 14.51 | 13.60 | Y |
| 3C 263 | 0.32567ⁱ | Y | 24, 11 | 2 | 1 | 34 | 17±2 | 15.23 | 14.02 | N |
| 3C 263 | 0.44672 | Y | 71, 32 | 2 | 1 | 44 | 9±3 | 14.17 | 13.87 | Y |

Notes

ᵃ Maximum velocity offset of the H I absorption components. When only one H I component is detected, Δv = 0 is listed.

ᵇ O IV absorbers are considered to be aligned with H I if |v(O VI) – v(H I)| $\leq$ 10 km s⁻¹ within the 1σ error.

[c] The results for the system at z = 0.01026 toward MRK 290 are from Narayanan et al. (2010b) and updated in this paper.

[d] The errors for the properties of the BLA are so uncertain we do not classify the O VI in this absorber as aligned.

[e] The results for the system at z = 0.22497 toward H 1821+643 are mostly from Narayanan et al. (2010a) with updates from this paper.

[f] The results for the system at z = 0.22600 toward HE 0153-4520 are from Savage et al. (2011b) and updated in this paper.

[g] Although the O VI component at -54±1 km s$^{-1}$ is formally aligned with the H I at -63±1 km s$^{-1}$, the O VI absorber at v = -64±6 km s$^{-1}$ is more likely associated with the H I absorption. We therefore do not consider this O VI component as aligned with H I.

[h] The results for the system at z = 0.20701 toward HE 0226-4110 are from Savage et al. (2005, 2011a) and updated in this paper.

[i] The results for the system at z = 0.16716 toward PKS 0405-123 are from Savage et al. (2010) and updated in this paper.

[j] The results for the system at z = 0.49507 toward PKS 0405-123 are from Narayanan et al. (2011) and updated in this paper.

[k] The results for the system at z = 0.06342 and 0.14072 toward 3C 263 are from Savage et al. (2012).

[l] The results for the system at z = 0.32567 toward 3C 263 are from Narayanan et al. (2009, 2012) as updated in this paper.

Table 3. Summary of the Properties of the O VI Systems

| | |
|---|---|
| O VI redshift path | 3.52 |
| Number of O VI systems | 54 |
| Number of H I components | 133 |
| Number of O VI components | 85 |
| Number of O VI components / number O VI systems | 1.57 |
| Number of H I components / number O VI systems | 2.46 |
| Redshift range for the O VI systems | 0.01 to 0.50 |
| Median redshift for the O VI systems | 0.17 |
| dn/dz for O VI systems (O IV components) with $W_r > 30$ mÅ | $17 \pm 2$ $(24 \pm 3)$ |
| Number of O VI systems with no other metal lines detected | 26 (48%) |
| Number of O VI systems with other metal lines detected | 28 (52%) |
| Systems with 1 H I and 1 OVI components | 11 (20%) |
| Systems with 1 H I and 2 O VI components | 1 (2%) |
| Systems with 2 H I and 1 OVI components | 16 (30%) |
| Systems with 3 H I and 1 OVI components | 6 (11%) |
| Systems with 2 H I and 2 OVI components | 7 (13%) |
| Systems with 3 H I and 2 OVI components | 2 (4%) |
| Systems more than 2 O VI components | 5 (10%) |
| O VI component range for log N(O VI) | 13.00 to 14.59 |
| Median component value of log N(O VI) | 13.68 |
| H I component range for log N(H I) | 12.45 to 16.61 |
| Median component value of log N(H I) | 13.64 |
| O VI component range for b(O VI) | 5 to 79 |
| Median component value for b(O VI) | 27 |
| H I component range for b(H I) | 11 to 151 |
| Median component value for b(H I) | 32 |
| | |
| Number of aligned O VI & H I components[a] | 54 |
| Number of temperature estimates in aligned components | 45 |
| Number with log T < 4.8 implying PI in cool absorbers | 31 |
| Number with log T > 5 implying warm absorbers | 14 |
| Range of log T for the warm absorbers | 4.99 to 6.14 |
| Median value of log T for the warm absorbers. | 5.24 |
| Range of log N(H) for warm absorbers with log T > 5.0 | 18.38 to 20.38 |
| Median of log N(H) for warm absorbers with log T > 5.0 | 19.35 |
| {log [N(H)/N(H I)]}$_{MEDIAN}$ for warm absorbers with log T > 5.0 | 5.7 |
| Range of [O/H] for the warm absorbers with log T > 5.4 | -1.93 to 0.03 |
| [O/H]$_{MEDIAN}$ for warm CI absorbers with log T > 5.4 | -1.03 |
| Range of $b_{NT}$ for PI absorbers (km s$^{-1}$) | 5 to 55 |
| Median $b_{NT}$ for PI absorbers (km s$^{-1}$) | 23 |
| Range of $b_{NT}$ for CI absorbers (km s$^{-1}$) | <10 to 56 |
| Median $b_{NT}$ for CI absorbers (km s$^{-1}$) | 29 |

[a] Aligned H I and O VI components are defined to have |v(O VI) – v(H I)| $\leq$ 10 km s$^{-1}$ within the 1$\sigma$ error.

Table 4. Properties of components with aligned H I and O VI absorption observed at high S/N[a]

| QSO | z (v comp) | log N(O VI) | b(O VI) | log N(H I) | b(H I) | log T | $b_{NT}$ | log N(H)[b] | [O/H][c] | Origin[d] |
|---|---|---|---|---|---|---|---|---|---|---|
| MRK 290 | 0.01026 (0) | 13.65±0.10 | 31±10 | 14.38±0.01 | 53±2 | 5.07 (+0.13, -0.19) | 29±11 | 19.33 (+0.33, -0.56) | … | CI/PI |
| MRK 876 | 0.00315 (0) | 13.38±0.12 | 35±13: | 14.00±0.02 | 58±2 | 5.14 (+0.16, -0.25) | 33±15 | 19.12 (+0.36, -0.83) | … | CI/PI |
| 3C 273 | 0.00336 (0) | 13.39±0.08 | 43±11 | 13.54±0.10 | 64±7 | 5.16 (+0.20, -0.38) | 41±12 | 18.72 (+0.52, -1.17) | … | CI/PI |
| 3C 273 | 0.12007 (0) | 13.38±0.02 | 10±2 | 13.50±0.01 | 23±1 | 4.44 (+0.06, -0.07) | 8±3 | … | … | PI |
| PG 1116+215 | 0.05897 (0) | 13.49±0.05 | 27±16 | 13.57±0.01 | 33±2 | 4.36 (+0.54, - 4.36) | 27±17 | … | … | PI |
| PG 1116+215 | 0.13850 (0) | 13.81±0.01 | 36±2 | 14.18±0.05 | 46±2 | 4.72 (+0.11, -0.15) | 35±2 | … | … | PI/CI |
| PG 1116+215 | 0.16553 (0) | 13.93±0.09 | 19±2 | 13.39±0.03 | 34±2 | 4.71 (+0.07, -0.10) | 18±2 | … | … | PI/CI |
| PG 1116+215 | 0.16553(178) | 13.38±0.04: | 51±6: | 13.38±0.10 | 14±2 | b(H I) < b(O VI) | … | … | … | UNK |
| PHL 1811 | 0.07773 (0) | 13.53±0.08 | 43±7 | 13.60±0.07 | 82±10 | 5.49 (+0.13, -0.19) | 39±8 | 19.47 (+0.23, -0.37) | -1.93 (+1.32, -0.11) | CI |
| PHL 1811 | 0.13280 (-133) | 13.74±0.04 | 30±3 | 14.67±0.11 | 29±4 | < 4.36 | <33 | … | … | PI |
| PHL 1811 | 0.13280 (0) | 13.68±0.03 | 23±3 | 12.82±0.21 | 21±8 | < 4.45 | <26 | … | … | PI |
| PHL 1811 | 0.15785 (0) | 13.68±0.02 | 31±3 | 13.25±0.04 | 97±10 | 5.73 (+0.09, -0.11) | 20±6 | 19.51(+0.13, -0.18) | -0.90 (+0.20, -0.32) | CI |
| PG 0953+414 | 0.00212 (0) | 13.60±0.06 | 43±8 | 13.15±0.03 | 38±3 | < 4.47 | <44 | … | … | PI |
| PG 0953+414 | 0.06808 (0) | 14.29±0.03 | 12±1 | 14.34±0.05 | 19±1 | 4.14 (+0.08, -0.10) | 11±1 | … | … | PI |
| PG 0953+414 | 0.14231 (0) | 14.09±0.01 | 19±1 | 13.57±0.01 | 26±1 | 4.31 (+0.08, -0.10) | 18±1 | … | … | PI |
| PG 0953+414 | 0.14231 (78) | 13.60±0.03 | 29±4 | 13.47±0.01 | 29±1 | < 4.25 | <31 | … | … | PI |
| H 1821+643 | 0.02443 (0) | 13.42±0.09 | 23±7 | 14.24±0.01 | 29±1 | 4.30 (+0.31, -4.30) | 23±8 | … | … | PI |
| H 1821+643 | 0.12141 (0) | 13.69±0.04 | 58±7 | 13.80±0.05 | 82±7 | 5.33 (+0.15, -0.24) | 56±8 | 19.36 (+0.29, -0.54) | … | CI/PI |
| H 1821+643 | 0.17036 (-95) | 13.20±0.22 | 31±12 | 13.65±0.01 | 54±2 | 5.10 (+0.14, -0.22) | 29±14 | 18.67 (+0.35, -0.65) | … | CI/PI |
| H 1821+643 | 0.21329 (0) | 13.49±0.02 | 26±2 | 14.42±0.01 | 40±1 | 4.77 (+0.06, -0.07) | 25±2 | … | … | PI/CI |
| H1821+643 | 0.22479 (0) | 14.25±0.01 | 45±1 | 13.94±0.17 | 95±11 | 5.65 (+0.11, -0.15) | 39±2 | 20.07 (+0.18, -0.25) | -1.17 (+0.34, -0.37) | CI |
| H 1821+643 | 0.24531 (0) | 13.76±0.01 | 27±1 | 13.08±0.05 | 36±5 | 4.56 (+0.22, -0.45) | 26±1 | … | … | PI |
| H 1821+643 | 0.26656 (0) | 13.61±0.03 | 24±2 | 13.64±0.02 | 46±2 | 4.99 (+0.05, -0.06) | 22±2 | 18.38 (+0.15, -0.19) | … | CI/PI |
| TON 236 | 0.19452 (0) | 14.03±0.03 | 44±5 | 14.05±0.01 | 49±1 | 4.47 (+0.29, -1.51) | 44±5 | … | … | PI |
| TON 236 | 0.39944 (0) | 13.74±0.03 | 61±18 | 14.10±0.01 | 45±1 | b(H I) < b(O VI) | … | … | … | UNK |
| HE 0153-4520 | 0.14887 (-113) | 13.82±0.02 | 50±3 | 13.34±0.03 | 35±3 | b(H I) < b(O VI) | … | … | … | UNK |
| HE 0153-4520 | 0.14887 (0) | 14.02±0.01 | 25±1 | 13.26±0.03 | 34±4 | 4.53 (+0.18, -0.32) | 24±1 | … | … | PI |
| HE 0153-4520 | 0.17090 (0) | 13.82±0.01 | 22±2 | 14.35±0.01 | 34±1 | 4.60 (+0.07, -0.09) | 22±2 | … | … | PI |
| HE 0153-4520 | 0.22600 (0) | 14.23±0.01 | 37±1 | 13.58±0.05 | 151±15 | 6.14(+0.08, -0.10) | <10 | 20.38 (+0.10, -0.13) | 0.03 (+0.10, -0.10) | CI |
| HE 0153-4520 | 0.29114 (0) | 13.85±0.03 | 19±2 | 14.22±0.02 | 20±1 | < 3.99 | <22 | … | … | PI |
| HE 0153-4520 | 0.40052 (-64) | 14.09±0.06 | 42±3 | 14.25-±0.01 | 42±1 | < 4.32 | <44 | … | … | PI |
| HE 0153-4520 | 0.40052 (0) | 13.53±0.12 | 24±4 | 14.27±0.01 | 19±1 | b(H I) < b(O VI) | … | … | … | UNK |
| HE 0226-4110 | 0.01749 (0) | 13.91±0.16 | 10±6 | 13.20±0.04 | 27±4 | 4.61 (+0.14, -0.22) | 8±8 | … | … | PI |
| HE 0226-4110 | 0.20701 (0) | 14.37±0.01 | 36±1 | 13.45±0.16 | 100±25 | 5.75 (+0.20, -0.37) | 27±6 | 19.73(+0.27, -0.63) | -0.31(+0.36, -0.10) | CI |
| HE 0226-4110 | 0.22005 (0) | 13.14±0.06 | 14±5 | 14.39±0.01 | 29±1 | 4.62 (+0.09, -0.12) | 12±6 | … | … | PI |
| HE 0226-4110 | 0.34034 (0) | 13.90±0.03 | 22±2 | 13.47±0.05 | 19±3 | < 3.73[e] | <22 | … | … | PI[d] |
| HE 0226-4110 | 0.35523 (0) | 13.33±0.03 | 10±2 | 13.70±0.02 | 34±2 | 4.83 (+0.05, -0.06) | 5±4 | … | … | PI/CI |
| PKS 0405-123 | 0.09658 (0) | 13.70±0.07 | 23±6 | 14.58±0.02 | 29±1 | 4.30 (+0.28, -1.02)[f] | 23±7 | … | … | PI[e] |
| PKS 0405-123 | 0.16716 (-135) | 13.89±0.04 | 29±2 | 13.35±0.05 | 12±1 | b(H I) < b(O VI) | … | … | … | UNK |
| PKS 0405-123 | 0.16716 (0) | 14.45±0.03 | 39±2 | 16.45±0.02 | 26±1 | b(H I) < b(O VI) | … | … | … | UNK |
| PKS 0405-123 | 0.18293(-71) | 13.64±0.02 | 30±2 | 14.71±0.01 | 31±1 | 3.59 (+0.51, -3.59) | 30±2 | … | … | PI |
| PKS 0405-123 | 0.18293 (0) | 13.84±0.01 | 18±1 | 14.06±0.02 | 32±1 | 4.65 (+0.04, -0.05) | 17±1 | … | … | PI |
| PKS 0405-123 | 0.29770 (0) | 13.60±0.02 | 55±3 | 13.82±0.02 | 61±2 | 4.65 (+0.20, -0.39) | 55±3 | … | … | PI |
| PKS 0405-123 | 0.36160 (-29) | 13.22±0.04 | 5±1 | 13.55±0.06 | 40±3 | b(H I)/b(O VI) = 8 | ... | ... | ... | UNK |
| PKS 0405-123 | 0.36335 (0) | 13.48±0.01 | 11±1 | 13.22±0.03 | 26±3 | 4.55 (+0.11, -0.15) | 9±1 | ... | ... | PI |
| PKS 0405-123 | 0.49507 (0) | 14.31±0.02 | 32±2 | 14.14±0.03 | 51±5 | 5.00 (+0.12, -0.18) | 30±2 | 18.91 (+0.34, -0.58) | ... | CI/PI |
| HE 0238-1904 | 0.40107 (0) | 13.47±0.03 | 22±2 | 14.93±0.01 | 31±1 | 4.49 (+0.09, -0.11) | 21±2 | ... | ... | PI |
| HE 0238-1904 | 0.42430 (0) | 14.16±0.02 | 23±1 | 14.57±0.01 | 24±1 | 3.48 (0.38, -3.48) | 23±1 | ... | ... | PI |
| HE 0238-1904 | 0.47024 (0) | 14.16±0.01 | 20±1 | 13.83±0.03 | 47±4 | 5.06 (+0.08, -0.10) | 17±1 | 18.77 (+0.21, -0.30) | ... | CI/PI |
| 3C 263 | 0.06342(0) | 14.59±0.07 | 39±6 | 14.30±0.36 | 19±17 | b(H I) < b(O VI) | ... | ... | ... | UNK |
| 3C 263 | 0.11389 (0) | 13.66±0.05 | 31±5 | 13.85±0.02 | 20±1 | b(H I) < b(O VI) | ... | ... | ... | UNK |
| 3C 263 | 0.14072 (0) | 13.60±0.09 | 33±12 | 13.47±0.10 | 87±15 | 5.62 (+0.15, -0.24) | 26±17 | 19.55(+0.23, -0.42) | -1.45 (+0.41, -0.14) | CI |
| 3C 263 | 0.44672 (0) | 13.87±0.03 | 32±3 | 14.17±0.03 | 32±1 | < 4.21 | <34 | ... | ... | PI |

Notes:

[a] Results are listed for components where H I and O VI are aligned in velocity to |v(O VI) – v(H I)| $\leq$ 10 km s$^{-1}$ within the 1σ error.

[b] Values of log N(H) are listed when log T $\geq$ 5 implying the plasma is warm.

[c] The oxygen abundance, [O/H], in the warm plasma can be reliably estimated when log T > 5.5 where the effects of non-equilibrium ionization of oxygen are relatively unimportant.

[d] The most likely ionization mechanism is listed based on the inferred temperature of the plasma. When the errors on the inferred temperature span the range consistent with either collisional ionization (CI) for log T > 4.7 or photoionization (PI) for log T < 5.3 both ionization mechanisms are listed. UNK for unknown is listed when it is not possible to estimate the temperature of the plasma or determine the origin of the O VI ionization. Most of these cases are for the situation where b(O VI) exceeds b(H I) implying it is unlikely the O VI and H I co-exist in the same plasma even though the absorbers are aligned in velocity.

[e] We list the results for the HE 0226-4110 z = 0.34034 absorber assuming the O VI exists in the narrow H I component with b = 19±3 km s$^{-1}$. If the O VI is instead associated with the BLA with b = 85±19 km s$^{-1}$, the difference in line widths implies log T = 5.64 (+0.17, -0.28) and an origin in collisionally ionized gas with log N(H) = 19.49 (+0.26, -0.51) and [O/H] = -1.02(+1.49, -0.10).

[f] We list the results here for the PKS 0405-123 z = 0.09658 absorber assuming the O VI exists in the narrow H I component with b = 29±1 km s$^{-1}$. However, if the O VI instead is associated with the BLA having b = 69±3 km s$^{-1}$, we obtain log T = 5.48±0.05 and b$_{NT}$ = 16±9 km s$^{-1}$ and an origin of the O VI is in warm collisionally ionized gas with log N(H) = 19.69 (+0.08, -0.11) and [O/H] = -1.98 (+0.11, -0.13) as discussed in the Appendix.

Table 5. Warm O VI Systems With log T ≥ 5.0 Ordered by Approximate Quality[a]

| QSO | z (v comp) | log T | log N(H) | [O/H] | Ioniz[b] |
|---|---|---|---|---|---|
| HE 0153-4520 | 0.22600 (0) | 6.14(+0.08, -0.10) | 20.38 (+0.10, -0.13) | 0.03 (+0.10, -0.10) | CI |
| MRK 290 | 0.01026 (0) | 5.07 (+0.13, -0.19) | 19.33 (+0.33, -0.56) | ... | CI/PI |
| 3C 263 | 0.14072 (0) | 5.62 (+0.15, -0.24) | 19.55(+0.23, -0.42) | -1.45 (+0.41, - 0.14) | CI |
| PKS 0405-123 | 0.49507 (0) | 5.00 (+0.12, -0.18) | 18.91 (+0.34, -0.58) | ... | CI/PI |
| H 1821+643 | 0.26656 (0) | 4.99 (+0.05, -0.06) | 18.38 (+0.15, -0.19) | ... | CI/PI |
| MRK 876 | 0.00315 (0) | 5.14 (+0.16, -0.25) | 19.12 (+0.36, -0.83) | ... | CI/PI |
| 3C 273 | 0.00336 (0) | 5.16 (+0.20, -0.38) | 18.72 (+0.52, -1.17) | ... | CI/PI |
| PHL 1811 | 0.15785 (0) | 5.73 (+0.09, -0.11) | 19.51(+0.13, -0.18) | -0.90 (+0.20, -0.32) | CI |
| HE 0226-4110 | 0.20701 (0) | 5.75 (+0.20, -0.37) | 19.73(+0.27, -0.63) | -0.31(+0.36, -0.10) | CI |
| H 1821+643 | 0.22479 (0) | 5.65 (+0.11, -0.15) | 20.07 (+0.18, -0.25) | -1.17 (+0.34, -0.37) | CI |
| H 1821+643 | 0.12141 (0) | 5.33 (+0.15, -0.24) | 19.36 (+0.29, -0.54) | ... | CI/PI |
| PHL 1811 | 0.07773 (0) | 5.49 (+0.13, -0.19) | 19.47 (+0.23, -0.37) | -1.93 (+1.32, -0.11) | CI |
| H 1821+643 | 0.17036 (-95) | 5.10 (+0.14, -0.22) | 18.67 (+0.35, -0.65) | ... | CI/PI |
| HE 0238-1904 | 0.47024 (0) | 5.06 (+0.08, -0.10) | 18.77 (+0.21, -0.30) | ... | CI/PI |
| Possible Warm O VI Systems[c] | | | | | |
| PKS 0405-123 | 0.09658 (0) | 5.43±0.05 | 19.69 (+0.08, -0.11) | -1.98 (+0.11, -0.13) | CI |
| 3C 263 | 0.32567 (0) | 5.61 (+0.07, -0.06) | 19.31 (+0.11, -0.13) | -1.21 (+0.20, -0.18) | CI |
| HE 0226-4110 | 0.34034 (0) | 5.64 (+0.17, -0.28) | 19.49 (+0.26, -0.51) | -1.02 (+1.49, -0.10) | CI |
| O VI Systems Possibly Tracing Delayed Recombination[d] | | | | | |
| PG 1116+215 | 0.16553 (0) | 4.71 (+0.07, -0.10) | ... | ... | PI/CI |
| PG 1116+215 | 0.13850 (0) | 4.72 (+0.11, -0.15) | ... | ... | PI/CI |
| H 1821+643 | 0.21329 (0) | 4.77 (+0.06, -0.07) | ... | ... | PI/CI |
| HE 0226-4110 | 0.35523 (0) | 4.83 (+0.05, -0.06) | ... | ... | PI/CI |

Notes

[a] We list the aligned warm systems from Table 4 ordered approximately by quality of the O VI and H I measurements. The systems are classified as warm when the observed value of log T ≥ 5.0.

[b] The origin of the ionization in the warm systems is likely to be dominated by CI when log T > 5.5 and a combination of CI and PI when log T ranges from 5.4 to 5 with the contribution from PI increasing as log T decreases. log N(H) can be estimated assuming CIE for all the warm systems. For log T from 5.4 to 5.0 the values will increase over those listed if the density is low enough for photoionization to contribute to the ionization. [O/H] can only reliably be estimated for the warm systems when log T > 5.4.

[c] Three possible warm systems are listed where the O VI is aligned with both narrow and broad H I absorption. We list here the properties of these systems assuming the O VI coexists in the gas traced by the BLA. Table 4 provided the alternate explanation that the O VI coexists with the narrower H I absorption and arises in cool photoionized gas. The detailed discussions of the alternate explanations listed here are found in the appendix. Both interpretations are equally possible. However, in the various plots and counting statistics these systems are considered to arise in cool photoionized gas.

[d] These systems have relatively high inferred temperatures. While it is possible the ionization is from photoionization in a low density plasma, the temperature is relatively high for photoionization heating. Therefore, it is possible that these systems are instead tracing the delayed recombination of O VI in gas that originally was collisionally ionized as discussed in Section 6.2. In the various plots and in the counting statistics of cool and warm absorbers, these systems are considered cool photoionized absorbers even though delayed recombination is also a distinct possibility.

Table 6. Galaxies Associated with the O VI Absorbers Observed at high S/N[a]

| AGN | $z_{abs}$ | $z_{gal}$ | total log N(H I) | total log N(O VI) | Name or RA(J2000)& DEC(2000) | Type[b] | $\Delta v$[c] | $L_{gal}$ (L*) | $L_{lim}$[d] (L*) | $\rho$ (kpc) | Ref.[e] |
|---|---|---|---|---|---|---|---|---|---|---|---|
| Mrk 290 | 0.01026 | 0.01032 | 14.56±0.01 | 13.65±0.10 | 15 35 14.3 +57 30 53.1 | E | 18 | 0.14 | 0.014 | 302 | 12 |
| Mrk 290 | " | 0.01000 | " | " | NGC 5987 | Sb | -78 | 1.3 | 0.014 | 434 | 1, 12 |
| Mrk 290 | " | 0.01104 | " | " | NGC 5965A | Sb | 234 | 0.16 | 0.014 | 499 | 12 |
| PKS 2155-304 | 0.05423 | 0.05404 | 14.02±0.05 | 13.53±0.06 | 21 58 23.8  -30 19 32.0 | ... | -54 | 3.0 | 0.087 | 544 | 3 |
| PKS 2155-304 | 0.05722 | 0.05704 | 14.59±0.01 | 13.65±0.04 | 21 58 40.7  -30 19 28.0 | A | -51 | 3.0 | 0.096 | 429 | 3 |
| MRK 876 | 0.00315 | 0.00304 | 14.08±0.02 | 13.38±0.12 | NGC 6140 | SBc | -34 | 0.23 | 0.018 | 188 | 1, 2, 12 |
| MRK 876 | " | 0.00333 | " | " | UGC 10369 | Scd? | 53 | 0.008 | 0.018 | 414 | 1, 2, 12 |
| MRK 876 | 0.01171 | 0.01173 | 14.01±0.05 | 13.26±0.23 | UGC 10294 | Sm? | 6 | 0.096 | 0.26 | 266 | 1 |
| 3C 273 | 0.00336 | 0.00302 | 14.32±0.03 | 13.39±0.08 | 12 28 15.9 +01 49 44.1 | E | -102 | 0.008 | 0.002 | 77 | 2, 13 |
| 3C 273 | " | 0.00433 | " | " | 12 27 46.5 +01 36 03.0 | ... | 290 | 0.016 | 0.002 | 141 | 2, 13 |
| 3C 273 | " | 0.00369 | " | " | 12 31 03.9 +01 40 34.4 | E | 97 | 0.022 | 0.002 | 155 | 9, 13 |
| 3C 273 | " | 0.00340 | " | " | 12 32 50.6 +02 47 50.3 | E | 13 | 0.008 | 0.002 | 300 | 9, 13 |
| 3C 273 | 0.09022 | 0.09004 | 13.56±0.02 | 13.11±0.10 | 12 28 51.9 +02 06 02.9 | A | -50 | 2.0 | 0.24 | 477 | 3 |
| 3C 273 | 0.12007 | 0.12020 | 13.50±0.01 | 13.38±0.02 | 12 28 17.8 +02 12 28.7 | A | 35 | 2.2 | 0.8 | 2009 | 9 |
| PG 1116+215 | 0.05897 | 0.0600 | 13.69±0.02 | 13.78±0.07 | 11 19 05.5 +21 17 33.3 | A | 292 | 0.10 | 0.053 | 124 | 3 |
| PG 1116+215 | " | 0.0594 | " | " | 11 19 03.1 +21 15 24 | A | 122 | 0.024 | 0.053 | 207 | 3 |
| PG 1116+215 | " | 0.05903 | " | " | 11 19 05.3 +21 15 37.7 | E | 16 | 1.3 | 0.053 | 258 | 3 |
| PG 1116+215 | 0.13850 | 0.13824 | 16.24±0.03 | 13.81±0.01 | 11 19 06.7 +21 18 28.7 | A | -70 | 3.4 | 0.33 | 139 | 2, 12 |
| PG 1116+215 | 0.16553 | 0.1660 | 14.68±0.02 | 14.25±0.12 | 11 19 12.2 +21 18 52.0 | A | 124 | 0.34 | 0.48 | 155 | 3 |
| PHL 1811 | 0.07773 | 0.07869 | 13.80±0.09 | 13.53±0.08 | 21 54 50.8  -09 22 33.2 | E | 267 | 0.15 | 0.095 | 234 | 12 |
| PHL 1811 | " | 0.07791 | " | " | 21 54 47.6  -09 22 53.9 | E | 50 | 0.48 | 0.095 | 308 | 12 |
| PHL 1811 | 0.13280 | 0.13253 | 14.71±0.11 | 14.24±0.04 | 21 55 06.5  -09 23 25.3 | E | 71 | 1.1 | 0.30 | 226 | 2, 3 |
| PHL 1811 | 0.15785 | 0.15820 | 13.25±0.04 | 13.80±0.02 | 21 55 05.1  -09 24 26.0 | E | 91 | 0.9 | (0.4) | 366 | 8 |
| PHL 1811 | 0.17651 | 0.17610 | 14.91±0.03 | 14.14±0.03 | 21 54 54.9  -09 23 31.0 | A | -104 | 7.5 | 0.6 | 351 | 2 |
| PG 0953+414 | 0.00212 | 0.00201 | 13.15±0.03 | 13.60±0.06 | NGC 3104 | E | -31 | 0.02 | 0.001 | 227 | 1 |
| PG 0953+414 | " | 0.00158 | " | " | 10 07 22.7  +38 58 10 | E | -163 | 0.00 | 0.001 | 483 | 1 |
| PG 0953+414 | 0.06808 | 0.06920 | 14.39±0.06 | 14.29±0.03 | 09 56 11.4 +41 14 50.3 | A | 315 | 1.9 | 0.072 | 607 | 6, 7 |
| PG 0953+414 | 0.14231 | 0.14326 | 13.94±0.01 | 14.21±0.02 | 09 56 40.5 +41 16 49.8 | E | 244 | 0.37 | 0.35 | 403 | 6, 7 |
| PG 0953+414 | " | 0.14266 | " | " | 09 56 38.9 +41 16 46.1 | Sb-Sc | 91 | 8.1 | 0.35 | 438 | 2, 6, 7 |
| H 1821+643 | 0.02443 | 0.02404 | 14.24±0.01 | 13.42±0.09 | 18 21 41.3 +63 51 37.0 | ... | -114 | 0.18 | 0.05 | 862 | 5 |
| H 1821+643 | 0.12141 | 0.12154 | 14.39±0.03 | 13.69±0.04 | 18 22 02.8 +64 21 39.0 | ... | 34 | 2.3 | (0.2) | 158 | 2, 12 |
| H 1821+643 | 0.17036 | 0.17086 | 14.15±0.01 | 13.81±0.07 | 18 21 36.7 +64 21 25.0 | ... | 128 | 2.4 | (0.4) | 417 | 5 |
| H 1821+643 | 0.21329 | ... | 14.42±0.01 | 13.49±0.02 | ... | ... | ... | < (0.6) | (0.6) | ... | 5 |
| H 1821+643 | 0.22497 | 0.22650 | 15.59±0.03 | 14.34±0.02 | 18 21 54.4 +64 20 09.3 | ... | 374 | 2.8 | (0.7) | 113 | 5 |
| H 1821+643 | " | 0.22650 | " | " | 18 21 38.9 +64 20 31.0 | ... | 374 | 2.3 | (0.7) | 435 | 5 |
| H 1821+643 | 0.24531 | 0.24435 | 13.08±0.05 | 13.76±0.01 | 18 21 56.3 +64 22 50.0 | ... | -231 | 1.9 | (0.8) | 521 | 5 |
| H 1821+643 | 0.26656 | 0.26669 | 13.64±0.02 | 13.61±0.03 | 18 22 10.2 +64 17 15.6 | Sc | 31 | 0.17 | (0.9) | 795 | 5 |
| HE 0226-4110 | 0.01783 | 0.01783 | 13.20±0.04 | 13.91±0.16 | NGC 954 | Sc | 101 | 3.0 | 0.16 | 596 | 1 |
| HE 0226-4110 | 0.20701 | 0.2065 | 15.27±0.07 | 14.37±0.01 | 02 28 14.5  -40 57 22.7 | E | -127 | 0.05 | 0.02 | 38 | 4 |
| HE 0226-4110 | " | 0.2078 | " | " | 02 28 13.5  -40 57 40.2 | E | 196 | 0.25 | 0.02 | 109 | 4 |
| HE 0226-4110 | " | 0.2077 | " | " | 02 28 10.5  -40 56 11.1 | A | 171 | 0.14 | 0.02 | 281 | 4 |
| HE 0226-4110 | 0.22005 | 0.2203 | 14.39±0.01 | 13.14±0.06 | 02 28 18.1  -40 52 34.2 | A | 61 | 0.20 | 0.06 | 999 | 4 |
| HE 0226-4110 | 0.34034 | 0.3416 | 13.88±0.06 | 13.91±0.02 | 02 28 19.0  -40 58 00.5 | E | 282 | 0.14 | 0.07 | 304 | 4 |
| HE 0226-4110 | 0.35523 | 0.3553 | 13.83±0.03 | 13.33±0.03 | 02 28 23.0  -40 57 18.8 | E | 15 | 0.17 | 0.08 | 437 | 4 |
| PKS 0405-123 | 0.09192 | 0.0923 | 14.58±0.01 | 13.85±0.07 | 04 07 49.4  -12 12 16 | E | 104 | 0.02 | 0.01 | 71 | 10 |
| PKS 0405-123 | " | 0.0908 | " | " | 04 07 43.2  -12 11 24 | A | -308 | 0.02 | 0.01 | 102 | 10 |
| PKS 0405-123 | " | 0.0914 | " | " | 04 07 43.2  -12 11 48 | A | -143 | 0.02 | 0.01 | 132 | 10 |
| PKS 0405-123 | " | 0.0917 | " | " | 04 07 40.2  -12 13 44 | A | -60 | 0.01 | 0.01 | 299 | 10 |
| PKS 0405-123 | " | 0.0908 | " | " | 04 07 54.4  -12 15 49 | E | -308 | 0.07 | 0.01 | 439 | 10 |
| PKS 0405-123 | 0.09658 | 0.0965 | 14.67±0.02 | 13.70±0.04 | 04 07 58.1  -12 12 24 | E | -22 | 0.15 | 0.01 | 269 | 10 |
| PKS 0405-123 | " | 0.0965 | " | " | 04 07 54.2  -12 14 45 | E | -22 | 0.15 | 0.01 | 371 | 10 |
| PKS 0405-123 | " | 0.0967 | " | " | 04 07 54.2  -12 14 50 | E | 33 | 1.2 | 0.01 | 380 | 10 |
| PKS 0405-123 | 0.16716 | 0.1669 | 16.49±0.03 | 14.78±0.04 | 04 07 48.4  -12 11 02 | E | -67 | 0.08 | 0.02 | 96 | 10 |
| PKS 0405-123 | " | 0.1672 | " | " | 04 07 51.2  -12 11 38 | E | 10 | 2.1 | 0.02 | 117 | 10 |
| PKS 0405-123 | 0.18293 | ... | 14.81±0.01 | 14.05±0.02 | ... | ... | ... | <0.04 | (0.04) | ... | 10, 14 |
| PKS 0405-123 | 0.29770 | 0.2978 | 13.98±0.03 | 13.60±0.02 | 04 07 50.6  -12 12 25 | E | -23 | 1.4 | 0.06 | 259 | 10 |
| PKS 0405-123 | 0.36160 | 0.3614 | 15.22±0.11 | 14.04±0.02 | 04 07 45.9  -12 11 09 | A | -88 | 4.5 | 0.08 | 236 | 10 |
| PKS 0405-123 | " | 0.3608 | " | " | 04 07 52.5  -12 11 56 | E | -176 | 0.08 | 0.08 | 316 | 10 |
| PKS 0405-123 | 0.36335 | 0.3617 | 13.22±0.03 | 13.48±0.01 | 04 07 45.0  -12 11 07 | ... | -363 | 6.5 | 0.1 | 237 | 12 |
| PKS 0405-213 | 0.49507 | 0.4942 | 14.26±0.03 | 14.44±0.03 | 04 07 49.1  -12 11 21 | E | -175 | 0.25 | 0.2 | 112 | 10 |
| 3C 263 | 0.06342 | 0.06322 | 15.45±0.10 | 14.59±0.07 | 175.02158 +65.80033 | E | -56 | 0.38 | 0.062 | 63 | 11 |
| 3C 263 | 0.11389 | 0.11370 | 14.33±0.02 | 13.66±0.05 | 174.98733 +65.74989 | E | -51 | 0.37 | 0.21 | 353 | 12 |
| 3C 263 | 0.14072 | 0.14087 | 14.55±0.03 | 13.60±0.09 | 175.03850 +65.73128 | A | 39 | 1.7 | 0.34 | 620 | 11, 12 |
| 3C 263 | 0.32567 | ... | 15.44±0.02 | 14.02±0.01 | ... | ... | ... | <2.2 | 2.2 | ... | 12 |
| 3C 263 | 0.44672 | ... | 14.30±0.03 | 13.87±0.03 | ... | ... | ... | <4.7 | 4.7 | ... | 12 |

[a] Galaxy redshift searches have not been performed for the fields surrounding TON 238, HE 0153-4520 and HE 0238-1904.

[b] Galaxy classification: E → emission line dominated, A → absorption line dominated.

[c] $\Delta v = v(galaxy) - v(absorber)$ in the rest fame of the absorber.

[d] Approximate estimates of the galaxy redshift survey luminosity limits are given. Values in parenthesis are very uncertain.

[e] References: (1) WS2009, (2) Stocke et al. (2013), (3) Prochaska et al. (2011a), (4) Chen & Mulchaey (2009), (5) Tripp et al. (1998), (6) Tripp et al. (2006), (7) Tripp & Savage (2000), (8) Jenkins et al. (2003),(9) Morris et al. (1993), (10) Johnson et al. (2013), (11) Savage et al. (2012), (12) Stocke et al. (2014, in prep) (13) The 3C 273 z = 0.00336 absorber occurs in the Virgo galaxy cluster. We have not included in the list above 13 additional galaxies with $300 < \rho < 500$ kpc and $0.002 < L/L^* < 0.38$ from Morris et al. (1993). (14) For the PKS 0405-123 z = 0.18283 absorber, Johnson et al. (2013) found no galaxies with $L > 0.04L^*$ for $\rho < 250$ kpc and $L > 0.3L^*$ for $\rho < 1$ Mpc. They suggest the absorber may lie in a galaxy void.

Table 7. Baryonic Content Estimates based on Observations

| Source | $\Omega_i$ | $\Omega_i/\Omega_b$ (%) | Note |
|---|---|---|---|
| Galaxies | 0.0032±0.0009 | 7±2 | 1 |
| Cold ISM | 0.00079±0.00019 | 1.7±0.4 | 1 |
| Hot Intercluster Medium | 0.0018±0.0007 | 4±1.5 | 1 |
| Warm gas traced by aligned O VI absorbers | 0.0020±0.0010 | 4.1±1.1 | 2 |
| Cool photoionized gas traced by narrow H I absorbers | 0.013±0.005 | 28±11 | 1 |
| Non-aligned O VI absorbers | ? | ? | 3 |
| Warm gas traced by BLAs without detected metals | (0.0065±0.0032) | (14±7) | 1,4 |
| Hot IGM | ? | ? | 5 |
| Total Observed | 0.0298±0.011 | 65±24 | |
| Total Expected | 0.0463±0.0024 | 100 | 6 |

Notes:
(1)These baryon content estimates and their errors are taken from Shull et al. (2012, appendix A) for those cases where the estimates appear to be relatively secure. (2) The estimate is from this paper based on the 14 aligned O VI absorbers tracing warm gas listed in Table 3. The error estimate includes the error in the estimated number density of these systems and the errors in the ionization corrections. The aligned O VI absorbers tracing cool photoionized gas are included as part of the cool photoionized IGM estimate listed above. (3) The 35 O VI components listed in Tables 2 for which temperature estimates are not possible represent 42% of the detected O VI components. These absorbers could be either photoionized or collisionally ionized. Without knowing the origin of the ionization or the abundance of oxygen in these components, any attempt to estimate their baryonic content is highly uncertain. (4) There are many BLAs for which metal lines have not been detected (Richter et al. 2004, 2006, Lehner et al. 2007, DS2008, Shull et al. 2012). Estimating the baryonic content of these absorbers requires an estimate of the temperature of the gas. However, without the measurement of a heavier ion in the absorber, it is not possible to separately determine the thermal and non-thermal contributions to the line broadening. Estimating the baryonic content of these absorbers therefore requires very uncertain assumptions about the relative amount of the non-thermal broadening. The value listed from Shull et al. (2012) only includes the BLAs without associated O VI absorption. The value is given in parenthesis because of the large and uncertain ionization correction. The listed value includes a 40% downward correction for an error made when calculating $\Omega$ in DS2008 and Danforth et al. (2010b). The error involved the calculation of $\Sigma X$ as discussed by Tilton et al. (2012). (5) The plasma with log T > 6 is predicted to exist at low z in most hydrodynamical simulations (see section 1). At z = 0, strong O VII and O VIII absorption detected by X-ray satellites is inconsistent with an IGM origin (Fang et al. 2006; Bregman & Lloyd-Davies 2007; Yao et al. 2010). Most of the absorption appears to be associated disk/halo of the Milky Way (Wang et al. 2005; Yao et al. 2009 ). Most of the claimed detections of O VII /O VIII at z > 0 are controversial with the exception of the z = 0.03 absorber toward H 2325-309 (Buote et al. 2009; Fang et al. 2010) that appears to be tracing the halo gas of galaxies located in the Sculptor wall (Williams et al. 2010, 2013). More sensitive and higher resolution X-ray observatories will be required to study the hot low density IGM. However, even when those observations become available it will be necessary to assume a value for [O/H] to estimate the baryonic content of the gas. (6) The analysis of the spectrum of the acoustic peaks in the cosmic microwave background obtained by the Wilkinson Microwave Anisotropy Probe found that baryons comprise a total fraction of the critical matter density of the Universe for $\Omega_b$ = 0.0463±0.0024 (Hinshaw et al. 2013).

Figure Captions

FIG. 1. COS and STIS observations of the O VI system at z = 0.14231 toward PG 0953+415 over the wavelength range from 1165 to 1190 Å. Both sets of observations have been binned to ~10 km s⁻¹ to allow a direct visual comparison of the S/N achieved. In this wavelength region the S/N per pixel in the COS observations is ~ 4 times larger than in the STIS observations when binned to the same pixel width.

FIG. 2. CIE curves from Gnat & Sternberg (2007) for a wide range of ions displayed as logN(X) vs log T for a slab of gas with [Z/H] = 0 and log N(H) = 19 with solar elemental reference abundances from Asplund et al. (2009).

FIG. 3. CIE and non-CIE curves from Gnat & Sternberg (2007) for isobaric cooling displayed as logN(X) vs log T for a slab of gas with log N(H) = 19 and [O/H] = 0 and -1 with the solar reference abundance from Asplund et al. (2009). The CIE curves are solid. The non-CIE curves are dashed. The curve for H I does not change between the CIE and non-CIE calculation. In contrast, substantial amounts of O VI are present in gas at log T < 5.3 in the non-CIE model because of delayed recombination of O VI into O V in the cooling gas. Note that the non-equilibrium effects become less significant as the abundance in the plasma decreases from [O/H] = 0 to -1.

FIG. 4. An ion ratio, ion ratio plot displaying log[N(Si IV)/N(C IV)] versus log[N(C IV)/N(O VI)] for the many different theoretical models for the warm (transition temperature) interstellar and intergalactic gas. The different models are displayed with different colored lines as follows: CIE from Gnat & Sternberg (2007) for log T ranging from 5.4 to 4.4 (Black line labeled CIE ). Non-equilibrium radiative cooling models from Gnat & Sternberg (2007) for log T from 5.4 to 4.4 for solar metallicity (Magenta line labeled static cooling). The line gets thicker at lower temperature where the gas spends more time. Shock models from Dopita & Sutherland (1996) for shock velocities of 200 and 300 km s⁻¹ (Orange line labeled shock heating). The conductive interface predictions of Borkowski et al. (1990) for the magnetic field perpendicular and parallel to the interface (Red lines labeled conductive interfaces). The model rapidly evolves along the red line to the large red dots where the ionic ratios stabilize. The non-equilibrium turbulent mixing layer model of Kwak & Shelton (2010) (Blue contours labeled NEI TML). The cooling flow model of Benjamin & Shapiro described in the appendix of Wakker et al. (2012) (Blue line labeled cooling flow). The model describes hot gas flowing through an interface with velocities of 20-30 km s⁻¹ (thicker line) and 12 to 42 km s⁻¹ (thinner part of the line). The Shelton (1998) halo gas model that employs old SN type Ia at large distances from the galactic plane for the heating of halo gas (Green lines labeled thick disk SNe). In all cases the model predictions have been adjusted to the solar abundances compiled by Asplund et al. (2009). The filled black circles are for the ionic ratios observed in Galactic high velocity clouds using the techniques described in Wakker et al. (2012). The open circles are for the ion ratios observed in damped Lyman α (DLA) systems by Fox et al. (2009).

FIG. 5. Examples of different O VI system profile types where only the observations for H I and O VI are shown. In most cases H I λ1215 and O VI λ1031 are shown. The full set of absorption profiles are displayed in the appendix. (a and b) Profiles showing sample cases where the O VI absorption is aligned with broad H I absorption implying a plasma with log T > 5. In these cases collisional ionization likely dominates the ionization of O VI. (c and d) Profiles showing sample

cases where the O VI absorption is aligned with a narrow H I absorption component implying a plasma with log T < 4.5 where photoionization is likely the dominant ionization process. (e and f) Profiles showing cases where some of the O VI and H I absorption components are not aligned in velocity as discussed in section 5. Blue profiles illustrate individual component fits to each absorber with the tick marks at the component velocities. Contaminating absorbers are indicated in each panel with the fitted green profiles and identified in the footnotes to the tables in the Appendix giving the fit results. Red profiles illustrate the total (components + contamination) fit to each absorber. Component tick marks for aligned H I and O VI components are displayed with a heavy line.

FIG. 6. Basic observed properties of the H I and O VI absorption components in all the O VI systems discussed in the appendix and listed in Table 2. (a) Observed distribution of redshift for the 54 O VI systems. (b) Distribution of the number of H I and O VI components in each O VI system. Note that simple systems with one or two H I components and one O VI component dominate the sample. (c) Distribution of the maximum velocity extent of the H I absorption components for the gas in the O VI systems. (d) Distribution of the minimum velocity offset between the O VI and H I absorption components, $|\Delta v(O\ VI - H\ I)|_{min}$. (e) Distribution of the Doppler parameters for the H I and O VI components in the O VI systems. (f) Distribution of the H I and O VI component column densities in the O VI systems. Detection incompleteness affects H I and O VI for log N < 13.2 and <13.5, respectively.

FIG. 7. Basic observed properties of the O VI components when aligned with H I absorption. The different symbols denote the likely origin of the ionization in each absorber (CI, PI or unknown). (a) Doppler parameters b(O VI) vs b(H I) for the aligned components. The solid line is for b(H I)/b(O VI) = 4 implying the line widths are dominated by thermal Doppler broadening. The dashed line is for b(H I)/b(O VI) = 1 implying the line widths are dominated by non-thermal broadening. (b) Histogram displaying the number distribution for the plasma temperatures determined for each aligned component. Gas with log T < 4.7 is probably photoionized. Gas with log T > 4.8 is probably collisionally ionized. Temperatures determined from the lower S/N TSB2008 STIS observations are shown with the dashed red histogram. High S/N is required to reliably detect the BLAs associated with O VI occurring in warm gas. (c) log N(H I) vs log N(O VI) for the aligned components. (d) log [N(H I)/N(O VI)] vs log N(H I) for the aligned components. (e) Histogram of the distribution of the non-thermal broadening, $b_{NT}$, for the warm (red) and cool photoionized (blue) absorbers. (f) Values of [O/H] in the collisionally ionized warm absorbers when 5.4 < log T < 6.2.

FIG. 8. Basic absorber properties versus the smallest associated galaxy impact parameter, ρ, from Table 5. (a) Number of absorbers versus ρ for all absorbers (Solid line) and for absorbers where ρ is determined from observations extending to a depth L < 0.3L* (red dashed line). (b) log N(O VI, total) versus impact parameter, ρ. (c) log N(H I, total) versus ρ. The circles plotted on b and c are from this paper with the circle size indicating the value of L* for the associated galaxy. Largest circles have L > 2.5L*, large circles have 2.5L* > L > 1.0L*. Medium circles have 1.0L* > L > 0.1L*, smallest circles have L < 0.1L*. Error bars are shown when they are larger than the plotted symbol size. Filled circles are from galaxy redshift observations extending to a depth of L < 0.3L*. For the open circles the search depth is comparable to the luminosity of the detected galaxy. The open blue squares show the O VI and H I absorption in the halos of targeted luminous star-forming galaxies from the COS-halo survey of Tumlinson et al. (2011b, 2013). The amount of H I and O VI

close to galaxies revealed by the COS-halo survey is much larger than for the more common O VI systems revealed in our blind survey.  The horizontal dashed lines in b and c indicate the approximate 3σ column density detection limits for O VI and H I in our survey spectra with average S/N.

FIG. 9.  Aligned absorber properties versus the smallest associated galaxy impact parameter, ρ, from Table 5.  The different symbols denote warm absorbers (red squares),  cool photoionized absorbers (blue diamonds) and absorbers where the origin of the ionization is unknown (black circles).  The size of these symbols indicates the luminosity of the associated galaxy.   Symbol size is coded with the galaxy luminosity as for Fig.8.   Filled symbols are from galaxy redshift observations extending to a depth of L < 0.3L*. For the open circles, the galaxy search depth is comparable to the luminosity of the detected galaxy. Error bars are shown when they are larger than the symbol. (a) Total (H I + H II) hydrogen column density, log N(H), versus ρ for the warm absorbers.  Very large H column densities extend to very large distances from the associated galaxies.  (b) log N(H I) versus ρ for  the aligned absorbers.  (c) log N(O VI) versus ρ for  the aligned absorbers.  The horizontal dashed lines in b and c indicate the approximate 3σ column density detection limits for O VI and H I in spectra with average S/N.

FIG. 10.  Absorption line profiles for all the O VI systems including the derived Voigt profile fits.  The QSO name and O VI system redshift are listed at the top of each figure.  The instrument producing the observations (COS, STIS, or FUSE) is indicated in each panel along with the ion and its rest-frame wavelength.   Blue profiles illustrate individual component fits to each absorber with the tick mark at the component velocity.  Contaminating absorbers are indicated in each panel with the fitted green profiles with green tick marks and identified in the footnotes to Tables A1-A14 giving the fit results.  Red profiles illustrate the total (components + contamination) fit to each absorber.   Component tick marks for aligned H I and O VI components are displayed with a heavy line.  The profile fit figures and tables are ordered by the increasing redshift of the QSO (see Table 1).   For each QSO the figures are ordered by increasing redshift of the absorption system (see Table 2).

APPENDIX

Properties of the Observed O VI Systems

In the following section we discuss the observed properties of the O VI absorption line systems detected toward each QSO. Whenever possible we attempt to determine the origin of the ionization of the O VI in each system. A number of the QSOs have been previously observed with FUSE and STIS. However, the new COS observations of these sightlines provide a substantial improvement in the S/N over the wavelength region from ~ 1140 to 1780 Å. In a number of cases, the improved S/N allows the detection of BLAs associated with the O VI absorption. The presence or absence of BLAs or narrow H I absorption aligned with the O VI absorption is crucial for determining the temperature of the gas and the origin of the ionization. The results presented here are ordered by the redshift of the QSO as listed in Table 1. Note that the errors listed for the component velocities in the Tables of this appendix are from the fit errors determined by VPFIT. The fit parameters marked with : indicate a very uncertain value. Additional velocity calibration errors range from 5 to 15 km s$^{-1}$ with the smaller error appropriate for the COS observations calibrated with the STIS observations of the same QSO. The O VI system velocity plots illustrated in this appendix are displayed with a pixel velocity sampling of ~2 km s$^{-1}$ for COS and ~ 5 km s$^{-1}$ for FUSE even though the resolutions are ~ 18 km s$^{-1}$ and ~ 20-25 km s$^{-1}$, respectively. The high sampling frequency sometimes makes it possible to see the effects of residual fixed pattern noise in the delay line detectors. References to many of the papers reporting analysis results for the FUSE and STIS IGM observations are given in the footnotes to Table 1.

Contaminating IGM and ISM absorption features are identified in the velocity plots of Figure 10 for each O VI absorption system and fitted using information from lines at other wavelengths produced by the same absorber. The identifications of the contaminating features are given in the footnotes to each table. Contaminating ISM H$_2$ absorption features are marked on the figures as simply H$_2$. The precise identification is given in the footnotes to the tables with the following notation: -14 km s$^{-1}$, H$_2$ L 6-0 P(4) $\lambda$1035.182, where we give the velocity of the contamination in the rest-frame of the O VI system, the H$_2$ band system, the vibrational quantum number change, the rotational transition designation, and the rest wavelength of the H$_2$ line. The properties of the H$_2$ contamination are from Wakker (2006) with occasional adjustments.

The statistical significance of an absorption feature can be inferred from the logarithmic errors on the column density determined from the profile fit code. We note that logarithmic column density errors of 0.10, 0.14, and 0.18 dex correspond to absorption lines with 5.0, 3.6, and 3.0σ detection significance, respectively. Footnotes to the table comment on the marginal detections.

### A1. O VI Absorption System toward MRK 290

The low redshift AGN MRK 290 with $z_{em}$ = 0.030 has a BLA and O VI absorber at z = 0.01026 first detected by Wakker & Savage (2009) with FUSE and STIS measurements and subsequently observed by Narayanan et al. (2010b) with COS at much higher S/N.

*The MRK 290 O VI absorber at z = 0.01026.* Narayanan et al. (2010b) present the full analysis of this absorber which contains two H I components and one O VI component with the COS LSF from Ghavamian et al. (2009). No other metal lines are detected in the system. Using our improved extraction of the MRK 290 COS observations, we re-fitted the absorbers in this system using the new COS LSF from Kriss (2011) with the results listed in Table A1 and shown in the velocity plot of Figure 10. The new fit parameters are in good agreement with those of Narayanan et al. (2010b). The O VI with log N(O VI) = 13.65±0.10 and b(O VI) = 31±10 km s$^{-1}$ is well aligned with the stronger H I component which is a BLA with log N(H I) = 14.38±0.01 and b(H I) = 53±2 km s$^{-1}$.

The difference in H I and O VI b values implies log T = 5.07 (+0.13, -0.19) and $b_{NT}$ = 29±11 km s$^{-1}$ if the two species exist in the same gas. That assumption is reasonable given the good alignment of H I and O VI with |Δv| = 3±7 km s$^{-1}$. The total baryonic content of the absorber, log N(H) = 19.33(+0.33,-0.56), follows from the temperature and log N(H I) = 14.38±0.01 assuming CIE for the H I. The absorber is tracing warm gas but the low temperature implies the effects of non-CIE are important for determining the O VI abundance. Narayanan et al. (2010b) argue that the ionization mechanisms operating include the effects of the non-equilibrium ionization in cooling gas along with photoionization from the EUV background to produce the O VI and also explain the limit on C IV in the absorber.

### A2. O VI Absorption Systems toward PKS 2155-304

The two low redshift O VI absorption systems are found toward the BL Lac object PKS 2155-304 with $z_{em}$ = 0.117. The new COS observations with S/N ~ 70 -110 greatly improve on the quality of the H I Ly α measurements associated with the O VI recorded by FUSE. The COS observations were obtained after July 23, 2012 when the positions of the spectra on the COS detector were shifted to Lifetime Position 2 resulting in a modified version of the LSF as described on the STScI web site (www.stsci.edu/hst/cos/performance/spectral_resolution). The LSF at Lifetime Position 2 varies according to the central wavelength by up to ~15% in the wing. However, the final fit parameters for non-saturated absorbers differ by less than 1% between the central wavelength setup 1291 and 1327Å LSFs. We used the LSF for the central wavelength setup 1309Å.

*The PKS 2155-304 O VI absorber at z = 0.05423.* The O VI observations for this system shown in Figure 10 are from FUSE. The individual H I absorber parameters have large uncertainties given the strong overlap between the two components observed by COS at -76, -34 and 29 km s$^{-1}$. Although the H I at -34±27 km s$^{-1}$ is aligned with the O VI at 0±3 km s$^{-1}$ to within the errors, it is not possible to reliably determine the properties of the H I that might be associated with the O VI absorption because of the very large uncertainties. Therefore, we do not list these components as aligned in Tables 2 and 4. The total H I column density of the system, log N(H I)= 14.07±0.10, is more reliably determined than the errors on the individual components might suggest because the component errors are strongly correlated.

*The PKS 2155-304 O VI absorber at z = 0.05722.* The COS observations reveal two strong BLAs at v = -176±1 and -38±1 km s$^{-1}$ with b = 50±1 and 68±1 km s$^{-1}$ and log N(H I) = 14.32±0.01 and 14.09±0.01. The BLAs are not aligned with the observed O VI absorption at v = 0±3 km s$^{-1}$. Therefore the origin of the O VI ionization is uncertain. There is possibly weak O VI absorption associated with the BLA at -177 km s$^{-1}$. However, the significance of the absorption is < 3σ. The feature may be associated with a detector edge effect.

### A3. O VI Absorption Systems toward MRK 876

The two O VI absorption systems in the spectrum of the UV bright AGN MRK 876 with $z_{em}$ = 0.129. The new COS observations yield much higher S/N observations of H I than the earlier STIS observations (see Table 1) and allow a much better determination of the properties of the H I associated with the O VI that is seen in FUSE spectra.

*The MRK 876 O VI absorber at z = 0.00315.* There is strong broad H I λ1215 absorption at v = -14±2 km s$^{-1}$ well recorded in the high S/N COS spectrum. The absorption is asymmetric revealing a second component redshifted by 48±3 km s$^{-1}$ from the broader primary component. The H I λ1025 absorption recorded by FUSE is strongly contaminated with ISM H$_2$. O VI λ1037 is not detected. However, O VI λ1031 is detected but blended with ISM H$_2$. The parameters derived for O VI use the H$_2$ absorption model of Wakker (2006). The O VI absorption with b = 35:±13: km s$^{-1}$ is reasonably well aligned with the BLA with b = 58±2 km s$^{-1}$ with |Δv| = 14±9 km s$^{-1}$. Using the O VI and BLA b values we obtain log T = 5.14 (+0.16, -0.25) and $b_{NT}$ = 33±15 km s$^{-1}$. The temperature determined from the line width differences implies the detection of collisionally ionized gas. However, the gas is in a temperature range where non-equilibrium ionization effects are important for oxygen so the inferred oxygen abundance is very uncertain. For log T = 5.14(+0.16, -0.25), the total baryonic content of the absorber is log N(H) = 19.12 (+0.36, -0.83), assuming the H I is in CIE. The uncertain O VI line width has resulted in large negative errors for the values of log T and for log N(H).

*The MRK 876 O VI absorber at z = 0.01171.* The COS absorption profiles shown in Figure 10 reveal narrow and broad components of the H I λ1215 absorber. The H I λ1025 absorption from FUSE is severely contaminated with ISM O VI λ1037. O VI λ1031 absorption likely extends from ~ -40 to 30 km s$^{-1}$. The O VI absorption is not well aligned with either the narrow or the broad H I absorption so it is not possible to determine the physical properties of the O VI bearing gas.

## A4. O VI Systems Toward 3C 273

The COS spectra of 3C 273 have the highest S/N of any QSO so far observed and allow a careful evaluation of the H I absorption in three O VI systems at z = 0.00336, 0.09022, and 0.12007 observed with FUSE and COS. The O VI in the system at z = 0.00336 is probably collisionally ionized. The O VI in the system at z = 0.12007 is likely photoionized. The O VI in the z = 0.09022 system is of uncertain origin since it does not align with the H I absorption.

*The 3C 273 O VI absorber at z = 0.00336.* This absorber is an example of an O VI absorber where the O VI is probably arising in collisionally ionized gas since the COS observations reveal the clear detection of a BLA superposed on the narrower H I absorption. The COS and FUSE observations of H I λλ1215, 1025 and FUSE observations of OVI λλ1031,1037 are shown in Figure 10. The fit parameters are given in Table A4. C III λ977 is not detected, O VI λ1037 is severely blended with ISM H$_2$. O VI λ1031 is partially blended with ISM H$_2$. The derived O VI fit parameters use the H$_2$ absorption model of Wakker (2003) to correct for the extra blended absorption. The H I λ1025 absorption is contaminated by ISM H$_2$. The H I λ1215 absorption has broad wings implying the existence of two H I absorbing components as shown by the model fit. The BLA was tentatively identified by TSB2008 but the parameters of the BLA are much better established with the higher S/N COS observations. The O VI absorption could be associated with the narrow H I or with broad H I absorption since the alignment is satisfactory for both components. However, the O VI absorption line does appear to be marginally broader than the narrow H I absorption component, suggesting the O VI absorption more likely is associated with the BLA. The difference in line widths implies log T = 5.16 (+0.20, - 0.38) and $b_{NT}$ = 41±12 km s$^{-1}$. The O VI is probably collisionally ionized although the negative temperature error is so large we can not exclude photoionization. The implied value of log N(H) = 18.72 (+0.43, -1.17) with the negative error uncertain because of the large negative error on log T.

*The 3C 273 O VI absorber at z = 0.09022.* This absorber contains well defined broad (b = 48±3) and narrow (b = 33±2) H I absorption. The O VI absorption is displaced 122±5 km s$^{-1}$ and 30±5 km s$^{-1}$ to positive velocity from the broad and narrow H I absorption. Since the O VI and H I absorption are not aligned, it is not possible to determine the origin of the ionization of the O VI bearing gas.

*The 3C 273 O VI absorber at z = 0.12007.* This absorber is an excellent example of an O VI absorber where the O VI is very likely arising in photoionized gas. The COS line profiles for O VI $\lambda\lambda$1031, 1037 and H I $\lambda\lambda$1025, 1215 along with the FUSE observation of the region of C III $\lambda$977 are shown in Figure 10. The profile fit results are given in Table A4. C III $\lambda$977 is not detected since the line observed near v = 0 km s$^{-1}$ is dominated by ISM H$_2$ absorption. The O VI and H I absorption lines are very well described by single Doppler broadened components. There is no evidence for a BLA superposed on the narrower H I absorption. The O VI and H I absorption are well aligned with |$\Delta$v| = 7±2 km s$^{-1}$. The difference in the H I and O VI Doppler line widths imply log T = 4.44 (+0.06, -0.07) and b$_{NT}$ = 8 ±3 km s$^{-1}$. The gas is consistent with the absorber being in photoionization equilibrium with the extragalactic background radiation. The COS results for this absorber agree with those from STIS presented by TSB2008.

## A5. O VI Absorption Systems toward PG 1116+215

The very high S/N COS spectrum of PG 1116+215 with $z_{em}$ = 0.177 supplements the very good quality FUSE and STIS observations. A complete analysis of the FUSE and STIS observations is found in Sembach et al. (2004). In this paper we consider the higher quality COS observations of the three O VI systems at z = 0.05897, 0.13850 and 0.16553. The system at z = 0.16553 would formally be considered an associated system since it lies within 4000 km s$^{-1}$ of the QSO. However, it is clearly associated with an enhancement in the density of galaxies near z = 0.165 which is well separated from the galaxies at z = 0.177 (see Fig. 22 in Sembach et al. 2004), so we include it in the group of intervening systems.

Sembach et al. (2004) noted two other possible O VI absorption systems that are associated with BLAs at z = 0.04125 and 0.06244. The new COS observations reveal a BLA at z = 0.04121 with b = 110±4 and log N(H I) = 13.41±0.01 while Sembach et al. (2004) report b = 105±18 km s$^{-1}$ and log N(H I) = 13.24±0.15 . For the absorber at z = 0.06244 we obtain b = 67±6 and log N(H I) = 12.96±0.03 at z = 0.06251, while Sembach et al. (2004) report b = 79±10 km s$^{-1}$ and log N(H I) = 13.18±0.09. The O VI features seen in the FUSE observations analyzed by Sembach et al. (2004) for these BLAs have less than 3$\sigma$ significance. We have not included these two possible O VI systems in our list of claimed O VI detections for the line of sight to PG 1116+215. These are two potentially very important O VI absorbers since the presence of a BLA and narrower O VI aligned in velocity with the BLA provides good evidence that the broadening of the BLA is mostly thermal.

*The PG 1116+215 O VI absorber at z = 0.05897.* The FUSE observations of O VI reveal two absorbing components with one at 0±12 km s$^{-1}$ and the other at 83± 3 km s$^{-1}$. The O VI $\lambda$1031 component at 0 km s$^{-1}$ has uncertain parameters because it is partially blended with ISM H$_2$ absorption at -57 km s$^{-1}$ and the corresponding O VI $\lambda$1037 is too weak to detect. The C IV $\lambda\lambda$1548, 1550 absorption detected at 82±2 km s$^{-1}$ is likely associated with the O VI at 83±3 km s$^{-1}$. The H I $\lambda$1215 line is well fitted with three components. The stronger H I component at -3±1 km s$^{-1}$ with b = 33±2 km s$^{-1}$ is well aligned with the O VI absorption at 0±12 km s$^{-1}$ with b = 27±16 km s$^{-1}$. If the H I and O VI coexist in the same gas, the difference in line width implies log T = 4.36 (+0.54, -4.36). This O VI absorber is probably photoionized. The large errors on log T are from the large error for the O VI line width.

The O VI and C IV near 83 km s$^{-1}$ differ in velocity by 24±4 km s$^{-1}$ from the weak H I absorption at 107±2 km s$^{-1}$. With this large a misalignment it is unlikely the O VI, C IV and H I co-exist in the same gas.   There is no way to reliably estimate the temperature or the origin of the ionization in the O VI and C IV components near 83 km s$^{-1}$.

*The PG 1116+215 O VI absorber at z = 0.13850.*  Our extraction of the H I 1215 absorption in this system considerably differs from the extraction obtained using the Danforth et al. (2010) software.  Because of wavelength misalignment problems when combining spectra and the presence of strong fixed pattern noise structures, the Danforth extraction code causes the BLA in the system to appear broader than we record.

This is a complex multi-phase absorption system containing H I, O VI, O III,  C IV, C III, C III, Si IV, Si III, Si II, N III, and N II (see Sembach et al. 2004).  The low ion Si II absorption recorded by STIS reveals two narrow components separated by 10 km s$^{-1}$.  Using the Si II absorption as a guide to the H I velocity structure, the H I absorption is well modeled with two narrow components  having comparable column densities of log N(H I) = 15.91±0.03 and 15.95±0.02.  The b values in these H I components of 14±1 and 19±1 km s$^{-1}$ imply the detection of cool H I and its associated set of photoionized low ionization metal absorption lines.  The O VI absorption with b(O VI) = 36±2 km s$^{-1}$ is much broader than all of the lower ionization absorption lines implying  the O VI exists in a different phase of a complex multiphase absorption system.   A BLA with b = 46±2 km s$^{-1}$ is detected with a velocity that suggests it could be associated with the O VI absorption.  If the BLA and O VI exist in the same gas, the difference in b values implies log T = 4.72(+0.07, -0.14) and b$_{NT}$ =35±2 km s$^{-1}$.  This is a relatively high temperature for gas in photoionization equilibrium although we cannot rule out that possibility.  However, the O VI and H I may exist in some type of collisionally ionized non-equilibrium cooling structure where the recombination of O VI has been delayed.

*The PG 1116+215 O VI absorber at z = 0.16553.*  This is a kinematically complex absorption system with the H I absorption containing 7 components spanning the velocity range from -113 to 435 km s$^{-1}$.   The 7 component fit includes a BLA with v = 117±7 km s$^{-1}$ and b = 100±7 km s$^{-1}$.  The O VI absorption reveals components at -131, -38 and 0 km s$^{-1}$ and a weak component with uncertain properties at  178 km s$^{-1}$.

 The strongest O VI component at 0±1 km s$^{-1}$ with b = 19±2 km s$^{-1}$ is reasonably well aligned with the H I component at v = -12±1 km s$^{-1}$ with b = 34±2 km s$^{-1}$.   If the O VI and H I in these components coexist in the same gas, the difference in b values implies log T = 4.71 (+0.07, -0.10) and b$_{NT}$ =18±2 km s$^{-1}$.  Gas this hot could be created by photoionization if the ionization parameter is large.  However, non-equilibrium collisional ionization in a cooling structure with delayed O VI recombination is also possible.

The broad marginally observed O VI component at 178±5 km s$^{-1}$ with b = 51:±6: km s$^{-1}$ is well aligned with the narrow H I component at 182±3 km s$^{-1}$ with b = 14±2 km s$^{-1}$.  The two absorbers can not exist in the same plasma because b(O VI)  >  b(H I).  The origin of the ionization for this O VI component is unknown.

The origin of the ionization in the other two O VI components at -131±2 and -38±14 km s$^{-1}$ is unknown because the O VI absorption is not well aligned with the H I absorption.

A6. O VI Absorption Systems Toward PHL 1811

Four intervening O VI absorption systems are detected in the COS/FUSE observations of PHL 1811 at z = 0.07773, 0.13280, 0.15785, and 0.17651.   The systems at z = 0.07773 and 0.15785 are clearly revealing O VI/BLA absorption associated with warm gas.

*The PHL 1811 O VI absorber at z = 0.07773.*  This system is detected in O VI λ1031, C III λ977, Si III λ1206, Si II λ1260,  and H I λλ1215 to 949.  The O VI λ1031 absorption is contaminated at -30 km s$^{-1}$ by ISM Fe II $\lambda$1112. The strength of the contamination is well modeled with reference to the other Fe II line in the FUSE and COS spectra. The H I absorption reveals narrow and broad components with b(H I) = 19±1 and 82±10 km s$^{-1}$. Lacki & Charlton (2010) have studied the properties of the C III, Si III, and C II in this absorber using the FUSE+STIS observations and fit the observations with a photoionization model. The O VI cannot arise in the photoionized gas producing the narrow H I absorption, C III, Si III and C II. Although the parameters for the O VI absorption are uncertain the absorption is definitely broader than that arising in the cool photoionized gas.

The O VI with b(O VI) = 43±7 km s$^{-1}$ is very likely associated with the well aligned BLA having b(H I) = 82±10 km s$^{-1}$. If the O VI and the BLA trace the same gas the difference in line width implies log T = 5.49(+0.14, -0.19) and $b_{NT}$ =39±8 km s$^{-1}$. The O VI absorber is tracing warm collisionally ionized cooling gas with log N(H) = 19.47 (+0.23, -0.37). The temperature and log N(H I) = 13.60±0.07 imply log N(H) = 19.47 (+0.23, -0.37) for H I in CIE. The temperature is large enough for the effects of non-CIE to be relatively small for O VI. Assuming CIE we obtain [O/H] = -1.93 (+1.32, -0.11).

*The PHL 1811 O VI absorber at z = 0.13280.* This O VI system has four components to the O VI $\lambda\lambda$ 1031, 1037 absorption spanning a velocity range of 218 km s$^{-1}$. The O VI component at -133±1 km s$^{-1}$ is well aligned to the H I component also at -133±1 km s$^{-1}$. The H I and O VI b values of 29±4 and 30±3 km s$^{-1}$ imply cool photoionized gas with log T < 4.36. The O VI absorption at -81±3 km s$^{-1}$ is not aligned with H I absorption. The temperature and origin of the ionization in this component are unknown. The O VI absorption at 0±2 km s$^{-1}$ with b = 23±3 km s$^{-1}$ is well aligned with H I absorption at 1±4 km s$^{-1}$ with b = 21±8 km s$^{-1}$ implying cool photoionized gas with log T < 4.45.

The O VI absorption well detected at 85±2 km s$^{-1}$ with log N(O VI) = 13.57 ±0.03 and b = 25±3 km s$^{-1}$ is possibly associated with the very uncertain BLA at v = 74:±15: km s$^{-1}$ with b = 49:±13: km s$^{-1}$ and log N(H I) = 12.65:±0.15:. If the two absorbers trace the same plasma the difference in line widths imply log T = 5.06 (+0.24, -0.56) and $b_{NT}$ =23±4 km s$^{-1}$. However, the properties of this BLA are too uncertain to clearly associate it with the O VI absorption and we report the O VI as not aligned with H I in Table 2. The O VI is either tracing relatively high metallicity photoionized gas with a very large ionization parameter or high metallicity collisionally ionized gas.

*The PHL 1811 O VI absorber at z = 0.15785.* O VI $\lambda\lambda$1031, 1037 are clearly detected in this absorber with b(O VI) = 31±3 and log N(O VI) = 13.68±0.03 at v = 0±2 km s$^{-1}$. The H I $\lambda$1215 absorption is contaminated by strong narrow O I $\lambda$1302 absorption in the Lyman limit system at z = 0.08092 studied by Jenkins et al. (2005). The broad absorption seen in the figure is a BLA with v = -7±7 km s$^{-1}$, b(H I) = 97±10 km s$^{-1}$, and log N(H I) = 13.25±0.04. If the O VI and the BLA do trace the same gas, log T = 5.73 (+ 0.09, –0.11) and $b_{NT}$ = 20±6 km s$^{-1}$. The evidence for warm gas is relatively strong and the temperature is large enough for CIE to be a good approximation for both H I and O VI. Assuming CIE, the values of log T and logN(H I) imply log N(H) = 19.51 (+0.13, -0.18) and [O/H] = -0.90 (+0.20, -0.32).

The O VI absorber at -87±3 km s$^{-1}$ is not detected in H I. The temperature and origin of the ionization of the O VI are unknown.

*The PHL 1811 O VI absorber at z = 0.17651.* O VI, C III and two-component H I are detected in this absorber. The O VI is only seen in the $\lambda$1037 line since the $\lambda$1031 line is lost in Geocoronal Ly$\alpha$ emission. C III $\lambda$977 exhibits a narrow and broad component with the narrow component lined up with the O VI absorption. However, the width of the O VI component at v ~ 0 km s$^{-1}$ is ~2.5 times larger than the width of the C III absorption. O VI and C III clearly trace different phases of the gas even though they share the same velocity. The strong narrow H I absorption at v = 15±1 km s$^{-1}$ does not line up with the OVI or C III absorption near 0 km s$^{-1}$. The H I component at −31±6 km s$^{-1}$ does not line up with the broad C III component at -31±10 km s$^{-1}$. Given the mis-alignments of all the absorbers it is not possible to determine useful information about the origin of the ionization of the O VI in this multiphase system.

### A7. O VI Absorption Systems Toward PG 0953+414

The COS observations of PG 0953+414 provide much higher S/N observations of H I and other ions in the three O VI systems at z = 0.00212, 0.06808 and 0.14231 previously studied with STIS and FUSE observations. We do not confirm the absorber at z = 0.05885 identified by TSB2008 with log N(O VI) = 13.74 (+0.10, -0.13) based on the O VI λ1031 line. The absorber identified as O VI is actually ISM H$_2$ L 1-0 R(1) λ1092.732 absorption that occurs in two components separated by 26 km s$^{-1}$ (see Wakker 2006).

*The PG 0953+414 O VI absorber at z = 0.00212.* This is a simple one component aligned H I and O VI absorber. With b(H I) = 38±3 km s$^{-1}$ and b(O VI) = 43± 8 km s$^{-1}$ the implied temperature in the absorber is low with log T < 4.47, obtained from b(H I) < 41 km s$^{-1}$ and b(O VI) > 35 km s$^{-1}$. The O VI in this absorber is likely produced by photoionization with the line broadening dominated by turbulence.

*The PG 0953+414 O VI absorber at z = 0.06808.* This simple absorber contains H I, O VI, N V, C IV, and C III. All the metal lines are well aligned with the narrow H I absorption at -2±1 km s$^{-1}$ with b(H I) = 19± 1 km s$^{-1}$. The second broader H I component at v = 32±23 km s$^{-1}$ was not seen in the lower S/N STIS observations. The parameters of this second H I component are uncertain because of its strong overlap with the much stronger but narrow H I absorption component. The photoionization origin of the O VI, N V, and C IV in this absorber was first discussed by Savage et al. (2002). The new COS observations are consistent with that interpretation. Note that the large line width for C III λ977 implies it arises in another gas phase. With b(H I) = 19±1 km s$^{-1}$ and b(O VI) = 12±1 km s$^{-1}$, log T = 4.14 (+0.11, -0.14) and b$_{NT}$ =11±1 km s$^{-1}$. Slightly different temperatures are obtained from the line width comparisons of H I/ C IV and H I/N V. This is probably the result of slight temperature changes within the absorbing structure. The O VI in this absorber is very likely photoionized.

*PG 0953+414 O VI absorber at z = 0.14231.* The COS observations of H I and O VI in this absorber have much higher S/N than the earlier STIS observations analyzed by Savage et al. (2002) and TSB2008. The COS H I λ1215 absorption reveals four components extending from -138 to 211 km s$^{-1}$. The two strongest H I components at 0 ±1 and 82±1 km s$^{-1}$ align well with the two detected O VI components at 0±1 and 78±3 km s$^{-1}$. These two narrow H I lines are slightly wider than the corresponding O VI lines implying log T = 4.21 (+0.08, -0.10) for the O VI absorber at 0 km s$^{-1}$ and log T < 4.25 for the O VI absorber at 78 km s$^{-1}$. Given the evidence for cool gas in each absorber, the O VI in both absorbers is likely produced by photoionization. However, the absorbers may be more kinematically and physically complex than this simple analysis assumes. Note that the C III absorption near -4 km s$^{-1}$ is narrower than the O VI absorption at 0 km s$^{-1}$ and the C III absorption at 96±2 km s$^{-1}$ is misaligned with the H I and O VI absorption near 80 km s$^{-1}$ While multiple gas zones are required to explain the H I, O VI, and C III absorption it appears reasonable to assume that photoionization is the dominant ionizing process. Higher resolution and higher S/N observations would be required to unravel the true complexity of the multiphase absorption extending from 80 to 100 km s$^{-1}$.

A8. O VI Absorption Systems Toward H 1821+643

The high S/N COS observations of H 1821+643 were obtained with the G130M grating. The 530 sec G160M COS observation has only moderate S/N. However, the STIS observations have relatively high S/N so we can combine the COS results for ions detected with the G130M grating with the STIS results at longer wavelengths. We report on the six O VI absorption systems near z = 0.02443, 0.12141, 0.17036, 0.21329, 0.22497, 0.24531, and 0.26656.

*The H 1821+643 O VI absorber at z = 0.02443.* This O VI system has single component H I and O VI absorption. The FUSE observation of the O VI absorption is based on the O VI λ1031 line since the O VI λ1037 line is strongly contaminated by ISM Fe II λ1063. According to our velocity calibration of the FUSE observations v(O VI) = 0±5 km s$^{-1}$ while for H I from the COS observations v(H I) = -12±1 km s$^{-1}$. The velocity from COS should be reliable because the COS observations were cross-correlated with the STIS observations which have a more reliable velocity calibration. The velocity errors listed above are only the profile fit errors. We estimate an additional 8 km s$^{-1}$ error in the FUSE velocity calibration for this object. Therefore, within the errors the H I and O VI absorbers are aligned. With b(H I) = 29±1 km s$^{-1}$ and b(O VI) = 23±7 km s$^{-1}$, we estimate log T = 4.30 (+0.31, -4.30). The gas in this absorber is cold and likely photoionized. There is no evidence for a BLA superposed on the narrow H I absorption component.

*The H 1821+643 O VI absorber at z = 0.12141.* This absorber contains 3 H I components and a single broad O VI component detected with COS in the O VI λ1031 line. The O VI λ1037 line is blended with IGM H I λ949 at z = 0.22504 and 0.22480. All three H I components have well determined Doppler widths with b > 40 km s$^{-1}$ implying the detection of 3 BLAs. The O VI absorption with v = 0±6 km s$^{-1}$ and b = 58±7 km s$^{-1}$ is aligned within the errors with the BLA at v = 16±9 km s$^{-1}$ having b = 82±7 km s$^{-1}$. The difference in Doppler parameters imply the detection of warm gas with log T = 5.33 (+0.16, -0.24) and $b_{NT}$ = 56±8 km s$^{-1}$. The H I ionization may be close to CIE with log N(H) = 19.36 (+0.29, -0.54). The O VI very likely is produced by collisional ionization in cooling gas and is present because of non-equilibrium delayed recombination. Tripp et al. (2001) first discussed the properties of the BLA in this system near v = 16±9 km s$^{-1}$ and its associated O VI absorption. However, the TSB2008 estimate of b(O VI) = 76 (+13, -11) from the lower quality STIS observations is substantially larger than the more accurate value from COS with b(O VI) = 58±7 km s$^{-1}$. Therefore, the derived temperature for the warm plasma based on the COS observations of log T = 5.33 (+0.16, -0.23) is larger by 0.31 dex than the value presented by TSB2008.

*The H 1821+643 O VI absorber at z = 0.17036.* O VI λ1031 absorption is detected by COS extending from ∼ -120 to 80 km s$^{-1}$. For v < -120 km s$^{-1}$ the absorption is contaminated by H I λ930 (ζ) at z = 0.29686. The corresponding O VI λ1037 absorption is contaminated by Geocoronal H I Lyα emission and ISM absorption. The component structure of the O VI absorption is uncertain. The fit results shown in the line profile plot and listed in Table A8 assumes two components. The H I λ1215 absorption in this system has 3 components spanning the velocity range from -179 to 33 km s$^{-1}$. The H I component at -87±2 km s$^{-1}$ is a BLA with b = 54±2 km s$^{-1}$. It is well aligned with the O VI component at v = -95±8 km s$^{-1}$ with b = 31±12 km s$^{-1}$. If these absorbers arise in the same gas the difference in Doppler parameters implies the detection of warm gas with log T = 5.10 (+0.14, -0.22), $b_{NT}$ = 29±14 km s$^{-1}$, and log N(H) = 18.67 (+0.35, -0.65). The O VI at these velocities likely has been collisionally ionized and is situated in a cooling plasma under non-CIE conditions. For a full discussion of the properties of gas with these parameters see the discussion of the physics of the O VI absorber at z = 0.01026 toward MRK 290 of Narayanan et al. (2010b).

The broad O VI absorber in this system centered at v = 0±15 km s$^{-1}$ with b = 79:±17: km s$^{-1}$ overlaps in velocity with the H I absorption but is not aligned with the H I absorption component velocities. The origin of the ionization in this O VI absorption is unknown.

*The H 1821+643 O VI absorber at z = 0.21329.* This absorber contains a single component to the O VI absorption which is well detected in the COS observations of O VI λ1031. The corresponding O VI λ1037 absorption is blended with ISM high velocity S II λ1259 absorption originating in the Milky Way's outer arm. The H I λ1215 absorption observed by COS is relatively noisy since it is the result of one 530 s extraction with the G160m grating. However, the H I parameters are well determined implying the detection of a BLA with b = 40±1 km s$^{-1}$ at v = -6±1 km s$^{-1}$. The BLA aligns well with the O VI absorption at v = 0± 2 km s$^{-1}$ with b = 26±2 km s$^{-1}$. There is also possible O IV λ787 absorption recorded by FUSE although its parameters are very uncertain. If the O VI and H I BLA are tracing the same gas, the difference in Doppler parameters implies log T = 4.77 (+0.06, -0.07) and $b_{NT}$ = 25±2 km s$^{-1}$. The temperature of this gas is below that commonly associated with warm gas in the IGM. However, it is difficult to produce gas this hot in the photoionized IGM unless the ionization parameter is very large. The absorption may be tracing delayed O VI recombination in a cooling plasma that was originally collisionally ionized. We list PI/CI for the origin of the O VI ionization in Table 4.

*The H 1821+643 O VI absorber at z = 0.22497.* This is a complex multi-phase O VI system previously studied by Narayanan et al.(2009) using observations from FUSE and STIS. The new COS observations provide much higher S/N G130M measurements from 1135 to 1460 Å yielding very high quality measurements of H I λλ1025, 972, 949, O VI λλ1031, 1037, and C III λ977. The new COS observations do not improve on the quality of the observations of H I λ1215 because of the short COS integration time for the G160M grating. Narayanan et al. (2009) propose the very strong broad O VI absorption at v = 0±2 km s$^{-1}$ with b = 45±1 km s$^{-1}$ is most likely tracing the warm gas in an interface between the cool photoionized gas in the multi-phase absorber (traced by H I, C II, C III, Si II, Si III, O III) and a hot unseen exterior medium. This absorber has the properties of Galactic high velocity clouds.

The new COS observations listed in Table A8 do improve on the STIS measurements of O VI which has components at 0±1, 60±1 and 343±1 km s$^{-1}$. The strongest O VI component at 0±1 km s$^{-1}$ is broad with b = 45±1 km s$^{-1}$ and log N(O VI) = 14.25±0.01. The H I observations shown in the velocity plot for H I λλ1215, 1025 are from the high quality STIS observations. Our multicomponent fit to the H I observations down to H I λ930 in the FUSE observations reveals the five components listed in Table A8. A broad H I component with at v = -23±17 km s$^{-1}$ with b = 95±11 km s$^{-1}$ is required to fit the wings of H I λ1215. Narayanan et al. (2010a) did not attempt to fit a BLA to the H I absorption. TSB2008 fitted the same STIS H I observations and instead of finding the BLA fitted the positive velocity wing of the H I λ1216 profile with a narrower feature at v = 71±1 km s$^{-1}$, b = 30(+15, -10) km s$^{-1}$ and log N(H I) = 13.05±0.21. This difference reveals the ambiguities that can arise when fitting complex overlapping multiphase absorption systems.

The BLA obtained from our profile fit is within the errors well aligned with the strong O VI component at v = 0±1 km s$^{-1}$. The BLA we find with b = 95± 11 km s$^{-1}$ may be tracing the warm highly ionized H I that co-exists with the O VI absorber. If the BLA and the O VI co-exist in the same plasma, the difference in the line widths implies log T = 5.65 (+0.11, -0.15) and $b_{NT}$ = 39±2 km s$^{-1}$. H I at this temperature should be relatively well described by CIE conditions implying log N(H )

= 20.07 (+0.18, -0.25). The O VI should also be well described by CIE allowing a determination of the oxygen abundance [O/H] = -1.17 (+0.34, -0.37).

The highly displaced O VI $\lambda\lambda 1031$, 1037 absorption at 343±2 km s$^{-1}$ with b = 16±1 km s$^{-1}$ and log N(O VI) = 13.48±0.01 is not detected in H I with a 3$\sigma$ limit of < 12.70 derived by Narayanan et al. (2010a), implying log [N(H I)/N(O VI)] < -0.8. In order to explain the O VI in this absorber by photoionization, a very large ionization parameter log U ~ -0.1 to -0.5 is required (see Table 4 in Narayanan et al. 2010a) implying very low gas densities of (2-4)x10$^{-6}$ cm$^{-3}$ and very large path lengths for [O/H] < -0.3. The large path lengths appear to be incompatible with the narrow O VI line width. Therefore, Narayanan et al. (2010a) argued a collisional ionization origin is more likely for this absorber. However, if the oxygen abundance in the absorber approaches solar, path lengths as small as ~20 kpc could explain the absorption for log U = -0.5. A photoionization origin for the O VI at 343 km s$^{-1}$ is unlikely although it cannot be completely ruled out.

*The H 1821+643 O VI absorber at z = 0.24531.* This simple single component H I and O VI absorber is well detected in O VI and H I. The O VI and H I absorption are well aligned in velocity. The O VI and H I line width difference implies log T = 4.56 (+0.22, -0.45) and $b_{NT}$ = 26±1 km s$^{-1}$. With this temperature, the O VI is very likely created in photoionized gas.

*The H 1821+643 O VI absorber at z = 0.26656.* For this simple single component O VI and H I absorber the O VI and H I are well aligned in velocity. The H I absorption is a BLA with b = 46±2 km s$^{-1}$. The difference in the O VI and H I line widths implies the detection of warm gas with log T = 4.99 (+0.05, -0.06) and $b_{NT}$ = 22±2 km s$^{-1}$. The O VI is probably experiencing delayed recombination in cooling gas. However, the ionization of H I will be relatively close to CIE provided log U < - 2. In that case log N(H) = 18.38 (+0.15, -0.19).

A9. O VI Absorption System Toward TON 236

There are no STIS observations of TON 236 and the FUSE observations have very low S/N and are not useful. TON 236 with $z_{em}$ = 0.447 contains only two O VI absorption systems at z = 0.19452 and 0.39944 over the redshift range from 0.100 to 0.427. The system at z = 0.39944 is tentative since the ID is only based on the O VI $\lambda 1031$ line. A possible two component narrow O VI absorber separated by 60 km s$^{-1}$ is at z = 0.25895. However, only two extractions cover the O VI $\lambda 1031$ line at 1299 Å with one showing the two absorbers and the other showing no absorption. The two observed features could be FPN. The line of sight is unusual since only two O VI systems are detected while four systems are expected for dn/dz = 21±3 from TSB2008.

*The TON 236 O VI absorber at z = 0.19452.* Our extraction of this absorber reveals two components to the H I absorption with the broader H I component with b(H I) = 49±1 km s$^{-1}$ aligned wit the broad O VI absorption with b(O VI) = 44±5 km s$^{-1}$. Our extraction of this absorber using the Danforth et al. (2010a) code reveals only one H I component with b = 60±1 km s$^{-1}$ and log N(H I) = 14.05±0.01. The blurring due to velocity misalignment of the combined spectra when using the Danforth et al. code likely explains the difference.

If the aligned H I and O VI exist in the same plasma, the line width difference implies log T = 4.47 (+0.30, -1.51) and $b_{NT}$ = 44±5 km s$^{-1}$. The O VI is likely produced by photoionization in cool gas.

*The TON 236 O VI absorber at z = 0.39944.* Possible broad O VI $\lambda 1031$ absorption with b(O VI) = 61±18 km s$^{-1}$ is detected that might be associated with H I $\lambda 1215$ absorption with log N(H I) = 14.10±0.01 and b(H I) = 45±1 km s$^{-1}$. With a velocity difference of 16±21 km s$^{-1}$ the two absorbers could be associated. However, it is unlikely they exist in the same plasma since b(O V) > b(H I), although the errors on b(O VI) are large. Because of severe contamination of the O VI

λ1037 line with H I λ1216 at z = 0.19451 it is possible the line identified as O VI λ1031 is instead H I λ1215 at z = 0.01387. The origin of the ionization in this absorber is unknown although within the errors it is marginally consistent with photoioniztion in cool gas.

### A10. O VI Absorption Systems Toward HE 0153-4520

Six O VI absorption systems are detected in the COS observation of HE 0153-4520 with $z_{em}$ = 0.450. The FUSE observations exist but the data have low S/N and are not useful. O VI systems are found at z = 0.14887, 0.17090, 0.22203, 0.22600, 0.29114, and 0.40052. The multi-phase Lyman limit system at z = 0.22600 containing cool photoionized gas and hot gas containing O VI and a very broad BLA has been discussed by Savage et al. (2011b).

*The HE 0153-4520 O VI absorber at z = 0.14887.* This absorber has two components for O VI and two for H I. The 0 km s$^{-1}$ O VI component is clearly detected in the O VI λ1031 line but contaminated by H I λ972 at z = 0.22596 in the O VI λ1037 line. Possible C IV λλ1548, 1550 absorption occurs at v = -10±1 km s$^{-1}$ and is significantly narrower than the O VI at 0±1 km s$^{-1}$. The possible C IV is well aligned with the H I component at v = -11±3 km s$^{-1}$. The C IV and H I likely trace photoionized gas given the small line width of the H I absorption. The O VI component at 0±1 km s$^{-1}$ may contain substructure and within the errors is also aligned with the H I absorption at v =-11±3 km s$^{-1}$. Assuming the O VI and H I coexist in the same gas the difference in line widths implies log T = 4.53 (+0.18, -0.32) and $b_{NT}$ =24±1 km s$^{-1}$. The low temperature implies photoionization explains the origin of the O VI.

The O VI component at -113±3 km s$^{-1}$ with b = 50±3 km s$^{-1}$ is relatively well aligned with the narrower H I component at -102.±3 km s$^{-1}$ with b = 35±3 km s$^{-1}$ suggesting the O VI occurs in a different gas phase. The origin of the ionization in the O VI absorber at -102 km s$^{-1}$ cannot be determined from these observations.

*The HE 0153-4520 O VI absorber at z = 0.17090.* The O VI λ1031 observation reveals a single line at v = 0±1 km s$^{-1}$ that is relatively well aligned with H I absorption at v =-8±1 km s$^{-1}$ detected in H I λλ1215, 1025, 972. O VI λ1037 is hidden by ISM Lyα absorption. The H I absorption also reveals a BLA at v = -40±2 km s$^{-1}$ with b(H I) = 108±2 km s$^{-1}$ and no associated O VI. If the O VI and H I absorption near 0 km s$^{-1}$ traces the same gas the difference in H I and O VI line widths implies log T = 4.60(+0.07, -0.09) and $b_{NT}$ =22 ±2 km s$^{-1}$. The temperature is consistent with an origin for the O VI in photoionized gas.

*The HE 0153-4520 O VI absorber at z = 0.22203.* The two component O VI absorption in this system is detected in the weaker λ1037 line. O VI λ1031 is very strongly contaminated by high velocity ISM Si II 1260 absorption associated with weak H I emission in the Magellanic Stream. C III λ977 is contaminated by high velocity ISM Si II λ1193 absorption. The H I λλ1215-937 absorption is well modeled with components at 19±1 and 31±1 km s$^{-1}$. The component at 31 km s$^{-1}$ is a BLA with b(H I) = 53±1 km s$^{-1}$. The O VI absorption components at 0±2.and 99±2 km s$^{-1}$ are not aligned with the H I absorption so it in not possible to obtain information about the temperature of the O VI bearing gas. The origin of the O VI ionization is unknown.

*The HE 0153-4520 O VI absorber at z = 0.22600.* This is an extremely interesting absorption system fully analyzed by Savage et al. (2011b). The absorber is a multiphase partial Lyman limit system tracing both cool and warm gas. The cool gas containing H I, C III, C II, N III, N II, Si III, and Si II is well modeled by equilibrium photoionization with log U = -2.8±0.1, log N(H) = 19.35±0.18 and log $n_H$ = -2.9±0.2 and [X/H] = -0.8 (+0.3, -0.2). The H I absorption in this cool gas obtained from the Lyman limit absorption observed by FUSE has log N(H I) = 16.61±0.15.

The fit results listed in Table A10 and shown in Figure 10 are based on the new extraction of the COS observations of the QSO. The new fit parameters for H I and O VI are in excellent agreement with those obtained by Savage et al.(2011b).

The fit to the COS and FUSE observations to H I $\lambda\lambda$1215 to 926 yields b(H I) = 28±1 km s$^{-1}$. The O VI absorption with b(O VI) = 37±1 km s$^{-1}$ does not arise in the cool gas but is instead associated with a very broad BLA implying the detection of warm gas. The broad H I absorption has log N(H I) = 13.58±0.05 and b(H I) = 151±15 km s$^{-1}$.

The O VI absorption is well aligned with the BLA. If the BLA and O VI exist in the same gas, the line width difference implies log T = 6.14 (+0.08, -0.10). The temperature and column densities of O VI and H I along with the assumption of CIE yield the oxygen abundance, [O/H] = 0.03±0.10, and the very large baryonic column of log N(H ) = 20.38 (+0.10, -0.13) in the hot collisionally ionized absorber. The high oxygen abundance suggests the absorber is probably closely associated with the enriched plasma from a galaxy.

*The HE 0153-4520 O VI absorber at z = 0.29114.* This simple system is well fitted with two O VI components in the strong and weak line of the doublet and two H I components. C III $\lambda$977 shown as a contaminating line in the O VI 1031 panel of the z = 0.22203 absorber is strongly contaminated by high velocity ISM Si II $\lambda$1260. The O VI component at v= 0±1 km s$^{-1}$ with b(O VI) = 19±2 is very well aligned with the H I component at v = -1±1 km s$^{-1}$ with b(H I) = 20±1 km$^{-1}$. It is very likely the O VI and H I occur in the same gas. However, the nearly identical line widths imply most of the broadening is non-thermal and that a possible thermal contribution to the broadening of the H I is small. Using b(H I) = 20±1 = 21 km s$^{-1}$ and b(O VI) =19-2 = 17 km s$^{-1}$ we derive an approximate 2 σ upper limit to the temperature of the gas of log T < 3.99 implying the absorber is cold for a IGM cloud in photoionization equilibrium.

The second O VI component at v = -53±3 km s$^{-1}$ is not well aligned with the BLA at v = -29±2 km s$^{-1}$ with b(H I) = 61±2 km s$^{-1}$ and probably is tracing a different phase of the gas. However, it is possible the true H I absorption is more complex than adopted in our simple two component fit to the observations. The origin of the ionization of the O VI in this component is unknown.

*The HE 0153-4520 O VI absorber at z = 0.40052.* This system has three components to the well observed O VI $\lambda\lambda$1031, 1037 absorption and three to the H I $\lambda\lambda$1215-949 absorption.

Narrow and broad O VI absorption with b(O VI) = 11±2 km s$^{-1}$ and 42±3 km s$^{-1}$ occur at –54 and -64 km s$^{-1}$. The broad O VI is well aligned with a BLA at -63 km s$^{-1}$ with b(H I) = 42±1 km s$^{-1}$. If these absorbers co-exist in the same gas the difference in line widths implies a 2σ upper limit to the temperature of log T < 4.32 and a photoionization origin to the O VI. The weak broad C III $\lambda$977 absorption at -79±5: with b(C III) = 41±7: km s$^{-1}$ may also exist in this cold gas.

Narrow C III $\lambda$977 absorption near 0 km s$^{-1}$ is well aligned with the H I component near 0 km s$^{-1}$ and the difference in line widths implies the C III is created by photoionization in cool gas. The O VI component at 0±6 km s$^{-1}$ with b = 24±4 km s$^{-1}$ is broader than the C III and H I absorption near 0 km s$^{-1}$ suggesting the O VI occurs in a different gas phase where the origin of the O VI ionization is unknown.

A11. O VI Absorption Systems Toward HE 0226-4110

Five O VI absorption systems are detected in the COS, FUSE, and STIS spectra of HE 0226-4110 with $z_{em}$ = 0.493. A complete analysis of the FUSE and STIS observations can be found in Lehner et al. (2006). The COS observations provide 3 to 4 times higher S/N than the STIS observations and yield improved information on the properties of the O VI and H I absorption. The existence of the STIS observations allows for a reliable wavelength calibration of the COS observations. Five intervening O VI systems are detected at z = 0.01749, 0.20701, 0.22005, 0.34034, and 0.35523. The multi-phase system at z = 0.20701 traces warm gas containing O VI, Ne VIII, and a BLA. It has been studied by Savage et al. (2006, 2011a). The COS observations for the other four systems are discussed for the first time in this paper.

*The HE 0226-4110 OVI absorber at z = 0.01749.* The system is clearly detected in O VI λ1031 and H I λ1215. O VI λ1037 is strongly contaminated with O IV λ787 from an O VI system at z = 0.34034. There is possible C IV λλ1548, 1550 absorption but the derived parameters are very uncertain due to strong fixed pattern noise. The O VI λ1031 line is unresolved in the FUSE observations and we obtain b = 10±6 km s$^{-1}$. The H I and O VI lines are well aligned and the difference in their b values implies log T = 4.61 (+0.14, - 0.22) and b$_{NT}$ = 8±8 km s$^{-1}$. The temperature implies the plasma is likely photoionized.

*The HE 0226-4110 OVI absorber at z = 0.20701.* This is the multiphase Ne VIII/ O VI/ BLA absorber analyzed by Savage et al. (2006) and Savage et al. (2011a). The moderately ionized ions observed including C III, O III, N III and Si III analyzed by Savage et al. (2006) are well modeled with a single slab of cool photoionized gas with log U ~ -1.85, log N(H) ~ 18.7 , log n$_H$ ~ - 4.6, and [X/H] = -0.5±0.2. The absorption system also contains Ne VIII, O VI and a BLA that do not arise in the photoionized gas.

New fit results for of H I and O VI based on our new COS data extraction methods are listed in Table A11. The O VI with b = 36±1 km s$^{-1}$ at v = 0±1 km s$^{-1}$ is broader than the narrow H I component at v = 18±9 km s$^{-1}$ and b(H I) = 28±4 km s$^{-1}$. The O VI is likely associated with the BLA having v = 16±14 km s$^{-1}$ and b(H I) = 100±25 km s$^{-1}$. Using the difference in the BLA and O VI line widths we obtain b$_{NT}$ = 27± 6 km s$^{-1}$ and log T = 5.75 (+0.20, -0.37) which is close to the temperature, log T = 5.73, obtained from the ratio of Ne VIII to O VI (see Savage et al. 2006) and somewhat larger than the value log T = 5.41 (+0.18, -0.11) obtained by Savage et al. (2011a) in an earlier analysis of the COS observations using the earlier extraction methods.

At a temperature as high as log T = 5.75(+0.20, -0.37), CIE should be a good approximation for determining H$^+$/H. Therefore, log N(H) = 19.73 (+0.27, -0.63). Assuming an oxygen ionization correction from CIE we obtain [O/H] = -0.31 (+0.36, -0.10). The absorber probably occurs in the halo of a foreground L = 0.25L* disk galaxy with an impact parameter of 109 kpc identified by Mulchaey & Chen (2009).

*The HE 0226-4110 OVI absorber at z = 0.22005.* The narrow H I component with v = -2±1 km s$^{-1}$ and b = 29±1 km s$^{-1}$ is well aligned with the O VI absorption with v = 0±4 km s$^{-1}$ and b = 14±5 km s$^{-1}$. The difference in the H I and O VI Doppler parameters implies the detection of cool gas with log T = 4.62 (+0.09, -0.12) and b$_{NT}$ = 12±6 km s$^{-1}$. The O VI in this absorber is very likely photoionized.

*The HE 0226-4110 OVI absorber at z = 0.34034.* The H I λ1215 absorption reveals 4 components extending from -214 to -5 km s$^{-1}$. H I λ1025 only reveals the stronger H I component at -5 km s$^{-1}$. BLAs appear to be present at -214±7 and -21±13 km s$^{-1}$ with b = 52±8 and 85±19 km s$^{-1}$, respectively. The O VI absorption at v = 0±2 km s$^{-1}$, b = 22±2 km s$^{-1}$, and log N(O VI) = 13.90±0.03 is well detected in both lines of the doublet. Weak O VI absorption at -46 km s$^{-1}$ is tentative and not aligned with H I absorption.

O IV λ787 is present although it is blended with O VI λ1031 at z = 0.01749 and therefore has uncertain parameters. The COS observations of C III λ977 have low S/N and are confused by fixed pattern noise. C III absorption appears to be present over the velocity range from -100 to 20 km s$^{-1}$ but the properties of the absorption are very uncertain and not listed in the Table.

The O VI absorption at 0±2 km s$^{-1}$ could be aligned with the BLA at v = -21±13 km s$^{-1}$ or with the narrow H I absorption at v = - 5±2 km s$^{-1}$. If the O VI exists in the gas traced by the narrow H I component the difference in line width implies cool gas with log T < 3.73 where the O VI is probably produced by photoionization. This is the case we list in Table 4 that summarized the properties of the aligned absorbers. However, if the O VI and the BLA trace the same gas, the difference in line widths implies log T = 5.64 (+0.17, -0.28). In this case the O VI is probably created by collisional ionization in a warm plasma. The estimated temperature is large enough for the ionization of H I and O VI to be reasonably well explained by CIE. Assuming CIE, log N(H) = 19.49 (0.26, -0.51) and [O/H] = -1.02 (+1.49, -0.10).

*The HE 0226-4110 OVI absorber at z = 0.35523* The H I λ1215 absorption reveals 3 components extending from -154 to -1 km s$^{-1}$ with a BLA with b(H I) = 45±15 km s$^{-1}$ at -154±13 km s$^{-1}$. Single components to O VI and O IV are detected and aligned with the H I component at v = -1±2 km s$^{-1}$ with b (H I) = 34±2 km s$^{-1}$.

If the narrow O VI with b = 10±2 km s$^{-1}$ is associated with the aligned H I with b = 34±2 km s$^{-1}$, the difference in Doppler parameters implies log T = 4.83 (0.10, -0.13) and b$_{NT}$ = 5±4 km s$^{-1}$. The narrow O VI and O IV likely occur in photoionized gas.

## A12. O VI Absorption Systems Toward PKS 0405-123

The relatively low S/N FUSE and STIS spectra of PKS 0405-123 with z$_{em}$ = 0.573 have been supplemented with the very high quality COS observations. The existence of the STIS observations permits improved wavelength calibration of the COS observations. Eight O VI absorption systems are detected at z = 0.09192, 0.09658, 0.16716, 0.18293, 0.29770, 36160, 0.36335 and 0.49507. COS results for the multi-phase systems at z = 0.16716 and 0.49507 containing cool and warm gas have been presented by Savage et al. (2010) and Narayanan et al. (2012) and updated in this paper.

*The PKS 0405-123 O VI absorber at z = 0.09192.* Relatively broad O VI absorption is detected with b = 39±3 km s$^{-1}$ and log N(O VI) = 13.85±0.07 in the FUSE spectrum. However, the strong H I absorption is displaced in velocity by -21±5 km s$^{-1}$ from the O VI absorption. This H I absorption appears to be associated with the C III λ977 absorption at v = -2±3 km s$^{-1}$. The O VI must occur in a different gas phase with little or no associated H I absorption because there is no evidence in the high quality H I profile for a component near the 0 ±5 km s$^{-1}$ velocity of O VI. This is a good example of misaligned O VI and H I absorption. The H I and C III near -21 km s$^{-1}$ probably arise in photoionized gas.

*The PKS 0405-123 O VI absorber at z = 0.09658.* The single component of O VI absorption recorded in the FUSE spectrum with v = 0±4 km s$^{-1}$ and b = 23±6 km s$^{-1}$ is either associated with the narrow (b = 29±1 km s$^{-1}$) H I absorption at v = -2±1 km s$^{-1}$ or with the broad (b = 69±3 km s$^{-1}$) H I with v = 9±2 km s$^{-1}$. Although the alignment appears better with the narrow H I absorption, the FUSE to COS velocity calibration error of ~10 km s$^{-1}$ means that the association of the absorption with the BLA is equally likely. If the O VI is associated with the narrow H I absorption the difference in the H I and O VI line widths yields log T = 4.30 (+0.28, -1.02) and b$_{NT}$ = 23±7 km s$^{-1}$ which is consistent with a origin of the O VI in cool photoionized gas. However, if the O VI is instead associated with the BLA the difference in line width yields log T = 5.43±0.05 and b$_{NT}$ = 16±9 km s$^{-1}$ and an origin of the O VI is in warm collisionally ionized gas. With such a large value of log T, CIE is a good assumption for H I and O VI. Therefore, if the warm gas interpretation is correct, log N(H) = 19.69 (+0.08, -0.11) and log N(O) = 14.41±0.07 implying [O/H] = -1.98 (+0.11, -0.13). In the ionization summary Table 3 we list this absorber as a cool photoionized absorber but note in the footnote to the table that the warm absorber interpretation is equally likely.

*The PKS 0405-123 O VI absorber at z = 0.16716.* Savage et al. (2010) have discussed the implications of the COS observations of this Lyman limit system with log N(H I) = 16.45±0.05. The system has complex multiple component H I absorption extending over 212 km s$^{-1}$ with at least six components along with multi component O VI absorption and the detection of many lower states of ionization including O III, O I, C IV, C III, C II, N V, N III, N II, Si IV, Si III, Si II, Fe III, and Fe II. Photoionizaiton modeling of the cool gas traced by the low ions implies log T ~ 3.9, log U ~ -3.1, $n_H$ ~ $10^{-2.7}$ cm$^{-3}$, L ~ 1 kpc, P/k ~ 40 cm$^{-3}$ K, log N(H) ~18.9, with abundances [O/H] = 0.1±0.2, [Si/H] = -0.4±0.2, [N/H] = 0.4±0.2, and [Fe/H] = 0.0±0.2. The photoionized gas is relatively cool and relatively confined. The O VI is not produced in the cool plasma.

Using the new extraction of the PKS 0405-123 observations in this paper we produced new profile fit results for H I, O VI, C III and Si II listed in Table A12. The new extraction has higher S/N and higher resolution than the older extraction used by Savage et al. (2010). The H I absorption is so complex it was necessary to fix some of the H I component velocities in order to get the H I fit process to converge. The component structure observed in C III and Si II was used to fix the H I velocities at -89, -39, -1, and +45 km s$^{-1}$. In all six H I components were required spanning the velocity range from -167 to +45 km s$^{-1}$. The H I fit is not unique even though the low ionization metal lines do provide additional insight about where H I components should exist. Some of the H I component properties are more uncertain than the formal fit error would suggest because of the strong overlap among the different components.

O VI exhibits four relatively strong components. The component at -65 km s$^{-1}$ does not align with H I absorption. The components at -135 and 0 km s$^{-1}$ do align with H I absorption. However, in both cases the O VI absorption is broader than the H I absorption and therefore the O VI does not arise in the same gas. The origin of the ionization in these three O VI components is unknown.

The O VI component at -278±2 km s$^{-1}$ is unusual in that it is very broad and has no associated H I. With log N(O VI) = 13.85±0.01 and log N(H I) < 12.7 for b = 60 km s$^{-1}$ the value of N(H I)/N(O VI) < 0.071. Although this O VI absorber could be created by photoionization in gas with [O/H] ~ 0, L = 180 kpc and log $n_H$ ~ -5.5, it appears more likely that the absorber is tracing warm/hot gas. If the thermal and non-thermal contributions to the O VI broadening are the same $b_{TH}$(O VI) = 37 km s$^{-1}$ implying log T ~ 6.11 and log N(H) ~ 19.9 for [O/H] ~ 0.

*The PKS 0405-123 O VI absorber at z = 0.18293.* O VI has two absorption components at -71±2 and 0±1 km s$^{-1}$. It is possible the component at -71 km s$^{-1}$ is actually two components at -86±2 and − 52±2 km s$^{-1}$ (see footnote 7 to Table A12). The H I absorption is well described by four components spanning a velocity range of 280 km s$^{-1}$. C III is detected at -81±5 km s$^{-1}$ but its parameters are uncertain due to strong blending with other absorbers.

The well defined O VI component at v = 0 ±1 km s$^{-1}$ with b = 18±1 km s$^{-1}$ is probably associated with the H I component at v = -10±1 km s$^{-1}$ with b = 32±1 km s$^{-1}$. The difference in the H I and O VI line widths implies the detection of photoionized gas with log T = 4.65 (+0.04, -0.05) and $b_{NT}$ = 17±1 km s$^{-1}$. The O VI absorption at -71±2 km s$^{-1}$ is likely associated with the H I absorption at -78±1 km s$^{-1}$. In this case the small difference in the b values implies the O VI arises in cool photoionized gas with log T = 3.59 (0.51, -3.59).

This absorption system is very interesting because there are no known associated galaxies to L < 0.04L* for ρ < 250 kpc( see Table 5 and Johnson et al. 2013).

*The PKS 0405-123 O VI absorber at z = 0.29770*. This simple absorber contains two H I and one O VI components. The broad H I component with b = 61±2 km s$^{-1}$ is well aligned with the broad O VI component with b = 55±3 km s$^{-1}$ implying an origin in cool photoionized gas with log T = 4.65 (+0.20, -0.39) and a large amount of non-thermal broadening with b$_{NT}$ = 55±3 km s$^{-1}$.

*The PKS 0405-123 O VI absorber at z = 0.36160*. This is a kinematically complex absorption system with 5 H I components spanning a velocity range of 185 km s$^{-1}$. There are possibly 6 O VI components extending over 330 km s$^{-1}$, although the four weak components with v < -50 km s$^{-1}$ are only detected in the O VI $\lambda$1031 line because of blending in the O VI $\lambda$1037 absorber with H I 1215 at z = 0.16118, 0.16149 and 0.16175.

The strongest O VI component at v = 0±1 is not aligned with the nearest H I component at v = -25±12 km s$^{-1}$ although it is possible that the H I absorption component has a more complex structure. The narrow (b = 5±1 km s$^{-1}$) and weak O VI component at v = -29±1 km s$^{-1}$ is aligned with the broad (b = 40±3 km s$^{-1}$) H I absorption at v = -25±12 km s$^{-1}$. However, the difference in the line widths is too large for the absorbers to occur in the same plasma although the narrow line width for the O VI implies log T < 4.67 and that the O VI is probably photoionized.

The strongest H I absorption is in two components at v = - 182±3 and -171±3 km s$^{-1}$. Weak O VI absorption occurs at v = -179±2 km s$^{-1}$ with stronger absorption by O IV, O III, N III and Si III near -171 to -174 km$^{-1}$ and C III absorbs at -180±1 km s$^{-1}$. The velocity differences among these different ions are so small it is difficult to determine which of the two H I components they are associated with. However, both H I components have b values small enough to imply that the metal lines near -171 to -179 km s$^{-1}$ including O VI arise in photoionized gas. If the O VI at -179 km s$^{-1}$ arises in the H I absorber at v = - 181 km s$^{-1}$ the difference in line width between the H I and O VI implies photoionized gas with log T = 4.02(+0.48, -4.02) and and b$_{NT}$ = 26±3 km s$^{-1}$.

Therefore, for the six O VI lines in this system, two are probably produced in photoionized gas while the origins of the other four components can not be determined.

*The PKS 0405-123 O VI absorber at z = 0.36355*. This is a simple one component O VI, C III, and H I system. The H I $\lambda$1215 absorption is blended with ISM C I* $\lambda$1657.37. The strength of the blending, estimated from the C I* 1560.682 absorption is relatively small (see Footnote 19 to Table A12).

N V $\lambda\lambda$1238, 1242 is possibly detected. However the absorption is situated in segment A of the G160M grating which has low sensitivity and has fixed pattern noise structures similar in strength to the possible N V absorption.

The 11±2 km s$^{-1}$ velocity offset between O VI and H I is relatively small given the systematic calibration errors. Therefore, the O VI , C III and H I probably exist in the same cool photoionized gas with log T = 4.55(+0.11,-0.15) and b$_{NT}$= 9±1 km s$^{-1}$ based on the difference in the O VI and H I line widths.

*The PKS 0405-123 O VI absorber at z = 0.49507*. The COS observations of this multiphase absorber, analyzed by Narayanan et al. (2011), provide clear detections of H I $\lambda$1025, 972, Ne VIII $\lambda\lambda$770,780, O VI $\lambda\lambda$1031, 1037, O IV $\lambda$787, O III $\lambda$832, and C III $\lambda$977 in a complex multi-phase system. FUSE observations also provide measures of O V $\lambda$630. Table A12 lists our new measurements of H I, O VI and C III in this system based in our updated extraction techniques.

The lack of coverage of H I $\lambda$1215 except with low resolution FOS observation prevents a sensitive study of the complexity of the H I absorption although the H I $\lambda$1025 absorption reveals two components at v = 4±4 and 103±9 km s$^{-1}$. The component at 4 km s$^{-1}$ is a BLA with b = 51±5 km s$^{-1}$ and log N(H I) = 14.14±0.03. O VI has two components at v = 0±2 and 50±2 km s$^{-1}$. The O VI component at v = 0±2 km s$^{-1}$ is well aligned with the BLA and their different line widths imply gas with log T = 5.00 (+0.12, -0.18) if both species co-exist in the same plasma. With this assumption, the origin of the O VI would be in cooling collisionally ionized gas with delayed recombination where log N(H) = 18.91 (+0.34, -0.58) assuming CIE for the H I. However, this is a complex multi-phase absorber with cool and warm gas where NeVIII has been detected and the true origin of the ionization in the multiple gas phases is likely much more complex. The analysis of Narayanan et al. (2011) suggested the O VI and Ne VIII could arise in warm gas with log T ~ 5.7. However, the BLA associated with gas that hot would have b ~ 100 km s$^{-1}$ rather than the value of 51 km s$^{-1}$ for the BLA actually detected. The system also contains cool photoionized gas traced by H I, C III, O III, and O IV. Some of the O VI could also arise in the cool gas. With the kinematical overlap of the absorption produced by so many gas phases, it is difficult to separately determine the properties of the gas in any individual phase.

The second O VI absorber at v = 50±2 km s$^{-1}$ with log N(O VI) = 13.84±0.05 is not aligned with an absorption component of H I. This O VI component is seen in O V, weakly in O IV and possibly in Ne VIII. It is probably tracing warm collisionally ionized gas although photoionization in a very low-density plasma with near solar abundances is also possible.

### A13. O VI Absorption Systems Toward HE 0238-1904

Three O VI intervening absorption systems at z = 0.40107, 0.42430, and 0.47204 are detected in the COS spectrum of HE 0238-1904 with z$_{em}$ = 0.631. No STIS observations exist for this QSO so the wavelength calibration of the COS observations is uncertain. The FUSE observations have relatively low S/N and was not used to search for low z O VI.

*The HE 0238-1904 O VI absorber at z = 0.40107.* This is a single component O VI system detected in O VI λ1031. O VI λ1037 is strongly blended with possible H I λ1215 absorption at z = 0.19598. A single H I absorption component is well fitted to the H I λλ1215-972 absorption. The H I absorption with b = 31±1 km s$^{-1}$ is well aligned with the O VI absorption with b = 22±2 km s$^{-1}$. Possible C III λ977 absorption is strongly blended with H I λ1025 at z = 0.33433 so we only report a limit to log N(C III). The difference in line widths for the well-aligned H I and O VI absorption implies log T = 4.49 (+0.09, -0.11) and b$_{NT}$ = 21±2 km s$^{-1}$ if the two species exist in the same gas. The O VI in this system is likely produced by photoionization in cool gas.

*The HE 0238-1904 O VI absorber at z = 0.42430.* This O VI system reveals two well modeled components to the lines of H I λλ1215-949, O VI λ1031, 1037, O IV λ787, and O III λ 832. The absorption by C III λ977 suggests three components. O VI λ1031 is weakly blended with H I λ1215 at z = 0.20843. The O VI at v = 0±1 km s$^{-1}$ with b = 23± 1 km s$^{-1}$ is well aligned with H I at 0±1 km s$^{-1}$ with b = 24±1 km s$^{-1}$. O IV, O III and C III also likely arise in this gas with absorption components at 7±10, 1±3 and -5±3 km s$^{-1}$. If the H I and O VI coexist in the same plasma the implied temperature is log T = 3.48 (+0.38, -3.48) and b$_{NT}$ = 23±1 km s$^{-1}$. Somewhat higher temperatures are obtained from the line widths of H I/ O IV and H I/ O III implying a structure more complex than a uniform temperature slab in photoionization equilibrium. However, given the good alignment of these ions and the H I b values suggesting cool gas it is likely the O VI is produced by photoionization in a cool plasma.

The origin of the O VI at -85±2 km s$^{-1}$ with b = 48±2 km s$^{-1}$ and log N(O VI) = 14.36±0.01 is unknown since it is not aligned with H I absorption.

*The HE 0238-1904 O VI absorber at z = 0.47204.* Two O VI components are clearly detected in this system at 0±1 and 54±1 km s$^{-1}$. However, only one COS extraction covers H I λ1215 with S/N ~ 12. H I components are detected at v = -134±5 and 14±4 km s$^{-1}$. The component at v = 14±4 km s$^{-1}$ with b = 47±4 km s$^{-1}$ is approximately aligned with the O VI component at 0±1 km s$^{-1}$ with b = 20±1 km s$^{-1}$. If these parameters are correct and the O VI and H I co-exist in the same plasma the difference in line widths implies log T = 5.06 (+0.08, -0.10) suggesting that the O VI is produced by collisional ionization in cooling gas with log N(H) = 18.77 (+0.21, -0.30) assuming CIE for the H I. Unfortunately, given the low S/N of the H I 1215 observation, it is not possible to determine if the H I component at 14±4 km s$^{-1}$ is actually two H I components with velocity separations similar to that seen in the O VI absorption.

The O VI component at 54±1 km s$^{-1}$ with b = 13±1 km s$^{-1}$ and log N(O VI) = 13.81±0.02 is not aligned with H I absorption. The O VI line width only implies log T < 5.3. Therefore, the O VI could be produced either by photo or collisional ionization.

### A14. O VI Absorption Systems Toward 3C 263

Five O VI intervening absorption systems are detected in the FUSE and COS spectra of 3C 263 with z$_{em}$= 0.652. Results for the systems at z = 0.06342, 0.14072 and 0.32567 have been presented by Narayanan et al. (2009, 2012) and Savage et al. (2012) and are briefly summarized below. The systems at z = 0.11389 and 0.44672 are presented for the first time in this paper. Although there is a complex O VI system at 0.52683 containing H I, O VI, O IV, O III, and C III, we do not consider it in this paper because the crucial H I λ1215 measurement is outside the COS wavelength coverage making it very difficult to evaluate the behavior of H I and to search for a BLA.

*The 3C 263 O VI absorber at z = 0.06342.* The multi-phase system at z = 0.06342 studied by Savage et al. (2012) traces cool photoionized gas and warm collisionally ionized gas associated with a L ∼ 0.31L* compact spiral emission line galaxy with an impact parameter of 63 kpc. The cool photoionized gas in the absorber is detected in H I, C II, Si II, and Si III and is well modeled with log U ∼ -2.6, log N(H) ∼17.8, log n(H) ∼ -3.3 and [Si/H] = -0.14±0.23. The collisionally ionized gas containing C IV and O VI probably arises in cooling shock heated transition temperature gas with log T ∼ 5.5. The H I associated with the gas containing O VI is not detected so the warm gas properties are uncertain. The absorber is likely tracing halo gas enriched by gas ejected from the spiral emission line galaxy.

*The 3C 263 O VI absorber at z = 0.11389.* The system at z = 0.11389 is detected in H I, O VI, C IV and C III. The H I λ1215 absorption traces two broad and one narrow absorption components at -219, -44, and 1 km s$^{-1}$. The properties of the BLA at -44 km s$^{-1}$ are uncertain because the component structure may be more complex than we have assumed. The BLAs at -219 and -44 km s$^{-1}$ are not detected in the metal lines. The O VI and C IV absorption is well aligned with the narrow H I absorption at 1 km s$^{-1}$ with b(H I) = 20±1 km s$^{-1}$. However, the O VI absorber with b = 31±5 km s$^{-1}$ is considerably broader than the C IV with b = 11±1 km s$^{-1}$ implying the O VI and C IV trace different phases of the gas. The C IV is likely associated with the H I component having b = 20±1 km s$^{-1}$. The difference in the C IV and H I line widths implies log T = 4.26 (+0.09, -0.11) and b$_{NT}$ = 10 (+1, -2) km s$^{-1}$ in the plasma containing C IV. The C IV is likely produced by photoionization. The origin of the ionization of the O VI is unknown. We do not understand the large width of the C III λ977 absorption with b = 50±8 km s$^{-1}$ recorded at relatively low S/N by FUSE.

*The 3C 263 O VI absorber at z = 0.14072.* The system at z = 0.14072 studied by Savage et al. (2012) only contains O VI and broad and narrow H I. The O VI with log N(O VI) = 13.60±0.09 and b(O VI) = 33±12 km s$^{-1}$ is likely associated with the well aligned BLA with log N(H I) = 13.47±0.10 and b (H I) = 87±15 km s$^{-1}$. The difference in Doppler parameters between O VI and H I implies the detection of a very large column of warm gas with log T = 5.62(+0.15, -0.24), log N(H) = 19.55(+0.23, -0.42)  and [O/H] = -1.45 (+0.41, -0.14).

*The 3C 263 O VI absorber at z = 0.32567.*  Narayanan et al (2010) detected Ne VIII, O IV, O III, C IV and N IV in this absorber and Narayanan et al. (2012) subsequently obtained high quality COS observations of the system revealing absorption by H I λλ1215 to 923, O VI, C III, C II,  N III,  and Si III.   The fit results for the updated extraction of our investigation are given in Table A14 and illustrated in Figure 10 for H I, O VI and C III.

 H I has two strong absorption components at -17±2 and 17±1 km s$^{-1}$ with b(H I) = 24±1 and 11±1 km s$^{-1}$.   C III is detected in both components while the other low ions are only detected in the 17 km s$^{-1}$ component.  Strong O VI is detected with v = 0±2 km s$^{-1}$, log N(O VI) = 14.01±0.01 and b(O VI) = 34±2 km s$^{-1}$.  The O VI absorption is not aligned with the narrow H I absorption components or with the low ionization metal line absorption.

All metal ions with the exception of O VI and Ne VIII are consistent with an origin in cool gas photoionized by the extragalactic EUV background.  The O VI and Ne VIII favor an origin in warm collisionally ionized gas with log T ~ 5.7.  The BLA associated with this absorption was only marginally detected by Narayanan et al. (2012) at 0 km s$^{-1}$ with log N(H I) = 13.25±0.17: and b = 86±6: km s$^{-1}$.  Assuming the O VI and BLA trace the same warm gas we obtain log T = 5.61(+0.07, -0.08)  and b$_{NT}$ =26±3 km s$^{-1}$.  Assuming CIE for the H I and O VI, log N(H) = 19.31(+0.11, -0.13) and [O/H] = -1.12 (+0.20, - 0.18).   Since the BLA detection is marginal, we do not list it in Tables 2, 4, and A14.

*The 3C 263 O VI absorber at z = 0.44672.*  This is a kinematically simple system detected in H I, O VI, O IV, O III and C III.   The lines of H I λλ1215, 1025 reveal two absorption components.  A BLA is detected at v = -35±12 with b = 71±5 km s$^{-1}$ and log N(H I) = 13.72±0.07.  No metal line absorption is observed at this velocity.  The H I absorption component at v = 9±1 km s$^{-1}$ and b = 32±1 km s$^{-1}$ is well aligned with the absorption produced by O VI, O IV, O III, and C III.  The b values of  O VI, O IV and O III, 32±3, 41±6, and 25±5 km s$^{-1}$ are the same at the ±2σ level.  It appears all three oxygen ions may exist in the gas producing the narrow H I absorption.  The gas is likely photoionized since the temperature constraint provided by the H I and O VI line widths is log T < 4.21 with b$_{NT}$ < 34 km s$^{-1}$.

Table A1
Profile Fit Results for MRK 290 ($z_{em}$ = 0.030)

| Ion | $\lambda s$ (Å) | v (km s$^{-1}$) | b (km s$^{-1}$) | log N(cm$^{-2}$) | Note |
|---|---|---|---|---|---|
| MRK 290 , z = 0.01026 (also see Narayanan et al. 2010b) | | | | | |
| H I | 1215 | -3±1 | 53±2 | 14.38±0.01 | 1, 2 |
| H I | 1215 | 110±1 | 32±1 | 14.08±0.01 | 1, 2 |
| O VI | 1031, 1037 | 0±7 | 31±10 | 13.65±0.10 | 1, 3 |

Notes: (1) This system has been carefully studied by Narayanan et al. (2010b). The fit results listed are based on our new extraction with the COS LSF from Kriss (2011). (2) H I λ1025 is strongly contaminated by ISM C II λ1036 and is not shown. (3) O VI λ1031 is contaminated at -240 km s$^{-1}$ by O I* λ1304.85 airglow emission; at 252 km s$^{-1}$ by ISM H$_2$ L 5-0 P(3) λ1043.503; from 100 to 200 km s$^{-1}$ by FPN. O VI λ1037 is contaminated at -40 km s$^{-1}$ by ISM Ar I λ1044 and at 283 km s$^{-1}$ by H$_2$ L 4-0 R(0) λ1049.367.

Table A2
Profile Fit Results for PKS 2155-305 ($z_{em}$ = 0.117)

| Ion | $\lambda s$ (Å) | v (km s$^{-1}$) | b (km s$^{-1}$) | log N(cm$^{-2}$) | Note |
|---|---|---|---|---|---|
| PKS 2155-304 , z = 0.05423 | | | | | |
| H I | 1215 | -76±8 | 28±4 | 13.68±0.41 | 1. 2 |
| H I | 1215 | -34±27 | 39±29 | 13.67±0.46 | 1, 2 |
| H I | 1215 | 29±10 | 21±14 | 12.76±0.47 | 1, 2 |
| H I | 1215 | 88±4 | 23±6 | 12.70±0.07 | 1 |
| O VI | 1031, 1037 | 0±3 | 24±5 | 13.53±0.06 | 1, 3 |
| PKS 2155-304, z = 0.05722 | | | | | |
| H I | 1215 | -176±1 | 50±1 | 14.32±0.01 | 1 |
| H I | 1215 | -38±1 | 68±1 | 14.09±0.01 | 1 |
| O VI | 1031 | 0±3 | 27±3 | 13.65±0.04 | 1, 4 |

Notes. (1) The new LSF for lifetime position 2 at central wavelength 1309 Å is used. (2) The large column density and velocity errors on the COS fitted components at -76, -34 and 29 km s$^{-1}$ result from the very strong overlap of two relatively broad features at the COS resolution. (3) The O VI λ1037 line is affected by bad instrumental pixels near line center. The fit result is based on the O VI λ1031 line. (4) The O VI λ1037 line in this system is contaminated at -29 km s$^{-1}$ by ISM Fe II λ1096.

Table A3
Profile Fit Results for MRK 876 ($z_{em}$ = 0.129)

| Ion | $\lambda$s (Å) | v (km s$^{-1}$) | b (km s$^{-1}$) | log N(cm$^{-2}$) | Note |
|---|---|---|---|---|---|
| | | MRK 876 , z = 0.00315 | | | |
| H I | 1215, 1025 | -14±2 | 58±2 | 14.00±0.02 | 1 |
| H I | 1215, 1025 | 34±2 | 25±4 | 13.30±0.11 | 1 |
| O VI | 1031, 1037 | 0±9 | 35:±13: | 13.38:±0.12: | 2 |
| | | MRK 876, z = 0.01171 | | | |
| H I | 1215 | -30±1 | 22±1 | 13.86±0.04 | 3 |
| H I | 1215 | -26±2 | 48±4 | 13.48±0.09 | 3 |
| O VI | 1031 | 0±15 | 22±9 | 13.26:±0.23: | 4 |

Notes. (1) The H I $\lambda$1215 is asymmetric and is well fitted by a narrow and broad component. (2) O VI $\lambda$1031 is contaminated at -14 km s$^{-1}$ with $H_2$ L 6-0 P(4) $\lambda$1035.182. O VI $\lambda$1037 is contaminated by $H_2$ as follows: -253 km s$^{-1}$, $H_2$ L 6-0 P(5) $\lambda$1040.058; -197 km s$^{-1}$, $H_2$ L 5-0 P(2) $\lambda$1040.367; -165 km s$^{-1}$, $H_2$ L 5-0 P(2) $\lambda$1040.367; 31 km s$^{-1}$, $H_2$ L 5-0 R(3) $\lambda$1041.158; and 63 km s$^{-1}$, $H_2$ L 5-0 R(3) $\lambda$1041.158. O VI $\lambda$1031 absorption is present but its fit parameters are uncertain. (3) The H I $\lambda$1025 absorption which is not plotted is strongly contaminated by ISM O VI $\lambda$1037. The H I $\lambda$1215 absorption requires a narrow and broad component to achieve an acceptable fit. (4) The O VI $\lambda$1031 absorption is contaminated by $H_2$ absorption at -194 km s$^{-1}$, $H_2$ L 5-0 P(3) $\lambda$1043.503; -162 km s$^{-1}$, $H_2$ L 5-0 P(3) $\lambda$1043.503; 104 km s$^{-1}$, $H_2$ L 5-0 R(4) $\lambda$1044.543; and 136 km s$^{-1}$, $H_2$ L 5-0 R(4) $\lambda$1044.543. The O VI $\lambda$1037 absorption is contaminated by $H_2$ absorption at -162 km s$^{-1}$, $H_2$ L 4-0 R(0) $\lambda$1049.367; -131 km/sec, $H_2$ L 4-0 R(0) $\lambda$1049.367; 6 km s$^{-1}$, $H_2$ L 4-0 R(1) $\lambda$1049.959; and 38 km s$^{-1}$, $H_2$ L 4-0 R(1) $\lambda$1049.959. There is a low significance detection of the O VI $\lambda$1031 line. Its parameters are very uncertain. The O VI does not align with the H I components.

Table A4
Profile Fit Results for 3C 273 ($z_{em}$ = 0.159)

| Ion | $\lambda$s (Å) | v (km s$^{-1}$) | b (km s$^{-1}$) | log N(cm$^{-2}$) | Note |
|---|---|---|---|---|---|
| | | 3C 273, z = 0.00336 | | | |
| H I | 1215 | -3±2 | 64±7 | 13.54±0.10 | 1 |
| H I | 1215 | 4±1 | 31±1 | 14.24±0.02 | 1 |
| O VI | 1031 | 0±7 | 43±11 | 13.39±0.08 | 2 |
| | | 3C 273, z = 0.09022 | | | |
| H I | 1215, 1025 | -122±2 | 48±3 | 13.24±0.02 | |
| H I | 1215, 1025 | -30±1 | 33±2 | 13.28±0.02 | |
| O VI | 1031, 1037 | 0±5 | 16±12 | 13.11±0.10 | 3 |
| C III | 977 | ... | ... | ... | 4 |
| | | 3C 273, z = 0.12007 | | | |
| H I | 1215, 1025 | -7±1 | 23±1 | 13.50±0.01 | |
| O VI | 1031, 1037 | 0 ±1 | 10±2 | 13.38±0.02 | |
| C III | 977 | ... | ... | ... | 5 |

Notes: (1) H I $\lambda$1215 requires a two component fit. (2) O VI $\lambda$1037 is blended with ISM $H_2$ at -215 km s$^{-1}$, $H_2$ L 5-0 P(2) $\lambda$1040.367; -177 km s$^{-1}$ $H_2$ L 5-0 P(2) $\lambda$1040.367; 11 km s$^{-1}$, $H_2$ L 5-0 R(3)

λ1041.158;  49 km s$^{-1}$, H$_2$ L 5-0 R(3) λ1041.158.   O VI λ1031 is blended at  -31 km s$^{-1}$, with H$_2$ L 6-0 P(4) λ1035.182.  This feature is very weak and probably an upper limit.  (3) O VI λ1031 is contaminated at 95 km s$^{-1}$  and 138 km s$^{-1}$  with ISM Fe II  λ1125. (4) C III λ977 is contaminated at  -167 km/sec by H$_2$ L 3-0 P(1) λ1064.605;  at  -129 km/sec by H$_2$ L 3-0 P(1) λ1064.605; at -58 km s$^{-1}$ by H$_2$ L 3-0 R(2) λ1064.994; and at  -20 km s$^{-1}$ H$_2$ L 3-0 R(2)  λ1064.994.  There is no evidence for C III in the O VI system. (5) C III λ977 is contaminated at -96 km s$^{-1}$ by  H I λ1025 at z = 0.06654;  at  -84 km s$^{-1}$ by H$_2$ L 1-0 P(1) λ1094.052;  at -46 km s$^{-1}$  with H$_2$ L 1-0 P(1)  λ1094.052; and at 6 km s$^{-1}$ with H$_2$ L 1-0 R(2) λ1094.244.  C III is not detected in the O VI system.

Table A5
Profile Fit Results for PG 1116+215 (z$_{em}$ = 0.177)

| Ion | λs (Å) | v (km s$^{-1}$) | b (km s$^{-1}$) | log N(cm$^{-2}$) | Note |
|---|---|---|---|---|---|
| PG 1116+215, z = 0.05897 | | | | | |
| H I | 1215 | -65±5 | 22±3 | 12.73±0.07 | |
| H I | 1215 | -3±1 | 33±2 | 13.57±0.01 | |
| H I | 1215 | 107±2 | 25:±3: | 12.78.±0.03: | 1 |
| O VI | 1031, 1037 | 0±12 | 27±16 | 13.49±0.05 | 2 |
| O VI | 1031, 1037 | 83±3 | 8±9 | 13.47±0.11 | 2 |
| C IV | 1548, 1550 | 82±2 | 12±3 | 12.73±0.05 | |
| PG 1116+215, z = 0.13850 | | | | | |
| H I | 1215 to 917 | -10f | 14±1 | 15.91±0.03 | 3, 4, 5 |
| H I | 1215 to 917 | 0f | 19±1 | 15.95±0.02 | 3, 4, 5 |
| H I | 1215 to 917 | 4±1 | 46±2 | 14.18±0.05 | 3, 4, 5 |
| O VI | 1031, 1037 | 0±1 | 36±2 | 13.81±0.01 | |
| C IV | 1548, 1550 | 3±3 | 20±4 | 13.14±0.05 | |
| C III | 977 | -4±1 | 7±1 | 15.33±0.53 | 6, 7 |
| C II | 1335 | -10f | 6±1 | 13.73±0.16 | 4, 8 |
| C II | 1335 | 0f | 5±3 | 13.41±0.12 | 4, 8 |
| N III | 989 | 1±2 | 16±3 | 14.01±0.05 | 7 |
| N II | 916, 1084 | -10f | 6±3 | 13.45±0.20 | 4, 8 |
| N II | 916, 1084 | 0f | 14±4 | 13.51±0.15 | 4, 8 |
| Si IV | 1393, 1402 | -10±1 | 3±1 | 13.07±0.23 | |
| Si III | 1206 | -10f | 3±1 | 12.68±0.27 | 4, 8 |
| Si III | 1206 | 0f | 4±2 | 13.18±0.64 | 4, 8 |
| Si II | 1260 | -10±1 | 5±1 | 12.63±0.10 | 8 |
| Si II | 1260 | 0±3 | 6±3 | 12.11±0.16 | 8 |
| PG 1116+215, z = 0.16553 | | | | | |
| H I | 1215, 1025 | -113±7 | 43±10 | 12.53±0.07 | 9 |
| H I | 1215, 1025 | -12±1 | 34±2 | 13.39±0.03 | |
| H I | 1215, 1025 | 117±7 | 100±7 | 13.55±0.01 | |
| H I | 1215, 1025 | 150±1 | 17±1 | 14.53±0.02 | |
| H I | 1215, 1025 | 182±3 | 14±2 | 13.38±0.10 | |
| H I | 1215, 1025 | 346±1 | 43±1 | 13.71±0.01 | 10 |
| H I | 1215, 1025 | 435±3 | 21±4 | 12.45±0.06 | |

| Ion | | v (km s⁻¹) | b (km s⁻¹) | log N(cm⁻²) | |
|-----|-----|-----|-----|-----|-----|
| O VI | 1031, 1037 | -131±2 | 18±3 | 13.17±0.04 | |
| O VI | 1031, 1037 | -38±14 | 40±9 | 13.72±0.16 | |
| O VI | 1031, 1037 | 0±1 | 19±2 | 13.93±0.09 | |
| O VI | 1031, 1037 | 178±5 | 51:±6: | 13.38:±0.04: | 11 |

Notes: (1) Only two COS integrations cover H I λ1215. The fit parameters for this weak absorber are uncertain. (2) The O VI λ1031absorption is contaminated at -204 km s⁻¹ with H₂ L 1-0 R(0) λ1092.195; at -121 km s⁻¹ with H I λ930.748 at z = 0.17361; at -57 km s⁻¹ with H₂ L 1-0 R(1) λ1092.732; and at 220 km s⁻¹ with H I λ937.803 at z = 0.16611. (3) STIS and COS observations simultaneously fitted. (4) Absorption line velocities indicated with f are fixed to the velocities measured from the two component Si II λ1260 STIS observations. (5) H I λ949 is contaminated at - 49 km s⁻¹ by H₂ L 2-0 P(2) λ1081.265, at 75 km s⁻¹ by H₂ L 2-0 R(3) λ1081.711, and at 123 and 154 km s⁻¹ by ISM Fe II λ1081. (6) C III λ977 is contaminated at -119 km s⁻¹ by ISM Fe II λ1112 and at - 1 km s⁻¹ by H₂ L 0-0 P(2) λ1112.495. (7) Fit to FUSE observations. (8) Fit to STIS observations. (9) H I λ1215 is contaminated at -196 km s⁻¹ by Si III λ1206 at z = 0.17361. (10) H I λ1025 is contaminated at 378 km/sec with ISM Mn II λ1197. (11) O VI λ1031 at 178 km s⁻¹ has uncertain parameters because of continuum placement uncertainty and blending with H I λ1025 at z = 0.17344 which is centered at 220 km s⁻¹. H I λ1025 at z = 0.17361 produces the strong line at 264 km s⁻¹ in the O VI λ1031 panel. The line at 381 km s⁻¹ in the O VI λ1037 panel is O VI λ1031 at z = 0.17344.

Table A6
Profile Fit Results for PHL 1811 (z_em= 0.201)

| Ion | λs (Å) | v (km s⁻¹) | b (km s⁻¹) | log N(cm⁻²) | Note |
|-----|-----|-----|-----|-----|-----|
| | | | PHL 1811, z = 0.07773 | | |
| H I | 1215 to 949 | 3±4 | 82±10 | 13.60±0.07 | 1 |
| H I | 1215 to 949 | 8±1 | 19±1 | 15.80±0.09 | 1 |
| O VI | 1031, 1037 | 0±9 | 43±7 | 13.53±0.08 | 2 |
| C IV | 1548 | … | … | … | 3 |
| C III | 977 | -19±4 | 4±13 | 13.19±2.44 | 4 |
| C III | 977 | 14±1 | 11±3 | 13.63±0.14 | 4 |
| C II | 1334 | 17±3 | 25±3 | 13.29±0.02 | |
| Si III | 1206 | 7±1 | 14±1 | 12.58±0.02 | 5 |
| Si II | 1260 | 4±2 | 18±2 | 12.34±0.03 | |
| | | | PHL 1811, z = 0.13280 | | |
| H I | 1215, 1025 | -133±1 | 29±4 | 14.67±0.11 | |
| H I | 1215, 1025 | -117±24 | 81±40 | 13.57±0.25 | 6 |
| H I | 1215, 1025 | 1±4 | 21±8 | 12.82±0.21 | 7 |
| H I | 1215, 1025 | 74:±15: | 49:±13: | 12.65:±0.15: | 8 |
| O VI | 1031, 1037 | -133±1 | 30±3 | 13.74±0.04 | 9 |
| O VI | 1031, 1037 | -81±3 | 27±6 | 13.51±0.06 | 9 |
| O VI | 1031, 1037 | 0±2 | 23±3 | 13.68±0.03 | |
| O VI | 1031, 1037 | 85±2 | 25±3 | 13.57±0.03 | 8 |
| | | | PHL 1811, z = 0.15785 | | |
| H I | 1215 | -7±7 | 97±10 | 13.25±0.04 | 10 |
| O VI | 1031, 1037 | -87±3 | 17±4 | 13.19±0.06 | 11 |
| O VI | 1031, 1037 | 0±2 | 31±3 | 13.68±0.02 | 11 |

| PHL 1811, z = 0.17651 | | | | | |
|---|---|---|---|---|---|
| H I | 1215, 972 | -31±6 | 26±3 | 13.74±0.11 | 12 |
| H I | 1215, 972 | 15±1 | 22±1 | 14.88±0.03 | 12 |
| O VI | 1037 | 0±2 | 20±2 | 14.14±0.03 | 13 |
| C III | 977 | -31±10 | 47±7 | 13.13±0.08 | |
| C III | 977 | 2±1 | 8±4 | 13.57±0.41 | 14 |

Notes: (1) H I $\lambda$1215 is contaminated at 324 km s$^{-1}$ with H I $\lambda$1215 at z = 0.07773 and 0.07890. H I $\lambda$972 is contaminated at 16 km s$^{-1}$ by ISM Ar I $\lambda$1048. The fitted line strength is determined from the Ar I $\lambda$1066 line. H I $\lambda$972 is also contaminated at -172 km s$^{-1}$ by H$_2$ L 5-0 P(4) $\lambda$1047.551; at 169 and 209 km s$^{-1}$ by C III $\lambda$977 at z = 0.07339 and 0.07354; at >240 km s$^{-1}$ by very strong H$_2$ L 4-0 R(0) $\lambda$1049.367. (2) O VI $\lambda$1031 is contaminated at -30 km s$^{-1}$ by ISM Fe II $\lambda$1112; at 87 km s$^{-1}$ by H$_2$ L 0-0 P(2) $\lambda$1112.495; at 111 km s$^{-1}$ by H$_2$ L 0-0 R(3) $\lambda$1112.583. O VI $\lambda$1037 is contaminated at -222 km s$^{-1}$ by H I $\lambda$949 at z = 0.17657. (3) C IV $\lambda$1548 is not detected. The contamination at 191and 222 km s$^{-1}$ is from ISM Al II $\lambda$1670. (4) C III $\lambda$977 is contaminated at 82 km s$^{-1}$ by H$_2$ L 4-0 P(2) $\lambda$1053.284; at 279 km s$^{-1}$ by H$_2$ L 4-0 R(3) $\lambda$1053.976. Lacki & Charlton (2010) studied C III and used 15 km s$^{-1}$ for the FUSE LSF width rather than the value 25 km s$^{-1}$ adopted here. The very large errors on the C III column densities are the result of line saturation effects produced by the large errors on the b values. If we instead adopt a resolution of 15 km s$^{-1}$ the C III column densities we obtain are consistent with those of Lacki & Charlton (2010). (5) Si III $\lambda$1206 is contaminated at 223 km s$^{-1}$ by H I $\lambda$1215 at z = 0.07040. (6) The properties of the BLA at -117±24 km s$^{-1}$ are very uncertain because of the strong overlap with the much stronger H I absorption at -133±1 km s$^{-1}$. (7) The weak H I absorption at 1±4 km s$^{-1}$ has uncertain properties because of overlap with the BLA at -117±24 km s$^{-1}$. (8) There is only marginal evidence for H I absorption associated with the O VI absorption at 85±2 km s$^{-1}$. The H I at 74:±15: km s$^{-1}$ with log N(H I) = 12.65:±0.15: is possibly present. (9) The COS observations clearly reveal two O VI $\lambda$1031 absorbers in the velocity range from -40 to -170 km s$^{-1}$ with the strongest absorption centered near -133 km s$^{-1}$. However, the derived parameters for these absorbers are uncertain because of their strong overlap. A 3$\sigma$ upper limit to the amount of H I at that velocity with b <50 km s$^{-1}$is log N(H I) < 13.0. We do not consider the feature listed as H I at 74±15 km s$^{-1}$ with log N = 12.65±0.15 a 3$\sigma$ detection. (10) The strong narrow line at the center of this absorption system at -5 km s$^{-1}$ is from O I $\lambda$1302 in a LLS at z = 0.08092. The profile fit of the O I $\lambda$1302 absorption includes a simultaneous fit to the O I $\lambda\lambda$988, 1039 lines observed by FUSE. (11) The absorption from -60 to -150 km s$^{-1}$ in the O VI $\lambda$1031 panel is probably O VI at -87±3 km s$^{-1}$. O VI $\lambda$1031 is contaminated at -215 and -138 km s$^{-1}$ by C I $\lambda$1193.995 and C I* $\lambda$1194.301, respectively. The O VI $\lambda$1037 observation is contaminated at -84 km s$^{-1}$ by ISM Mn II $\lambda$1201 and at <-140 km s$^{-1}$ by ISM N I $\lambda$1200. The strength of the Mn II $\lambda$1201 is determined from Mn II $\lambda$1197. (12) H I $\lambda$1025 is not plotted because it is strongly contaminated by ISM Si III $\lambda$1205. H I $\lambda$972 is contaminated at -261 and 187 km s$^{-1}$ by Fe II $\lambda$1143, 1144, respectively. (13) O VI $\lambda$1031 is contaminated by H I $\lambda$1215 geocoronal emission and ISM absorption. (14) The C III column density for the narrow component is very uncertain due to line saturation effects given the large error for the small b value. C III $\lambda$977 is contaminated at -40 km s$^{-1}$ by H I $\lambda$1025 at z = 0.12050 and at -68 km s$^{-1}$ by Fe II $\lambda$1063 at z = 0.08092.

Table A7
Profile Fit Results for PG 0953+414 ($z_{em}$= 0.239)

| Ion | $\lambda$s (Å) | v (km s$^{-1}$) | b (km s$^{-1}$) | log N(cm$^{-2}$) | Note |
|---|---|---|---|---|---|
| | | PG 0953+414, z = 0.00212 | | | |
| H I | 1215 | -1±2 | 38±3 | 13.15±0.03 | |
| O VI | 1031,1037 | 0±6 | 43±8 | 13.60±0.06 | 1 |
| | | PG 0953+414, z = 0.06808 | | | |
| H I | 1215, 1025 | -2±1 | 19±1 | 14.34±0.05 | |
| H I | 1215, 1025 | 32±23 | 37±15 | 13.41±0.30 | 2 |
| O VI | 1031, 1037 | 0±1 | 12±1 | 14.29±0.03 | |
| N V | 1238, 1242 | -4±1 | 8 ±1 | 13.42±0.02 | 3 |
| C IV | 1548, 1550 | -5±1 | 9±1 | 13.91±0.05 | |
| C III | 977 | -3±2 | 19±3 | 13.20±0.04 | 4 |
| | | PG 0953+414, z = 0.14231 | | | |
| H I | 1215 | -138±3 | 20±5 | 12.59±0.06 | |
| H I | 1215 | 0±1 | 26±1 | 13.57±0.01 | |
| H I | 1215 | 82±1 | 29±1 | 13.47±0.01 | |
| H I | 1215 | 211±3 | 58±5 | 13.21±0.02 | |
| O VI | 1031, 1037 | 0±1 | 19±1 | 14.09±0.01 | |
| O VI | 1031, 1037 | 78±3 | 29±4 | 13.60±0.03 | |
| C III | 977 | -4±2 | = 9 | 12.51±0.07 | 5 |
| C III | 977 | 96±1 | 9±2 | 12.93±0.04 | |

Notes. (1) O VI $\lambda$1037 absorption is contaminated at -314 km s$^{-1}$ by H I $\lambda$972 at z = 0.06807, at -226 and -181 km s$^{-1}$ by ISM O I $\lambda$1039 absorption, at -130 km s$^{-1}$ by geocoronal O I $\lambda$1039 emission, and at 135 and 166 km s$^{-1}$ by H$_2$ L 5-0 P(2) $\lambda$1040.365. If the O VI $\lambda$1031 absorption is fitted alone we obtain v = 0±7 km s$^{-1}$, b = 41±10 km s$^{-1}$ and log N = 13.54±0.08. (2) This weak absorber has uncertain properties because of its strong blending with the much stronger H I absorption near -2±1 km s$^{-1}$. (3) N V $\lambda$1242 is contaminated at 105 km s$^{-1}$ by H I $\lambda$1215 at z = 0.09230 and at 290 km s$^{-1}$ by H I $\lambda$1215 at z = 0.09316. (4) C III $\lambda$977 is contaminated at -29 km s$^{-1}$ and 3 km s$^{-1}$ by weak H$_2$ L 5-0 P(3) $\lambda$1043.503 absorption. (5) The b value for C III $\lambda$977 at -4±2 km s$^{-1}$ was fixed at 9 km s$^{-1}$ to obtain a reasonable error on log N(C III).

Table A8
Profile Fit Results for H 1821+643 ($z_{em}$ = 0.297)

| Ion | $\lambda$s (Å) | v (km s$^{-1}$) | b (km s$^{-1}$) | log N(cm$^{-2}$) | Note |
|---|---|---|---|---|---|
| | | H 1821+643, z = 0.02443 | | | |
| H I | 1215 | -12±1 | 29±1 | 14.24±0.01 | 1 |
| O VI | 1031 | 0±5 | 23±7 | 13.42±0.09 | 2 |
| | | H 1821+643, z = 0.12141 | | | |
| H I | 1215 | -62±1 | 42±1 | 14.23±0.02 | |
| H I | 1215 | 16±9 | 82±7 | 13.80±0.05 | |
| H I | 1215 | 207±2 | 44±2 | 13.16±0.02 | |
| O VI | 1031 | 0±6 | 58±7 | 13.69±0.04 | 3 |
| H 1821+643, z = 0.17036 | | | | | |
| H I | 1215 | -179±1 | 33±1 | 13.86±0.01 | |
| H I | 1215 | -87±2 | 54±2 | 13.65±0.01 | |
| H I | 1215 | 33±1 | 30±1 | 13.39±0.01 | |
| O VI | 1031 | -95±8 | 31±12 | 13.20±0.22 | 4 |
| O VI | 1031 | 0±15 | 79:±17: | 13.69±0.08 | 4 |
| H 1821+643, z = 0.21329 | | | | | |
| H I | 1215, 1025 | -6±1 | 40±1 | 14.42±0.01 | 5 |
| O VI | 1031 | 0±2 | 26±2 | 13.49±0.02 | 6 |
| O IV | 787 | 10±7 | 10±13 | 14.12±0.33 | 7 |
| H 1821+643, z = 0.22497 (also see Narayanan et al. 2010) | | | | | |
| H I | 1215 to 930 | -103±8 | 35±6 | 13.97±0.13 | 8 |
| H I | 1215 to 930 | -40±2 | 26±2 | 15.24±0.03 | 8 |
| H I | 1215 to 930 | -23±17 | 95±11 | 13.94±0.17 | 8 |
| H I | 1215 to 930 | 18±2 | 18±1 | 15.28±0.03 | 8 |
| H I | 1215 to 930 | 293±3 | 58±4 | 13.51±0.02 | 8 |
| O VI | 1031, 1037 | 0±1 | 45±1 | 14.25±0.01 | |
| O VI | 1031, 1037 | 60±1 | 5±2 | 13.11±0.04 | |
| O VI | 1031, 1037 | 343±1 | 16±1 | 13.48±0.01 | 9 |
| H 1821+643, z = 0.24531 | | | | | |
| H I | 1215 | -1±5 | 36±5 | 13.08±0.05 | 10 |
| O VI | 1031, 1037 | 0±1 | 27±1 | 13.76±0.01 | |
| H 1821+643, z = 0.26656 | | | | | |
| H I | 1215 | 4±2 | 46±2 | 13.64±0.02 | 10 |
| O VI | 1031, 1037 | 0±2 | 24±2 | 13.61±0.03 | |

Notes. (1) H I λ1215 is contaminated at -216 km s$^{-1}$ by strong H I λ1025 at z = 0.21326. (2) The feature at -209 km s$^{-1}$ in the O VI λ1031 panel is H$_2$ L 4-0 P(3) λ1056.471. O VI λ1037 is strongly contaminated with ISM Fe II λ1063.176 at -86 km s$^{-1}$ and 27 km s$^{-1}$, Fe II λ1062.151 at -261 km s$^{-1}$, Fe II λ1063.971 at 251 km s$^{-1}$, H$_2$ L 3-0 R(0) λ1062. 882 at -45 km s$^{-1}$, and H$_2$ L 3-0 R(1) λ1063.460 at 118 km s$^{-1}$. The O VI column density is based on O VI λ1031. (3) O VI λ1037 is strongly contaminated at -90 km s$^{-1}$ and -33 km s$^{-1}$ with H I λ949 at =0.22480 and 0.22504. The O VI column density is based on the O VI λ1031 absorption. (4) O VI λ1037 is strongly contaminated by ISM Lyα absorption and geocoronal emission. The column densities are based on O VI λ1031. O VI λ1031 is contaminated with H I λ930 at z = 0.29686 and 0.29668 at -166 km s$^{-1}$ and -209 km s$^{-1}$. For v < -220 km s$^{-1}$ ISM Si III λ1206 is the major contaminant. Two O VI absorbers are identified. Their properties are uncertain because of the strong overlap of the absorption lines. (5) Only one extraction with a 530 s integration time contributes to the COS observation of H I λ1215. There could be additional H I λ1215 absorption at ~80 km s$^{-1}$. The feature at 198 km s$^{-1}$ in the H I λ1025 panel is H I λ1215 at z = 0.02439. (6) O VI λ1037 is contaminated by ISM S II λ1259 at -30 to 175 km s$^{-1}$, and by ISM Si II 1260 at >175 km s$^{-1}$. (7) The O IV λ787 absorption is contaminated at -226, -25, and 248 km s$^{-1}$ by H$_2$ L 13-0 R(1) λ955.065, H$_2$ L 13-0 P(1) λ955.708, and H$_2$ L 13-0 R(2) λ956.579 absorption, respectively. The feature at -78 km s$^{-1}$ is unidentified. (8) Profile fits to FUSE H I λλ1025 to 930 observations and the high quality STIS observations of H I λ1215. The COS observations of H I λ1215 are of lower S/N than the STIS observations because only one short integration was obtained with the COS G160M grating. (9) The O VI component at 343±1 km s$^{-1}$ is a good example of an O VI absorber at a velocity where H I absorption is not evident. (10) STIS observations are used for H I λ1215 because the short COS G160M integration has lower S/N.

Table A9
Profile Fit Results for TON 236 ($z_{em} = 0.447$)

| Ion | λs (Å) | v (km s$^{-1}$) | b (km s$^{-1}$) | log N(cm$^{-2}$) | Note |
|-----|--------|-----------------|------------------|-------------------|------|
| | | TON 236, $z = 0.19452$ | | | |
| H I | 1215, 1025 | 1±1 | 49±1 | 14.05±0.01 | 1 |
| H I | 1215, 1025 | 83±3 | 16±5 | 12.89±0.08 | 1 |
| O VI | 1031, 1037 | 0±4 | 44±5 | 14.03±0.03 | 2 |
| | | TON 236, $z = 0.39944$ | | | |
| H I | 1215 | 16±1 | 45±1 | 14.10±0.01 | |
| O VI | 1031 | 0±21 | 61±18 | 13.74±0.03 | 3 |

Notes: (1) H I λ1215 is contaminated at 134 km s$^{-1}$ by H I λ1025 at z = 0.41624. H I λ1025 is contaminated at -210 km s$^{-1}$ by H I λ972 at z = 0.25896. (2) O VI λ1037 is contaminated at -213 km s$^{-1}$ by λ1215 at z = 0.01883 and at -126 km s$^{-1}$ by H I λ1025 at z = 0.20786. (3) O VI λ1037 is strongly contaminated at 10, 93 and 120 km s$^{-1}$ by H I λ1215 at z = 0.19451 and z = 0.19484 and H I λ1025 at z = 0.41624. The line identified as O VI λ1031 at z = 0.39944 could instead be H I λ1215 at z = 0.01387.

Table A10
Profile Fit Results for HE 0153-4520 ($z_{em}$ = 0.450)

| Ion | λs (Å) | v (km s$^{-1}$) | b (km s$^{-1}$) | log N(cm$^{-2}$) | Note |
|---|---|---|---|---|---|
| \multicolumn HE 0153-4520, z = 0.14887 | | | | | |
| H I | 1215 | -102±2 | 35±3 | 13.34±0.03 | 1 |
| H I | 1215 | -11±3 | 34±4 | 13.26±0.03 | 1 |
| O VI | 1031,1037 | -113±3 | 50±3 | 13.82±0.02 | 2 |
| O VI | 1031 | 0±1 | 25±1 | 14.02±0.01 | 2 |
| C IV | 1548, 1550 | -9±1 | 13±2 | 13.20±0.03 | 3 |
| HE 0153-4520, z = 0.17090 | | | | | |
| H I | 1215 to 972 | -40±2 | 108±2 | 13.88±0.01 | 4 |
| H I | 1215 to 972 | -8±1 | 34±1 | 14.35±0.01 | 4 |
| O VI | 1031 | -96±2 | 20±3 | 13.32:±0.07: | 5, 6 |
| O VI | 1031 | 0±1 | 22±2 | 13.82±0.01 | 6 |
| HE 0153-4520, z = 0.22203 | | | | | |
| H I | 1215 to 937 | 19±1 | 33±1 | 14.95±0.03 | 7 |
| H I | 1215 to 937 | 31±1 | 53±1 | 14.64±0.05 | 7 |
| O VI | 1037 | 0±2 | 34±4 | 13.69±0.04 | 8 |
| O VI | 1031,1037 | 99±2 | 26±3 | 13.71±0.03 | 8 |
| C III | 977 | 0±1: | 17±3: | 12.98±0.04: | 9 |
| HE 0153-4520, z = 0.22600  (see Savage et al. 2011) | | | | | |
| H I | 1215 to 926 | -12±1 | 28±1 | 16.61f±0.15 | 10 |
| H I | 1215 to 926 | -6±4 | 151±15 | 13.58±0.05 | 10 |
| O VI | 1031, 1037 | 0±1 | 37±1 | 14.23±0.01 | 11 |
| HE 0153-4520, z = 0.29114 | | | | | |
| H I | 1215 to 949 | -29±2 | 61±2 | 13.91±0.02 | |
| H I | 1215 to 949 | -1±1 | 20±1 | 14.22±0.02 | |
| O VI | 1031, 1037 | -53±3 | 31±3 | 13.78±0.04 | 12 |
| O VI | 1031, 1037 | 0±1 | 19±2 | 13.85±0.03 | 12 |
| HE 0153-4520, z = 0.40052 | | | | | |
| H I | 1215 to 949 | -63±1 | 42±1 | 14.25±0.01 | 13 |
| H I | 1215 to 949 | 3±1 | 19±1 | 14.27±0.01 | 13 |
| H I | 1215 to 949 | 66±5 | 24±7 | 12.67±0.10 | 13 |
| O VI | 1031, 1037 | -64±6 | 42±3 | 14.09±0.06 | |
| O VI | 1031, 1037 | -54±1 | 11±2 | 13.82±0.06 | |
| O VI | 1031, 1037 | 0±6 | 24±4 | 13.53±0.12 | |
| C III | 977 | -79±5: | 40±7: | 12.70±0.05: | 14 |
| C III | 977 | 1±1 | 5±3 | 12.92±0.12 | |

Notes: (1) H I λ1215 is contaminated at 199 km s$^{-1}$ by H I λ1215 at z = 0.14964.  (2) O VI λ1031 is weakly contaminated at -144 and -32 km s$^{-1}$ by H I λ1025 at z = 0.15527 and 0.15570, respectively. O VI λ1037 at -280 and -210 km s$^{-1}$ is contaminated by ISM Si II λ1190,  at 50 km s$^{-1}$ by H I λ972 at z = 0.22596,  at >240 km s$^{-1}$ by ISM Si II λ1193.  The properties of the component at 0±1 km s$^{-1}$ are based on the O VI λ1031 line alone. The properties of the component at -113±3 km s$^{-1}$ are determined

from both lines of the doublet.   (3) The C IV $\lambda\lambda$1548, 1550 identification is tentative. (4) H I $\lambda$1025 is contaminated by ISM N I $\lambda\lambda$1200.223, 1200.709 at v = -177 and -56 km s$^{-1}$, respectively, and at 180 km s$^{-1}$ by a combination of H I $\lambda$930 at z = 0.29114 and H I $\lambda$949 at z = 0.26540.  The H I profile fit is mostly determined from the H I $\lambda\lambda$ 1215, 972 measurements. (5) Although this absorber is affected by FPN, it appears real and the only reasonable ID is O VI.  (6) O VI $\lambda$1037 is contaminated by ISM H I $\lambda$1215 absorption.  O VI $\lambda$1031 is contaminated at <-200 km s$^{-1}$ by ISM Si III $\lambda$1206.  (7) H I $\lambda$1025 is contaminated with ISM S II $\lambda$1253 for v = 60 to 140 km s$^{-1}$.  H I $\lambda$949 is contaminated by H I $\lambda$1025 at z = 0.13205 for v from 80 to 200 km s$^{-1}$.  H I $\lambda$937 is contaminated at -270 km s$^{-1}$ with ISM Fe II $\lambda$1144. (8) O VI $\lambda$1031 is contaminated by ISM Si II $\lambda$1260 over the velocity range from -210 to 62 km s$^{-1}$, and at 88 km s$^{-1}$ by C III $\lambda$977 at z = 0.29109.  The O VI fit parameters are determined from the O VI $\lambda$1037 absorption and partially by O VI $\lambda$1031 since the ISM Si II $\lambda$1260 can be constrained by the weaker Si II $\lambda\lambda$1526, 1190, 1193 lines.  (9) C III $\lambda$977 is contaminated by ISM Si II $\lambda$1193 over the velocity range from -220 to 70 km s$^{-1}$.  The properties of the C III absorption are very uncertain.  (10) The H I column density for the narrow component was  fixed based on the careful analysis by Savage et al. (2011) of the properties of the partial Lyman limit system observed by FUSE.  The resulting fit results for the H I component velocities and b values are in excellent agreement with the values in Savage et al. (2011). The error on b(H I) and log N(H I) for the BLA was obtained by allowing log N(H I) for the narrow component to extend from 16.44 to 16.73, the range allowed by the fit to the partial Lyman limit absorption.  (11) We only list results for O VI.  Savage et al.  (2011) give the full details of the results for other metal lines detected  including Si IV, Si III, Si II, C III, C II, N III, and N II. (12) O VI $\lambda$1031 is contaminated at 152 and 213 km s$^{-1}$ by H I $\lambda$1215 at z = 0.09655 and 0.09677, respectively.  (13) H I $\lambda$972 is contaminated by H I $\lambda$1215 at z = 0.11981 for v < -100 km s$^{-1}$ and at  6 km s$^{-1}$ with H I $\lambda$1215 at z = 0.12043.  (14) The existence of C III $\lambda$977 at v = -79±5 km s$^{-1}$ is tentative.

Table A11
Profile Fit Results for HE 0226-4110 ($z_{em}$ = 0.493)

| Ion | $\lambda s$ (Å) | v (km s$^{-1}$) | b (km s$^{-1}$) | log N(cm$^{-2}$) | Note |
|---|---|---|---|---|---|
| | | HE 0226-4110, z = 0.01749 | | | |
| H I | 1215 | -8±2 | 27±4 | 13.20±0.04 | 1 |
| O VI | 1031 | 0±2 | 10±6 | 13.91±0.16 | 2 |
| C IV | 1548, 1550 | -11±7: | 32±5: | 13.31±0.06: | 3 |
| C IV | 1548, 1550 | 53±3: | 10±5: | 12.98±0.09: | 3 |
| | | HE 0226-4110, z = 0.20701   (See Savage et al. 2005, 2011a) | | | |
| H I | 1215 to 926 | -18±3 | 21±1 | 15.13±0.07 | |
| H I | 1215 to 926 | 16±14 | 100±25 | 13.45±0.16 | |
| H I | 1215 to 926 | 18±9 | 28±4 | 14.67±0.15 | |
| O VI | 1031, 1037 | 0±1 | 36±1 | 14.37±0.01 | 4 |
| | | HE 0226-4110, z = 0.22005 | | | |
| H I | 1215, 1025 | -2±1 | 29±1 | 14.39±0.01 | 5 |
| O VI | 1031, 1037 | 0±4 | 14±5 | 13.14±0.06 | 6 |
| HE 0226-4110, z = 0.34034 | | | | | |
| H I | 1215 | -214±7 | 52±8 | 13.19±0.05 | 7 |
| H I | 1215 | -130±3 | 9±5 | 12.76±0.11 | 7 |
| H I | 1215 | -21±13 | 85±19 | 13.39±0.06 | 7 |
| H I | 1215, 1025 | -5±2 | 19±3 | 13.47±0.05 | 7 |
| O VI | 1031, 1037 | -46±10 | 17±10 | 13.00±0.21 | 8 |
| O VI | 1031, 1037 | 0±2 | 22±2 | 13.90±0.03 | 8 |
| O IV | 787 | 0±17 | 26±18 | 13.89±0.23 | 9 |
| HE 0226-4110, z = 0.35523 | | | | | |
| H I | 1215 | -154±13 | 45±15 | 12.92±0.10 | 10 |
| H I | 1215 | -75±4 | 16±5 | 13.00±0.08 | |
| H I | 1215 | -1±2 | 34±2 | 13.70±0.02 | |
| O VI | 1031, 1037 | 0±1 | 10±2 | 13.33±0.03 | 11 |
| O IV | 787 | 3±10 | 10f | 13.74±0.20 | 12 |

Notes: (1) H I λ1215 is contaminated at 256 km s$^{-1}$ by H I λ1215 at z = 0.20694. (2) O VI λ1037 is contaminated at 12 km s$^{-1}$ with O IV λ787 at z = 0.34034 and by ISM Fe II 1055 at -132 km s$^{-1}$. The listed column density is based on the O VI λ1031 observation. The O VI line is unresolved. The large error for b causes the large column density error due to line saturation effects. (3) The C IV λλ1548, 1550 lines occur in a region of the spectrum strongly affected by fixed pattern noise. The existence of these features is questionable. We only list the measured parameters for completeness. (4) O VI λ1031 is contaminated at -78 km s$^{-1}$ by very weak H I λ972 at z = 0.28039 and at 218 km s$^{-1}$ by H I λ1215 at z = 0.02533. O VI λ1037 is contaminated at -295 km s$^{-1}$ by H I λ1215 at z = 0.02921, at −238 km s$^{-1}$ by H I λ1025 at z = 0.22004, at − 7 km s$^{-1}$ by H I λ1025 at z = 0.22098, and at 43 km s$^{-1}$ by H I λ1025 at z = 0.22119. (5) The continuum placement for H I λ1215 is uncertain. H I λ1215 is contaminated at 230 km s$^{-1}$ by H I λ1215 at z = 0.22098 and at 279 km s$^{-1}$ by H I λ1215 at z = 0.22119. H I λ1025 is contaminated at -192 km s$^{-1}$ by ISM S II 1250, at -60 km s$^{-1}$ by possible H I λ1215 at z = 0.02921, at 236 km s$^{-1}$ by O VI λ1037 z = 0.20701. (6) O VI λ1031 is contaminated at 236 km s$^{-1}$ by ISM S II λ1259. (7) Four H I components are detected in H I λ1215. The strongest narrow component at v = -5±2 km s$^{-1}$ is also detected in H I λ1025. (8) O VI λ1031 is contaminated at -168 km s$^{-1}$ by H I λ926 at z = 0.49246, at 140 km s$^{-1}$ by H I λ972 at z = 0.42286, and at 145 km s$^{-1}$ by H I λ1215 at z = 0.13831. O VI λ1037 is contaminated at -356 km s$^{-1}$ by H I λ930 at z = 0.49246, at -223 km s$^{-1}$ by H I λ1025 at z = 0.35489, and at -148 km s$^{-1}$ by H I λ1025 at z = 0.35521. (9) O IV λ787 is probably detected but it has very uncertain parameters because of strong blending at -11 km s$^{-1}$ with O VI λ1037 at z = 0.01749. O IV is also contaminated at -143 km s$^{-1}$ by ISM Fe II λ1055. (10) The H I λ1215 absorption at -154 km s$^{-1}$ is weak and has uncertain properties. (11) O VI λ1037 is contaminated at -290 km s$^{-1}$ with H I λ1215 at z = 0.15552 and at  -89 km s$^{-1}$ with possible H I λ1215 at z = 0.15639. O VI λ1031 is contaminated at 244 km s$^{-1}$ by H I λ937 at z = 0.49246 and at 280 km s$^{-1}$ by H I λ1215 at z = 0.15171. (12) The O IV λ787 b value was constrained to agree with the value observed for O VI. O IV λ787 is contaminated at -233 km s$^{-1}$ by ISM Ar I λ1066.

Table A12
Profile Fit Results for PKS 0405-123 ($z_{em} = 0.573$)

| Ion | λs (Å) | v (km s$^{-1}$) | b (km s$^{-1}$) | log N(cm$^{-2}$) | Note |
|------|--------|------|------|------|------|
| PKS 0405-123, z = 0.09192 | | | | | |
| H I | 1215, 1025 | -21±1 | 36±1 | 14.57±0.01 | |
| H I | 1215, 1025 | 50±6 | 23±6 | 12.78±0.13 | |
| O VI | 1031, 1037 | 0±5 | 39±3 | 13.85±0.07 | 1 |
| C III | 977 | -22±3 | 16±5 | 13.41±0.12 | 2 |
| PKS 0405-123, z = 0.09658 | | | | | |
| H I | 1215, 1025 | -2±1 | 29±1 | 14.58±0.02 | 3 |
| H I | 1215, 1025 | 9±2 | 69±3 | 13.93±0.04 | 3 |
| O VI | 1031, 1037 | 0±4 | 23±6 | 13.70±0.07 | |
| PKS 0405-123, z = 0.16716 (also see Savage et al. 2010) | | | | | |
| H I | 1215 to 916 | -167±19 | 54±13 | 13.11±0.16 | 4 |
| H I | 1215 to 916 | -136±1 | 12±1 | 13.35±0.05 | 4 |
| H I | 1215 to 916 | -89f | 27±5 | 13.56±0.08 | 4 |
| H I | 1215 to 916 | -39f | 16±1 | 15.41±0.06 | 4 |
| H I | 1215 to 916 | -1f | 26±1 | 16.45±0.02 | 4 |
| H I | 1215 to 916 | 45f | 41±3 | 13.90±0.08 | 4 |
| O VI | 1031, 1037 | -278±2 | 56±2 | 13.85±0.01 | 5 |
| O VI | 1031, 1037 | -135±2 | 29±2 | 13.89±0.04 | 5 |
| O VI | 1031, 1037 | -65±3 | 38±5 | 14.23±0.07 | 5 |
| O VI | 1031, 1037 | 0±3 | 39±2 | 14.45±0.03 | 5 |
| C III | 977 | -130±4 | 22±4 | 13.25±0.07 | |
| C III | 977 | -89±3 | 16±6 | 13.06±0.11 | |
| C III | 977 | -39f | 14 | 13.52±0.08 | |
| C III | 977 | -1f | 16 | 15.35±0.54 | |
| C III | 977 | 45±6 | 10±6 | 13.21±0.12 | |
| Si II | 1260 | -39±1 | 9 | 12.44±0.02 | 6 |
| Si II | 1260 | -1±1 | 10 | 13.28±0.01 | 6 |
| PKS 0405-123, z = 0.18293 | | | | | |
| H I | 1215, 1025 | -145±3 | 18±2 | 12.82±0.08 | |
| H I | 1215, 1025 | -78±1 | 31±1 | 14.73±0.01 | |
| H I | 1215, 1025 | -10±1 | 32±1 | 14.06±0.02 | |
| H I | 1215, 1025 | 135±2 | 21±2 | 12.63±0.03 | |
| O VI | 1031, 1037 | -71±2 | 30±2 | 13.64±0.02 | 7 |
| O VI | 1031, 1037 | 0±1 | 18±1 | 13.84±0.01 | 7 |
| C III | 977 | -81±5 | 20±4 | 13.13±0.12 | 8 |
| PKS 0405-123, z = 0.29770 | | | | | |
| H I | 1215, 1025 | -13±1 | 20±2 | 13.47±0.04 | 9 |
| H I | 1215, 1025 | 9±2 | 61±2 | 13.82±0.02 | 9 |
| O VI | 1031, 1037 | 0±2 | 55±3 | 13.60±0.02 | 10 |
| PKS 0405-123, z = 0.36160 | | | | | |
| H I | 1215 to 972 | -210±57 | 58±23 | 13.46±0.76 | 11 |
| H I | 1215 to 972 | -182±3 | 29±5 | 14.78±0.13 | |
| H I | 1215 to 972 | -171±3 | 15±2 | 14.96±0.11 | |
| H I | 1215 to 972 | -112±7 | 32±10 | 13.83±0.13 | |
| H I | 1215 to 972 | -25±12 | 40±3 | 13.55±0.06 | |
| O VI | 1031 | -330±4 | 38±5 | 13.20±0.04 | 12, 13 |

| | | | | |
|---|---|---|---|---|
| O VI | 1031 | -248±2 | 24±3 | 13.27±0.04 | 12, 13 |
| O VI | 1031 | -179±2 | 26±3 | 13.28±0.04 | 12,13 |
| O VI | 1031, 1037 | -85±7 | 37±9 | 13.01±0.07 | 12, 13 |
| O VI | 1031, 1037 | -29±1 | 5±1 | 13.22±0.04 | 12 |
| O VI | 1031, 1037 | 0±1 | 17±1 | 13.69±0.01 | 12 |
| O IV | 787 | -178±3 | 15±4 | 14.26±0.08 | 14 |
| O III | 832 | -174±2 | 12±3 | 14.23±0.06 | 15 |
| N III | 989 | -174±2 | 11±2 | 13.05±0.04 | 16 |
| C III | 977 | -242±13 | 5 (+12,-4) | 12.03±0.18 | 17 |
| C III | 977 | -180±1 | 18±1 | 13.63±0.01 | 17 |
| C III | 977 | -134±3 | 9±5 | 12.21±0.10 | 17 |
| C III | 977 | -18±4 | 23±4 | 12.38±0.06 | 17 |
| Si III | 1206 | -171±1 | 13±1 | 12.61±0.02 | 18 |
| PKS 0405-123, z = 0.36335 | | | | | |
| H I | 1215, 1025 | 11±2 | 26±3 | 13.22±0.03 | 19 |
| O VI | 1031, 1037 | 0±1 | 11±1 | 13.48±0.01 | 20 |
| C III | 977 | 0±1 | 14±1 | 12.65±0.02 | 21 |
| PKS 0405-123, z = 0.49507 (also see Narayanan et al. 2011) | | | | | |
| H I | 1025 | 4±4 | 51±5 | 14.14±0.03 | 22 |
| H I | 1025 | 103±9 | 40±5 | 13.64±0.06 | 22 |
| O VI | 1031 | 0±2 | 32±2 | 14.31±0.02 | 23 |
| O VI | 1031 | 50±2 | 17±2 | 13.84±0.05 | 23 |
| C III | 977 | -33±1 | 3.4±2.7 | 12.58±0.14 | 24 |
| C III | 977 | 2±1 | 11±1 | 12.96±0.01 | |
| C III | 977 | 56±2 | 13±2 | 12.39±0.03 | |

Notes: (1) O VI $\lambda$1037 is contaminated at 121 km s$^{-1}$ by O III $\lambda$832 at z = 0.36080 and at 199 km s$^{-1}$ by ISM Fe II $\lambda$1133. (2) The C III absorption is contaminated at -104 km s$^{-1}$ by H I $\lambda$972 at z = 0.09657, at -25 km s$^{-1}$ by ISM Ar I $\lambda$1066, at 41 km s$^{-1}$ by H$_2$ L 3-0 P(2) $\lambda$1066.900, at 80 km s$^{-1}$ by H I $\lambda$914.2 at z = 0.16715, at 174 km s$^{-1}$ by H I $\lambda$914.5 at z = 0.16515, at 199 km s$^{-1}$ by H$_2$ L 3-0 R(3) $\lambda$1067.478, and at 287 km s$^{-1}$ by H I $\lambda$914.9 at z = 0.16715. (3) H I $\lambda$1215 is contaminated at -238 by C III $\lambda$977 at z = 0.36334, at 163 km s$^{-1}$ by H I $\lambda$1215 at z = 0.09718, at > 200 km s$^{-1}$ by ISM C II $\lambda$1334. H I $\lambda$1025 is contaminated at 198 km s$^{-1}$ by ISM Fe II $\lambda$1125. (4) The H I component structure is complex for v between 0 and -100 km s$^{-1}$. In order for the fit process to converge, it was necessary to fix the H I component velocities denoted with 'f' to agree with the C III and Si II component structure. The following contamination affects the H I absorption: H I $\lambda$1215 at 228 km s$^{-1}$ by H I $\lambda$949 at z = 0.49510. H I $\lambda$926 at 79 km s$^{-1}$ by H$_2$ L 2-0 P(2) $\lambda$1081.266, at 199 km s$^{-1}$ by H$_2$ L 2-0 R(3) $\lambda$1081.711, and at 250 km s$^{-1}$ by ISM Fe II 1081. H I $\lambda$919 at -312 km s$^{-1}$ by O IV $\lambda$787 at z = 0.36079 and at -399 km s$^{-1}$ by H I $\lambda$926 at z = 0.16715. H I $\lambda$918 at -391 km s$^{-1}$ by H$_2$ L 3-0 P(3) $\lambda$1070.140, at -348 km s$^{-1}$ by H I $\lambda$917 at z = 0.16700, at -310 km s$^{-1}$ by H I $\lambda$917 at z = 0.16715, and at 85 km s$^{-1}$ by O IV $\lambda$787 at z = 0.36079. (5) O VI $\lambda$1037 is contaminated at -370 km s$^{-1}$ by C II $\lambda$1334 at z = 0.16715. (6) Si II $\lambda$1260 is contaminated at 103, 151 and 256 km s$^{-1}$ by H I $\lambda$1215 at z = 0.21054, 0.21073, and 0.21116, respectively. (7) O VI $\lambda$1037 is contaminated at -129 km s$^{-1}$ by H I $\lambda$1215 at z = 0.00924. We list here a 2-component fit to the O VI absorption. A 3 component fit improves the value of $\chi^2$ from 1.089 to 0.877 and separates the component near -71 km s$^{-1}$ into two components at v = -86±2 km s$^{-1}$ and -52±2 km s$^{-1}$ with b = 11±2 km s$^{-1}$ and 15±4 km s$^{-1}$ with log N(O VI) = 13.38±0.05 and 13.30±0.07. The fit parameters for the O VI absorption at v = 0 hardly change. (8) The properties of this C III absorber are uncertain because of contamination at -136 and -123 km s$^{-1}$ with N III $\lambda$989 at z = 0.16713 and 0.16718 and by contamination at -108 km s$^{-1}$ by Si II $\lambda$989 at z = 0.16715. (9) H I $\lambda$1025 is contaminated at -191 and 211 km s$^{-1}$ by C III $\lambda$977 at z = 0.36151 and

0.36334, respectively. (10) O VI λ1031 is contaminated at -236 km s$^{-1}$ by H I 949 at z = 0.40888. O VI λ1037 is contaminated at 91 km s$^{-1}$ by N III λ989 at z = 0.36080. (11) This H I absorber at -210 ±57 km s$^{-1}$ has very uncertain parameters because it lies in the wing of the much stronger H I absorption at v = -182±3 km s$^{-1}$. (12) O VI λ1031 is contaminated at -473 and -423 km s$^{-1}$ by ISM Si IV λ1402. O VI λ1037 is contaminated at -255, -176, -108, and 217 km s$^{-1}$ by H I λ1215 at z = 0.16118, 0.16149, 0.16175, and 0.16301, respectively. (13) The O VI lines at -330, -248, -179, -85 km s$^{-1}$ are only detected in the O VI λ1031 line. (14) O IV λ787 is contaminated at -303 and -265 km s$^{-1}$ by H I λ918 at z = 0.16700 and 0.16715, respectively, and at 95 and 133 km s$^{-1}$ by H I λ919 at z = 0.16700 and 0.16715, respectively. (15) O III λ832 is contaminated at -295 km s$^{-1}$ by O VI λ1037 at z = 0.09192, at -96 km s$^{-1}$ by ISM Fe II λ1133 and FPN, and at 12, 35, 77 and 101 km s$^{-1}$ by ISM N I λλ1134.165, 1134.414, and FPN. (16) N III λ989 is contaminated at -265 km s$^{-1}$ by O VI 1037 at z = 0.29770. (17) C III λ977 is contaminated at -360 km s$^{-1}$ by H I λ1215 at z = 0.09298, at -310 km s$^{-1}$ by ISM C I λ1328.833, at -251 km s$^{-1}$ by ISM C I* λ1329.100, and at 160 km s$^{-1}$ by H I λ1025 at z = 0.29764. (18) Si III λ1206 is contaminated at -73, 42, 109 and 174 km s$^{-1}$ by H I λ1215 at z = 0.35100, 0.35152, 0.35182, and 0.35211, respectively. (19) H I λ1215 is contaminated at -180, -61, 20 and 116 km s$^{-1}$ by C I* λ1656.267, C I λ1656.928, C I* λ1657.379, and C I* λ1657.907, respectively. The contaminated is estimated from the fit to the stronger C I and C I* absorption lines recorded in the COS observations. (20) O VI λ1031 is contaminated at 238 and 277 km s$^{-1}$ by Si III λ1206 at z = 0.16700 and 0.16716, respectively. O VI λ1037 is contaminated at -166 km s$^{-1}$ by H I λ1215 at z = 0.16301. (21) C III λ977 is contaminated at -223, -202 km s$^{-1}$ by H I λ1025 at z = 0.29764 and 0.29773, respectively, and at 236 km s$^{-1}$ by H I λ1215 at z = 0.09657. (22) H I λ1025 is contaminated at -241 km s$^{-1}$ by H I λ1215 at z = 0.26045. (23) O VI λ1037 is contaminated at -96 km s$^{-1}$ by ISM C IV λ1550. (24) This C III λ977 absorber properties are uncertain because of the narrow line width.

Table A13
Profile Fit Results for HE 0238-1904 ($z_{em}$ = 0.631)

| Ion | $\lambda$s (Å) | v (km s$^{-1}$) | b (km s$^{-1}$) | log N(cm$^{-2}$) | Note |
|---|---|---|---|---|---|
| HE 0238-1904, z = 0.40107 | | | | | |
| H I | 1215 to 972 | -1±1 | 31±1 | 14.93±0.01 | |
| O VI | 1031 | 0±2 | 22±2 | 13.47±0.03 | 1 |
| C III | 977 | -4±8 | 21±4 | <12.36 | 2 |
| HE 0238-1904, z = 0.42430 | | | | | |
| H I | 1215, 1025, 950 | -70±1 | 35±1 | 14.72±0.01 | |
| H I | 1215, 1025, 950 | 0±1 | 24±1 | 14.57±0.01 | |
| O VI | 1031, 1037 | -85±2 | 48±2 | 14.36±0.01 | 3 |
| O VI | 1031, 1037 | 0±1 | 23±1 | 14.16±0.02 | |
| O IV | 787 | -81±13 | 39±14 | 14.65±0.11 | 4 |
| O IV | 787 | 7±10 | 22±10 | 14.81±0.33 | 4 |
| O III | 832 | -67±3 | 30±4 | 14.05±0.03 | |
| O III | 832 | 1±3 | 19±3 | 13.93±0.04 | |
| C III | 977 | -104±10 | 16±4 | 13.12±0.18 | 5 |
| C III | 977 | -71±5 | 21±3 | 13.50±0.07 | 5 |
| C III | 977 | -5±3 | 16±1 | 13.52±0.03 | 5 |
| HE 0238-1904, z = 0.47204 | | | | | |
| H I | 1215 | -134±5 | 38±5 | 13.62±0.05 | 6 |
| H I | 1215 | 14±4 | 47±4 | 13.83±0.03 | 6 |
| O VI | 1031, 1037 | 0±1 | 20±1 | 14.16±0.01 | 7 |
| O VI | 1031, 1037 | 54±1 | 13±1 | 13.81±0.02 | |

Notes: (1) O VI $\lambda$1037 is blended at 30 km s$^{-1}$ with possible H I $\lambda$1215 at z = 0.19598. (2) C III $\lambda$977 is contaminated at -49 km s$^{-1}$ with H I $\lambda$1025 at z = 0.33433. (3) The O VI $\lambda$1031 blends at -145 km s$^{-1}$ with H I $\lambda$1215 at z = 0.20843. (4) The FUSE spectrum of O IV $\lambda$787 has S/N ~ 3. The continuum is very uncertain. The fit errors are large. O IV $\lambda$787 is contaminated at 23 and 142 km s$^{-1}$ by ISM Fe II $\lambda$1121 and ISM Fe III $\lambda$1122. (5) C III $\lambda$977 has the following blending: at -265 km s$^{-1}$ H I $\lambda$1025 at z = 0.35547, at 83 km s$^{-1}$ H I $\lambda$1215 at z = 0.14501, at 144 km s$^{-1}$ H I $\lambda$1025 at z = 0.35732, and at 217 km s$^{-1}$ H I $\lambda$1025 at z = 0.35765. (6) Only one COS extraction covers H I $\lambda$1215 with S/N ~12. (7) O VI $\lambda$1037 is contaminated at -200 to -50 km s$^{-1}$ by ISM Si II $\lambda$1526 and FPN, and at 163 km s$^{-1}$ by H I $\lambda$1215 at z = 0.25712.

Table A14
Profile Fit Results for 3C 263 $(z_{em} = 0.652)$[a]

| Ion | $\lambda s$ (Å) | $v$ (km s$^{-1}$) | $b$ (km s$^{-1}$) | log N(cm$^{-2}$) | Note |
|---|---|---|---|---|---|
| 3C 263, z = 0.06342 (from Savage et al. 2012) | | | | | |
| H I | 1215 to 972 | -151±3 | 28±2 | 13.92±0.05 | 1 |
| H I | 1215 to 972 | -40f | 45±8 | 15.04±0.10 | 2 |
| H I | 1215 to 972 | -9f | 19±18 | 14.30±0.36 | 2 |
| H I | 1215 to 972 | 25f | 21±2 | 15.15±0.13 | 2 |
| H I | 1215 | 96±4 | 13±10 | 12.70±0.14 | 1 |
| O VI | 1031, 1037 | 0±4 | 39±6 | 14.59±0.07 | 3 |
| C IV | 1548, 1550 | -40±8 | 22±8 | 13.57±0.19 | |
| C IV | 1548, 1550 | -9±3 | 10±7 | 13.45±0.22 | |
| C IV | 1548, 1550 | 25±1 | 15±2 | 13.86±0.03 | |
| Si III | 1206 | -48±2 | 22±3 | 12.65±0.03 | |
| Si III | 1206 | 24±1 | 13±1 | 12.90±0.03 | |
| Si II | 1260, 1193 | 17±3 | 12±6 | 12.26±0.09 | |
| C II | 1334 | 22±2 | 17±3 | 13.37±0.04 | |
| 3C 263, z = 0.11389 | | | | | |
| H I | 1215 | -219±1 | 47±1 | 14.05±0.01 | |
| H I | 1215 | -44±6 | 71±5 | 13.49±0.05 | |
| H I | 1215 | 1±1 | 20±1 | 13.85±0.02 | |
| O VI | 1031, 1037 | 0±4 | 31±5 | 13.66±0.05 | 4 |
| C IV | 1548, 1550 | 6±1 | 11±1 | 13.17±0.03 | |
| C III | 977 | 23±6 | 50±8 | 13.35±0.05 | 5 |
| 3C 263, z = 0.14072 (from Savage et al. 2012) | | | | | |
| H I | 1215, 1025 | 7±1 | 28±1 | 14.51±0.03 | |
| H I | 1215, 1025 | -7±7 | 87±15 | 13.47±0.10 | |
| O VI | 1031, 1037 | 0±6 | 33±12 | 13.60±0.09 | |
| 3C 263, z = 0.32567 (also see Narayanan et al. 2012) | | | | | |
| H I | 1215 to 937 | -17±2 | 24±1 | 15.23±0.02 | 6 |
| H I | 1215 to 937 | 17±1 | 11±1 | 15.02±0.02 | 6 |
| O VI | 1031, 1037 | 0±2 | 34±2 | 14.02±0.01 | 7 |
| O III | 831 | -3±7 | 41±8 | 14.27±0.07 | |
| C III | 977 | -21±1 | 12±1 | 13.43±0.02 | |
| C III | 977 | 14±1 | 9±2 | 13.65±0.15 | |
| 3C 263, z = 0.44672 | | | | | |
| H I | 1215, 1025 | -35±12 | 71±5 | 13.72±0.07 | 8 |
| H I | 1215, 1025 | 9±1 | 32±1 | 14.17±0.03 | |
| O VI | 1031, 1037 | 0±3 | 32±3 | 13.87±0.03 | 9 |
| O IV | 787 | 12±7 | 41±6 | 14.51±0.06 | |
| O III | 832 | 0±4 | 25±5 | 13.49±0.05 | 10 |
| C III | 977 | 2±1 | 23±1 | 13.12±0.02 | 11 |

[a]3C 263 does not have a STIS spectrum. Therefore the wavelength calibration is uncertain, particularly for $\lambda$ < 1190 Å.

Notes: (1) Two extra components at negative and positive velocity not seen in C IV are required to fit the H I absorption. The component at 96 km s$^{-1}$ has very uncertain properties. Although fitted by a narrow feature, this component might instead represent the positive velocity extent of a BLA associated with warm gas in the absorber. (2) The principal H I three component structure was

assumed to have the same velocities as the well defined C IV component structure. The three H I components are strongly saturated in the lines of H I $\lambda\lambda$1215, 1025. Therefore, the H I column densities in the three components are mostly derived from the weaker H I $\lambda$972 absorption which is not strongly saturated. H I $\lambda$949 is strongly affected by an instrumental glitch. (3) A single component fit to the broad O VI absorption is listed. The column density error may be larger if the actual O VI component structure is more complex. O VI $\lambda$1031 is contaminated at -196 and -156 km s$^{-1}$ by ISM Fe II $\lambda$1096. O VI $\lambda$1037 is contaminated at 207 km s$^{-1}$ by O III $\lambda$832 at z = 0.32567. (4) The O VI $\lambda\lambda$1031, 1037 at z = 0.11389 is at 1149.5 and 1155.8 Å. There is H I $\lambda$1025 absorption for z = 0.12228 at 1151.15 Å. The wavelength region of the O VI doublet was adjusted to make H I $\lambda\lambda$1025, 1215 in the z = 0.12228 system agree. In the combined spectra there are two integrations for O VI $\lambda$1031 and three for O VI $\lambda$1037. (5) The C III z = 0.11389 absorption in the low S/N FUSE spectrum is unusually broad. The displayed spectrum has 0.04Å sampling. The unbinned observations suggest that C III could be blended with an unidentified line in the positive velocity wing. (6) There could be a BLA but only two COS integrations cover H I $\lambda$1215. Weak ISM Fe II $\lambda$1611 contaminates H I $\lambda$1215 at -131 and -91 km s$^{-1}$. (7) O VI $\lambda$1037 is contaminated at 153 km s$^{-1}$ by H I 1215 at z = 0.13208. (8) H I $\lambda$1025 at z = 0.44672 is contaminated as follows: 205 km s$^{-1}$, H I $\lambda$972 at z = 0.52688; 262 km s$^{-1}$, H I $\lambda$972 at z = 0.52717. (9) The O VI $\lambda$1031 absorption at z = 0.44672 is contaminated as follows: -285 km s$^{-1}$, H I $\lambda$1025 at z = 0.45387; -222 km s$^{-1}$, C III $\lambda$977 at z = 0.52688; -176 km s$^{-1}$, C III $\lambda$977 at z = 0.52712; -17 km s$^{-1}$, C III $\lambda$977 at z = 0.52793; 62 km s$^{-1}$, C III $\lambda$977 at z = 0.52833. The O VI properties are mostly determined by the O VI $\lambda$1037 absorption. (10) O III at z = 0.44672 is contaminated from 100 to 300 km s$^{-1}$ by ISM Si III 1206 and at -275 km s$^{-1}$ by O IV $\lambda$787 at z = 0.52835. (11) C III $\lambda$977 in the z = 0.44672 system is contaminated as follows: -181 km s$^{-1}$, H I $\lambda$1025 at z = 0.37719; -142 km s$^{-1}$, H I $\lambda$1025 at z = 0.37737; 28 km s$^{-1}$, H I $\lambda$972 at z = 0.45352; 99 km s$^{-1}$, H I $\lambda$972 at z = 0.45387; 161 km s$^{-1}$, H I $\lambda$926 at z = 0.52688.

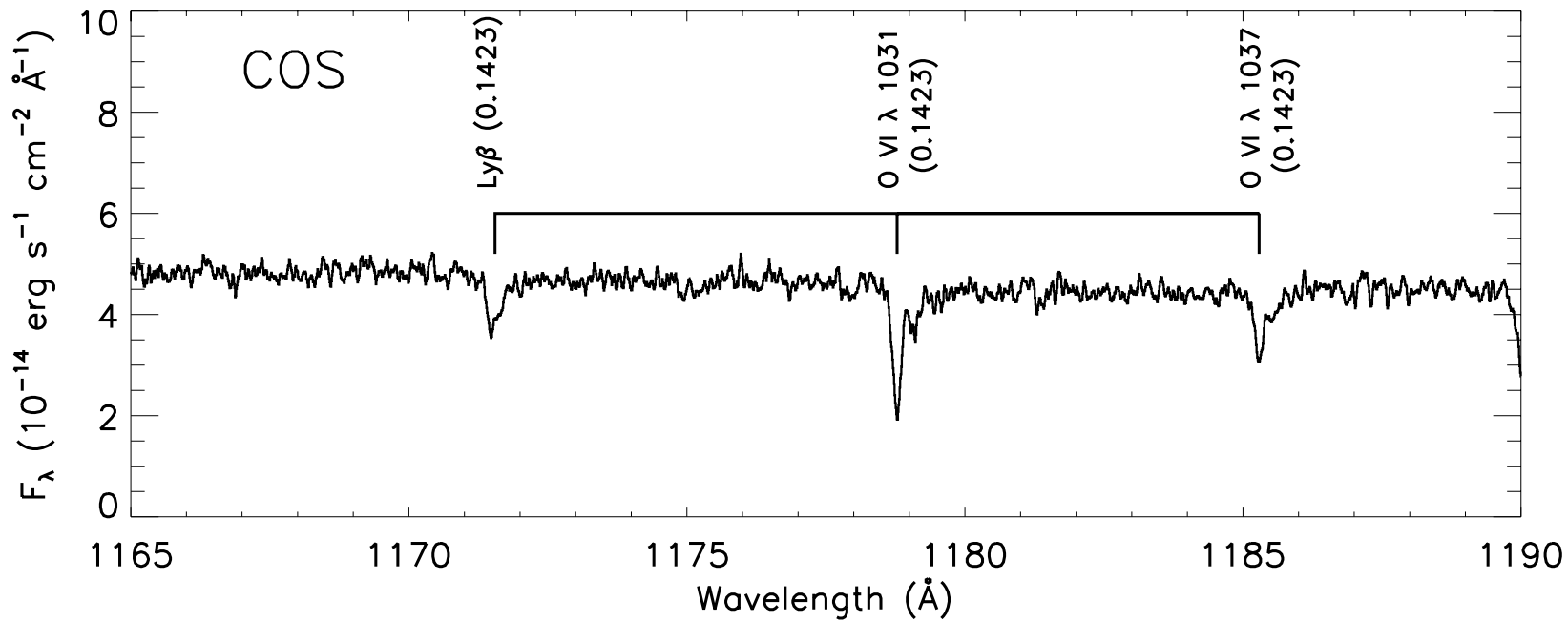

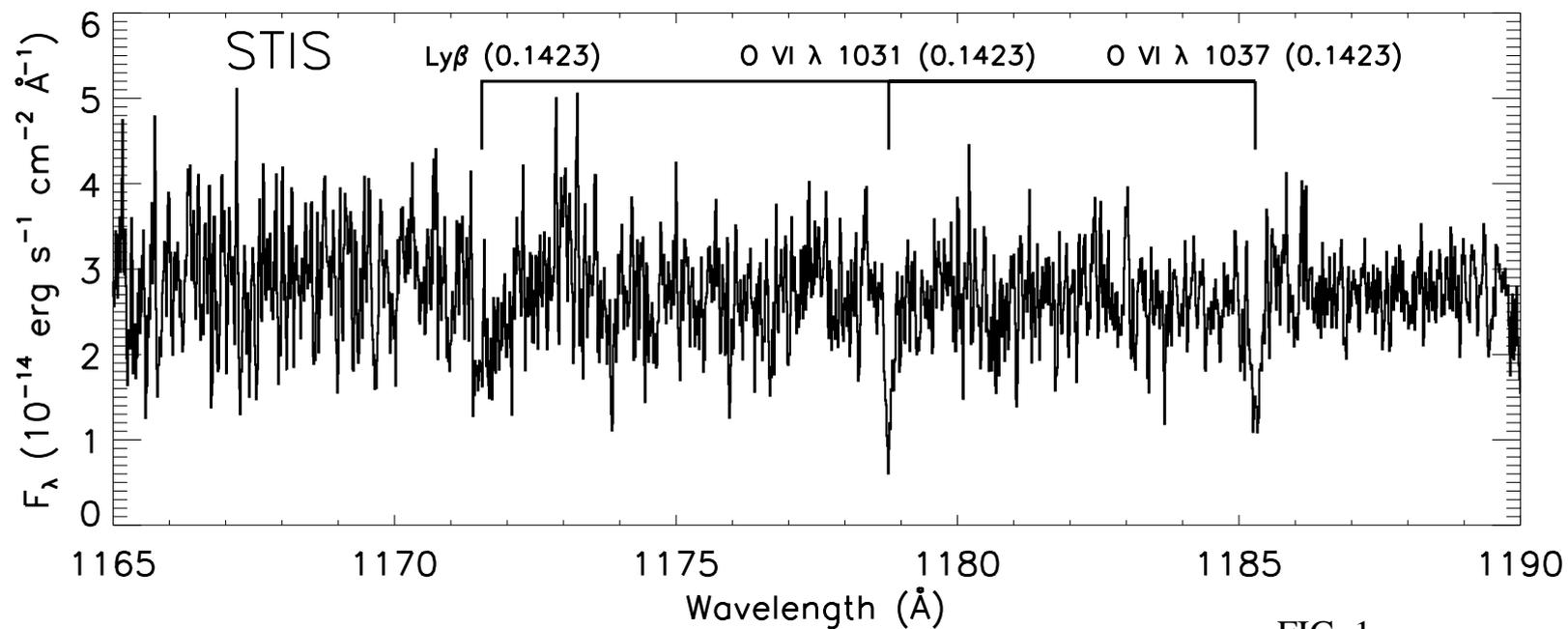

FIG. 1

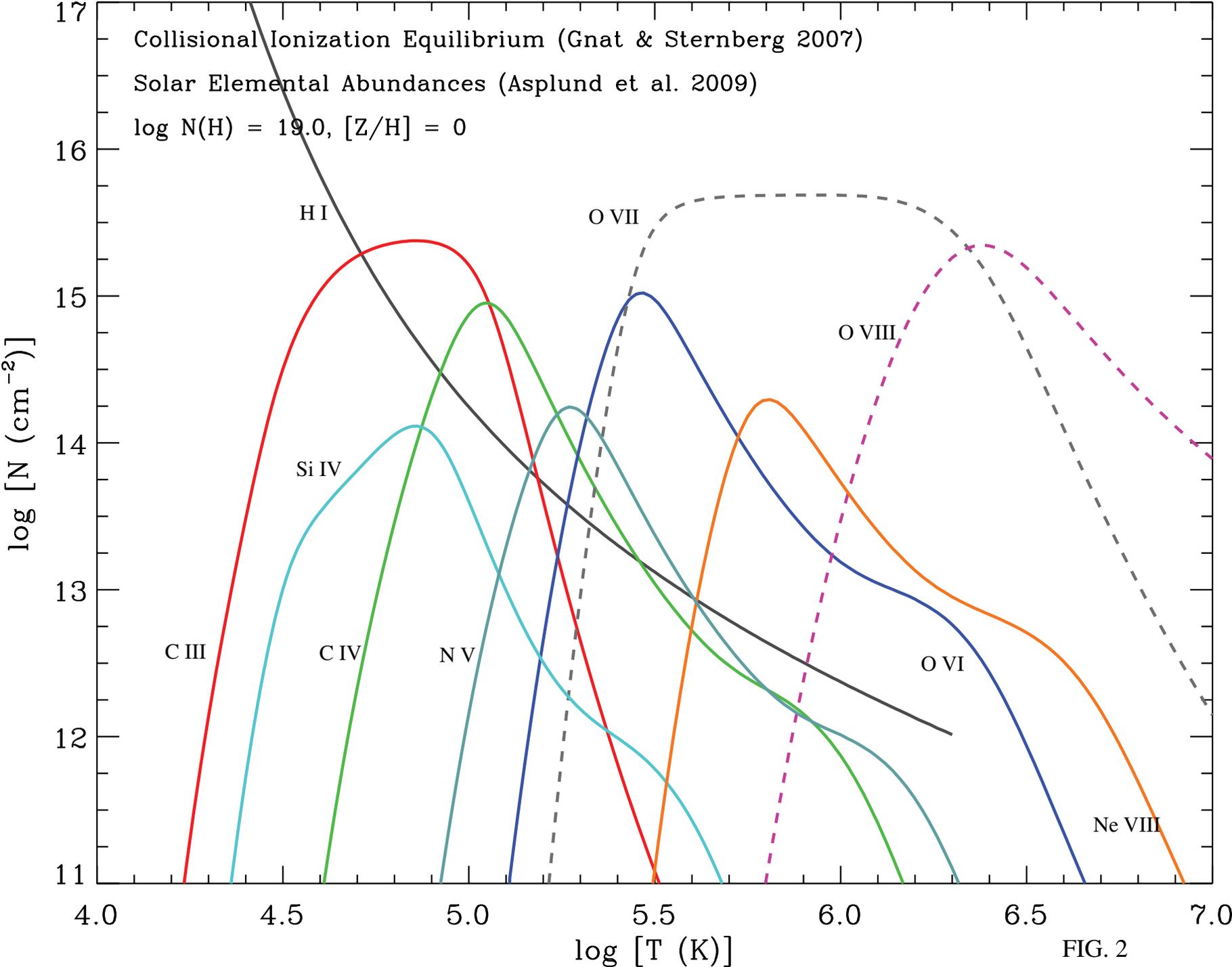

Collisional Ionization Equilibrium (Gnat & Sternberg 2007)

Solar Elemental Abundances (Asplund et al. 2009)

log N(H) = 19.0, [Z/H] = 0

FIG. 2

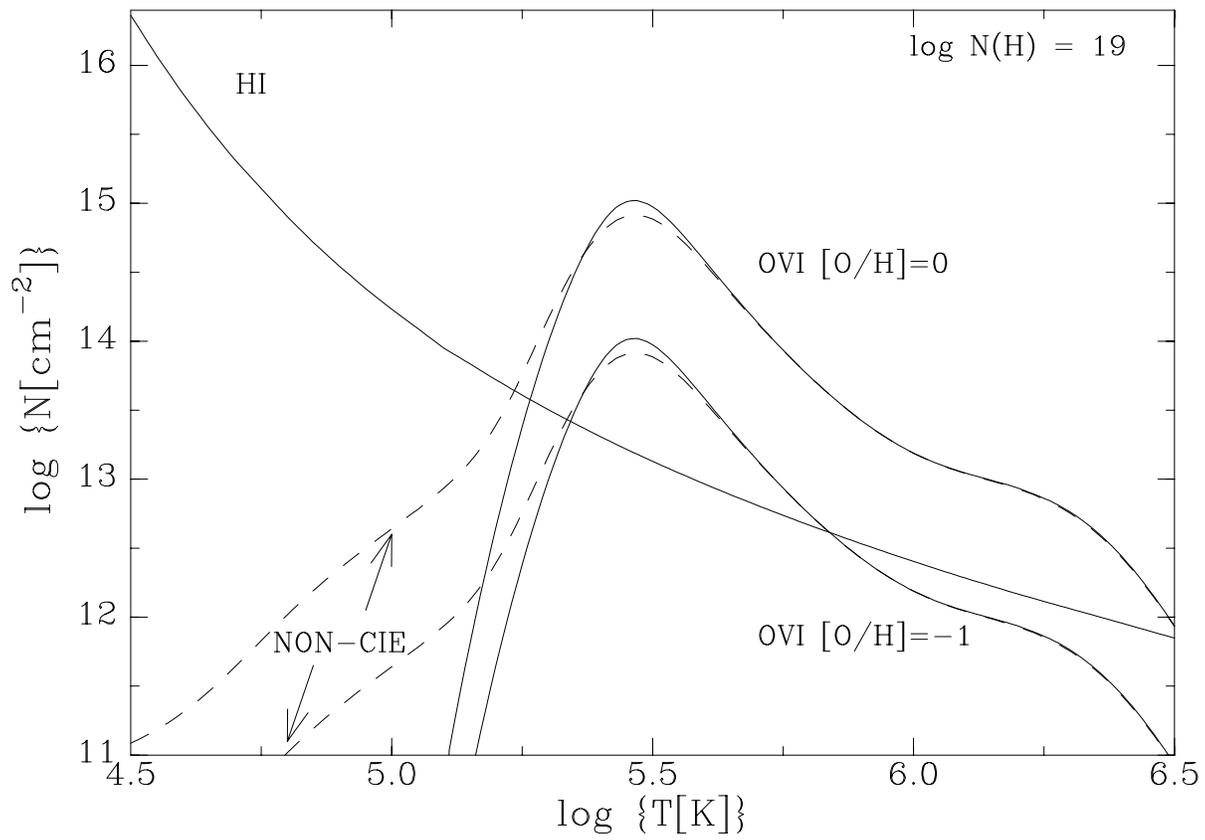

FIG. 3

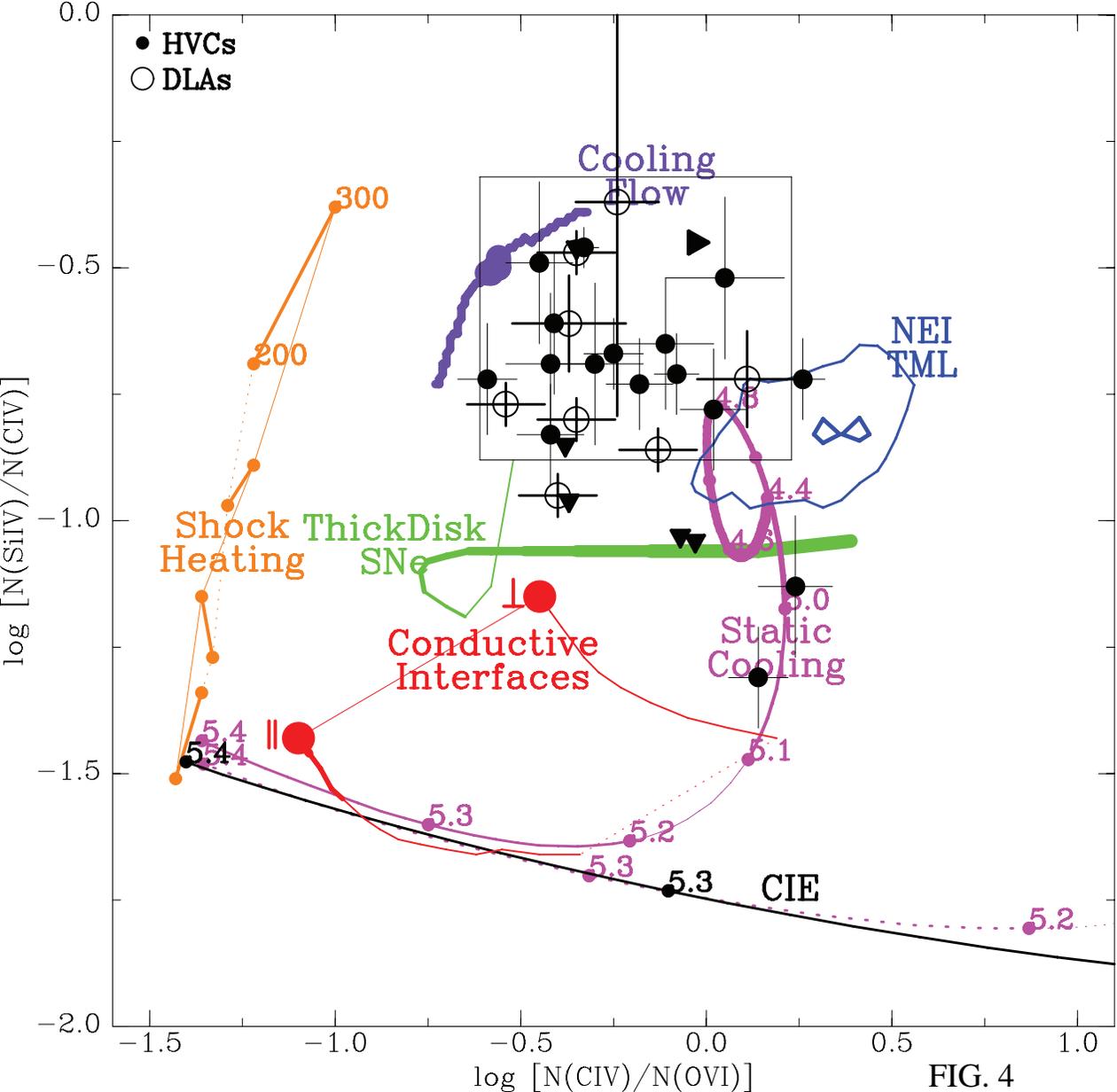

FIG. 4

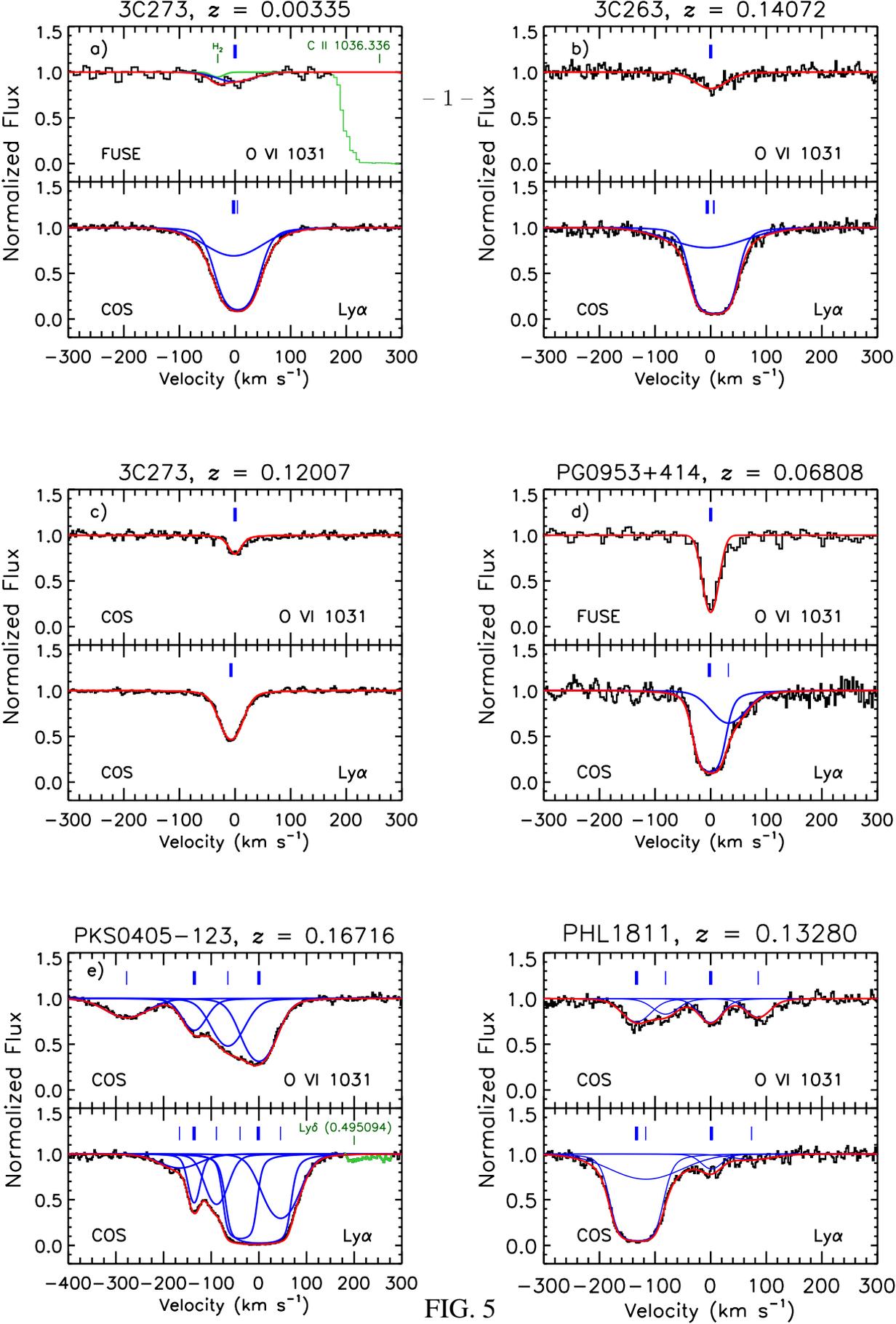

FIG. 5

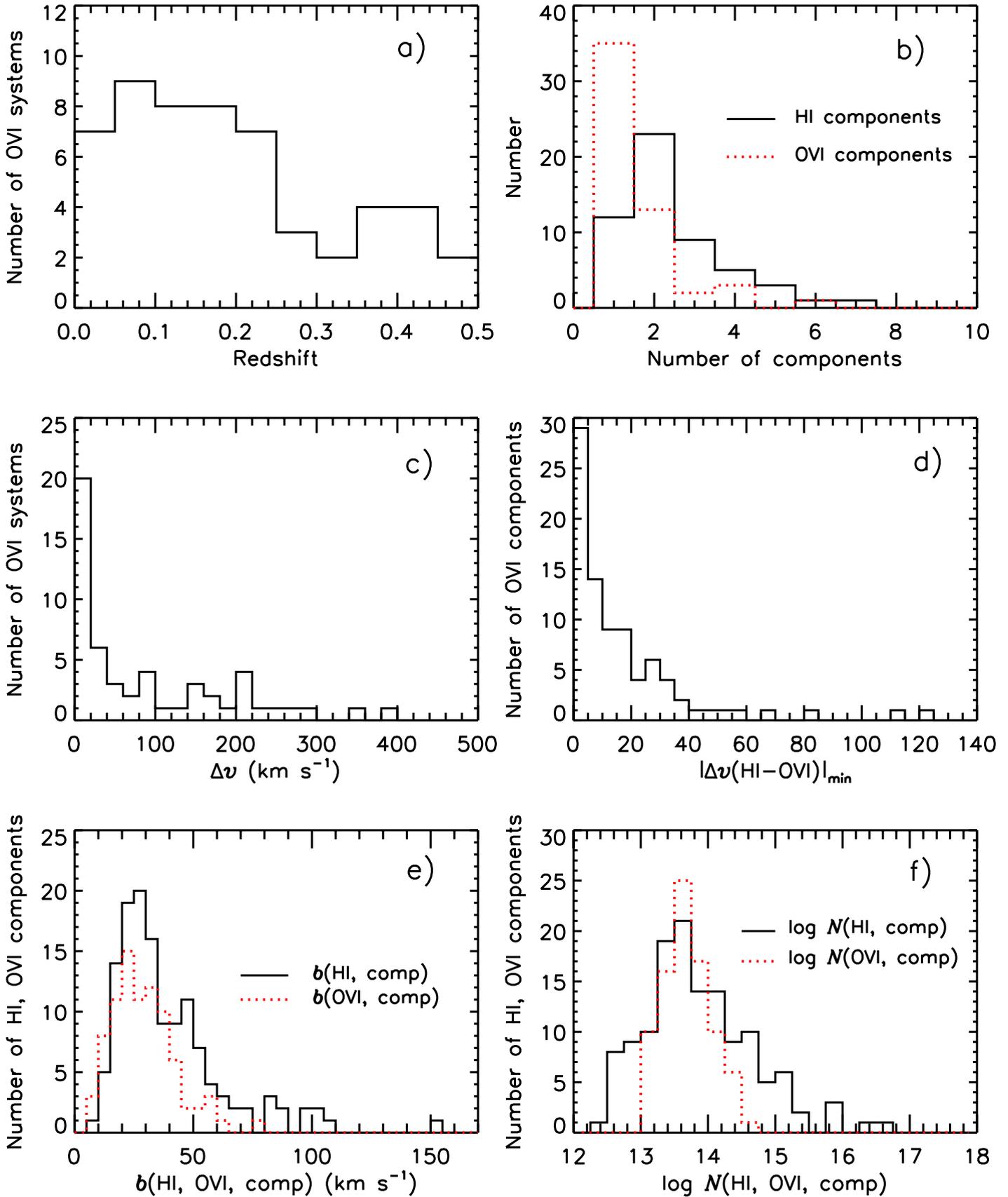

FIG. 6

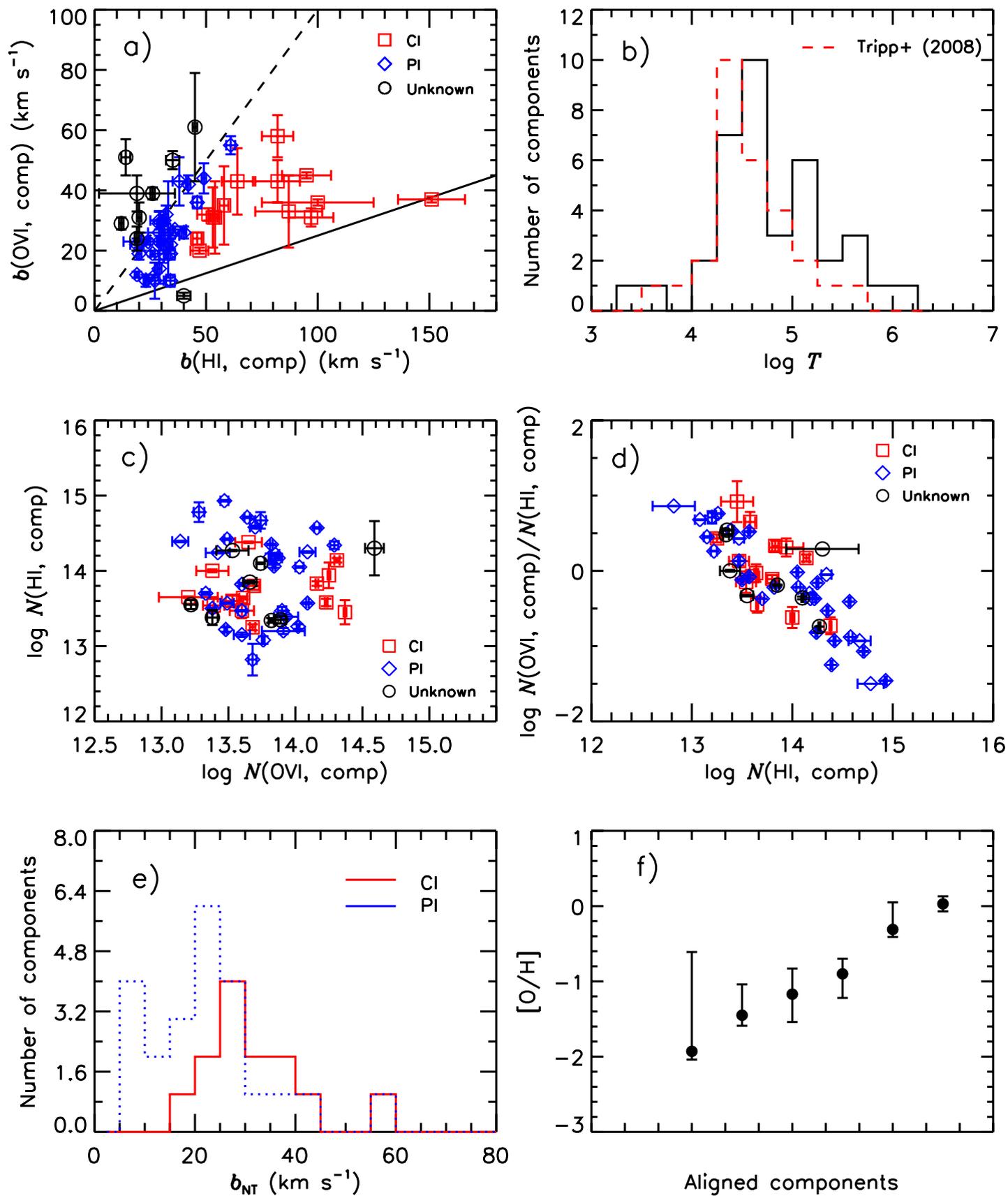

FIG. 7

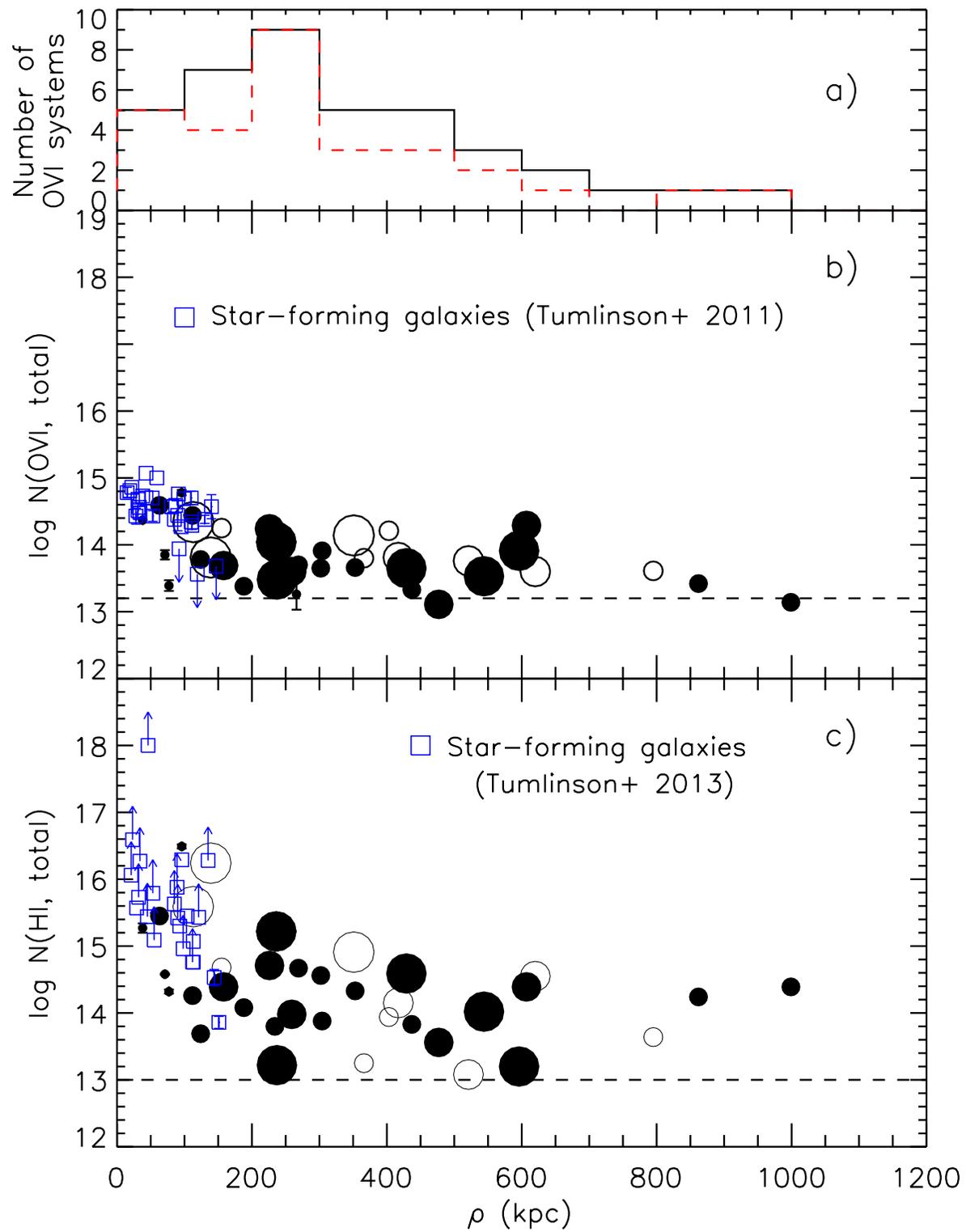

FIG. 8

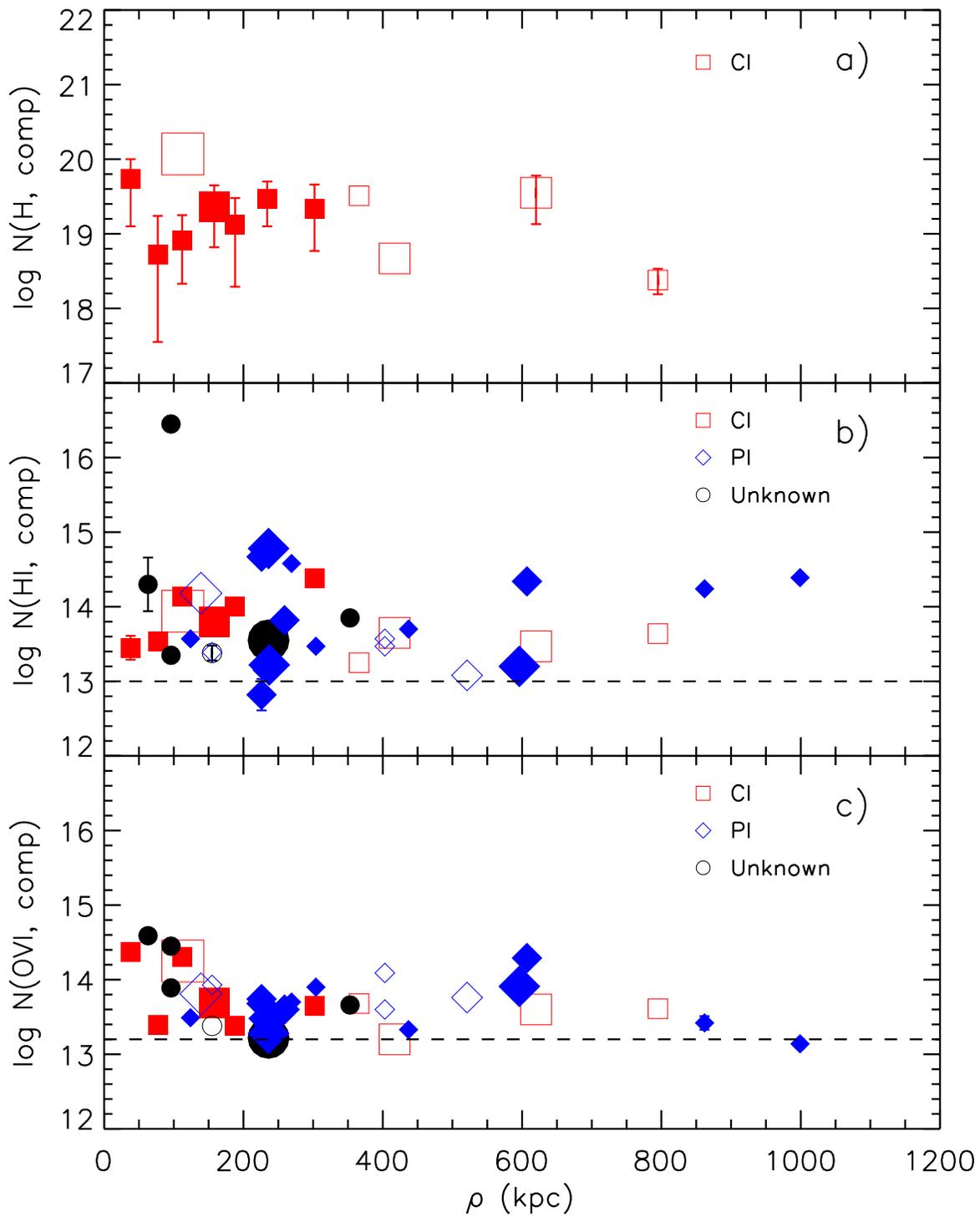

FIG. 9

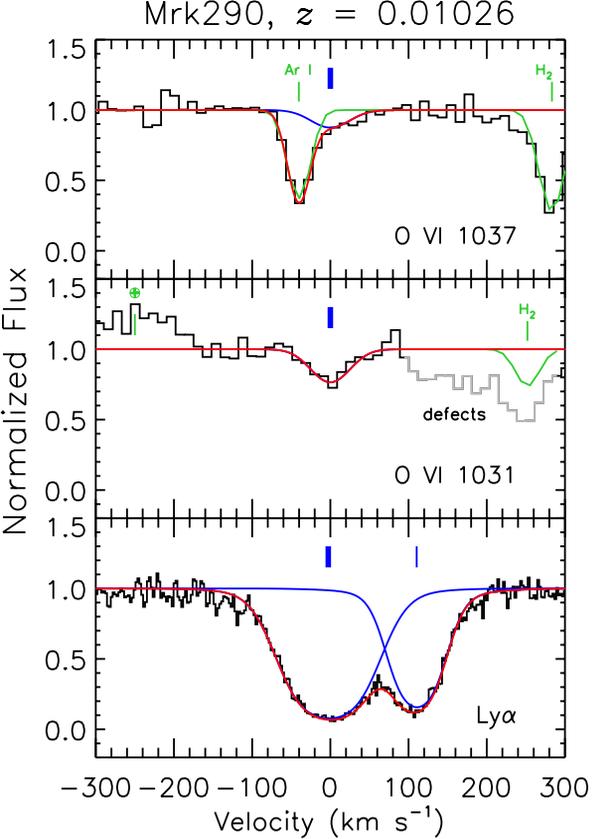
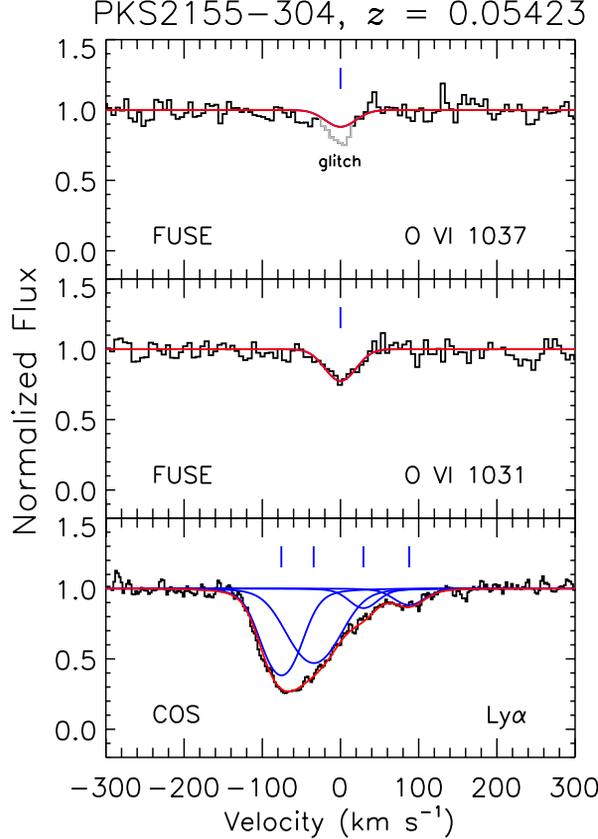
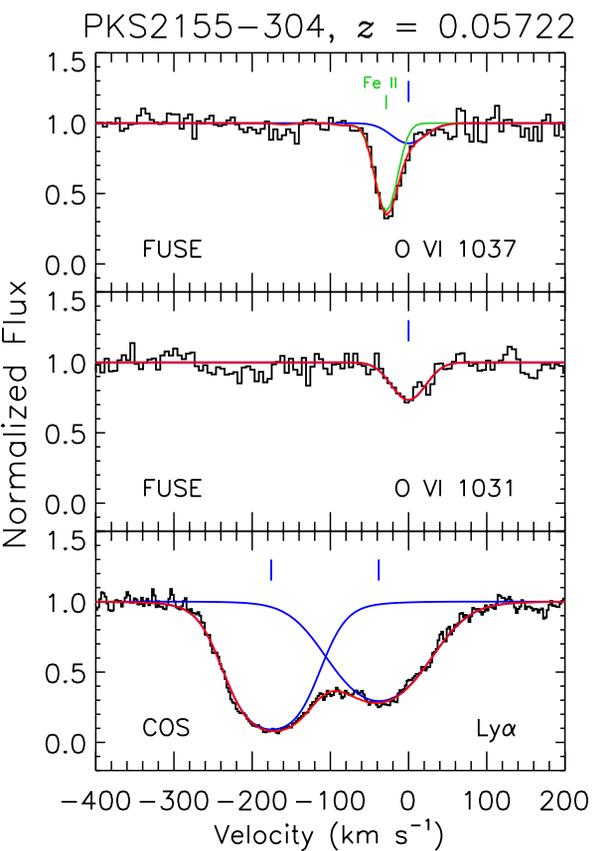
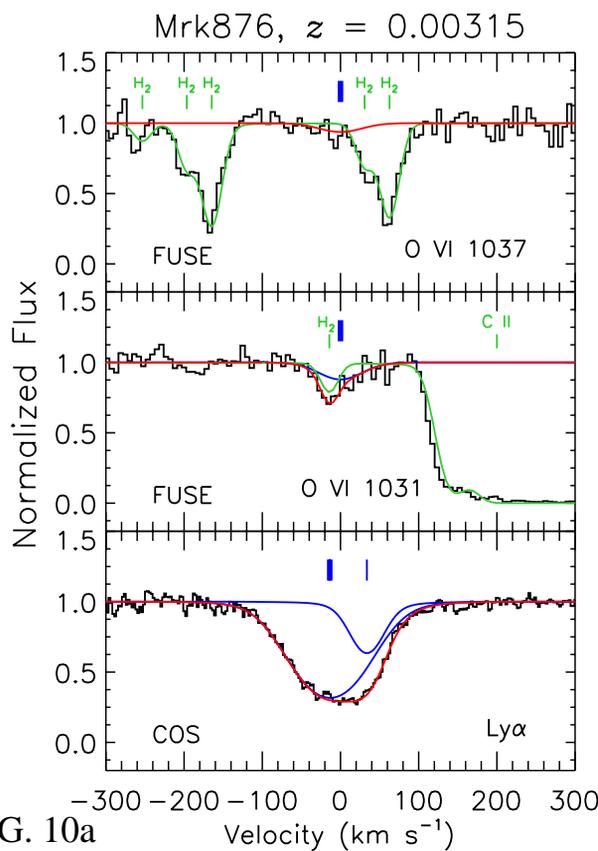

FIG. 10a

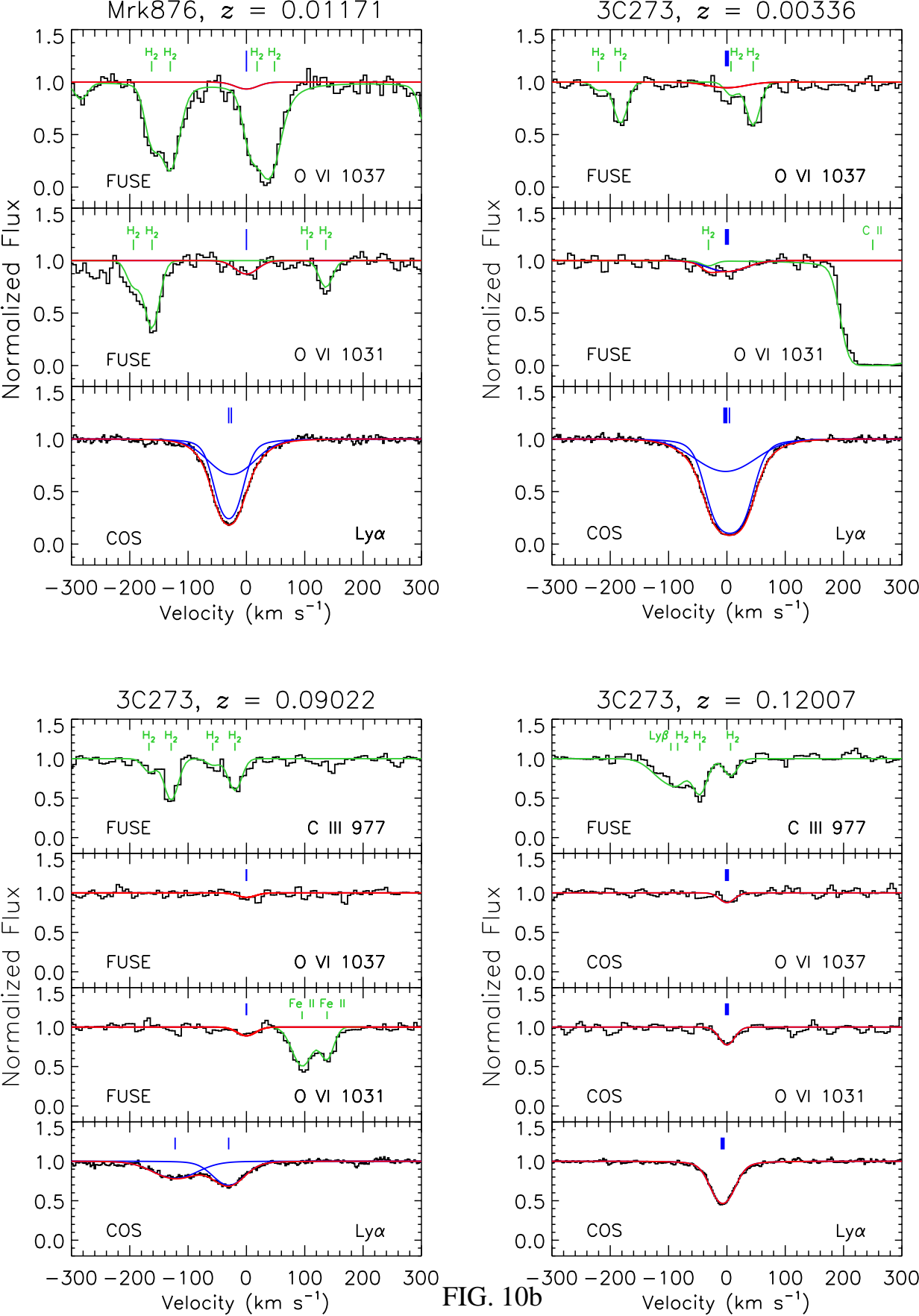

FIG. 10b

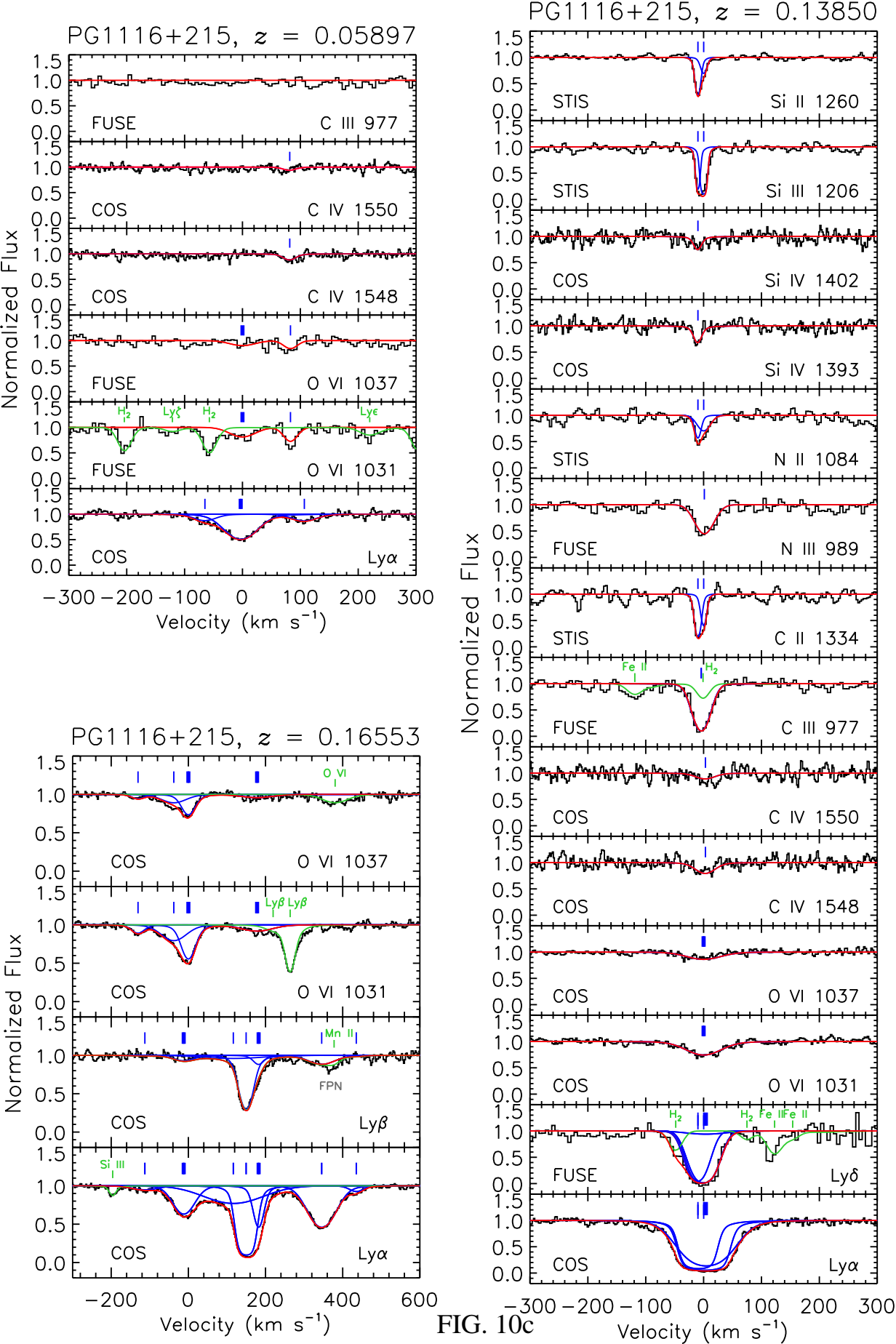

FIG. 10c

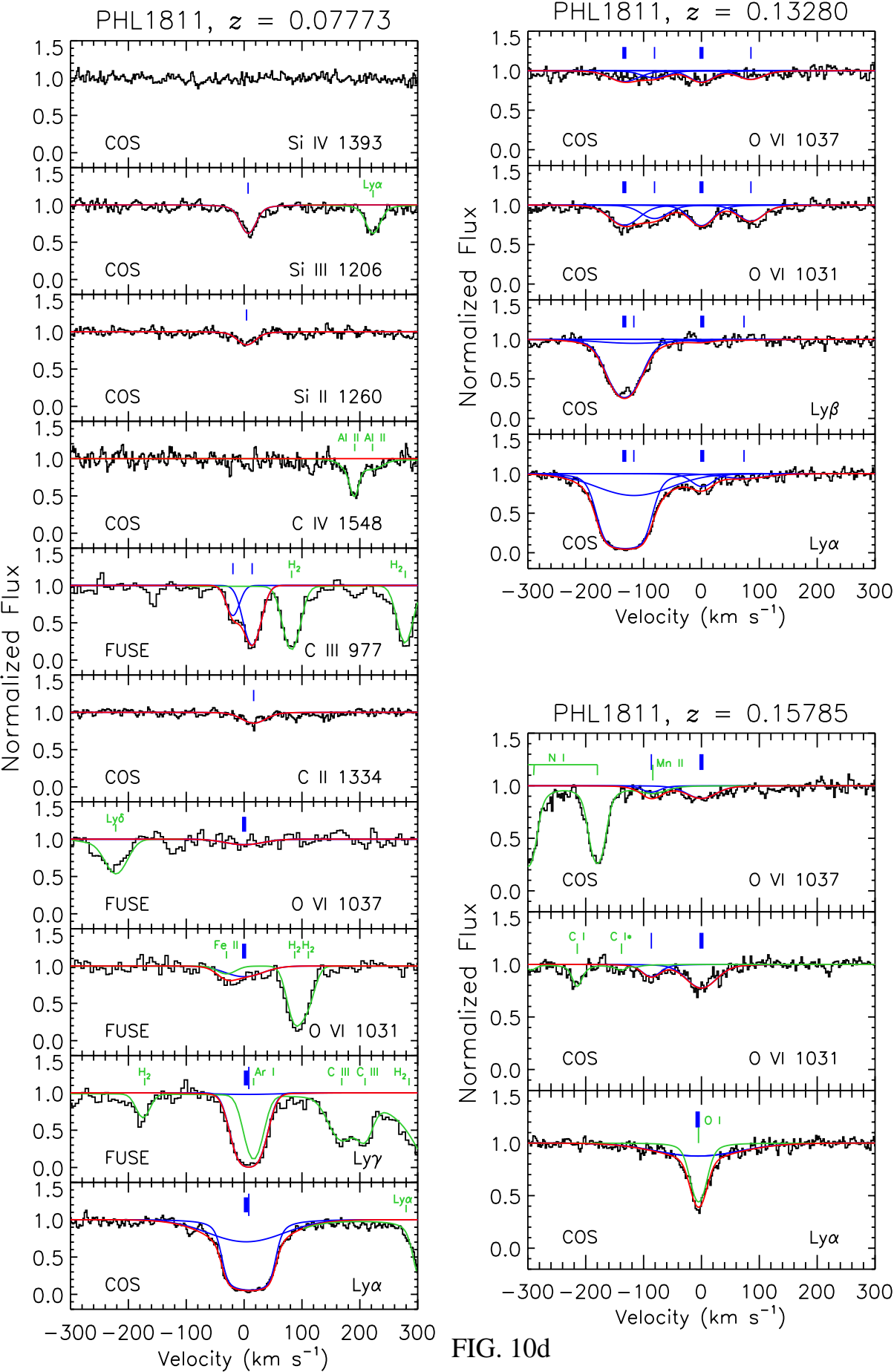

FIG. 10d

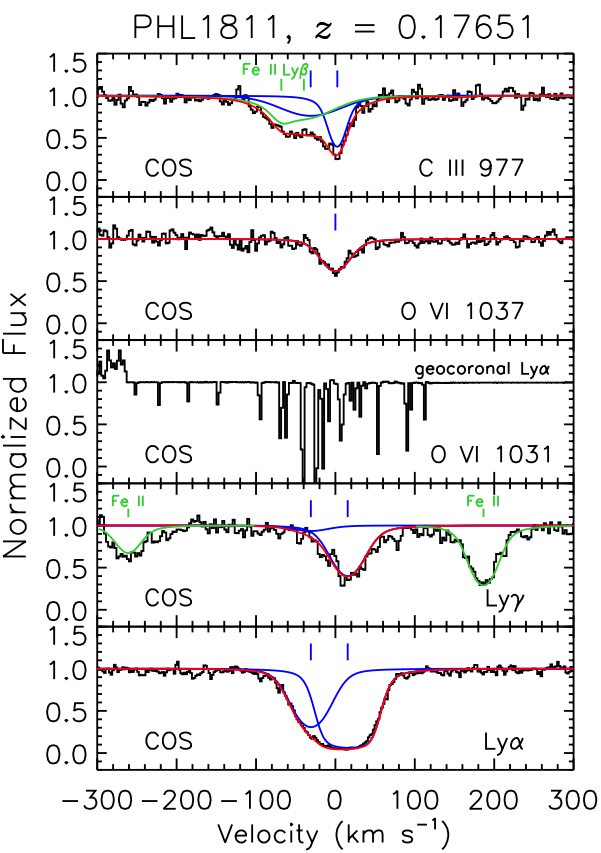

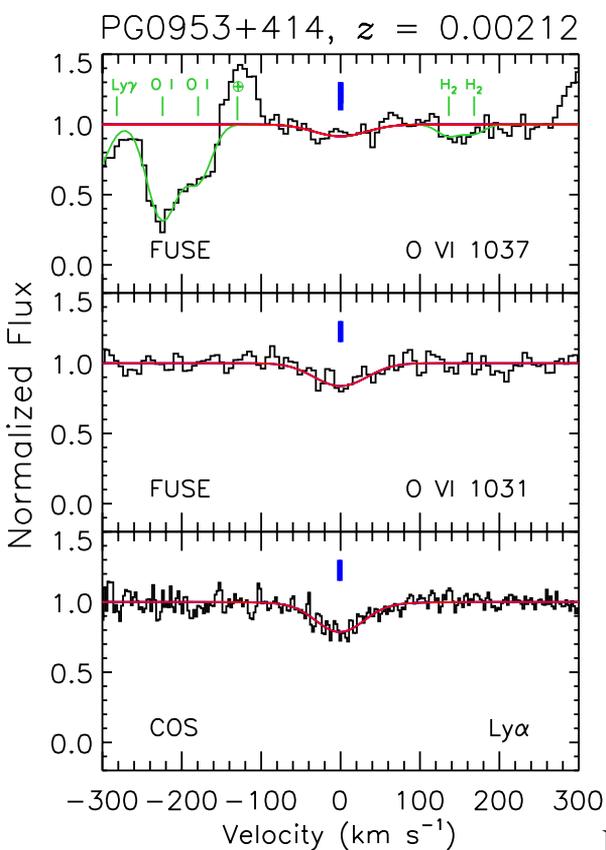

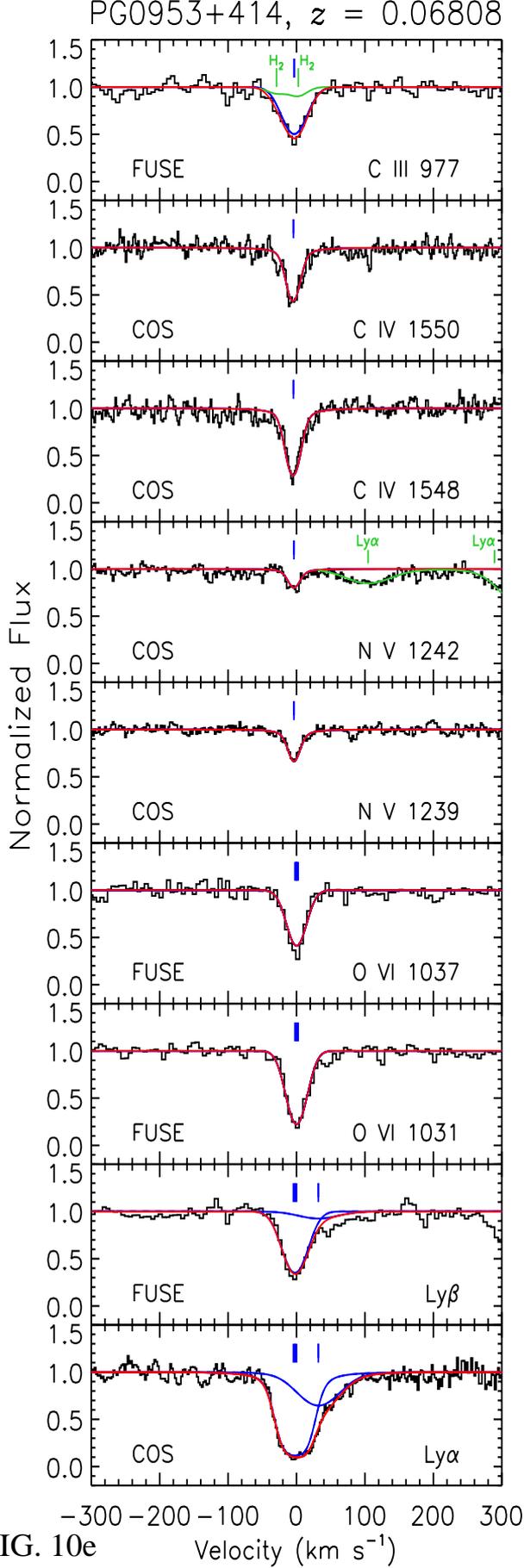

FIG. 10e

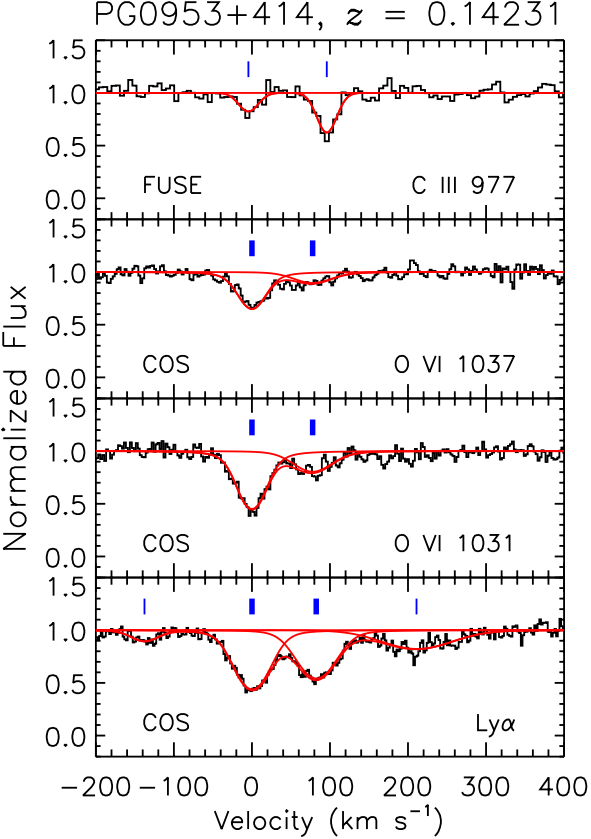

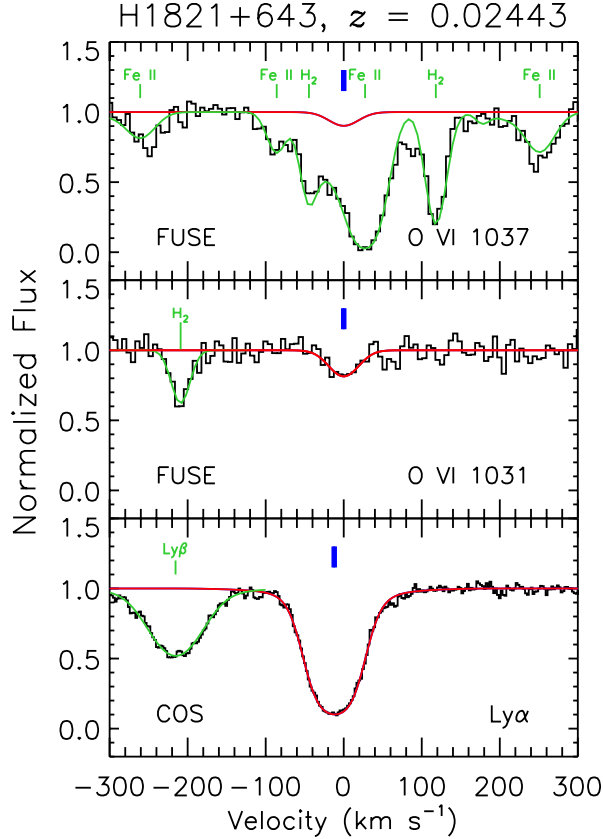

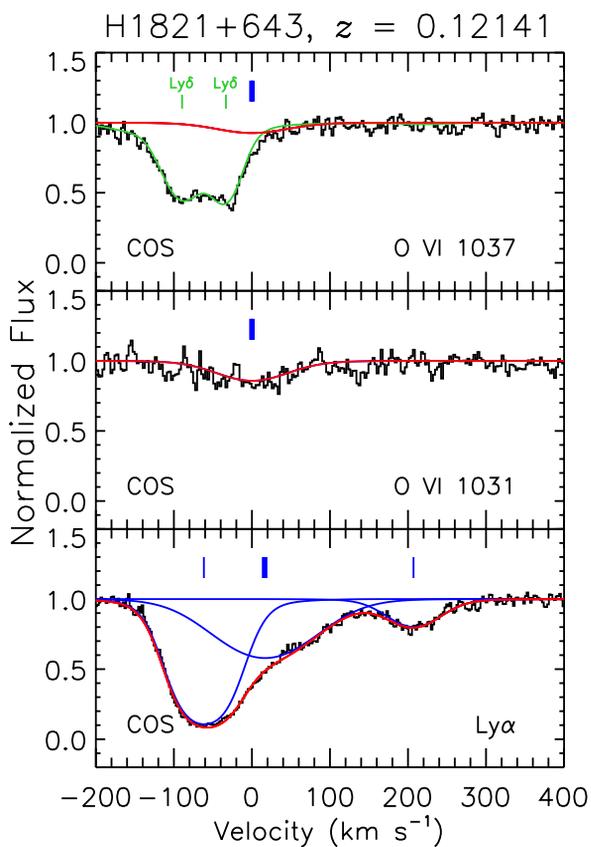

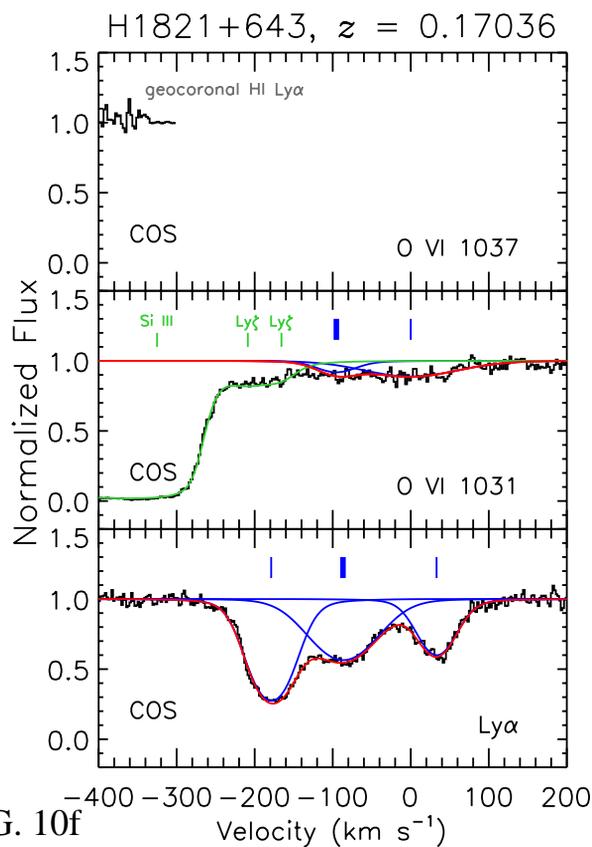

FIG. 10f

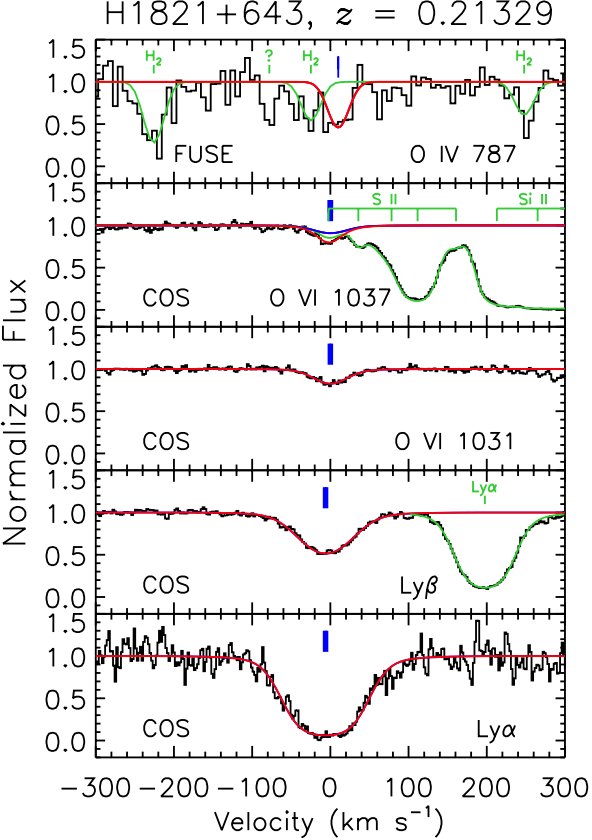

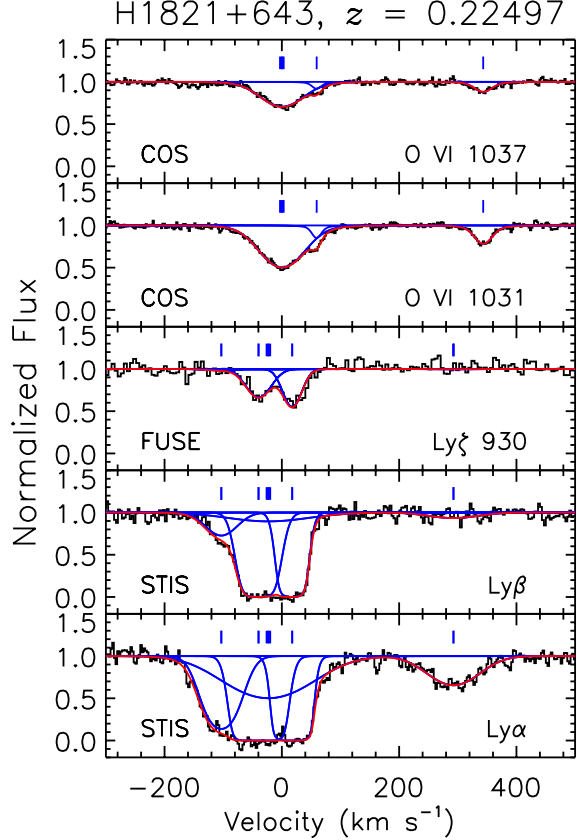

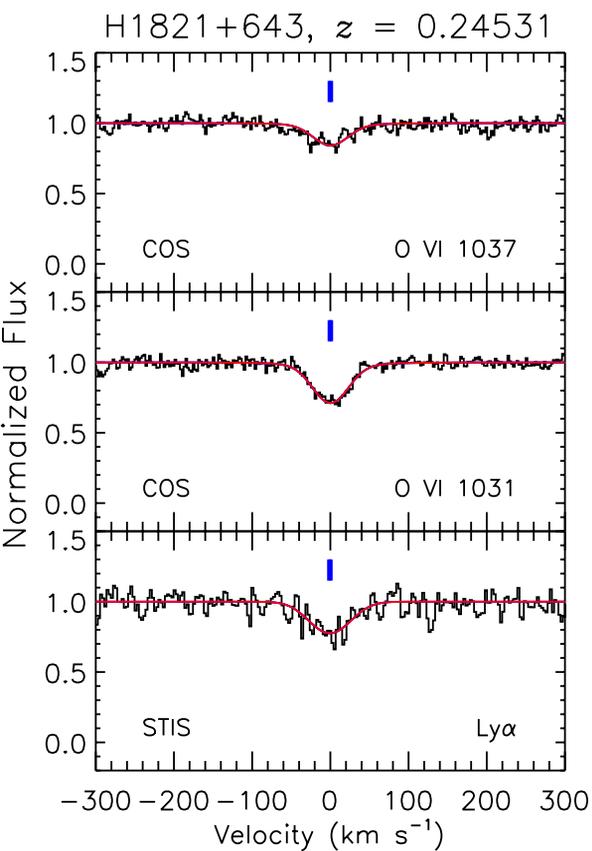

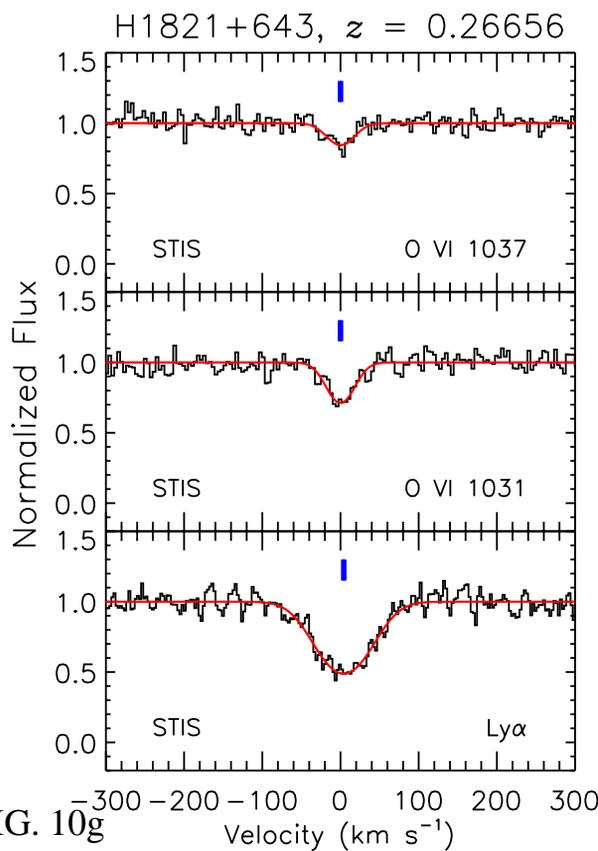

FIG. 10g

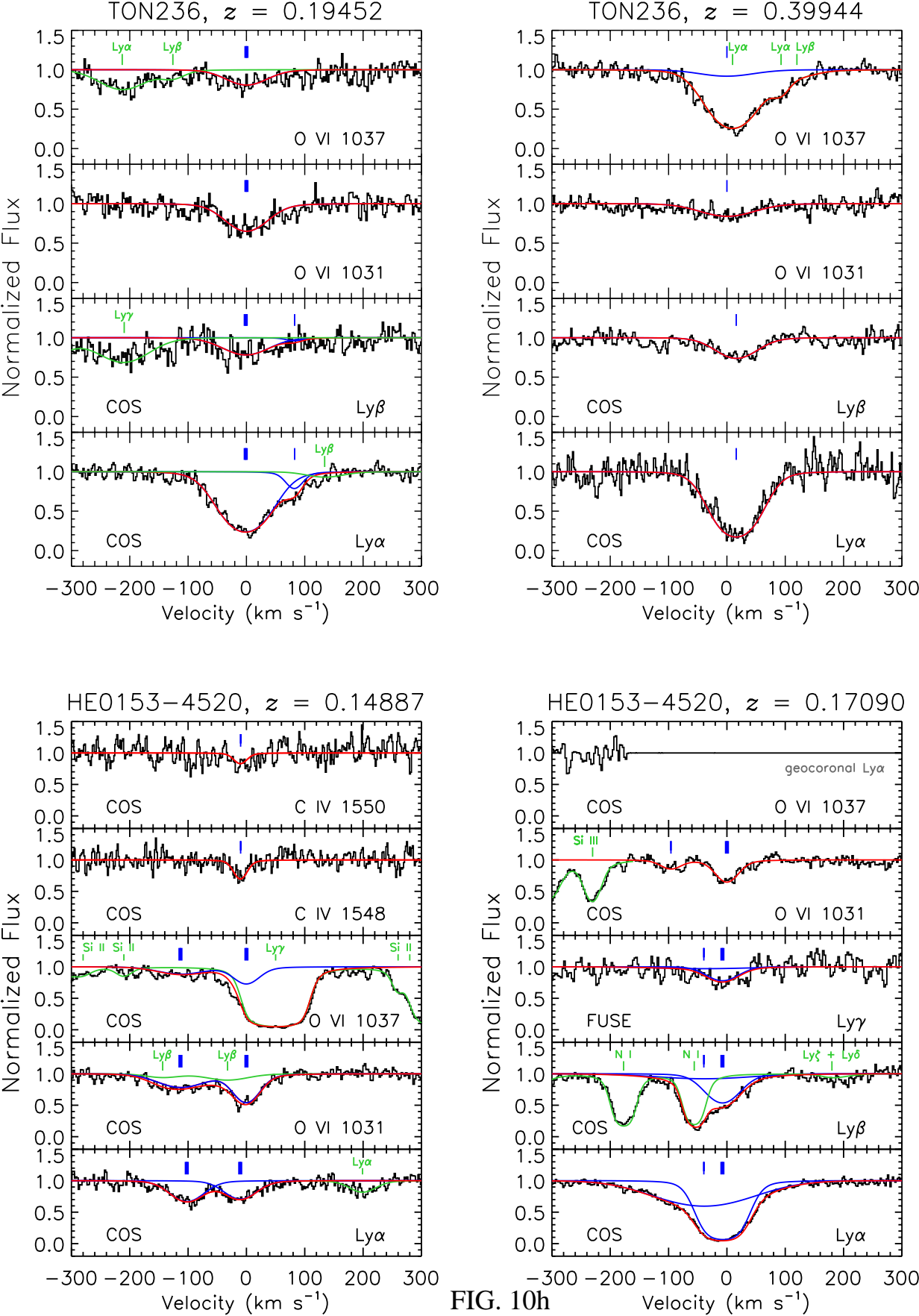

FIG. 10h

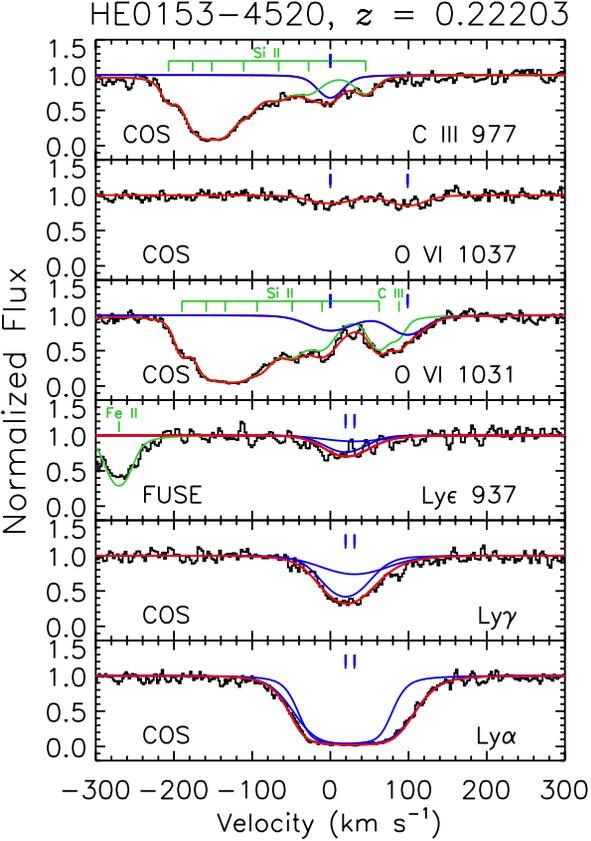

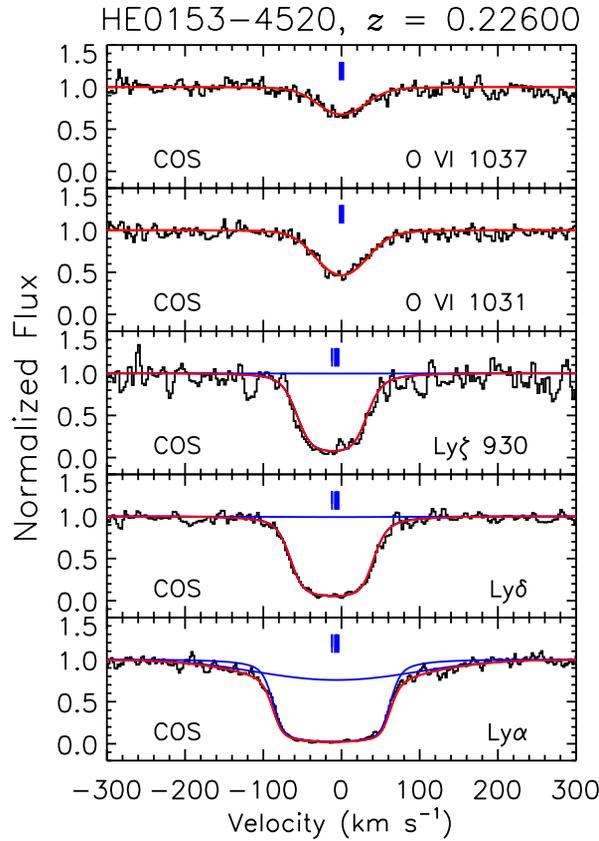

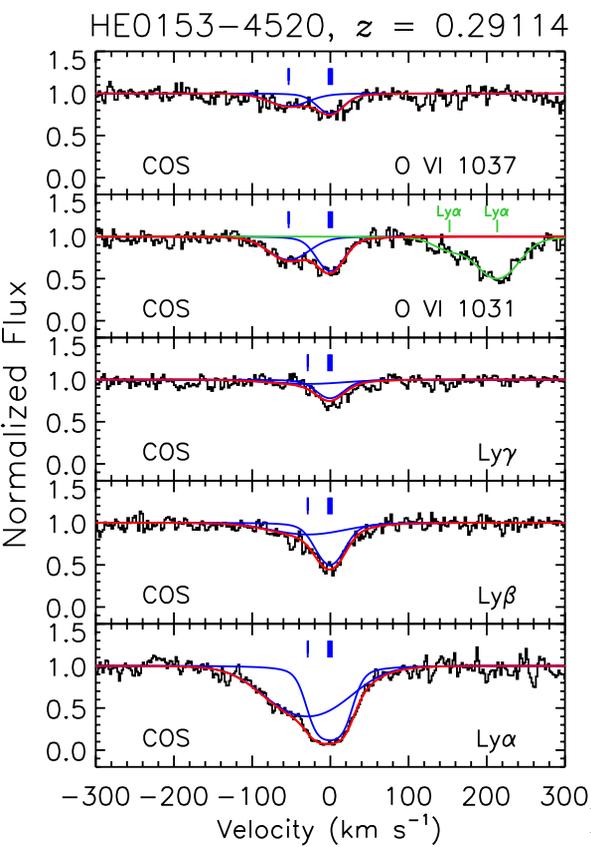

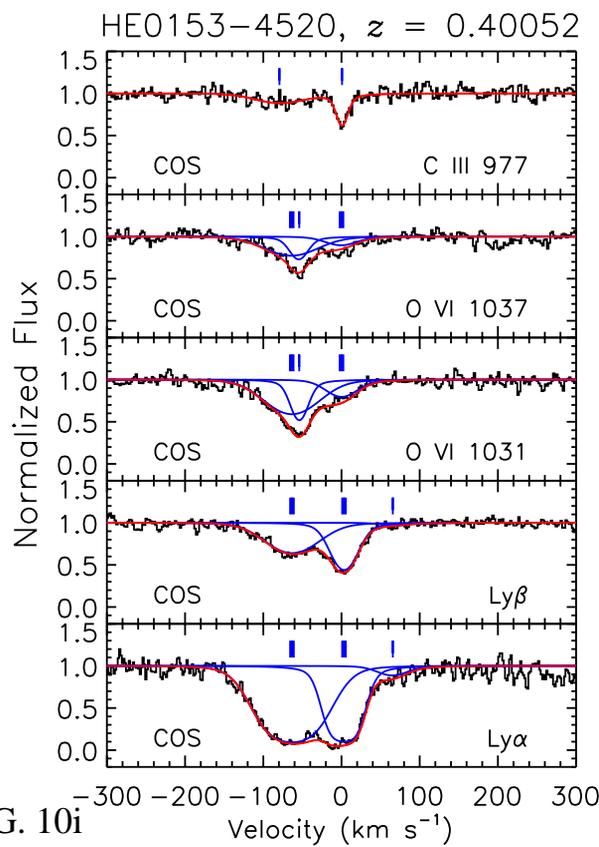

FIG. 10i

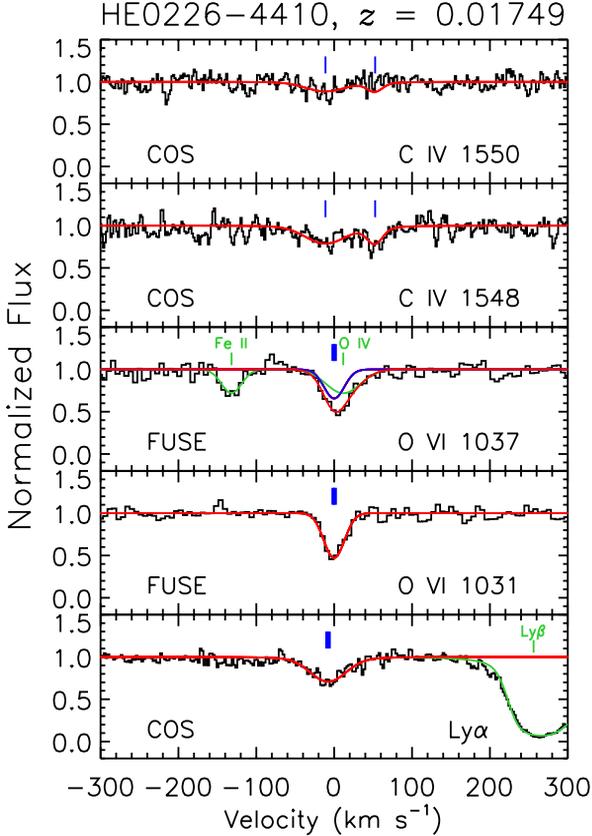
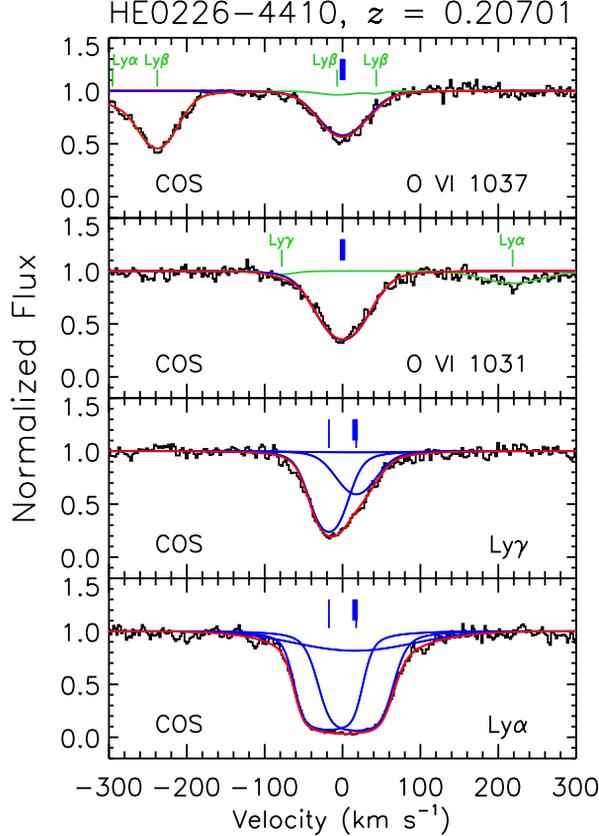
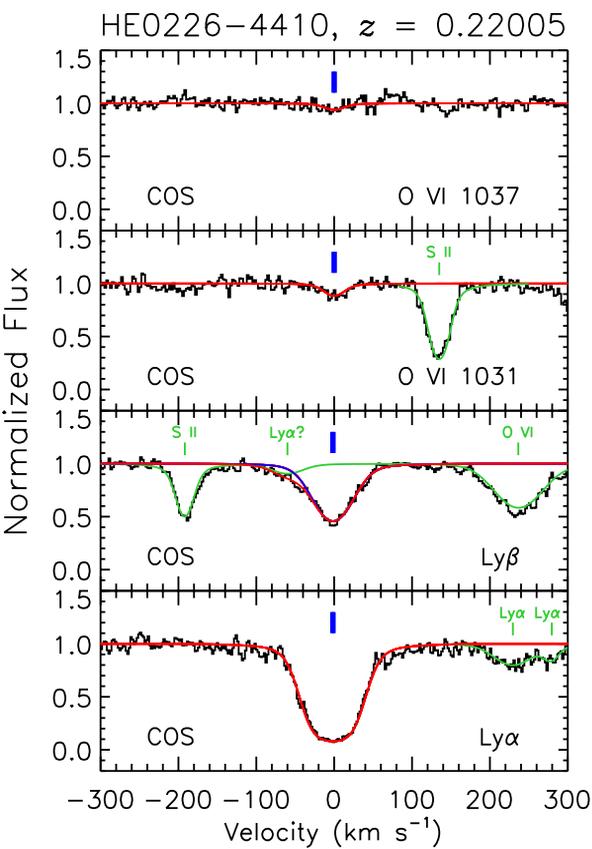
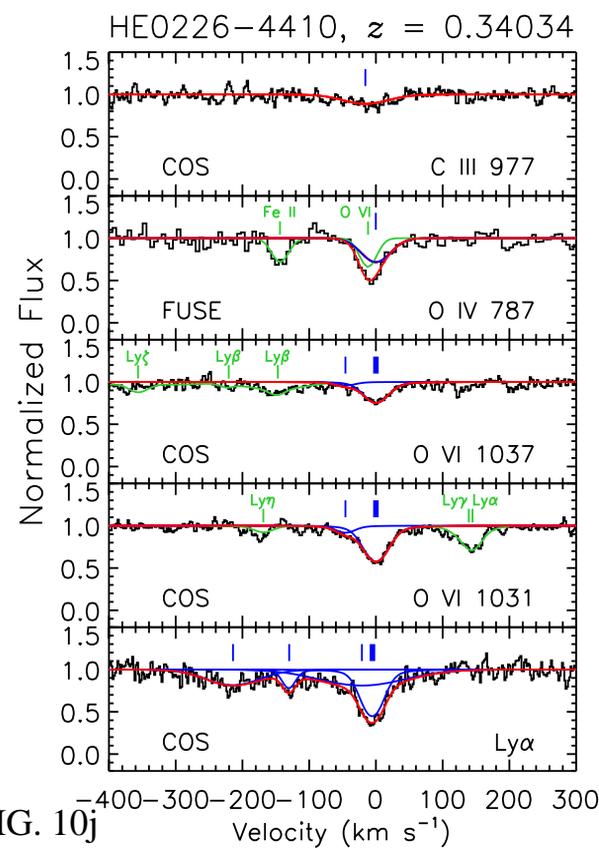

FIG. 10j

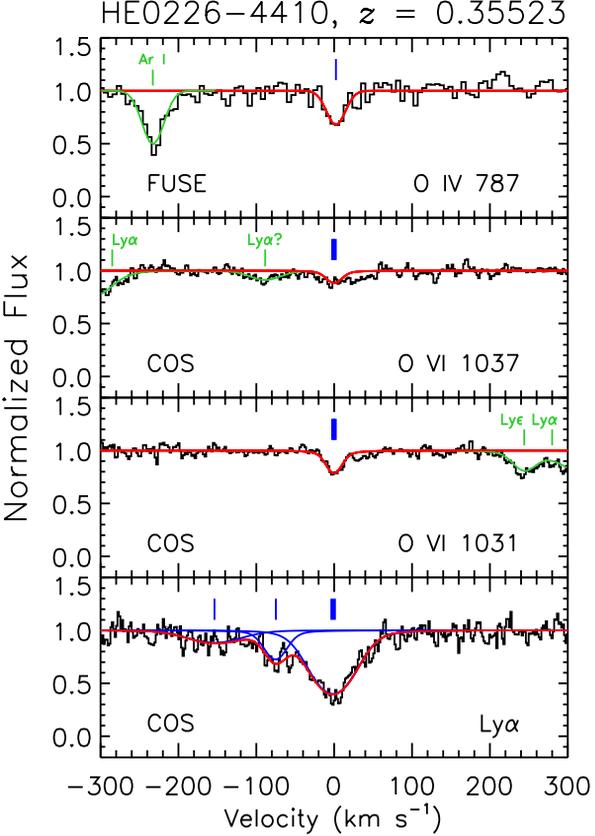

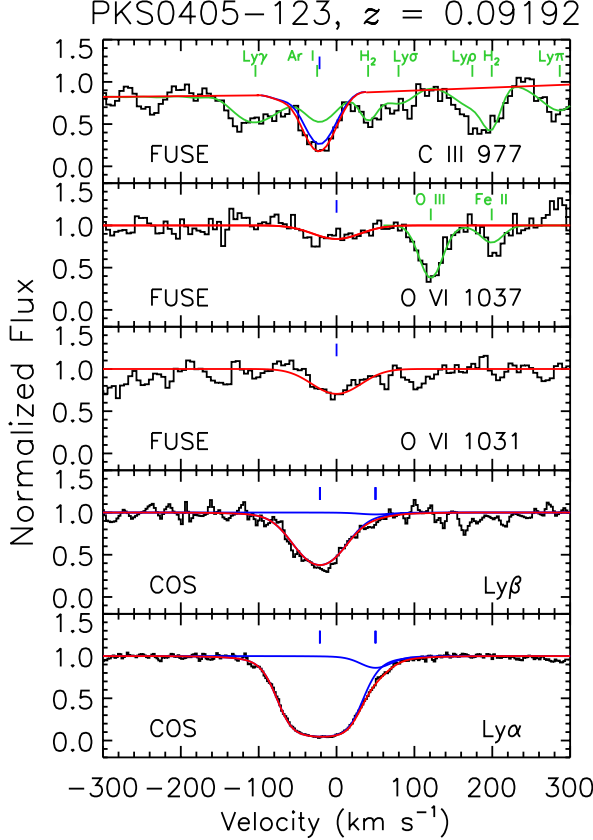

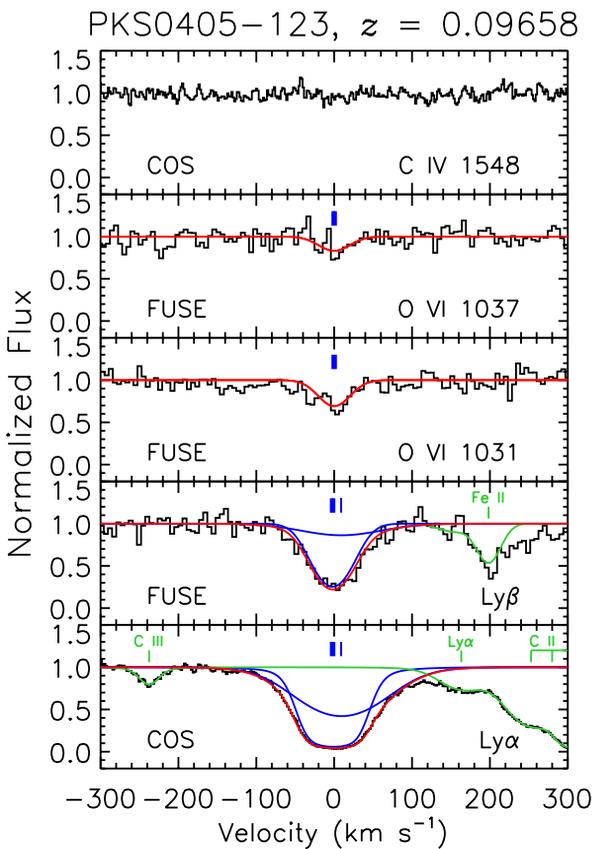

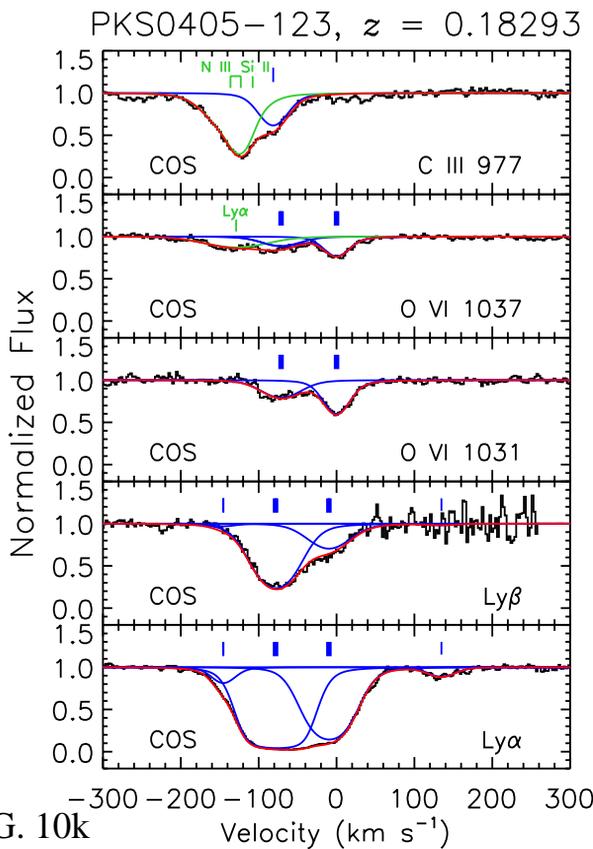

FIG. 10k

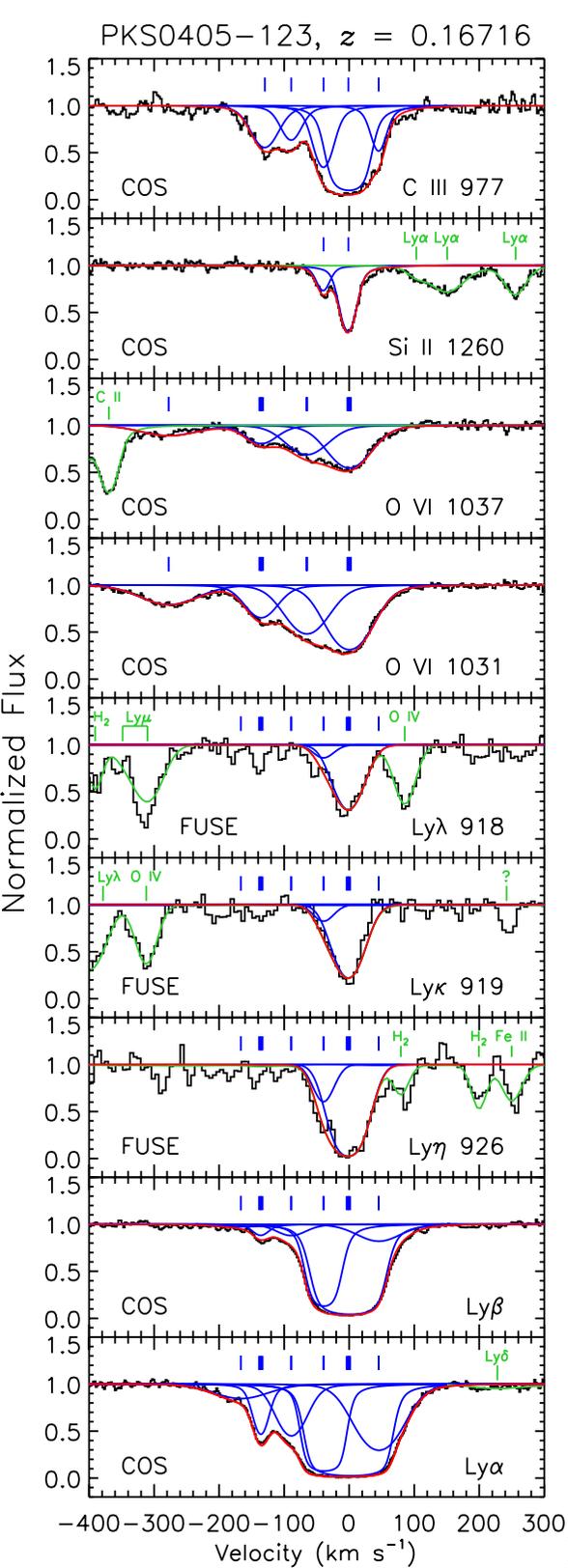
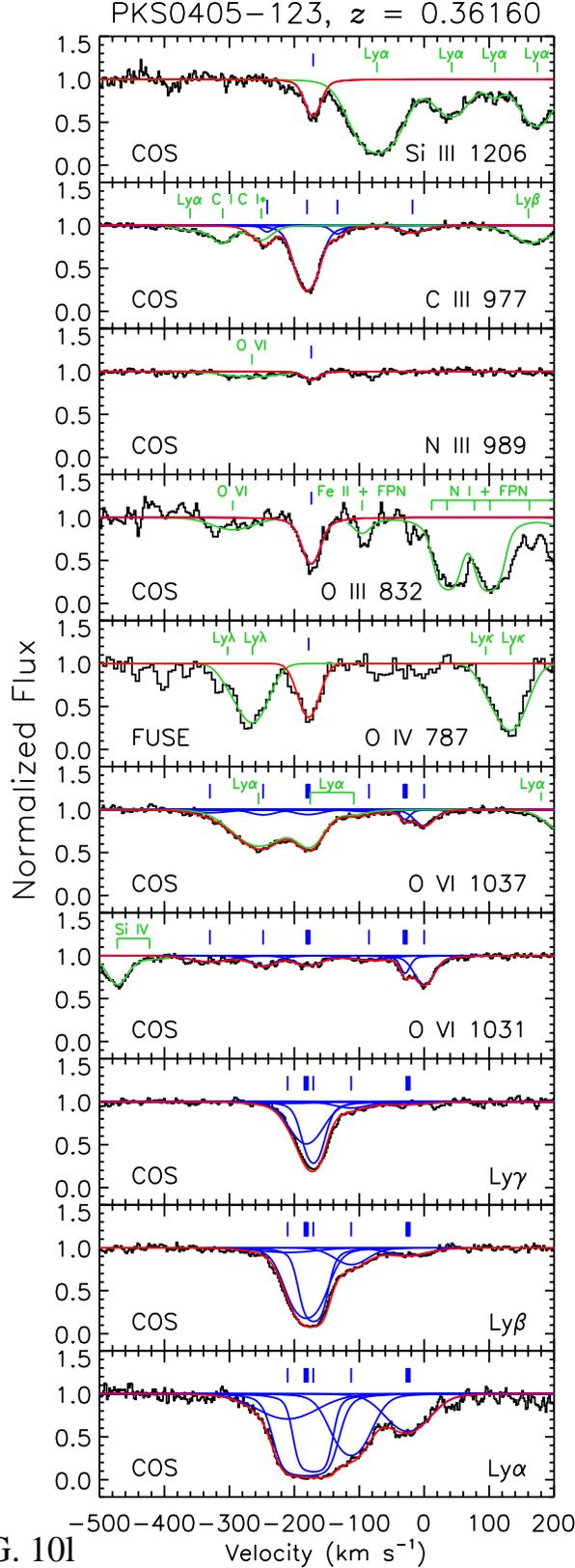

FIG. 10l

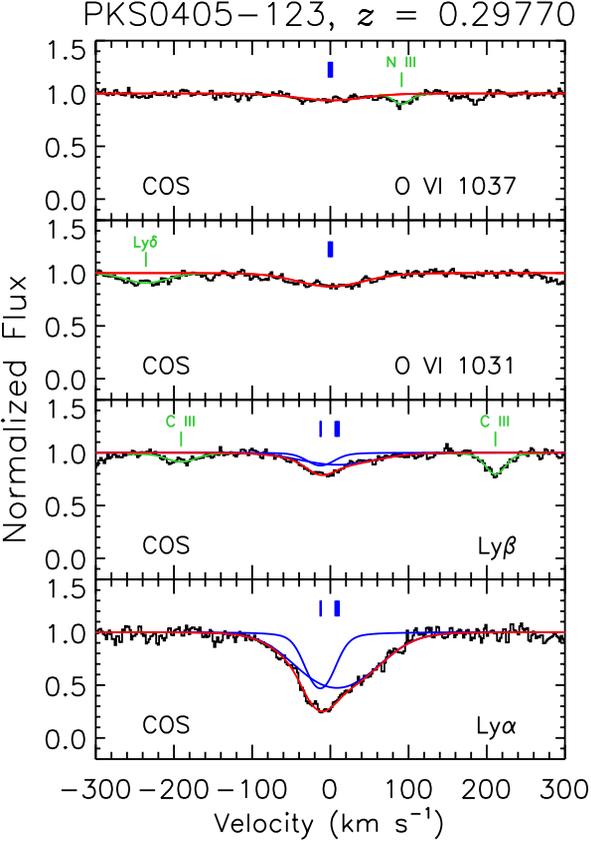
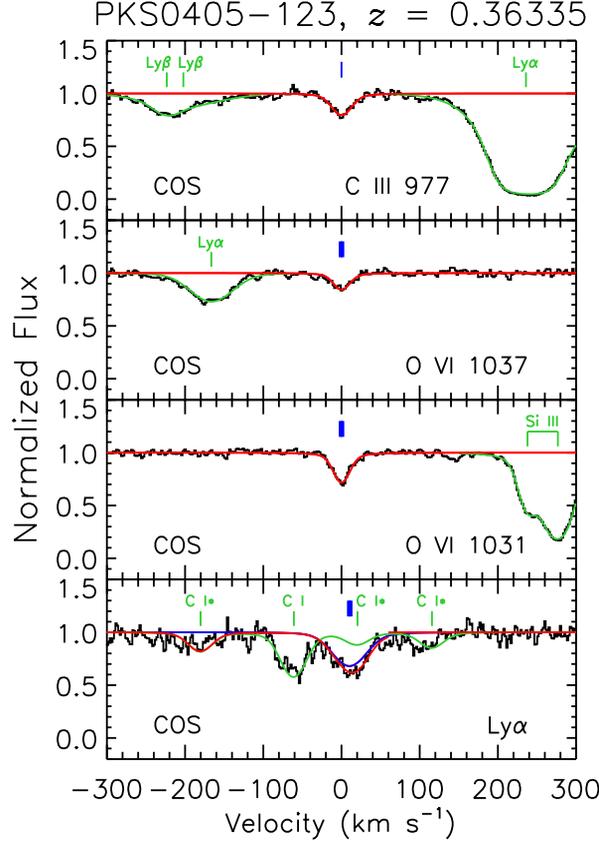
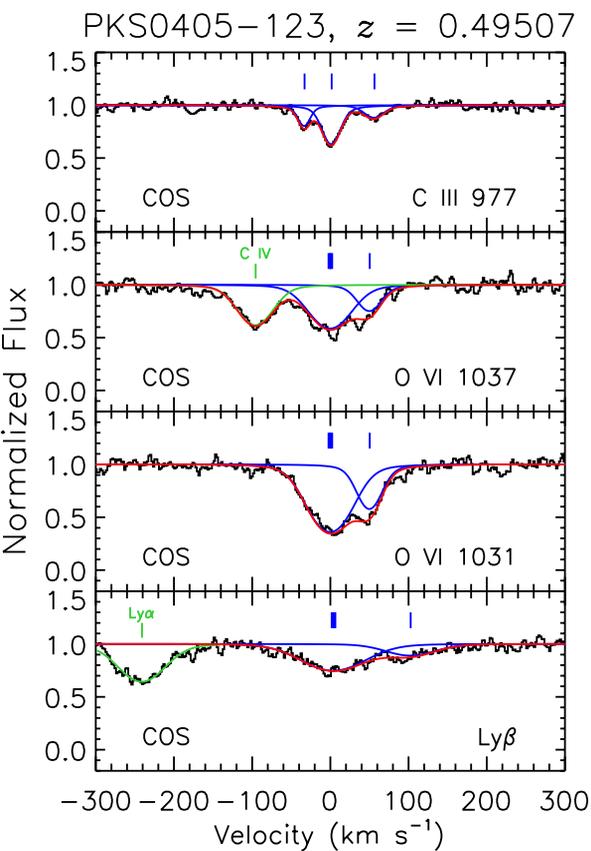
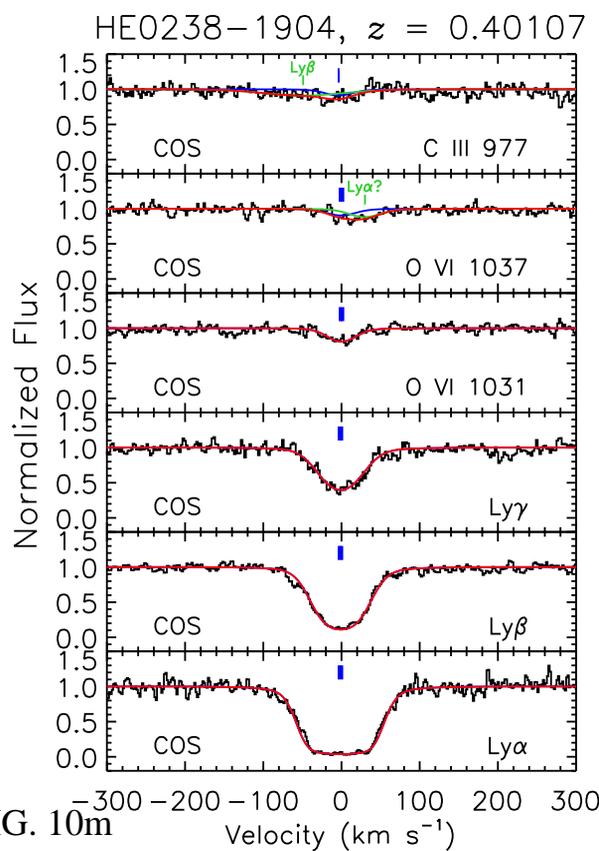

FIG. 10m

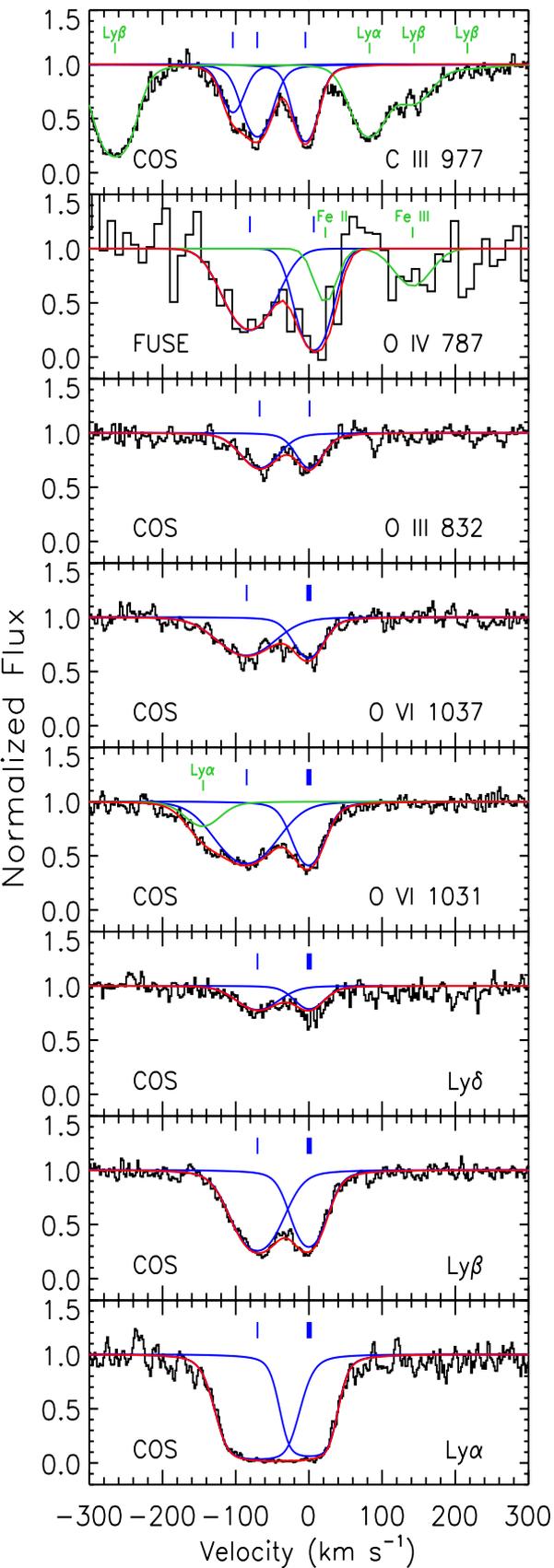

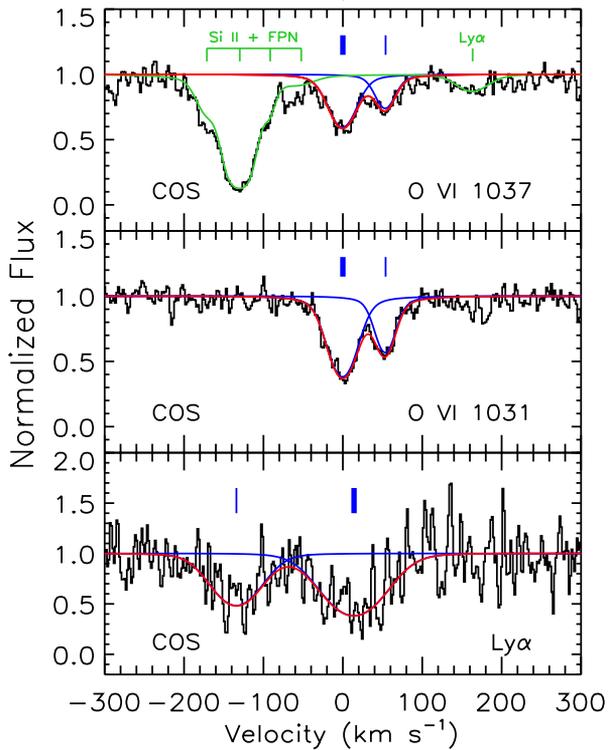

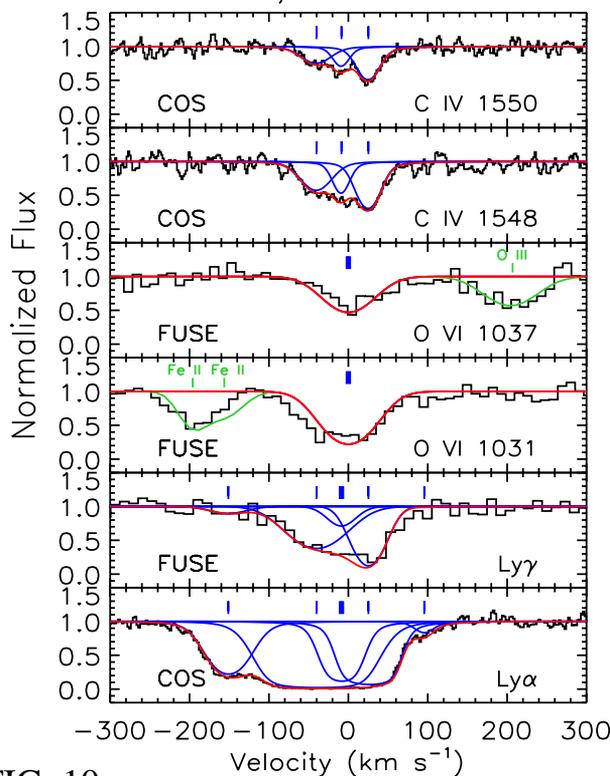

FIG. 10n

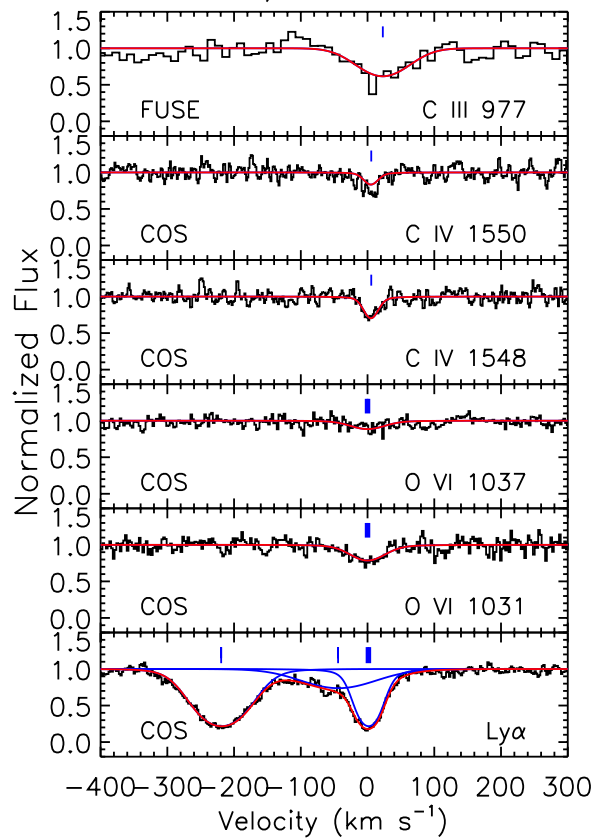

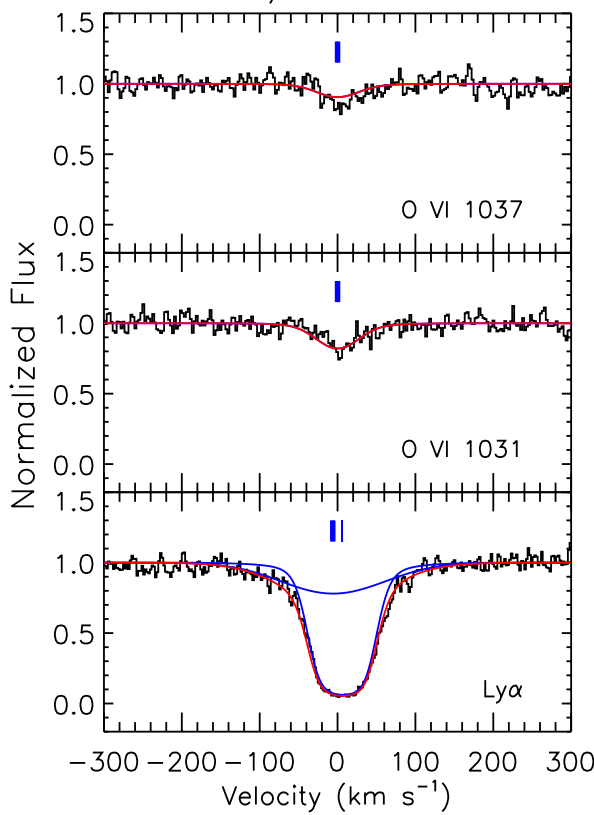

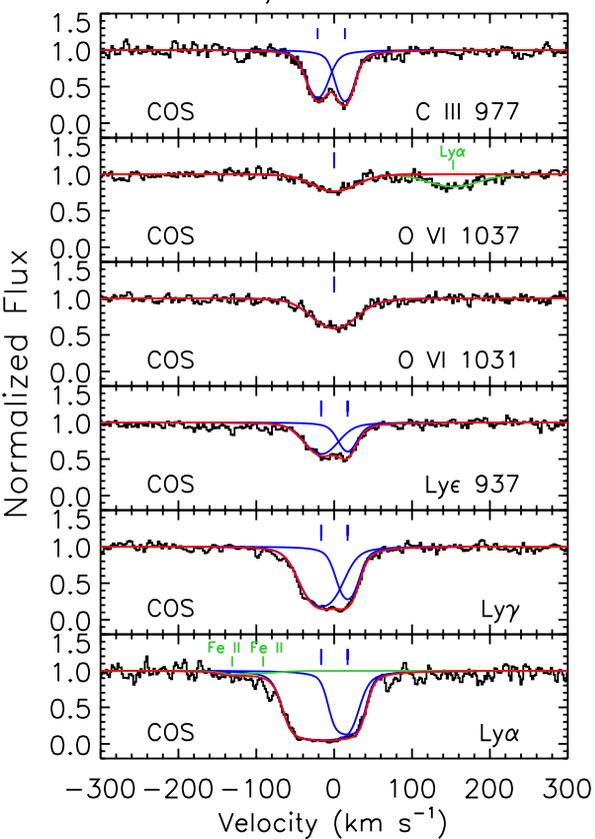

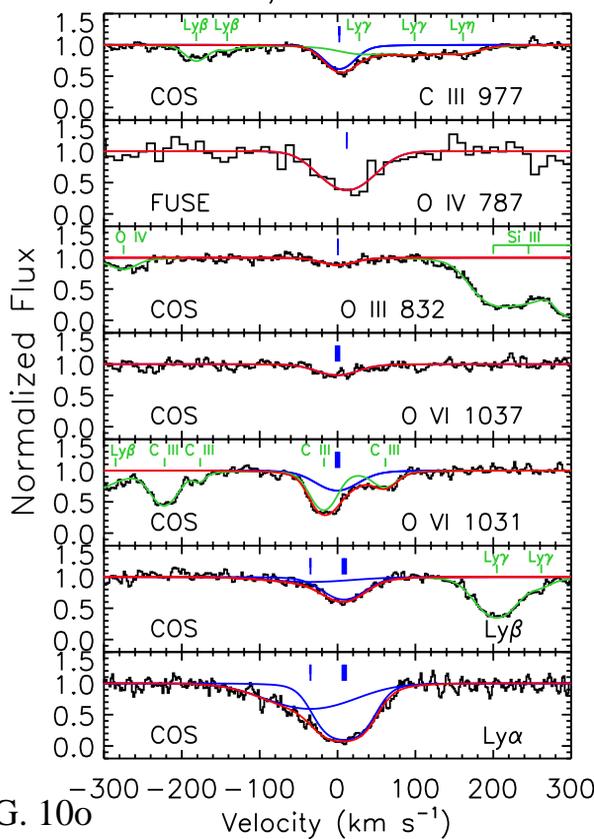

FIG. 10o